\documentclass[12pt]{iopart}
\usepackage{comment}
\usepackage{epsfig}
\usepackage{latexsym}
\usepackage{graphicx}
\usepackage{amssymb}
\usepackage{amsfonts}
\usepackage{epstopdf}
\usepackage{color}
\usepackage{caption}
\usepackage{hyperref}
\usepackage[perpage]{footmisc}
\usepackage{etoolbox}
\usepackage{stackengine}
\stackMath

\def\lsim{\mathrel{\hbox{\rlap{\hbox{\lower4pt\hbox{$\sim$}}}\hbox{$<$}}}}
\def\gsim{\mathrel{\hbox{\rlap{\hbox{\lower4pt\hbox{$\sim$}}}\hbox{$>$}}}}

\def\mP{m_{\mathrm{Pl}}}

\newcommand{\be}{\begin{equation}}
\newcommand{\ee}{\end{equation}}
\newcommand{\bea}{\begin{eqnarray}}
\newcommand{\eea}{\end{eqnarray}}
\newcommand{\Eqref}[1]{Eq.~(\ref{#1})}
\newcommand{\eqref}[1]{(\ref{#1})}

\newcommand{\bk}{{\bf k}}
\newcommand{\bx}{{\bf x}}

\newcommand{\HH}{H}

\newcommand{\vev}[1]{\langle #1 \rangle}

\newcommand{\uetcT}{\mathcal{U}}

\newcommand{\rhogw}{\rho_{\text{GW}}}
\newcommand{\Omgw}{\Omega_{\text{GW}}}
\def\Mpl{M_{\mathrm{ Pl}}}
\def\TT{{\rm TT}}
\newcommand{\nn}{\nonumber}

\newcommand{\gws}{gravitational waves}
\newcommand{\GW}{{_{\rm GW}}}
\newcommand{\boldsymbol}[1]{\bf #1}
\newcommand{\text}[1]{\rm #1}
\newcommand{\msun}{M_\odot}
\newcommand{\mn}{{\mu\nu}}
\newcommand{\ab}{{\alpha\beta}}
\newcommand{\lambdabar}{{\mkern0.75mu\mathchar '26\mkern -9.75mu\lambda}}

\makeatletter
\def\@mkboth#1#2{}
\newlength\appendixwidth
\preto\appendix{\addtocontents{toc}{\protect\patchl@section}}
\newcommand{\patchl@section}{%
  \settowidth{\appendixwidth}{\textbf{Appendix }}%
  \addtolength{\appendixwidth}{1.5em}%
  \patchcmd{\l@section}{1.5em}{\appendixwidth}{}{\ddt}%
}
\makeatother

\begin{document}

\title{Cosmological Backgrounds of Gravitational Waves}

\newcommand{\addressAPC}{Laboratoire Astroparticule et Cosmologie, CNRS UMR 7164, Universit\'e Paris-Diderot, 10 rue Alice Domon et L\'eonie Duquet, 75013 Paris, France.}
\newcommand{\addressCERN}{Theoretical Physics Department, CERN, Geneva, Switzerland}
\newcommand{\addressEPFL}{Laboratory of Particle Physics and Cosmology Institute of Physics (LPPC), \'Ecole Polytechnique F\'ed\'erale de Lausanne (EPFL), CH-1015 Lausanne, Switzerland.}

\author{Chiara Caprini}
\address{\addressAPC}

\author{Daniel G. Figueroa}
\address{\addressEPFL}

\begin{abstract}
Gravitational waves (GWs) have a great potential to probe cosmology. We review early universe sources that can lead to cosmological backgrounds of GWs. We begin by presenting proper definitions of GWs in flat space-time and in a cosmological setting (Section 2). Following, we discuss the reasons why early universe GW backgrounds are of a stochastic nature, and describe the general properties of a stochastic background (Section 3). We recap current observational constraints on stochastic backgrounds, and discuss the basic characteristics of present and future GW detectors, including advanced LIGO, advanced Virgo, the Einstein Telescope, KAGRA, and LISA (Section 4). We then review in detail early universe GW generation mechanisms, as well as the properties of the GW backgrounds they give rise to. We classify the backgrounds in five categories: GWs from quantum vacuum fluctuations during standard slow-roll inflation (Section 5), GWs from processes that operate within extensions of the standard inflationary paradigm (Section 6), GWs from post-inflationary preheating and related non-perturbative phenomena (Section 7), GWs from first order phase transitions related or not to the electroweak symmetry breaking (Section 8), and GWs from general topological defects, and from cosmic strings in particular (Section 9). The phenomenology of these early universe processes is extremely rich, and some of the GW backgrounds they generate can be within the reach of near-future GW detectors. A future detection of any of these backgrounds will provide crucial information on the underlying high energy theory describing the early universe, probing energy scales well beyond the reach of particle accelerators. 
\end{abstract}

\tableofcontents


\section{Introduction. Gravitational waves, probe of the early universe}
\label{sec:intro}

The first detection of gravitational waves (GWs) 
by the LIGO/Virgo collaboration on Sept.~2015~\cite{Abbott:2016blz}, has happily ended 50 years of experimental effort towards a direct detection of GWs. At the same time, it has proven the existence of a quite unexpected source, binary systems with fairly massive stellar-origin black holes. It was a first hint of the great potential of GW detection for the exploration and understanding of the universe: further detections by the aLIGO interferometer first, and by the aLIGO and aVirgo network starting from summer 2017, have subsequently fully revealed this potential~\cite{Abbott:2016nmj,Abbott:2017vtc,Abbott:2017gyy,Abbott:2017oio}. The implications of these detections concern not only the discovery of new astrophysical objects, but extend also to powerful tests of fundamental physics and cosmology. Several aspects of General Relativity (GR) can be probed, such as for example the speed of propagation and polarisation of GWs. The first detection of the coalescence of two neutron stars, accompanied by the coincident detection of the same event in various electromagnetic bands~\cite{TheLIGOScientific:2017qsa,GBM:2017lvd}, has strongly constrained the GW propagation speed $|c_T-1| \leq 5\cdot 10^{-16}$. Focussing on cosmology this has, for example, important consequences for modified gravity scenarios candidates to explain the current acceleration of the universe, see e.g.~\cite{Creminelli:2017sry,Ezquiaga:2017ekz,Baker:2017hug,Sakstein:2017xjx}, and~\cite{Lombriser:2015sxa,Lombriser:2016yzn} for early works. Moreover, this observation has provided a measurement of the Hubble rate today~\cite{Abbott:2017xzu}, though not precise enough yet to help solving the tension between Cosmic Microwave Background (CMB)~\cite{Ade:2015xua} and local universe measurements~\cite{Riess:2016jrr}. The measurement of the Hubble rate will however improve consistently as new GW signals will be detected in the upcoming years.  

The aim of the present review is to show that the discovery potential of GW observations also concerns the cosmology of the early universe. On general grounds, due to the weakness of gravity, GWs are decoupled from the rest of matter and radiation components in the universe, upon production: comparing the rate of interaction of GWs with the Hubble rate, one gets qualitatively~\cite{Maggiore:1999vm}
\begin{equation}
	\frac{\Gamma(T)}{H(T)}\sim \frac{G^2\,T^5}{T^2/M_{Pl}}=\left(\frac{T}{M_{Pl}}\right)^3\,,
\end{equation}
where $M_{Pl}$ denotes the Planck mass, $G=1/M_{Pl}^2$ the Newton constant, $H(T)\sim T^2/M_{Pl}$ the Hubble rate in the radiation dominated era, and we have assumed a weak interaction with rate $\Gamma(T)=n\,\sigma \,v$, with the number density of particles $n\sim T^3$, cross-section $\sigma\sim G^2T^2$ and $v\sim1$. This estimate shows that the GW interaction rate is smaller than the Hubble parameter, essentially at any temperature in the universe $T < M_{Pl}$ for which our present knowledge about gravitation holds. In other words, GWs propagate freely in the early universe, immediately after they are generated. This means that GWs carry unique information about the processes that produced them, and therefore about the state of the universe at epochs and energy scales unreachable by any other means. The energy scales that GWs can probe extend far beyond the reach of presently available observational probes of the universe, mostly based on electromagnetic emission. 
Furthermore, GWs can provide information on particle physics theories, in a complementary way to the Large Hadron Collider or future particle colliders. In this review we present the majority of mechanisms proposed in the literature, typically based on theories beyond the Standard Model of particle physics, that generate GWs in the early universe. These mechanisms can occur within a broad range of energies, from the QCD scale $\sim \mathcal{O}(10^2)$ MeV, all the way up to the inflationary scale, bounded as $E_{\rm inf}\lesssim 10^{16}$ GeV. 

The potential of GW detection to improve our knowledge of the early universe, is in principle comparable to the one of the CMB detection at its dawn, which marked the beginning of modern cosmology. Still, the GW signals must have sufficient amplitude to be captured by current and future GW detectors. In general, this requires the production of a substantial amount of tensor 
metric fluctuations. 
Fortunately, as we will see in detail, there are a number of well motivated mechanisms that can generate cosmological GW backgrounds within the reach of some GW experiments. Furthermore, the number of present and planned GW detectors is increasing, and presently include: the network of terrestrial interferometers, currently composed by aLIGO and aVirgo, but to be complemented by KAGRA in the near future, and subsequently by LIGO India; the space-based interferometer LISA, which has been accepted by the European Space Agency with a predicted launch date around 2034; and Pulsar Timing Arrays, which in the future will reach an extreme sensitivity with the Square Kilometre Array. The present review aims at presenting in detail an updated view of our current understanding of potential sources from the early universe, that may generate a cosmological GW background. Our review updates and completes in this way, previous reviews on the topic, such as~\cite{Allen:1996vm,Hogan:1998dv,Hogan:2006va,Maggiore:1999vm,Buonanno:2003th}, or the more recent ones~\cite{Guzzetti:2016mkm,Cai:2017cbj}. 

The review is organised as follows. In Section~\ref{sec:GWdef} we provide a definition of GWs, initially on a flat space-time, and then focussing on the relevant case of a Friedmann-Lema\^itre-Robertson-Walker (FLRW) background, together with some noteworthy solutions of the GW equation of motion. In Section~\ref{sec:general}, after discussing why any GW background produced in the early universe is of stochastic nature, we present a general characterization of a statistically homogeneous and isotropic stochastic gravitational wave background (SGWB). In Section~\ref{sec:bounds} we review the present constraints on SGWBs, and discuss some of the characteristics of current and future GW detectors. In the rest of the manuscript, Sects.~\ref{sec:inf} - \ref{sec:CosmicDefects}, we consider proposed mechanisms of GW generation during the early universe. In Section \ref{sec:inf} we discuss the irreducible SGWB arising from quantum vacuum fluctuations in standard inflation. In Section \ref{sec:infII} we review production mechanisms that operate within extensions of the standard inflationary scenario,  such the occurrence of particle production during inflation, or enhancement mechanisms of the inflationary SGWB at high frequencies, due to extra features beyond the standard single field slow-roll vanilla scenario. In section \ref{sec:PreheatingAndOthers} we concentrate on post-inflationary preheating mechanisms like parametric resonance or tachyonic instabilities, as well as on related non-perturbative processes, like the dynamics of flat-directions or oscillons. Section \ref{sec:FOPT} is dedicated to first order phase transitions beyond the standard model of particle physics, both related and unrelated to the electroweak symmetry breaking. Finally, in Section \ref{sec:CosmicDefects}, we present the GW generation from topological defects, including the irreducible SGWB from any network of cosmic defects, and the more specific background produced by the decay of cosmic string loops. In Section~\ref{sec:Conclusion} we summarize and conclude. 

{\bf Notations}. Throughout the review, unless otherwise specified, we use units $c=\hbar=1$. We will use interchangeably the Newton constant $G$, the full Planck mass $M_{Pl} \simeq 1.22\cdot 10^{19}$ GeV, or the reduced Planck mass $m_{Pl} \simeq 2.44\cdot 10^{18}$ GeV, related through $M_{Pl}^2 = 8\pi m_{Pl}^2 = 1/G$. Latin indices are reserved for spatial dimensions $i, j, k, ... = 1,2,3$, and Greek indices for space-time dimensions $\mu, \nu, \alpha, \beta, ... = 0,1,2,3$. We assume the Einstein convention so that repeated indices are interpreted as a sum over their values. We use a flat FLRW metric $ds^2 = -dt^2 + a^2(t) \, \delta_{ij} \, dx^i dx^j$ $= a^2(\eta)(d\eta^2 + \, \delta_{ij} \, dx^i dx^j)$, where $t$ denotes physical time, and $\eta$ (alternatively $\tau$ in some sections) denotes the conformal time. Comoving momenta are presented by $k$, the physical Hubble rate is denoted by $H$, and the conformal Hubble rate by $\mathcal{H}$. The critical density today is 
\be
\label{rhoc}
\rho_c^0 = \frac{3 H_0^2}{8 \pi G}\,,
\ee
and, unless otherwise specified, cosmological parameters are fixed to the CMB values given in \cite{Ade:2015xua}. Our Fourier convention is given in section \ref{gweom}.

\section{Definition of gravitational waves}
\label{sec:GWdef}

In this section we review the basic concepts needed to define GWs. We first discuss the case of `linearized gravity' in Sect.~\ref{subsec:GWdef}, defining GWs as metric perturbations in globally-vacuum asymptotically flat space-times. We present a more general definition in Sect.~\ref{subsec:SVT}, including the case when sources are present, by decomposing the metric perturbation into  scalar-vector-tensor components, which are irreducible under three-dimensional rotations. We identify the GWs with the only gauge-invariant radiative part of the metric perturbation. After a discussion on the notion of GWs in arbitrary space-times (Sect.~\ref{subsec:GWcurved}), we focus on a cosmological context, introducing the GW equation of motion in an expanding FLRW universe, and discussing some of its solutions in vacuum (Sect.~\ref{gweom}). The GW evolution in the presence of a generic source is postponed to Sect.~\ref{sec:general}, after we have introduced the statistical characterization of cosmological SGWBs.

\subsection{Linearized theory in vacuum: the transverse-traceless gauge}
\label{subsec:GWdef}

A natural approach to introduce gravitational waves (GWs) is that of `linearised gravity', 
by which one considers a small perturbation over a fixed Minkowski background $\eta_{\mu\nu} \equiv {\rm diag}(-1,+1,+1,+1)$,
\begin{eqnarray}\label{eq:LinearGravity}
g_{\mu\nu}(x) = \eta_{\mu\nu} + h_{\mu\nu}(x)\,,~\hspace*{1cm} |h_{\mu\nu}(x)| \ll 1\,.
\end{eqnarray}
The condition $|h_{\mu\nu}(x)| \ll 1$ implies that one is allowing for $a)$ only weak gravitational fields, and $b)$ a restricted set of coordinate systems where Eq.~(\ref{eq:LinearGravity}) holds. General Relativity (GR) is invariant under general coordinate transformations $x^\mu \longrightarrow x'^{\mu}(x)$, under which the metric tensor transforms as $g'_{\mu\nu}(x') = {\partial x^\alpha\over\partial x'^{\mu}}{\partial x^\beta\over\partial x'^{\nu}}g_{\alpha\beta}(x)$. This implies that, under general infinitesimal coordinate transformations $x'^{\mu} \longrightarrow x^{\mu} + \xi^\mu$, with $\xi^{\mu}(x)$ an arbitrary infinitesimal vector field, the metric perturbation transforms as
\begin{eqnarray}\label{eq:TensorChange}
h'_{\mu\nu}(x') = h_{\mu\nu}(x) - \partial_\mu\xi_\nu - \partial_\nu\xi_\mu\,.
\end{eqnarray}
In order to preserve the functional form of Eq.~(\ref{eq:LinearGravity}) in the new system of coordinates, i.e.~$g'_{\mu\nu}(x') = \eta_{\mu\nu} + h'_{\mu\nu}(x'),~|h'_{\mu\nu}(x')| \ll 1$, we require $|\partial_\alpha\xi_\beta| \lesssim |h_{\alpha\beta}|$. Therefore, only slowly varying infinitesimal coordinate transformations are a symmetry of the linearised theory\footnote{Notice that under a Lorentz transformation $x'_\mu = \Lambda_\mu^{~\,\nu}x_\nu$, $g'_{\mu\nu}(x') = \Lambda_\mu^{~\,\alpha}\Lambda_\nu^{~\,\beta}g_{\alpha\beta}(x)$, preservation of  Eq.~(\ref{eq:LinearGravity}) requires $|\Lambda_\mu^{~\,\alpha}\Lambda_\nu^{~\,\beta}h_{\alpha\beta}(x)|\ll 1$, so that it remains true that $|h'_{\mu\nu}(x')| \ll 1$. Rotations do not spoil the condition $|h_{\mu\nu}(x)| \ll 1$, but boosts could, and therefore must be restricted to those that do not spoil such condition. As $h_{\mu\nu}(x)$ is invariant under constant displacements $x'^\mu \longrightarrow x^\mu + a^\mu$, linearised gravity Eq.~(\ref{eq:LinearGravity}) is also invariant under Poincar\'e transformations. 
}.

The {\it affine} connection ({\it Christoffel} symbols) to linear order in the tensor perturbation is 
\begin{eqnarray}
\hspace*{-1.2cm}\Gamma^\alpha_{~\,\mu\nu} \equiv {1\over2}g^{\alpha\beta}(\partial_\nu g_{\beta\mu} + \partial_\mu g_{\beta\nu} - \partial_\beta g_{\mu\nu}) = {1\over2}(\partial_\nu h^\alpha_{~\,\mu} + \partial_\mu h^\alpha_{~\,\nu} - \partial^\alpha h_{\mu\nu}) + \mathcal{O}(h_{**}^2)\,,
\end{eqnarray}
where $h^{\mu}_{\,~\nu} \equiv \eta^{\mu\alpha}h_{\alpha\nu} = h_{\nu}^{\,~\mu}$. Using this expression we can write, also to linear order, the Riemann tensor, Ricci tensor and Ricci scalar, as
\begin{eqnarray}
&& \hspace*{-1cm} R^\alpha_{~\,\mu\nu\beta} = \partial_\nu\Gamma^\alpha_{~\,\mu\beta} - \partial_\beta\Gamma^\alpha_{~\,\mu\nu} = {1\over2}(\partial_\mu\partial_\nu h^\alpha_{~\,\beta} + \partial_\beta\partial^\alpha h_{\mu\nu} - \partial_\nu\partial^\alpha h_{\mu\beta} - \partial_\beta\partial_\mu h^\alpha_{~\,\nu})\,,\\
&& \hspace*{-1cm} R_{\mu\nu} \equiv -R^\alpha_{~\,\mu\nu\alpha} = {1\over2}(\partial_\nu \partial^\alpha h_{\alpha\mu} + \partial_\mu \partial^\alpha h_{\alpha\nu} - \partial_\mu\partial_\nu h - \Box h_{\mu\nu})\,,\\
&& \hspace*{-1cm} R = R^\mu_{~\,\mu} = (\partial^\alpha\partial^\beta h_{\alpha\beta}-\Box h)\,,
\end{eqnarray}
where $h \equiv h^\alpha_{~\,\alpha}$ is the trace of the metric perturbation, $\partial^\alpha \equiv \eta^{\alpha\beta}\partial_\beta$ and $\Box \equiv \partial_\alpha\partial^\alpha = \eta^{\alpha\beta}\partial_\alpha\partial_\beta$. From these expressions we can construct the Einstein tensor, again to first order in the metric perturbation, as
\begin{eqnarray}\label{eq:EinsteinTensorLinTh}
\hspace*{-2cm} G_{\mu\nu} &\equiv& R_{\mn} - {1\over2}\eta_\mn R = {1\over2}(\partial_\nu \partial_\alpha h^\alpha_{\,~\mu} + \partial_\mu \partial^\alpha h_{\nu\alpha} - \partial_\mu\partial_\nu h - \Box h_{\mu\nu} - \eta_\mn\partial^\alpha\partial_\beta h_\alpha^{~\beta}+\eta_{\mn}\Box h) \nonumber\\
\hspace*{-2cm}  &=& {1\over2}(\partial_\alpha \partial_\nu {\bar h}^\alpha_{~\,\mu} + \partial^\alpha \partial_\mu {\bar h}_{\nu\alpha} - \Box {\bar h}_{\mu\nu} - \eta_\mn\partial_\alpha\partial^\beta {\bar h}^\alpha_{~\,\beta})\,,
\end{eqnarray}
where in the last line, for convenience, we have introduced a new metric perturbation 
\begin{eqnarray}\label{eq:hTraceRev}
{\bar h}_{\mn} \equiv h_\mn - {1\over2}\eta_{\mn}\,h\,.
\end{eqnarray}
As the trace of ${\bar h}_{\mn}$ has opposite sign to that of $h_\mn$, ${\bar h} = - h$, ${\bar h}_{\mn}$ is referred to as the {\it trace-reversed} metric perturbation. Writing $G_\mn$ in terms of ${\bar h}_\mn$ has the advantage that it eliminates the trace. 

The expression of $G_\mn$ in Eq.~(\ref{eq:EinsteinTensorLinTh}) can be further simplified, by exploiting the invariance of the linearised theory under slowly varying infinitesimal coordinate transformations. Under $x'^{\mu} \longrightarrow x^{\mu} + \xi^\mu$, the metric perturbation $h_\mn$ changes as in Eq.~(\ref{eq:TensorChange}), whereas the trace-reversed perturbation transforms as
\begin{eqnarray}\label{eq:TensorChangeTR}
{\bar h}'_{\mu\nu}(x') = {\bar h}_{\mu\nu}(x) + \xi_\mn(x)\,,~~~ \xi_\mn(x) \equiv \eta_\mn\partial_\alpha\xi^\alpha - \partial_\mu\xi_\nu - \partial_\nu\xi_\mu\,.
\end{eqnarray}
In light of the expression of $G_\mn$ in terms of ${\bar h}_\ab$, it seems convenient to make a coordinate transformation such that the metric perturbation verifies
\begin{eqnarray}\label{eq:LorentzGauge}
\partial^\mu{\bar h}_\mn(x) = 0\,.
\end{eqnarray}
We now demonstrate that this gauge choice, known as the {\it Lorentz gauge}, is always possible. 
Let us start with an arbitrary perturbation ${\bar h}_\mn$ for which $\partial^\mu{\bar h}_\mn \neq 0$. The Lorentz gauge condition, using Eq.~(\ref{eq:TensorChangeTR}), transforms as
\begin{eqnarray}
\partial'^\mu{\bar h}'_\mn(x') = \partial^\mu{\bar h}_\mn(x) - \Box \xi_\nu\,,
\end{eqnarray}
so that we can always demand that $\partial'^\mu{\bar h}'_\mn(x') = 0$, as long as
\begin{equation}\label{eq:LorentzCond}
\Box \xi_\nu = f_\nu(x)\,, ~~~~~~f_\nu(x) \equiv \partial^\mu{\bar h}_\mn(x)\,.
\end{equation}
One can always find solutions to the above Eq.~(\ref{eq:LorentzCond}), simply because the {\it d'Alembertian} operator $\Box$ is invertible. Therefore, one is free to exploit the invariance of the linearised theory under infinitesimal coordinate transformations, in order to pick the Lorentz gauge. 

The advantage of expressing $G_\mn$ in terms of the trace-reverse metric, becomes now manifest. Restricting the coordinate systems to those verifying the Lorentz-gauge condition Eq.~(\ref{eq:LorentzGauge}), leads to a very simple expression for the Einstein tensor,
\begin{eqnarray}\label{eq:EinsteinTensorLorentz}
G_\mn^{\rm (L)} = -{1\over2}\Box{\bar h}_\mn\,, 
\end{eqnarray}
where $^{\rm (L)}$ refers to the Lorentz gauge. One can always write $G_\mn$ as in Eq.~(\ref{eq:EinsteinTensorLorentz}), as long as one restricts oneself to the set of Lorentz coordinate systems verifying Eq.~(\ref{eq:LorentzGauge}). The amount of gauge freedom in Lorentz coordinate systems is of course reduced, compared to the full freedom in linearized gravity under general infinitesimal coordinate transformations given by Eqs.~(\ref{eq:TensorChange}), (\ref{eq:TensorChangeTR}). Once ${\bar h}_\mn$ verifies $\partial^\mu{\bar h}_\mn(x) = 0$, we can always make a further infinitesimal coordinate transformation $x'^{\mu} \longrightarrow x^{\mu} + \xi^\mu$, so that the new metric perturbation ${\bar h}'_\mn$ still verifies $\partial'^\mu{\bar h}'_\mn(x') = 0$, as long as Eq.~(\ref{eq:LorentzCond}) is satisfied with $f_\nu(x) = 0$. The gauge freedom within Lorentz coordinate systems amounts therefore, to an infinitesimal vector displacement characterized by four functions $\xi_\nu(x)$, which are not fully free, but rather restricted to satisfy the source-less wave equation\footnote{Alternatively, the gauge freedom in Lorentz coordinate systems, amounts to 8 free functions depending on the 3 spatial coordinates, determining the initial data hyper-surface.} $\Box\xi_\nu = 0$.

In the Lorentz gauge, linearized Einstein gravity reduces therefore to the equation
\begin{eqnarray}\label{eq:LinearizedEinsteinEqs}
\Box {\bar h}_\mn = -{2\over m_p^2}T_\mn\,,
\end{eqnarray}
which is nothing else but a wave equation with a source. The general {\it homogeneous} solution to the wave equation, based on the superposition of the linearly independent solutions, can be written as
\begin{eqnarray}\label{eq:HomSolWaveEq}
{\bar h}_\mn(x) = \int d^3k ~({\bar h}_{\mn}({\bf k})e^{ikx} + {\bar h}_{\mn}^*({\bf k})e^{-ikx})\,,~~~~ {\rm with}~~ k^\mu{\bar h}_\mn = 0\,,
\end{eqnarray}
where $kx \equiv k^\mu x_\mu = -\omega(|{\bf k}|)t + {\bf k}\cdot {\bf x}$, $\omega(|{\bf k}|) = |{\bf k}|$, and ${\bar h}_{\mn}({\bf k})$, are functions that depend solely on the wave vector ${\bf k}$. The latter are not free, but must rather satisfy $k^\mu{\bar h}_\mn = 0$, as it follows from Eq.~(\ref{eq:LorentzGauge}). The solution Eq.~(\ref{eq:HomSolWaveEq}) characterizes completely a gravitational wave background $h_\mn(x) \equiv {\bar h}_\mn(x) - {1\over2}\eta_\mn {\bar h}(x)$, at every space-time point $x = ({\bf x},t)$. 

In light of Eqs.~(\ref{eq:LinearizedEinsteinEqs}) and \eqref{eq:HomSolWaveEq}, it becomes clear why, in fact, we speak of {\it gravitational waves}. As the Lorentz gauge Eq.~(\ref{eq:LorentzGauge}) represents 4 constraints over the trace-reverse metric perturbation ${\bar h}_\mn$, one might be tempted to conclude that there are in total $10 - 4 = 6$ radiative degrees of freedom verifying Eq.~(\ref{eq:LinearizedEinsteinEqs}), and hence propagating at the speed of light. However, this is not the case: 
as discussed above, fixing the Lorentz gauge does not saturate completely the gauge freedom. In fact, from Eq.~(\ref{eq:TensorChangeTR}), we see that the trace-reversed metric changes under infinitesimal coordinate transformations, as ${\bar h}_\mn \longrightarrow {\bar h}_\mn + \xi_\mn$, with $\xi_\mn \equiv \eta_\mn\partial^\alpha\xi_\alpha - \partial_\mu\xi_\nu - \partial_\nu\xi_\mu$. Once in a Lorentz frame $\partial^\mu{\bar h}_\mn = 0$, to remain in a Lorentz frame after applying a new infinitesimal coordinate transformation, requires simply that $\Box \xi_\mu = 0$. From this we observe that also $\Box \xi_\mn = 0$ is verified. 

We now restrict to globally-vacuum spatially flat space-times, where the first condition implies $T_\mn(x) = 0$ at every point and time, and the second conditions requires $h_{\mn}(x) \longrightarrow 0$ as $|{\bf x}| \longrightarrow \infty$ (the case where a source is present is treated in the next section). It follows then, that the wave equation is invariant under Lorentz preserving infinitesimal coordinate transformations\footnote{The box operator $\Box$ also changes under a coordinate transformation, $\Box' = \eta^\mn{\partial\over\partial x'^\mu}{\partial\over\partial x'^\nu}$ = $\eta^\mn[(\delta^\alpha_\mu - {\partial\xi^\alpha\over\partial x^\mu}){\partial\over\partial x^\alpha}]$ $[(\delta^\beta_\nu - {\partial\xi^\beta\over\partial x^\nu}){\partial\over\partial x^\beta}]$ = $\eta^\mn{\partial\over\partial x^\mu}{\partial\over\partial x^\nu} + \mathcal{O}(\partial\xi)$ = $\Box + \mathcal{O}(\partial\xi)$. As $\mathcal{O}(|\partial\xi|) \lesssim \mathcal{O}(|h_{**}|)$, then $\Box' = \Box + \mathcal{O}(|h_{**}|)$, and hence $\Box' h_\mn = \Box h_\mn + \mathcal{O}(h^2_{**})$.}, as $\Box' {\bar h}'_\mn = \Box ({\bar h}_\mn+\xi_\mn) = \Box {\bar h}_\mn=0$. In light of this, we can consider taking 4 infinitesimal vector displacements $\xi_\mu$ (appropriately chosen so that $\Box \xi_\mu = \Box\xi_\mn = 0$ is verified), and use them to impose 4 conditions over a newly (Lorentz-gauge preserving) transformed metric perturbation ${\bar h}'_\mn(x') = {\bar h}_\mn(x) + \xi_\mn(x)$, hence eliminating 4 degrees of freedom. In other words, out of the seemingly 6 degrees of freedom that verify the wave equation Eq.~(\ref{eq:LinearizedEinsteinEqs}), we realize that, in reality, only two independent degrees of freedom are present. These are the truly physical propagating degrees of freedom, as any new Lorentz-gauge preserving coordinate transformation ${\bar h}_\mn \longrightarrow {\bar h}_\mn + \xi_\mn$, with $\Box \xi_\mu = \Box \xi_\mn = 0$, would not reduce further the number of independent degrees of freedom below 2.


Under these circumstances, one can exploit the residual gauge freedom to eliminate directly components of the metric perturbation.
For instance, using Eq.~(\ref{eq:TensorChangeTR}), we can make the trace and the spatial-temporal components to vanish, ${\bar h} = {\bar h}_{0i} = 0$. This implies that we do not need to differentiate any more between trace-reversed and normal perturbations, as they become equal ${\bar h}_\mn = h_\mn$. From the Lorentz condition we obtain that ${\dot h}_{00} = -\partial_ih_{i0} = 0$, and hence that the temporal-temporal component is only a function of the spatial coordinates $h_{00} = V({\bf x})$. This time independent term corresponds in fact to the static part of the gravitational interaction, i.e.~to the Newtonian potential. As GWs are only concerned with the time-dependent part of the gravitational interaction, we may very well set $h_{00} = 0$. So in summary, we have specialized the gauge to
\begin{eqnarray}\label{eq:TTgauge}
h_{\mu 0} = 0\,,~~~~~~ h = h^i_{~\,i} = 0\,,~~~~~~\partial_ih_{ij} = 0\,,
\end{eqnarray}
where the last condition follows from the Lorentz condition Eq.~(\ref{eq:LorentzGauge}). This is known as the {\it transverse-traceless} (TT) gauge. The counting of the degrees of freedom in the TT gauge becomes now more clear than before, as after having eliminated all temporal components $h_{\mu 0} = 0$, we are left with 6 degrees of freedom in the spatial components $h_{ij}$. Out of these 6 degrees of freedom, 3 are further eliminated from the 3 transversality condition(s) $\partial_ih_{ij} = 0$, and 1 more degree of freedom is eliminated from the trace-less condition $h = h^i_{~\,i} = 0$. Hence, we obtain that there are, finally, only $6 - 3 - 1 = 2$ degrees of freedom surviving. Once the TT gauge Eq.~(\ref{eq:TTgauge}) is adopted, gauge freedom is saturated. 

The TT gauge is therefore particularly convenient as it fixes completely the gauge freedom, so that the metric perturbation contains only the physical radiative degrees of freedom. This can be seen particularly clearly by considering a plane wave propagating in direction $\hat n = {\bf k}/|{\bf k}|$. Due to the transversality condition $\partial_ih_{ij} = 0$, we see that the tensor components parallel to the direction of propagation vanish, as $\hat n_ih_{ij} = 0$. Without loss of generality, we can fix $\hat z = \hat n$, so that only $h_{11}, h_{12}, h_{21}$ and $h_{22}$ are non-zero in such system of reference. As we also require the metric perturbation to be trace-less and symmetric, then we are left only with 2 independent components, which we call $h_{\times} \equiv h_{12} = h_{21}$ and $h_+ \equiv h_{11} = - h_{22}$. We find that the perturbed line element, due to the passing of a GW, is thus given by
\begin{eqnarray}
ds^2 = -dt^2 + dz^2 + (1+h_{+})dx^2 + (1-h_{+})dy^2 + 2h_{\times}dxdy\,,
\end{eqnarray}
where it is manifest that there are only 2 degrees of freedom, $h_{\times}$ and $h_{+}$.

Furthermore, in a globally vacuum space-time, all non-zero components of the Riemann tensor can be obtained from $R_{i0j0}$, which in the TT gauge reads
\begin{eqnarray}\label{eq:RiemannTT}
R_{i0j0} = -{1\over2}{\ddot h}_{ij}\,.
\end{eqnarray}
This simple relation between the Riemann tensor and metric perturbations in the TT-gauge, makes particularly simple the study of the response of a detector when a GW passes through it. We refer the reader to the elaborated discussion on this in section 1.3 of~\cite{MaggioreBook}.

\subsection{Linearized theory in matter: scalar-vector-tensor decomposition} 
\label{subsec:SVT}

We ended the previous section considering linearised gravity over a Minkowski background ($g_\mn = \eta_\mn + h_\mn$), in asymptotically-flat ($h_\mn \longrightarrow 0$ at infinity) and globally vacuum space-times (i.e.~with null stress-energy tensor everywhere $T_\mn = 0$). In this setting, we were able to demonstrate that GWs are characterised by only two physical degrees of freedom, $h_+$ and $h_\times$. However, this characteristic is a manifestation of the intrinsic nature of the gravitational interaction, 
mediated by the graviton, a spin-two massless field that has only two independent helicity states (see e.g.~chapter 2 of \cite{MaggioreBook}). The TT gauge, that can only be picked in vacuum, clearly exhibits the fact that GWs are characterised by only two physical degrees of freedom, corresponding to two polarisation states. However, this must be true in general, not only in globally vacuum space-times. In this section we develop a formalism that renders this fact manifest. 

We thus maintain, in the following, the conditions of linearised gravity over a Minkowski background ($g_\mn = \eta_\mn + h_\mn$, $|h_\mn| \ll 1$), but consider the more realistic situation where a non-vanishing stress-energy tensor is present, $T_\mn \neq 0$. 
The rest of this section is based on \cite{Flanagan:2005yc} (see also \cite{Carroll:2004st}), which presents the flat space-time limit of Bardeen's cosmological gauge-invariant perturbation theory \cite{Bardeen:1980kt}. We follow standard notation in cosmology (that deviates somewhat from that of the previous section): the background metric is $\bar g_{\mn}=\eta_\mn$, and we denote the first order metric perturbation as $\delta g_\mn$ (this corresponds to $h_\mn$ in the previous section). The main difference with the cosmological setting of \cite{Bardeen:1980kt} is that our metric background is Minkowski, and hence the energy momentum tensor $T_\mn$ at the background level, must vanish. 

We begin by decomposing the metric perturbation and the energy-momentum tensor into irreducible parts with respect to spatial rotations, 
\begin{eqnarray}\label{eq:MetricSVT}
	\delta g_{00} &=& -2\phi \,,\\
	\label{eq:MetricSVT2}
	\delta g_{0i} &=& \delta g_{i0} = \left(\partial_i B + S_i \right) \,,\\
	\label{eq:MetricSVT3}
	\delta g_{ij} &=& \delta g_{ji} = -2\psi\delta_{ij} + (\partial_i\partial_j - {1\over3}\delta_{ij}\nabla^2)E + \partial_i F_j + \partial_j F_i + h_{ij} \,,
\end{eqnarray}
and
\begin{eqnarray}\label{eq:TmunuSVT}
	T_{00} &=& \rho \,,\\
	\label{eq:TmunuSVT2}
	T_{0i} &=& T_{i0} = \partial_i u + u_i \,,\\
	\label{eq:TmunuSVT3}
	T_{ij} &=& T_{ji} = p\,\delta_{ij} + (\partial_i\partial_j - {1\over3}\delta_{ij}\nabla^2)\sigma + \partial_{i} v_{j} + \partial_{j} v_{i} + \Pi_{ij} \,.
\end{eqnarray}
By construction, the above functions can be classified as scalars, vectors and tensors, according to how they transform under the 3-dimensional Euclidean rotation group,
\begin{center}
\begin{tabular}{c||c||c}
\, & $\delta g_\mn$ & $T_\mn$\\ \hline
\vspace*{-5mm}
\, & \, & \, \\ \hline
Scalar(s) & ~~~$\phi$, $B$, $\psi$, $E$~~~ & ~~~$\rho$, $u$, $p$, $\sigma$~~~\\ \hline
Vector(s) & ~~~$S_i$, $F_i$~~~ & ~~~$u_i$, $v_i$~~~\\ \hline
Tensor(s) & ~~~$h_{ij}$~~~ & ~~~$\Pi_{ij}$~~~ 
\end{tabular}
\end{center}
In order not to overcount degrees of freedom, the vector and tensor parts must satisfy the following conditions
\begin{equation}\label{eq:MetricConstraintI}
	\partial_i S_i = 0 ~(1~{\rm constraint})\,, \quad \partial_i F_i = 0 ~(1~ {\rm constraint})\,,
\end{equation}
\begin{equation}\label{eq:MetricConstraintII}
	\partial_i h_{ij} = 0 ~(3~ {\rm constraints})\,, \quad h_{ii} = 0 ~(1~ {\rm constraint}) \,.
\end{equation}
and
\begin{equation}\label{eq:TmunuConstraintI}
	\partial_i u_i = 0 ~(1~ {\rm constraint})\,, \quad \partial_i v_i = 0 ~(1~ {\rm constraint})\,,
\end{equation}
\begin{equation}\label{eq:TmunuConstraintII}
	\partial_i \Pi_{ij} = 0 ~(3~ {\rm constraints})\,, \quad \Pi_{ii} = 0 ~(1~ {\rm constraint})\,,
\end{equation}
where we have implicitly assumed that all terms vanish $\rho, u, u_i, p, \sigma, v_i, \Pi_{ij} \longrightarrow 0$ at infinity. The total number of degrees of freedom is as follows. For the  metric perturbation, the total number of functions introduced in Eqs.~(\ref{eq:MetricSVT})-(\ref{eq:MetricSVT3}) is 16 = 4 scalars ($\phi$, $B$, $\psi$, $E$) + 6 vector components ($S_i, F_i$) + 6 tensor components of the $3\times3$ symmetric tensor $h_{ij}$. The total number of constraints in Eqs.~(\ref{eq:MetricConstraintI}), (\ref{eq:MetricConstraintII}) is 6, so the number of independent functions in the decomposition defined by Eqs.~(\ref{eq:MetricSVT})-(\ref{eq:MetricSVT3}) is $10 = 16 - 6$, as expected for a $4\times4$ symmetric tensor. Analogous consistent counting follows for the energy-momentum tensor, decomposed in Eqs.~(\ref{eq:TmunuSVT})-(\ref{eq:TmunuSVT3}), and subject to the constraint equations~(\ref{eq:TmunuConstraintI}), (\ref{eq:TmunuConstraintII}).

Given our assumptions about metric perturbations over a flat background $\eta_\mn$, and asymptotic flatness $\delta g_\mn \longrightarrow 0$, the decomposition of the metric perturbation into scalar, vector and tensor pieces, defined by Eqs.~(\ref{eq:MetricSVT})-(\ref{eq:MetricSVT3}), is actually unique. Given a metric perturbation $\delta g_\mn$, one can always solve unequivocally for $\phi$, $B$, $\psi$, $E$, $S_i$, $F_i$ and $h_{ij}$, as a function of $\delta g_\mn$. Similarly, the decomposition of the energy-momentum tensor by Eqs.~(\ref{eq:TmunuSVT})-(\ref{eq:TmunuSVT3}) is also unique. 

All functions introduced so far (scalars, vectors and tensors) are assumed to be arbitrary functions of space-time coordinates and, in general, they are not independent from each other. For instance, from the conservation of the stress-energy tensor $\partial^\mu T_\mn = 0$, it follows that the functions introduced in Eqs.~(\ref{eq:TmunuSVT})-(\ref{eq:TmunuSVT3}), must satisfy\footnote{In reality, Eqs.~(\ref{eq:TmunuConservII}) and (\ref{eq:TmunuConservIII}) do not follow immediately from $\partial^\mu T_\mn = 0$, as this rather leads to ${\dot u}_i - \nabla^2 v_i = \partial_i f(x)$, where $f(x) = {2\over3}\nabla^2 \sigma + p - \dot {u}$. Applying $\vec\nabla \vec u = \vec\nabla \vec v = 0$ this leads to $\vec\nabla^2 f = 0$, so that ${\dot u}_i - \nabla^2 v_i = \partial_i f(x)$ corresponds to two constraints and $\nabla^2 f = 0$ to another. As functions vanish at infinity, it follows that $f(x) = 0\,~\forall\,x$, i.e.~it vanishes everywhere always, and hence $\nabla^2 \sigma = {3\over2}(\dot {u}- p)$ [c.f.~Eq.~(\ref{eq:TmunuConservII})] and $\nabla^2 v_i = {\dot u}_i$ [c.f.~Eq.~(\ref{eq:TmunuConservIII})] must be also true.}
\begin{eqnarray}\label{eq:TmunuConserv}
\nabla^2 u &=& \dot {\rho}~{\rm ~(1~constraint)}\,,\\
\label{eq:TmunuConservII}
\nabla^2 \sigma &=& {3\over2}(\dot {u}- p)~{\rm ~(1~constraint)}\,,\\
\label{eq:TmunuConservIII}
\nabla^2 v_i &=& {\dot u}_i~{\rm ~(2~constraints)}\,.
\end{eqnarray}
Therefore, considering these 4 extra constraints, in the case of the energy-momentum tensor, out of the 10 seemingly independent functions $\rho, u, u_i, p, \sigma, v_i, \Pi_{ij}$, there are in reality only $6 = 10 - 4$ independent degrees of freedom. For instance $\rho, u_i, p, \Pi_{ij}$ can be set arbitrarily, whereas the remaining functions $u, \sigma, v_i$ can be derived from the latter, by solving the system of Eqs.~(\ref{eq:TmunuConserv})-(\ref{eq:TmunuConservIII}).

Similarly, from the conservation of the Einstein tensor $\partial^\mu G_\mn = 0$ (which amounts to 4 constraints), it follows that out of the 10 metric functions $\phi$, $B$, $S_i$, $\psi$, $E$, $F_i$, $h_{ij}$ in the metric decomposition Eqs.~(\ref{eq:MetricSVT})-(\ref{eq:MetricSVT3}), only $6 = 10 - 4$ functions are truly independent degrees of freedom. The relation among metric components is however more complicated than in the case of the stress-energy tensor components Eqs.~(\ref{eq:TmunuConserv})-(\ref{eq:TmunuConservIII}), as it is the metric perturbation $\delta g_\mn$ (and not the Einstein tensor $G_\mn$), that we have decomposed into scalar, vector, and tensor parts. In order to reduce the number of independent degrees of freedom in Eqs.~(\ref{eq:MetricSVT})-(\ref{eq:MetricSVT3}) from 10 to 6, it is more practical to exploit the invariance of linearised gravity, under arbitrary infinitesimal coordinate transformations $x_\mu \longrightarrow x_\mu + \xi_\mu$. Following the logic of the metric decomposition in Eqs.~(\ref{eq:MetricSVT})-(\ref{eq:MetricSVT3}), let us first express an arbitrary infinitesimal 4-vector displacement as
\begin{eqnarray}
\xi_\mu = (\xi_0,\xi_i) \equiv (d_0,\partial_i d + d_i)\,~~~~~~{\rm with}~~\partial_i d_i = 0\,,
\end{eqnarray}
where $d, d_0, d_i$ are general functions of the space-time coordinates $(t,{\bf x})$. As the metric perturbation transforms under an arbitrary infinitesimal diffeomorphism as $\delta g_\mn \longrightarrow \delta g_\mn - \partial_\mu\xi_\nu - \partial_\nu\xi_\mu$, see Eq.~(\ref{eq:TensorChange}), one obtains that scalar parts of the metric perturbation in Eqs.~(\ref{eq:MetricSVT})-(\ref{eq:MetricSVT3}), transform as
\begin{eqnarray}\label{eq:gaugeTransformScalar}
\phi \longrightarrow \phi + {\dot d}_0\,,~~~~~~~~~~~ B \longrightarrow B - d_0 - \dot d\,,\\
\label{eq:gaugeTransformScalarII}
\psi \longrightarrow \psi + {1\over3}\nabla^2d\,, ~~~~~~ E \longrightarrow E - 2 d\,,
\end{eqnarray}
the vector parts as
\begin{eqnarray}\label{eq:gaugeTransformVector}
S_i \longrightarrow  S_i - \dot d_i\,, ~~~~~~~~~~ F_i \longrightarrow \quad F_i - d_i\,,
\end{eqnarray}
and the tensor part as
\begin{eqnarray}\label{eq:gaugeTransformTensor}
h_{ij} \longrightarrow h_{ij}\,.
\end{eqnarray}
The 2 degrees of freedom encoded in the tensor perturbation $h_{ij}$ are therefore gauge invariant, i.e.~independent of the system of coordinates (as long as we preserve the infinitesimal condition $|\delta g_\mn| \ll 1$). Since we know that there should be only $6~(= 10 - 4)$ physical degrees of freedom, it must be possible to reduce the above scalar and vector perturbations (8 functions) to only $4~ (= 6 - 2)$ degrees of freedom. In light of Eqs.~(\ref{eq:gaugeTransformScalar}), (\ref{eq:gaugeTransformScalarII}) and (\ref{eq:gaugeTransformVector}), one can build new scalar and vector perturbations 
\begin{eqnarray}\label{eq:NewScalar}
\Phi &\equiv& \phi + \dot B - {1\over2}\ddot E\,,\\
\label{eq:NewScalarII}
\Theta &\equiv& -2\psi - {1\over3}\nabla^2E\,,\\
\label{eq:NewVector}
\Sigma_i &\equiv& S_i - {\dot F}_i\,,~~~~~~{\rm with}~~\partial_i\Sigma_i = 0\,,
\end{eqnarray}
which are directly invariant under arbitrary infinitesimal coordinate transformations. 

Note that for the energy momentum tensor, we did not need to perform a coordinate transformation to exhibit the gauge-invariant independent degrees of freedom, as we just did for the metric perturbations: they followed simply after imposing energy-momentum conservation, see discussion after Eqs.~\eqref{eq:TmunuConserv}-\eqref{eq:TmunuConservIII}. This is a consequence of the fact that, in our approach, the energy momentum tensor must be zero in the background, because of the assumption of linearisation around Minkowski. In fact, the first-order perturbation of a generic tensor $T_\mn=\bar T_\mn+\delta T_\mn$ transforms under an infinitesimal coordinate transformation as $\delta T_\mn \longrightarrow \delta T_\mn + L_\xi \bar T_\mn$, where $L_\xi \bar T_\mn$ denotes the Lie derivative of the background component $\bar T_\mn$, along the vector field ${\bf \xi}_\mu$ (which reduces to Eq.~\eqref{eq:TensorChange} in flat spacetime and for the metric tensor). Therefore, a tensor with $\bar T_\mn=0$ is automatically gauge invariant, i.e. it is invariant under arbitrary infinitesimal coordinate transformations. This is the so-called Stewart Walker lemma \cite{SWlemma,DurrerBook}.

As the set of variables $\Phi, \Theta, \Sigma_i$ and $h_{ij}$ are gauge invariant, and all together represent in total 6 degrees of freedom (= 1 ($\Phi$) + 1 ($\Theta$) + 2 ($\Sigma_i$) + 2 ($h_{ij}$)), we are certain that these variables represent the truly physical degrees of freedom of the metric. It must be possible therefore, to express the Einstein equations as a function exclusively of these variables. As a matter of fact, the Einstein tensor 
can be written purely in terms of such gauge invariant quantities, as
\begin{eqnarray}
&& \hspace{-1cm}G_{00} = -\nabla^2\Theta\,,
\label{eq:G00}\\
&& \hspace{-1cm}G_{0i} = -\frac{1}{2}\nabla^2\Sigma_i - \partial_i\dot\Theta\,,
\label{eq:G0i}\\
&& \hspace{-1cm}G_{ij} = -\frac{1}{2}\Box h_{ij} -
\partial_{(i}\dot\Sigma_{j)} - \frac{1}{2}\partial_i\partial_j\left(2\Phi
+ \Theta\right) + \delta_{ij}\left[\frac{1}{2}\nabla^2
\left(2\Phi + \Theta\right) - \ddot \Theta\right]\,.
\label{eq:Gij}
\end{eqnarray}
Introducing now Eqs.~(\ref{eq:G00})-(\ref{eq:Gij}) in the left hand side of the Einstein Equations $G_\mn = {1\over m_p^{2}}T_\mn$, and Eqs.~(\ref{eq:TmunuSVT})-(\ref{eq:TmunuSVT3}) in the right hand side, one finds, with the help of Eqs.~(\ref{eq:TmunuConserv})-(\ref{eq:TmunuConservIII}), that
\begin{eqnarray}
\begin{array}{lcl}
\nabla^2\Theta = -{1\over m_p^{2}}\rho\,, &\, & \nabla^2\Phi = {1\over 2 m_p^{2}}\left(\rho + 3 p - 3\dot u\right)\,,\vspace*{0.2cm}\\
\nabla^2\Sigma_i = -{2\over m_p^{2}} S_i\,, &\, & \Box h_{ij} = -{2\over m_p^{2}}\Pi_{ij}\,.
\end{array}
\label{eq:hijTTeqn}
\end{eqnarray}
It appears that only the tensor part of the metric $h_{ij}$ obeys a wave equation. The other variables $\Theta$, $\Phi$ and $\Sigma_i$, obey Poisson-like equations. Indeed, in a globally vacuum space-time, the above equations reduce to five Laplace equations and a wave equation,
\begin{eqnarray}
\begin{array}{lcl}
\nabla^2\Theta = 0\,, &\, & \nabla^2\Phi = 0\,,\vspace*{0.2cm}\\
\nabla^2\Sigma_i = 0\,, &\, & \Box h_{ij} = 0\,.
\end{array}
\end{eqnarray}
This demonstrates explicitly that, among the gauge-invariant degrees of freedom of the metric perturbation $\Theta, \Phi, \Sigma_i$ and $h_{ij}$, only the {\it tensor} part $h_{ij}$ (which has two independent components) represents radiative degrees of freedom that can propagate in vacuum. 

The above statement is actually independent of the system of reference, as long as the metric perturbation remains as such, i.e.~a perturbation $|\delta g_\mn| \ll 1$. In Sect.~\ref{subsec:GWdef} we found that the invariance under infinitesimal coordinate transformations of the linearised theory, allows to saturate the gauge freedom once one reduces the metric perturbations to only 2 degrees of freedom, in the transverse-traceless gauge and in vacuum. However, identifying correctly the truly gauge invariant and radiative degrees of freedom is not just a matter of a gauge choice. In some gauges, as e.g. the Lorentz one, it is possible to have all metric components satisfying a wave equation, but this is only a `gauge artefact', arising due to the choice of coordinates. Such gauge choices, although useful for calculations, may mistakenly led to identify pure gauge modes, with truly physical gravitational radiation. 

To summarise, in general, a metric perturbation $\delta g_{\mn}$ contains: $i)$ gauge spurious degrees of freedom, $ii)$ physical but non-radiative degrees of freedom, and $iii)$ physical radiative degrees of freedom. We have found that, using infinitesimal coordinate transformations, one can arrive to the result that only two physical radiative degrees of freedom are relevant. However, due to the presence of the physical non-radiative degrees of freedom, these cannot be made explicit, unless in vacuum: it is not possible in general to write the metric perturbation in the TT gauge, since usually we cannot eliminate the temporal components of the stress-energy tensor that do not vanish $T_{00},~T_{0i} \neq 0$. Nonetheless, here we have demonstrated that the linearised metric perturbation can be split up uniquely into scalar, vector and tensor parts, as in Eqs.~(\ref{eq:MetricSVT})-(\ref{eq:MetricSVT3}). This decomposition contains all type of degrees of freedom $i)-iii)$. From Einstein equations it appears clearly that the physical radiative degrees of freedom correspond only to the tensor piece of the metric perturbation, i.e.~to the piece that satisfies a wave equation and verifies the TT gauge conditions (often referred to as the TT piece), irrespective of the gauge choice. In vacuum, the TT-gauge happens to correspond to the set of coordinate systems where the whole metric perturbation reduces to the physical radiative degrees of freedom. In the presence of matter, there are instead four physical degrees of freedom on top of the TT ones. Yet, the latter are -- unmistakably -- the only physical degrees of freedom truly representing gravitational radiation, independently of the gauge choice, and/or the presence of matter.

\subsection{Gravitational waves in a curved background}
\label{subsec:GWcurved}

In section \ref{subsec:GWdef} we have presented the definition of GWs in the context of linearised gravity over a Minkowski background, in asymptotically-flat and globally vacuum space-times. In section \ref{subsec:SVT} we have demonstrated that GWs can be unequivocally defined also dropping the assumption of globally-vacuum space-times. The next step is to drop the assumption of linearised theory over Minkowski, and tackle the definition of GWs over a curved background. 

This step becomes indeed mandatory in order to define the GW energy momentum tensor \cite{Misner:1974qy,MaggioreBook,Flanagan:2005yc}. In GR, every form of energy contributes to the curvature of space-time. In order to find expressions for the energy and momentum carried by GWs, one has to explore in which sense GWs are themselves a source of space-time curvature. However, this simple statement is enough to conclude that one needs to go beyond linearised theory over Minkowski: sticking to it, one excludes from the beginning the possibility of generating any form of curvature in the space-time background, being by definition flat. Furthermore, in the rest of this review we will be dealing with GWs generation processes operating in the early universe, and hence it is crucial that we determine how to define GWs over a FLRW background, which naturally corresponds to a curved background. 

Going beyond linearised theory over Minkowski emerges as an outstanding necessity, but it is far from being a simple task. We need to generalize the theory to $g_{\mu\nu}(x) = \bar g_{\mu\nu}(x) + \delta g_{\mu\nu}(x)$, with $|\delta g_{\mu\nu}(x)| \ll |\bar g_\mn(x)|$ and $\bar g_\mn(x)$ a general metric. However, it is clear that in this setting it becomes non-obvious to define GWs, since it is non-trivial to distinguish the background from the fluctuation, as $\bar g_\mn(x)$ can contain space- and time-dependent components, due for instance to space- and time-varying Newtonian fields. The only way to define fluctuations representing GWs in this context, is to {\it exploit a possible separation of scales/frequencies}: if the background $\bar g_\mn(x)$ varies over a typical length-scale $L_B$ (or its time variation is characterised by a typical frequency $f_B$), and the GWs have typical reduced wavelength $\lambdabar=\lambda / 2\pi$ (or frequency $f=1/\lambda$), one can distinguish the GWs from the background provided that $L_B\gg \lambdabar$ ($f_B\ll f$). In this case, the GWs can be viewed as small perturbations on a smooth background (from `their' point of view), or rapidly varying perturbations over a slowly varying background. 

Let us make this more explicit with two examples. We will properly introduce the case of the FLRW metric later on, but let us anticipate that in this case, it is easy to see that the typical space and time variations of the background today correspond to the Hubble factor, $L_B\sim 1/f_B\sim 1/H_0$. For a GW production mechanism operating causally (i.e.~within the causal horizon) at a given time $t_*$ in the radiation or matter dominated eras, the typical wave-lengths/frequencies of the GW signal today would correspond to $\lambda=1/f \leq a_0/(a_* H_*)$. Since the universe is expanding it holds that $a_* H _*\gg a_0 H_0$, and therefore it is clear that the above conditions (in terms of length-scales and in terms of frequencies) are satisfied in this case. This is an anticipation of the fact that GWs can indeed be well defined in a cosmological context.  

In the case of GWs arriving on Earth, e.g.~from a compact binary coalescence, the situation is more complicated. 
Earth-based GW detectors have the best sensitivity to GWs for frequencies around $f\sim 100-1000$ Hz, corresponding to $\lambdabar \sim 500-50$ km: on these length scales, the Newtonian gravitational field of the Earth does have spatial variations, rendering it impossible to distinguish it from GWs solely based on the condition $L_B\gg \lambdabar$. Moreover, its amplitude is much bigger than the GW one, $\delta g_{00}^N\sim 10^{-9} \gg \delta g_{ij}^{\rm GW}\sim 10^{-21}$. On the other hand, the Earth gravitational field is almost static in the frequency window in which terrestrial interferometers operate: it varies mostly on typical frequencies $f_B\lesssim 0.1 $ Hz. Therefore, Earth-based interferometers can indeed perform GW measurements (as proven recently!), and distinguish GWs from the background, based on the condition $f_B\ll f$: i.e.~by maximizing their sensitivity in a frequency window which is clean from the time varying Newtonian gravitational field, Earthquakes and other seismic motions. 

Let us then start from the principle that GWs can be defined within the approximation that their wave-length (frequency) is much smaller (bigger) than the length (inverse time-scale) characterising the background space-time over which the waves are propagating. The method to implement this definition on a practical level, is to perform averages of physical quantities over a length-scale $\ell$ (time-scale $\tau$) such that $\lambdabar\ll \ell \ll L_B$ ($1/f \ll \tau \ll 1/f_B$). For proper covariant definitions of the averaging procedure, see \cite{Brill:1964zz}. 

Since an averaging is involved, it becomes clear that, in order to proceed consistently in the definition of GWs, it is necessary to {\it go to second order} in the expansion of the metric perturbation $|\delta g_\mn|\ll 1$. Averaging the first order contribution in fact gives zero, because of the oscillatory nature of the waves. So we need to look for the contribution of GWs to the background curvature, but linear quantities in $\delta g_\mn$ cannot influence the background, as they average to zero. On the other hand, the averaging of a second order quantity can mix two short wavelength (high frequency) modes, in such a way, that in total they contribute a long wavelength (low frequency) mode, as commonly experienced in convolutions. Therefore, second order quantities in $\delta g_{\mu\nu}$ can give rise to the corrections to the background metric at scales larger than the GW wavelength. A particularly relevant point, that will help to clarify these ideas, is the definition GW energy momentum tensor, which we discuss next.  

For a full analysis of perturbation theory over a generic background to second order in $|\delta g_\mn| \ll 1$, we refer the reader to the excellent treatments of Refs.~\cite{Misner:1974qy,MaggioreBook,Flanagan:2005yc}. Here, we only present the two main results that will be useful for us in the review: the definition of the GW energy momentum tensor, and the equation for the propagation of GWs on a curved background. Concerning the first point, the analysis of the Einstein equations at second order, shows that the effective energy momentum tensor of GWs is obtained by averaging the second order Ricci tensor. Denoting the metric perturbation as $h_{\mu\nu} = \delta g_{\mu\nu}$ and  generalizing the trace-reserve metric perturbation definition as $\bar{h}_{\mu\nu} = h_{\mu\nu} - {1\over 2}\bar g_{\mu\nu}\bar g^{\alpha\beta}h_{\alpha\beta}$ (where $\bar g_{\mu\nu}$ is the background metric), the result reads, in the Lorentz gauge $\nabla^\mu \bar{h}_{\mu\nu} = 0$ and once all spurious gauge modes have been removed, as
\begin{equation}\label{eq:GWenmontens}
	{T^{\rm GW}}_{\mu\nu}=\frac{\langle \nabla_\mu h_{\alpha \beta}\nabla_\nu h^{\alpha \beta} \rangle}{32\pi \,G}\,,
\end{equation}
where $\nabla$ denotes the covariant derivative with respect the background space-time metric, and $\langle ...\rangle$ an average over $\ell$ and/or $\tau$. The above GW energy momentum tensor must be inserted in the right hand side of the background Einstein equations, as any other form of matter. It contributes to the background space-time curvature as a term of order $\mathcal{O}(h^2/\lambdabar^2)$, which must satisfy $h^2/\lambdabar^2\lesssim 1/L_B^2$, where the $<$ sign applies if the background already contains a contribution from another (dominant) source of space-time curvature, whereas the $\simeq$ sign applies if the only contribution is due to the GWs. The consistency of the second order treatment is then manifest, since one has $|h_{\mn}| \lesssim \lambdabar/L_B$: the definition of GWs in the limit $\lambdabar/L_B\ll 1$ implies, therefore, that these are small perturbations, $|h_{\mn}|\ll 1$. From Eq.~\eqref{eq:GWenmontens} we see that the energy density of GWs becomes the well known expression 
\begin{equation}\label{eq:GWendensTT}
	\rho_{\rm GW}= T_{\rm GW}^{00}=\frac{\langle {\dot{h}}_{ij} \,{\dot{h}^{ij}} \rangle}{32\pi \,G}\,,
\end{equation}
which we have written in the TT gauge. Following the results presented in sections \ref{subsec:GWdef} and \ref{subsec:SVT}, TT gauge can be chosen either far away from the sources where one is almost in vacuum, in which case $\dot{h}_{ij}$ denotes the derivative with respect to the time variable of Minkowski metric; or in curved space-time, for example in FLRW, in which case $\dot{h}_{ij}$ denotes the derivative with respect to physical time, leading to Eq.~\ref{rhogw}. Note that, for FLRW at first order in perturbation theory, $\nabla_0{h}_{ij}=\dot{h}_{ij}$. 

To analyse the propagation of GWs on a curved background, on the other hand, one does not need to go to second order in the $h_\mn$ expansion. The Einstein equations at first order,  in the case of the expansion around a flat background, led to the propagation equation in \eqref{eq:hijTTeqn}. In the case of a curved background, this generalises to (see e.g.~\cite{Isaacson:1967zz,Isaacson:1968zza,
Pfenning:2000zf,Barausse:2013ysa})
\begin{eqnarray}\label{eq:GWeqNoGauge}
&-\frac{1}{2} \Box \bar{h}_{\mu\nu} + & R^\lambda{}_{\mu \nu}{}^\sigma \bar{h}_{\lambda \sigma} + \nabla_{(\nu} \nabla^\sigma \bar{h}_{\mu)\sigma} -\frac{1}{2} \bar{g}_{\mu\nu} \nabla^\alpha \nabla^\beta \bar{h}_{\alpha \beta} + \\
 & &  +{R}^{\alpha \beta} \left[\frac{1}{2} \bar{g}_{\mu\nu} \bar{h}_{\alpha \beta} -\frac{1}{2} \bar{h}_{\mu\nu} \bar{g}_{\alpha \beta} + \bar{g}_{\beta (\mu} \bar{h}_{\nu)\alpha}\right] = 8\pi\, G\, \delta T_{\mu\nu}\,, \nonumber
\end{eqnarray}
where round parentheses in a subscript denote symmetrisation, $\bar{h}_{\mn}$ is the trace reversed metric perturbation, $\Box $, $\nabla$, the Riemann and the Ricci tensors are defined with respect to the background metric $\bar{g}_{\mu\nu}$. We have decomposed the matter energy momentum tensor as $T_\mn = \bar T_\mn + \delta T_\mn$, where $\bar T_\mn$ is the background contribution, sourcing the background curvature of $\bar{g}_{\mu\nu}$, while $\delta T_\mn$ is the first order contribution, that can act as source of GWs. 
Note that we have written the above equation without fixing the gauge choice and in presence of a generic matter source, as in this way it can be directly adapted to the FRLW case. The above equation is most commonly written in vacuum and in the Lorentz gauge $\nabla^\alpha{\bar h}_{\alpha\beta} = 0$, where it takes the simpler form \cite{Flanagan:2005yc}
\begin{equation}
	\Box \bar{h}_{\mu\nu} -2 R^\lambda{}_{\mu \nu}{}^\sigma \bar{h}_{\lambda \sigma}=0\,.
\end{equation}
In the limit $L_B\gg \lambdabar$, the couplings to the background due to the $\Box$ term and the term proportional to the background Riemann tensor, have the effect of imprinting gradual changes on the properties of the GWs, e.g. in their amplitude and polarisations. This can be analysed by solving the above equation in the geometric optics limit $L_B\gg \lambdabar$, with the result that GWs propagate along null geodesics of the space-time background , with parallel-transported polarisation, orthogonal to the rays, see e.g.~\cite{MaggioreBook} for a discussion about this. Furthermore, one of the consequences of the geometric optics limit is a conservation law, that represents, in the quantum language, the conservation of the number of gravitons. In Section \ref{sec:inf} we will present however, an example of a situation, namely inflation, where the conservation of gravitons does not hold. During inflation one has that the GWs (sub-Hubble tensor metric perturbations) gradually exit the Hubble scale, breaking the condition $L_B\gg \lambdabar$. When this occurs, the dynamics of the GWs is strongly coupled to the dynamics of the background, and the graviton number is no longer conserved. We will explain this in detail in Section \ref{sec:inf}. The important point to remark here is that, even though in that case $L_B\simeq \lambdabar$, one can still distinguish the background from the metric fluctuations (albeit the tensor modes become GWs only after they have crossed back inside the Hubble scale). See Section~\ref{sec:PrimTensInf} in particular, for further details.

All of this finally brings us to the definition of GWs in the particular case when the background space-time is the FLRW metric. It is a remarkable case in which the splitting of the metric into a curved background component plus linear perturbations can be uniquely defined, even in the regime\footnote{Another example are e.g.~static space-times.} $L_B\simeq \lambdabar$. The reason is the symmetries of the FLRW background, experimentally well verified by CMB observations \cite{Ade:2013vbw}: the hyper-surfaces of constant time are homogeneous and isotropic. Therefore, there is no possible ambiguity between what pertains to $\bar g_\mn(x)$ and what to $h_\mn(x)$. 
Furthermore, because of homogeneity and isotropy, two-index tensors can be irreducibly decomposed on these hyper-surfaces under spatial translations (harmonic analysis) and rotations. One can therefore perform the same decomposition carried on in section \ref{subsec:SVT}, but on a time-evolving background. A general perturbation of the metric can be decomposed into scalar, vector and tensor modes, according to the way they transform under spatial rotations in the background space-time. For derivations of this decomposition in the case of a FLRW background, see e.g.~\cite{Bardeen:1980kt,Kodama:1985bj,DurrerBook,WeinbergCosmo}. 

As already mentioned in section \ref{subsec:SVT}, the main difference with the flat space-time case is the presence of a non-zero energy momentum tensor at the background level, which has to satisfy the symmetries of the FLRW space (admitted cases are for example an unperturbed perfect fluid or a scalar field that depends on time only). Consequently, in cosmological perturbation theory one has to build a set of gauge-invariant variables also for the components of the energy momentum tensor, which we did not need to do in section \ref{subsec:SVT}. Moreover, in cosmological perturbation theory, it is customary to also expand the metric perturbations into eigenfunctions of the Laplacian (i.e.~in Forurier modes), which represent irreducible components under translations. Other than this, the two analyses proceed in a very similar way.

In the context of cosmological perturbation theory one also finds that the two degrees of freedom of the tensor modes of the metric perturbations, $h_{ij}$, are the only radiative modes\footnote{Note that, in this context, one also finds a wave-like equation for the scalar Bardeen potential, which in the case of adiabatic perturbations of a perfect fluid and zero spatial curvature reads $\ddot{\Phi}+3\mathcal{H}(1+c_s^2)\dot \Phi+ [\mathcal{H}^2 (1+3 c_s^2)- \mathcal{H}^2 (1+3 w)+ k^2c_s^2] \Phi=0$, where $w$ is the background fluid equation of state, and $c_s$ its sound speed. However, these are sound waves and represent the perturbations in the matter-radiation fluid, coupled to the metric. They do not exist if the background fluid is not present, and their modes cannot propagate in vacuum: they have therefore an entirely different nature than the tensor mode.}, and therefore correspond to GWs: scalar and vector modes cannot propagate in vacuum. The symmetries of the FLRW background imply that $h_{ij}$ vanishes in the background: it is therefore gauge-invariant at first order by the Stewart Walker lemma (c.f.~section \ref{subsec:SVT}). Scalar, vector and tensor modes are decoupled from each other at linear order in perturbation theory \cite{DurrerBook}. 
GWs may then be represented by the tensor spatial perturbations $h_{ij}$ ($i, j = 1, 2, 3$) of the FLRW metric 
\be
\label{GWcosmo}
ds^2 = -dt^2 + a^2(t) \, (\delta_{ij} + h_{ij}) \, dx^i dx^j
\ee  
with
\be 
\label{TT}
\partial_i h_{ij} = h_{ii} = 0 \, .
\ee
Since $h_{ij}$ is symmetric, the transverse and traceless conditions (\ref{TT}), leave only two independent degrees of freedom, which correspond to the two GW polarizations. For most practical purposes there is no need to go beyond linear order in perturbations theory. The only exception in this review will be presented in sub-section \ref{sec:SecGWScal}, where we will consider second order scalar perturbations as a source of GWs.

\subsection{Propagation of gravitational waves in expanding backgrounds}
\label{gweom}

In this section we analyse the propagation of GWs in a cosmological context, and provide some useful definitions that will hold throughout the review. The GW equation of motion is given by the Einstein equations linearized to first order in $h_{ij}$, over a FLRW background. It can be deduced from Eq.~\eqref{eq:GWeqNoGauge}, by specifying the FLRW connection, Riemann and Ricci tensors, and keeping only the TT piece $h_{ij}$ of the metric perturbation. This leads to 
\be
\label{gweqx}
\ddot{h}_{ij}(\mathbf{x}, t) + 3 \, H \, \dot{h}_{ij}(\mathbf{x}, t) -  
\frac{\mathbf{\nabla}^2}{a^2} \, h_{ij}(\mathbf{x}, t) = 16\pi G \, \Pi_{ij}^{TT}(\mathbf{x}, t)\,,
\ee
where $\mathbf{\nabla}^2 = \partial_i \, \partial_i$ is the Laplacian associated to the comoving coordinates $x^i$ in \eqref{GWcosmo}, a dot denotes derivative with respect to $t$, $H = \dot{a} / a$ is the Hubble rate, and $\Pi_{ij}^{TT}$ is the transverse and traceless part of the anisotropic stress. 
The anisotropic stress is given by
\be
\label{Piij}
a^2 \, \Pi_{ij} = T_{ij} - p \, a^2 \, (\delta_{ij} + h_{ij})\,,
\ee
where $T_{ij}$ denotes the spatial components of the energy-momentum tensor of the source, and $p$ is the background pressure. In the RHS of Eq.~(\ref{Piij}), the term in $p \, \delta_{ij}$ is a pure trace that does not contribute to 
$\Pi_{ij}^{TT}$, while the term in $p\,h_{ij}$ cancels out with an identical term of opposite sign that emerges in the derivation of \Eqref{gweqx}.

The transverse and traceless part of a tensor is most easily extracted in Fourier space. Consider the spatial Fourier transform
\be
\Pi_{ij}(\mathbf{x}, t) = \int \frac{d^3 \mathbf{k}}{(2 \pi)^{3}}\,
\Pi_{ij}(\mathbf{k}, t)\,e^{- i \mathbf{k} \cdot \mathbf{x}} \, .
\ee
The transverse and traceless part of a symmetric tensor is then given by the projection (see e.g.~Ref.~\cite{Misner:1974qy})
\bea
\label{TTproj}
\hspace{-1cm}\Pi_{ij}^{\mathrm{TT}}(\mathbf{k}) = \mathcal{O}_{ij , lm}(\mathbf{\hat{k}}) \, \Pi_{lm}(\mathbf{k}) = 
\left[P_{il}(\mathbf{\hat{k}})\,P_{jm}(\mathbf{\hat{k}}) - \frac{1}{2}\,P_{ij}(\mathbf{\hat{k}})\,P_{lm}(\mathbf{\hat{k}})\right]
\Pi_{lm}(\mathbf{k})\,,
\eea
with
\be
\label{Pij}
P_{ij}(\mathbf{\hat{k}}) = \delta_{ij} - \hat{k}_i\,\hat{k}_j\,,
\ee
where $\mathbf{\hat{k}} = \mathbf{k} / k$ is the unit vector in the $\mathbf{k}$ direction. The operators $P_{ij}$ are projectors on the subspace orthogonal to $\mathbf{k}$, satisfying $P_{ij} k_i = 0$ and $P_{ij}\,P_{jl} = P_{il}$. From this it follows directly that $k_i\,\Pi_{ij}^{\mathrm{TT}} = \Pi_{ii}^{\mathrm{TT}} = 0$. 

The transverse and traceless perturbation $h_{ij}$ can be decomposed into the two polarization states $r = +, \times$, as
\be
\label{hrketa}
h_{ij}(\mathbf{x}, t) = \sum_{r = + , \times} \, \int \frac{d^3 \mathbf{k}}{(2 \pi)^{3}} \, h_r(\mathbf{k},t) \,
e^{- i \mathbf{k} \cdot \mathbf{x}} \, e_{ij}^{r}(\mathbf{\hat{k}})
\ee
where the two polarisation tensors $e_{ij}^{r}(\mathbf{\hat{k}})$ can be taken to be real and to satisfy 
$e_{ij}^{r}(\mathbf{- \hat{k}}) = e_{ij}^{r}(\mathbf{\hat{k}})$. The condition for $h_{ij}$ to be real is then 
$h_r^*(\mathbf{k}, t) = h_r(- \mathbf{k}, t)$. The two polarisation tensors depend only on the unit vector $\mathbf{\hat{k}}$ and are symmetric ($e^r_{ij} = e^r_{ji}$), transverse ($\hat{k}_i\,e^r_{ij} = 0$) and traceless ($e_{ii}^r = 0$). They can be written as
\bea
\label{eij}
e^+_{ij}(\mathbf{\hat{k}}) &=& \hat{m}_i\,\hat{m}_j - \hat{n}_i\,\hat{n}_j\,, \nonumber\\
e^{\times}_{ij}(\mathbf{\hat{k}}) &=& \hat{m}_i\,\hat{n}_j + \hat{n}_i\,\hat{m}_j\,, 
\eea
where $\mathbf{\hat{m}}$ and $\mathbf{\hat{n}}$ are two unit vectors that are orthogonal to $\mathbf{\hat{k}}$, and well as to each other. We then have the orthonormal and completeness relations
\bea
\label{orthonormal}
e_{ij}^{r}(\mathbf{\hat{k}}) \, e_{ij}^{r'}(\mathbf{\hat{k}}) &=& 2 \, \delta_{r r'}\,, \\
\label{completeness}
\sum_{r = + , \times} \, e_{ij}^{r}(\mathbf{\hat{k}}) \, e_{lm}^{r}(\mathbf{\hat{k}}) &=& 
P_{il} \, P_{jm} + P_{im} \, P_{jl} - P_{ij} \, P_{lm}\,,
\eea
where the projectors $P_{ij}$ are defined in Eq.~(\ref{Pij}), and can be written alternatively as 
$P_{ij} = \hat{m}_i\,\hat{m}_j + \hat{n}_i\,\hat{n}_j$. 

The space-time behavior of GWs is determined by Eq.~\eqref{gweqx}, with solutions depending on the particular source considered. Most examples of cosmological sources last however, only for a finite amount of time, and become eventually negligible. In linearised GR, once the source has stopped operating, GWs propagate freely through the FLRW space-time. It is therefore useful to derive the corresponding free solutions of Eq.~\eqref{gweqx}, particularly in the two regimes of interest in a cosmological setting: for wavelengths smaller and larger than the Hubble radius. 

It is convenient to work with conformal time $d\eta = dt / a(t)$, so that the metric (\ref{GWcosmo}) reads
\be
\label{GWconf}
ds^2 = a^2(\eta) \, \left[ - d\eta^2 + (\delta_{ij} + h_{ij}) \, dx^i dx^j \right] \, .
\ee 
Defining
\be
H_{ij}(\mathbf{k}, \eta) = a\,h_{ij}(\mathbf{k}, \eta) \,,
\ee
Eq.~(\ref{gweqx}) in Fourier space becomes
\be
\label{gweq2}
H_{ij}''(\mathbf{k}, \eta) + \left(k^2 - \frac{a''}{a}\right)\,H_{ij}(\mathbf{k}, \eta) = 
16\pi G \, a^3\,\Pi_{ij}^{TT}(\mathbf{k}, \eta) \,,
\ee
where primes denote derivatives with respect to $\eta$, and $k = |\mathbf{k}|$ is the comoving wave-number. Restricting ourselves to the case where the source is absent, $\Pi_{ij}^{TT}(\mathbf{x}, t)=0$ (the solution in the presence of a generic stochastic source is deferred to section~\ref{sec:spectrum_generic}), we are interested in solving the time-dependence of the Fourier amplitudes $h_r(\mathbf{k}, \eta)$ in \Eqref{hrketa}. These can be easily obtained from the equation
\be
\label{gweq3}
H_{r}''(\mathbf{k}, \eta) + \left(k^2 - \frac{a''}{a}\right)\,H_{r}(\mathbf{k}, \eta) = 0\,,
\ee
where we have set the source to zero, and defined $H_{r}(\mathbf{k}, \eta) = a\, h_{r}(\mathbf{k}, \eta)$. Let us focus on a generic scale factor with power law behaviour $a(\eta)=a_n\eta^n$, which covers the cases of radiation ($n=1$) and matter ($n=2$) domination, as well as of {\it de Sitter} inflation ($n = -1$), see e.g. \cite{DurrerBook}). The general solution is 
\begin{equation}
	h_r({\bf k},\eta)=\frac{A_r(\mathbf{k})}{a(\eta)}\eta\, j_{n-1}(k\eta)+\frac{B_r(\mathbf{k})}{a(\eta)}\eta\,y_{n-1}(k\eta)\,,
	\label{hsol_for_n}
\end{equation}
where $j_n(x), y_n(x)$ are the spherical Bessel functions, and $A_r(\mathbf{k})$ and $B_r(\mathbf{k})$ are dimensional constants, to be established from the initial conditions. 

Somewhat more explicit solutions can be obtained using the fact that, for a power law scale factor, $a''/a\propto \mathcal{H}^2$, where $\mathcal{H}=a'/a$ is the comoving Hubble factor. One can therefore solve approximately \Eqref{gweq3} in the limits of super-Hubble ($k \ll \mathcal{H}$) and sub-Hubble ($k\gg \mathcal{H}$) scales, which simply correspond to the solutions one obtains taking the limits $k\eta\ll 1 $ and $k\eta\gg 1 $ in Eq. \eqref{hsol_for_n}. 

For sub-Hubble scales, one neglects the term $a'' / a$ with respect the term $k^2$, in Eq. (\ref{gweq3}), and the solution becomes 
\be
\label{subHub}
h_r(\mathbf{k}, \eta) = \frac{A_r(\mathbf{k})}{a(\eta)}\,e^{i k \eta} + 
\frac{B_r(\mathbf{k})}{a(\eta)}\,e^{- i k \eta}\,, \hspace*{0.5cm} \mbox{for } k \gg \mathcal{H} \, .
\ee 
Again, $A_r(\mathbf{k})$ and $B_r(\mathbf{k})$ are dimensional constants [of different dimension than in Eq.~(\ref{hsol_for_n})], to be established from the initial conditions. For $h_{ij}(\mathbf{x}, \eta)$ to be real, they must satisfy the conditions $A_r(-\mathbf{k})=B_r^*(\mathbf{k})$ and $B_r(-\mathbf{k})=A_r^*(\mathbf{k})$. With the above solution for sub-Hubble modes, Eq.~(\ref{hrketa}) reduces to a superposition of plane waves with wave-vectors $\mathbf{k}$, and amplitude decaying as $1/a(\eta)$,
\begin{eqnarray}
	h_{ij}(\mathbf{x}, \eta) &=& \frac{1}{a(\eta)} \sum_{r = + , \times} \, \int \frac{d^3 \mathbf{k}}{(2 \pi)^{3}} \, e_{ij}^{r}(\mathbf{\hat{k}}) [A_r(\mathbf{k}) e^{ik\eta- i \mathbf{k} \cdot \mathbf{x}}+ {\rm c.c.}]=\label{eq:plane_wave_A}\\	
	&=& \frac{1}{a(\eta)}\sum_{r = + , \times} \, \int \frac{d^3 \mathbf{k}}{(2 \pi)^{3}} \, e_{ij}^{r}(\mathbf{\hat{k}}) [B_r(\mathbf{k}) e^{-ik\eta- i \mathbf{k} \cdot \mathbf{x}}+ {\rm c.c.}] \label{eq:plane_wave_B} \,,
\end{eqnarray}
where c.c.~stands for complex conjugate. These formulas will be useful at the end of subsection~\ref{sec:stochback2}. 

To solve \Eqref{gweq3} for super-Hubble scales, one neglects instead the term $k^2$ in \Eqref{gweq3}, and then we obtain
\be
\label{supHub}
h_r(\mathbf{k}, \eta) = A_r(\mathbf{k}) + B_r(\mathbf{k}) \, \int^{\eta} \, \frac{d \eta'}{a^2(\eta')} 
\hspace*{0.5cm} \mbox{for } k \ll \mathcal{H} \, ,
\ee
where the first term in the RHS is constant in time, and the second one decays with the expansion of the universe (again, $A_r(\mathbf{k})$ and $B_r(\mathbf{k})$ are arbitrary constants). As it will be discussed in Section~\ref{sec:inf}, Eq.~(\ref{supHub}) applies, in particular, to the super-Hubble modes generated from quantum fluctuations of the tensor metric perturbation during inflation. In that case, the decaying mode in Eq.~(\ref{supHub}) becomes quickly negligible due to the quasi-exponential expansion of the universe, so that $h_r(\mathbf{k}, \eta)$ is constant in time, for modes outside the Hubble radius. These tensor perturbations eventually re-enter the Hubble radius during the post-inflationary evolution, and then become standard GWs, behaving as in Eq.~(\ref{subHub}). 

Note that, in a radiation-dominated universe, the term $a'' / a$ vanishes identically, therefore strictly speaking one cannot perform the approximation based on neglecting $k^2$ vs $a'' / a$. The super-Hubble solution in that case, is simply \Eqref{subHub} in the limit $k\eta\ll 1$, which also reduces to a decaying and a constant mode.

\section{Cosmological (ergo stochastic) gravitational wave backgrounds}
\label{sec:general}
 
 
In this section we introduce general aspects of cosmological backgrounds of GWs. We review the reasons why GW backgrounds generated by cosmological sources are expected to have a stochastic character (sub-section~\ref{sec:stochback1}), and we detail how to characterize the spectrum of such stochastic backgrounds of GWs (sub-section~\ref{sec:stochback2}). Furthermore, we discuss the evolution of a cosmological background from the time of its production until the present epoch, as it redshifts with the expansion of the universe (sub-section~\ref{redshift}). Finally, we present a derivation of the GW spectrum by a generic stochastic source (sub-section~\ref{sec:spectrum_generic}).
 
\subsection{Stochastic nature of cosmological backgrounds}
\label{sec:stochback1}
 
Early universe sources typically lead to the production of stochastic backgrounds of GWs today. This means that the amplitude of the tensor perturbation $h_{ij}({\mathbf x},\eta)$ in Eq.~\eqref{hrketa} is a random variable, which can be characterised only statistically, by means of ensemble averages. In principle, to perform an ensemble average, many copies of the system should be available; obviously in the case under analysis this does not happen, as there is only one observable universe. What is customarily done in cosmology is to invoke the {\it ergodic} hypothesis, equating the ensemble average with either spatial and/or temporal averages. This implies that, by observing today large enough regions of the Universe, or a given region for long enough time, one has access to many realisations of the system. Two conditions must be met for this to hold. The first one is that the universe is almost homogeneous and isotropic, so that the `initial conditions' of the GW generating process are the same (even if only in a statistical sense) at every point in space. The second one is that a GW source fulfils causality, and operates at a time when the typical size of a region of causal contact in the universe was smaller than the causal horizon today\footnote{The case of inflation does not verify this condition, but we analyse this later on in the section.}.
Under these conditions, the GW signal from the early universe takes the form of a stochastic background and one can invoke the ergodic theorem to study its properties. Before showing in further detail why GWs from the early universe must be viewed as a stochastic field, let us remark that, under the ergodic hypothesis, the average over length- and/or time-scales introduced in section \ref{subsec:GWcurved}, necessary to define GWs, can be identified with the ensemble average needed to characterise the statistical properties of the GW signal. 
 
Because of causality, a cosmological GW source acting at a given time in the early universe, cannot produce a signal correlated at length/time scales larger than the cosmological horizon at that time. Denoting with a subscript $p$ the time of production, the (physical) correlation scale of the emitted GWs, must satisfy $\ell_p\leq H_p^{-1}$, while the GW signal can be correlated at best on a time scale $\Delta t_p\leq H_p^{-1}$. Here we have set the inverse Hubble factor $H^{-1}_p$ as the cosmological horizon, which is a good approximation for most of the cosmological evolution (except for inflation, a case that we discuss below). Since at the present time we have access to much larger length/time scales than today's redshifted scale associated to $H^{-1}_p$, the GW signal in the universe today is composed by the superposition of many signals, that are uncorrelated in time and space. The number of independent signals can be actually obtained, knowing the evolution of the universe and depending on the time of GW production. 

We start by comparing the size of the horizon today $H_0^{-1}$ with the correlation length scale redshifted to today $\ell_p^0=\ell_p (a_0/a_p)$ (for simplicity, we focus for now on the length-scale only and analyse the case of time-scales later on). This gives
\be\label{ellp}
\frac{\ell_p^0}{H_0^{-1}} = \frac{\ell_p}{H_0^{-1}} \frac{a_0}{a_p} \leq \frac{H^{-1}_p}{H_0^{-1}} \, \frac{a_0}{a_p} = \frac{a_0/a_p}{\sqrt{\Omega_{\rm mat}(z_p) + \Omega_{\rm rad}(z_p)+ \Omega_\Lambda}}\,,
\ee
where in the third equality $z$ denotes the redshift, and we have inserted one of the Friedmann equations in its form $H(z)=H_0\sqrt{\Omega_{\rm mat}(z) + \Omega_{\rm rad}(z)+ \Omega_\Lambda}$, where $H_0 = 100\,h\,\mathrm{km}\,\mathrm{sec}^{-1}\,\mathrm{Mpc}^{-1}$ is the Hubble constant today and $\Omega_*(z)=\rho_*(z)/\rho_c^0$, with $\rho_c^0 = 3 H_0^2/(8 \pi G)$ the critical energy density today, $\rho_*(z)$ denoting the energy densities associated with radiation $\rho_{\rm rad}(z)$, matter $\rho_{\rm mat}(z)$, and the cosmological constant $\rho_\Lambda \equiv \Lambda /(8\pi G)$.
 
Since we are interested in sources operating far in the radiation era, the term proportional to $\Omega_{\rm rad}(z)$ dominates. In order to rewrite Eq.~(\ref{ellp}), we first need to make a short digression. It is an excellent approximation to treat the expansion of the universe as adiabatic, so that it is governed until today by the conservation of entropy per comoving volume\footnote{The total entropy of the universe is very large and dominated by the relativistic species: extra entropy production due to known decoupling processes is sub-dominant with respect to the total entropy.} \cite{DurrerBook},
\be
\label{conss}
g_S(T)\,T^3\,a^3(t) = \mathrm{constant}
\ee
where $T$ is the photon temperature at time $t$, and $g_S$ is the effective number of entropic degrees of freedom at that time~\cite{Kolb:1990vq}. As the universe cools down, $g_S$ decreases when some species become non-relativistic. When this occurs, they release their entropy to the species that are still in thermal equilibrium, causing the temperature to decrease as $T \propto a^{-1}\,g_S^{-1/3}$, i.e. slower than the usual $T \propto a^{-1}$. The amount the universe has expanded between GW production at a temperature $T_p$, and today, is then characterized by the ratio
\be
\label{a0ap}
\frac{a_0}{a_p} = \left(\frac{g_S(T_p)}{g_S(T_0)}\right)^{1/3} \, \left(\frac{T_p}{T_0}\right) \simeq
1.25 \times 10^{13} \, \left(\frac{g_S(T_p)}{100}\right)^{1/3} \, \left(\frac{T_p}{\mathrm{GeV}}\right)\,,
\ee
where we have used $T_0 \simeq 2.35\times 10^{-13}$ GeV for the photon temperature today~\cite{Olive:2016xmw} and $g_S(T_0) \simeq 3.91$ for the Standard Model degrees of freedom with three light neutrino species~\cite{Kolb:1990vq}. Note that $g_S(T_0)$ must be evaluated in this calculation as if the neutrinos were still relativistic today. This is because they decouple from the thermal plasma when they are relativistic (at $T \sim$ MeV, while $m_{\nu} < 2$ eV \cite{Olive:2016xmw}) and do not release their entropy to the photons when they later become non-relativistic. In the Standard Model, the last decrease of $g_S$ occurs when electrons and positrons become non-relativistic (at $T \sim m_e \simeq 0.5$ MeV), and the photon temperature evolves simply as
$T \propto a^{-1}$ afterwards. In the radiation era, the total energy density is given by \cite{Kolb:1990vq}
\be
\label{rhorad}
\rho_{\mathrm{rad}} = \frac{\pi^2}{30}\,g_*(T)\,T^4\,,
\ee
where $g_*(T)$ is the effective number of relativistic degrees of freedom at temperature $T$. Combining Eqs.~\eqref{a0ap} and \eqref{rhorad}, one obtains
\begin{equation}\label{Omrad}
           \Omega_{\rm rad}(T)=\Omega_{\rm rad}^0 \left(\frac{g_S(T_0)}{g_S(T)}\right)^{4/3} \left(\frac{g_*(T)}{g_*(T_0)}\right) \left(\frac{a_0}{a}\right)^4 \,,
\end{equation}
where $\Omega_{\rm rad}^0= h^{-2} \, 2.47\times 10^{-5}$ is the radiation energy density today, and $g_*(T_0)=2$. Note that for numerical estimates we take $h=0.67$ \cite{Ade:2013zuv}. Saturating inequality \eqref{ellp}, i.e.~setting $\ell_p\equiv  H_p^{-1}$, and keeping only the dominant term during the radiation era, Eqs.~(\ref{a0ap}) and \eqref{Omrad} lead to (in the last equality we set $g_*(T_p) = g_S(T_p)$ for $T_p > 0.1$ MeV \cite{Kolb:1990vq})
\bea
\frac{\ell_p^0}{H_0^{-1}} & \simeq &  \frac{a_0/a_p}{\sqrt{\Omega_{\rm rad}(T_p)}}  =  \frac{1}{\sqrt{\Omega_{\rm rad}}} \left(\frac{g_S(T_p)}{g_S(T_0)} \right)^{1/3} \sqrt{\frac{g_*(T_0)}{g_*(T_p)}}\,\frac{T_0}{T_p} \nn \\
&\simeq & 1.3\times 10^{-11}\left(\frac{100}{g_*(T_p)}\right)^{1/6}\, \left(\frac{\rm GeV}{T_p}\right) \label{eq:number_horizons}\,,
\eea
which clearly shows that the correlation scale today of a GW signal from the early universe is tiny comparable to the present Hubble scale. 
 
The number of uncorrelated regions from which we are receiving today independent GW signals, can be found from calculating the angle $\Theta_p$ subtending the size $\ell_p$ at $z_p$:
\begin{equation}\label{eq:Thetap}
\Theta_p=\frac{\ell_p}{d_A(z_p)}     \,,
\end{equation}
where $d_A(z_p)$ is the angular diameter distance
\begin{equation}\label{eq:dA}
           d_A(z_p)=\frac{1}{H_0(1+z_p)}\int_0^{z_p} \frac{dz'}{\sqrt{\Omega_{\rm mat}(z') + \Omega_{\rm rad}(z')+ \Omega_\Lambda}}\,.
\end{equation}
In total, today one has access to $\sim d_A(z_p)^2/ \ell_p^2=\Theta_p^{-2}$ uncorrelated regions. 
Let us consider, as an example of a GW source, the electroweak (EW) phase transition at $T_{\rm EW} \sim \mathcal{O}(10^2)$ GeV, for which we can take $g_S(T_{\rm EW}) \sim 100$. The redshifted scale today corresponding to the 
horizon scale at the EW phase transition is $(a_0/a_{\rm EW})H_{\rm EW}^{-1}\simeq 2.7\times 10^{-4}$ pc. 
Inserting $h^2\Omega_{\rm mat}=0.12$ and $\Omega_\Lambda=1-\Omega_{\rm mat}$ in \Eqref{eq:dA}, from \Eqref{eq:Thetap} one gets $\Theta_{\rm EW}\simeq 2 \times 10^{-12}$ deg, meaning that the GW signal (as received today on Earth) due to a causal process operating at the EW epoch, is composed by the superposition of independent signals emitted by at least $ \sim 10^{24}$ uncorrelated regions (even more, if the inequality $\ell_{\rm EW}\leq H_{\rm EW}^{-1}$ is not saturated)\footnote{It is important to note that, even though these regions are not in causal contact, and hence are uncorrelated, the phase transition is happening everywhere at the same time, because the temperature is the same everywhere, as the the universe is homogeneous and isotropic.  Inflation is the leading mechanism to provide the right initial conditions for this to happen, see Sect.~\ref{sec:inf}.}. This indicates that the GW signal can only be described statistically. Thus, a GW signal from the early universe cannot possibly be resolved beyond its stochastic nature. In order to resolve individual realizations of the signal, a GW detector should have an angular resolution as good as $\Theta_p$. However, as shown for the example of a GW signal produced at the time of the EW transition, $\Theta_{\rm EW}$ corresponds to a tiny resolution unreachable by any realistic GW detector. 

It appears therefore entirely justified to consider the GW signal from sources operating in the early universe, as a stochastic background. Furthermore, since any signal is composed by the superposition of sources operating in regions that are not in causal contact during the GW generation, but in which the same physical process is taking place, it is completely justified to invoke the ergodic hypothesis, and assume that a spatial average corresponds to an ensemble average.
 
In fact, the above considerations apply even to much lower energy scales than the one of the EW transition. For instance, at the epoch of photon decoupling, at redshift $z_{\rm dec} \simeq 1090$, we have $\Theta_{\rm dec}\simeq 0.9$ deg. For the sake of the exercise, let us postulate (quite optimistically) the existence of a stochastic background with a high enough amplitude so that a future GW detector could observe it, say with an angular resolution of about 10 deg. The redshift at which the background should have been generated, in order to be correlated on an angular scale of 10 deg, should be, according to Eq.~\eqref{eq:Thetap}, $z_p\simeq 17$. Consequently, the property of stochasticity holds for any GW signal sourced until well into the matter dominated era.
 
One may ask whether the signal could be resolved in time, instead of in terms of its characteristic length-scale. The correlation time-scale of the GW signal is again given by $\Delta t_p\leq H_p^{-1}$. Saturating the inequality, for the EW phase transition at $T_{\rm EW}\sim 100$ GeV, one finds a time interval today $\Delta t_{\rm EW}^{(0)}\simeq 8$ hours, while for the QCD phase transition at $T_{\rm QCD}\sim 200$ MeV, it is $\Delta t_{\rm QCD}^{(0)}\simeq 9$ months. The correlation time of primordial sources looks therefore reasonable from a point of view of observational time. However, to resolve the signal, one would need a detector capable of pointing in the same direction within the above calculated angular size $\Theta_p$, for a time interval corresponding to $\Delta t_p^{(0)}$. This is clearly not possible, again because of the limited resolution of GW detectors.
 
Note that the above arguments remain valid also for causal GW sources that are not localised in time at a given moment $t_p$, but are continuously operating during several Hubble times. The paradigmatic example of this are topological defects, which we will introduce in section~\ref{sec:CosmicDefects}. For example, a network of cosmic strings, emits GWs continuously, all the way since the epoch of the phase transition that produced it, until today. As we will explain in section \ref{sec:CosmicDefects}, the GW signal in this case is the sum of two components. One is the irreducible component, given by GWs that are produced around the horizon at each time $t$, by the anisotropic stress of the network (presented in section \ref{sec:IrreduclibleGWCD}). The second contribution, is the superposition of the emission of GWs from sub-horizon cosmic string loops, at each time $t$ (presented in section \ref{sec:GWcosmicStringLoops}). The GW signal from these two components is a stochastic background, contributed by the superposition of the many horizons that at every moment, fit (redshifted) within today's horizon. Therefore, if we observe it today, it cannot be resolved beyond its stochastic nature, for the same reasons discussed above. As we shall see, the main difference between this continuously sourced background and the one arising from a source localised at a time $t_p$, is that the former extends over many frequencies, precisely due to the source operating during many Hubble times.
 
On the other hand, during inflation the causal horizon grows exponentially~\cite{DurrerBook}, and the above arguments do not apply. In this case, the reason why the inflationary GW signal is a stochastic background, resides in the intrinsic quantum nature of the generating process. As we will see in detail in section \ref{sec:inf}, the source of this GW background are quantum vacuum fluctuations of the metric during inflation. The tensor metric perturbations are therefore random variables with random phases. They become effectively classical, as the universe expands and the wave-numbers of the fluctuations become larger than the Hubble scale during inflation, leading to very large occupation numbers, see discussion e.g.~in \cite{Allen:1996vm} and \cite{lectures_lesgourgues,Senatore:2016aui}. This quantum-to-classical transition renders the metric perturbation, of quantum origin, equivalent to a stochastic variable after Hubble crossing. The perturbations re-enter progressively the Hubble radius during the radiation and matter dominated eras, leading to a GW signal which is intrinsically stochastic.
 
In general, the stochastic GW background from sources in the early universe is assumed to be {\it statistically homogeneous and isotropic}, {\it unpolarised} and {\it Gaussian}. The reasons behind these assumptions are easily understood, as we will explain next (for a thorough discussion see also \cite{Allen:1996vm}).
 
Statistical homogeneity and isotropy is inherited from the same property of the FLRW universe, be it during inflation or afterwards during the thermal era. It implies that the two-point spatial correlation function satisfies
\begin{equation}
           \langle h_{ij}(\mathbf{x}, \eta_1) \, h_{lm}(\mathbf{y}, \eta_2) \rangle = \xi_{ijlm}(|\mathbf{x}-\mathbf{y}|, \eta_1,\eta_2)\,,\label{h_two_point}
\end{equation}
where $h_{ij}(\mathbf{x}, \eta_1)$ is the tensor perturbation of Eq.~\eqref{GWconf}, and $\langle ... \rangle$ denotes the ensemble average (that becomes an average over volume/time under the ergodic hypothesis). In the case presented before about a phase transition operating during the radiation dominated era for example, even though the GW signal is given by the superposition of the signals emitted from many uncorrelated regions, the (statistical) homogeneity and isotropy of the universe causes these regions to have, essentially, the same characteristics, e.g.~the temperature and particle densities. Therefore the phase transition happens everywhere in the universe at the same time and with the same outcome, so that the produced GW background is statistically homogeneous and isotropic. The same holds for the irreducible GW background generated during inflation, because the tensor metric perturbations representing the GWs, are generated over the homogeneous and isotropic FLRW background.
 
The GW cosmological backgrounds are assumed typically to be unpolarised, as a consequence of the absence of a significant source of parity violation in the universe. If the process sourcing the GWs is based on interactions that are symmetric under parity, the outcome is a GW background for which the two polarisations $+,\,\times$, are uncorrelated. In terms of the Fourier amplitudes of Eq.~(\ref{hrketa}), this means $\langle h_+({\mathbf k},\eta)h_\times({\mathbf k},\eta)\rangle=0$. The connection with the parity symmetry is made more explicit by introducing the helicity basis $\epsilon_i^{\pm}(\hat{\mathbf k})=(\hat {\mathbf m}\pm i\,\hat{\mathbf n})_i/\sqrt{2}$, where $\hat {\mathbf m},\,\hat {\mathbf n}$ are the unit vectors used in Eq.~(\ref{eij}). Out of the usual $e_{ij}^{+,\times}$ polarisation tensors defined in Eq.~(\ref{eij}), one can construct a basis for the transverse-traceless tensor space representing the two independent helicity states $\pm 2$: $e_{ij}^{\pm 2}=(e_{ij}^{+}\pm i\,e_{ij}^{\times})/2$. The basis transforms as $e_{ij}'^{\pm 2}=e^{\pm 2i\theta}e_{ij}^{\pm 2}$ under rotation by an angle $\theta$ around the $\hat{\mathbf k}$ axis (see e.g. \cite{Rubakovbook,Hu:1997hp}). An arbitrary symmetric rank two transverse-traceless tensor is in general a mixture of both helicity states, and can be expressed as a linear combination in this basis, $h_{ij}=h_{+2}\,e_{ij}^{+2}+h_{-2}\,e_{ij}^{-2}$. Using these definitions, one can easily derive that $\langle h_{+2}({\mathbf k},\eta)h_{+2}({\mathbf k},\eta) - h_{-2}({\mathbf k},\eta)h_{-2}({\mathbf k},\eta)\rangle = \langle h_+({\mathbf k},\eta)h_\times({\mathbf k},\eta)\rangle = 0$, where the last equality holds if the background is unpolarised (see e.g. \cite{Caprini:2003vc}).
The absence of a net polarisation is therefore equivalent to the condition that the two independent helicity modes are produced, on the average, with the same amplitude, i.e.~with identical expectation values. If this is not the case, the GW background can be chiral and must arise from some parity-violating source. We will present an example of such chiral background in section~\ref{sec:SustainedPartProdInf}.
 
Gaussianity also follows straightforwardly in most cases of GW backgrounds formed by the emission of many uncorrelated regions. As discussed above, since the signal is composed by a large number of sources that were independent at the moment of the GW emission, by the central limit theorem one can expect the outcome signal given by the superposition of all independent signals, to have a Gaussian distribution \cite{Allen:1996vm}. Gaussianity also applies in the case of the irreducible background generated during inflation, again because of the quantum nature of this background: in the simplest scenarios, the tensor metric perturbation can be quantised as a free field, and hence with Gaussian probability distribution for the amplitudes\footnote{In reality, there is always a small degree of deviation from gaussianity in the inflationary perturbations, as they are created over a dynamical quasi-de Sitter background that also evolves (even if slowly) during inflation~\cite{Maldacena:2002vr}. In practice, the amount of non-gaussianity is `slow-roll suppressed'.}.
 
Note that, although the properties of statistical homogeneity and isotropy, gaussianity, and absence of net polarisation, are satisfied to a good approximation for most cosmological sources, there can be exceptions. For example, a certain level of large-scale anisotropy in the universe is allowed by present CMB constraints \cite{Ade:2013vbw}. A typical example of GW source from the early universe, leading to a statistically anisotropic GW background, is the excitation of a gauge field during inflation (although this has been studied mainly for the scalar mode, see e.g.~\cite{Bartolo:2012sd,Bartolo:2014hwa,Naruko:2014bxa}). The GW background generated by gauge field dynamics during inflation is also non-Gaussian, since the GW source is quadratic in the fields~\cite{Cook:2013xea}, and it can be polarised if the interaction between the gauge field and the inflaton is parity-breaking~\cite{Sorbo:2011rz,Anber:2012du} (in this case the inflaton is a pseudo-scalar). We will discuss precisely this later in Section~\ref{sec:SustainedPartProdInf}.
 
\subsection{Characterization of a stochastic gravitational wave background}
\label{sec:stochback2}
 
In the following we introduce different quantities that are used to characterize the power spectrum of a stochastic GW background. The Fourier amplitudes $h_r(\mathbf{k}, \eta)$ of Eq.~(\ref{hrketa}), are considered to be random variables. For a statistically homogeneous and isotropic, unpolarised and Gaussian GW background, their power spectrum can be written as
\be
\label{powerspec}
\langle h_r(\mathbf{k}, \eta) \, h^*_{p}(\mathbf{q}, \eta) \rangle = \frac{8\pi^5}{k^3} \,
\delta^{(3)}(\mathbf{k} - \mathbf{q}) \, \delta_{r p} \, h_c^2(k, \eta)\,,
\ee
where $h_c$ is dimensionless, real and depends only on the time $\eta$ and the comoving wave-number $k = |\mathbf{k}|$. The delta function in $\mathbf{k},\,\mathbf{q}$, and the fact that $h_c$ does not depend on the direction $\mathbf{\hat k}$, are consequences of statistical homogeneity and isotropy; the delta function in the polarisation states $r,\,p$ is a consequence of the absence of a net polarisation, and gaussianity implies that the above expectation value contains all the relevant information on the statistical distribution of the random variables $h_r(\mathbf{k}, \eta)$. We do not need to investigate therefore higher-point correlation functions, as for a Gaussian field even-point correlation functions can be rewritten in as powers of $h_c^2(k, \eta)$, while odd-point correlations are simply vanishing. The factor $8\pi^5$ in \Eqref{powerspec} has been chosen so that Eqs.~(\ref{hrketa}) and (\ref{orthonormal}) give
\be
\label{hcketa}
\langle h_{ij}(\mathbf{x}, \eta) \, h_{ij}(\mathbf{x}, \eta) \rangle = 2 \, \int_0^{+\infty} \frac{dk}{k} \, h_c^2(k, \eta)\,,
\ee
where the factor $2$ in the RHS is a convention motivated by the fact that the LHS involves contributions from two independent polarizations (we adopt here the same convention as \cite{MaggioreBook,Maggiore:1999vm}, while the one adopted in \cite{Romano:2016dpx} differs by a factor $2$). It appears from the above equation that $h_c(k, \eta)$ represents a characteristic GW amplitude per logarithmic wave-number interval and per polarization state, at a time $\eta$.
 
As discussed at the beginning of subsection \ref{sec:stochback1}, for free waves at sub-Hubble scales (those detectable today), the average in the LHS of Eq.~(\ref{powerspec}) can be taken both as a volume average over sufficiently large regions compared to the GW wavelengths, and a time average over several periods of oscillation (i.e.~the average under which GWs can be defined, following what presented in subsection \ref{subsec:GWcurved}). The time behaviour of the GW Fourier amplitudes for sub-Hubble modes is given by \Eqref{subHub}. Inserting this solution into \Eqref{powerspec}, and keeping in mind that the presence of the delta function imposes $k=q$, one can average out the terms that are oscillatory in time, and find
\begin{equation}
           \langle h_r(\mathbf{k}, \eta) \, h^*_{p}(\mathbf{q}, \eta) \rangle = \frac{1}{a^2(\eta)}[\langle A_r(\mathbf{k}) \, A^*_{p}(\mathbf{q})  \rangle+ \langle B_r(\mathbf{k}) \, B^*_{p}(\mathbf{q})\rangle]\,.
           \label{eq:h_sub_hub_AB}
\end{equation}
The above equation, together with \Eqref{powerspec}, shows that for free waves inside the Hubble radius, $h_c(k, \eta) \propto 1/a(\eta)$ after the oscillatory terms are averaged out.
 
Besides $h_c$, a quantity of prime interest to characterize a stochastic GW background, is the spectrum of GW energy density per logarithmic wave-number interval, $d \rho_{\rm GW} / d \mathrm{log} k$. The energy density in GWs is given by the 00-component of the energy-momentum tensor, see Eq.~\eqref{eq:GWendensTT}
\be
\label{rhogw}
\rho_{\rm GW} \, = \, \frac{\langle \dot{h}_{ij}(\mathbf{x}, t) \, \dot{h}_{ij}(\mathbf{x}, t) \rangle}{32 \pi G} \, = \, 
\frac{\langle h'_{ij}(\mathbf{x}, \eta) \, h'_{ij}(\mathbf{x}, \eta) \rangle}{32 \pi G \, a^2(\eta)} \, = \, 
\int_0^{+\infty} \frac{dk}{k}\,\frac{d \rho_{\rm GW}}{d \mathrm{log} k}\,,
\ee
where in the second equality we have converted the derivatives with respect to the physical time $t$ into derivatives with respect to the conformal time $\eta$, while the third equality defines
$d \rho_{\rm GW} / d \mathrm{log} k$. Again, we have seen from the discussion in subsection \ref{subsec:GWcurved} (c.f.~also what presented in \cite{Misner:1974qy,MaggioreBook}) that, even for a deterministic GW signal, the energy-momentum tensor of GWs cannot be localized inside a volume smaller than the GW typical wavelength, but can only be defined by performing an average over volume and/or time (over several wavelengths and/or frequencies for its Fourier components). For a stochastic background generated in the early universe, invoking the ergodic hypothesis, the average performed in Eq.~(\ref{rhogw}) corresponds to the usual ensemble average of Eq.~(\ref{powerspec}).
 
An expression for the GW energy density power spectrum $d \rho_{\rm GW} / d \mathrm{log} k$ valid for free waves inside the Hubble radius, can be found from \Eqref{rhogw}, inserting the time behaviour of the GW Fourier modes given by Eq.~(\ref{subHub}). The first step is to postulate that the same structure of Eq.~\eqref{powerspec} holds for the power spectrum of the conformal time derivatives of the Fourier modes $h_r(\mathbf{k},\eta)$,
\be
\label{powerspecprime}
\langle h_r'(\mathbf{k}, \eta) \, {h'_{p}}^*(\mathbf{q}, \eta) \rangle = \frac{8\pi^5}{k^3} \,
\delta^{(3)}(\mathbf{k} - \mathbf{q}) \, \delta_{r p} \, {h'_c}^{2}(k, \eta)\,,
\ee
where we have defined a new characteristic amplitude ${h'_c}^{2}(k, \eta)$, analogous to $h_c^{2}(k, \eta)$. One then substitutes solution (\ref{subHub}) in the above equation. Again, because of the delta function imposing $k=q$, it is straightforward to average out the terms that oscillate in time (as we did to derive \Eqref{eq:h_sub_hub_AB}). Besides, one can neglect the $\mathcal{H}^2$ term arising due to the conformal time derivative of \eqref{subHub}, with respect to the term $k^2$, since in the case under analysis, $k\gg \mathcal{H}$. One then finds a simple relation among the amplitudes:
\begin{equation}
           {h'_c}^{2}(k, \eta) \simeq k^2 \,h_c^{2}(k, \eta) \,.
\end{equation}
With this identity, one can evaluate \Eqref{rhogw} with the help of Eqs.~(\ref{hrketa}),~(\ref{orthonormal}) and (\ref{powerspecprime}), to find
\be
\label{drhodlogk}
\frac{d \rho_{\rm GW}}{d \mathrm{log} k} = \frac{k^2\,h_c^2(k, \eta)}{16 \pi G \, a^2(\eta)} \, .
\ee
Furthermore, we have seen before that $h_c(k, \eta) \propto 1/a(\eta)$ for sub-Hubble modes. Thus, as expected for massless degrees of freedom, the GW energy density is diluted as radiation with the expansion of the universe, $\rho_{\rm GW} \propto a^{-4}$.
 
In order to make connection with observations, it is necessary to evaluate the GW background today in terms of the present-day physical frequency $f = k / (2 \pi \, a_0)$, corresponding to the comoving wave-number $k$ redshifted to today (we remind that a subscript ``$0$'' indicates a quantity evaluated at the present time). The characteristic GW amplitude per logarithmic frequency interval today, is then given by
\be
\label{hcf}
h_c(f) = h_c(k, \eta_0)\,,
\ee
which corresponds to the definition given in Ref.~\cite{Maggiore:1999vm}. A stochastic background is often characterized also by its spectral density
\be
\label{Shf}
S_h(f) = \frac{h_c^2(f)}{2 f} \, ,
\ee
which has dimension $\mathrm{Hz}^{-1}$. This quantity is directly comparable to the noise in a detector, parametrised by $S_n(f)$. We will use the spectral density in section \ref{sec:interferometers}, when discussing the sensitivity of interferometric experiments to stochastic backgrounds. 
 
The spectrum of GW energy density per logarithmic frequency interval, can be conveniently normalized as
\be
\label{Omegagw}
\Omega_{\rm GW}(f) = \frac{1}{\rho_c}\,\frac{d \rho_{\rm GW}}{d \mathrm{log} f}\,,
\ee
where $\rho_c =3 H^2/(8 \pi G)$ is the critical energy density at time $t$. The quantity traditionally considered by cosmologists is $h^2\,\Omega_{\rm GW}^{(0)}$, because it is independent of the observational uncertainty on the value of $H_0$. Eqs.~(\ref{drhodlogk} - \ref{Shf}) with $f = k / (2 \pi \, a_0)$ give (note the factor two difference with respect to e.g. \cite{Romano:2016dpx})
\be
\Omega_{\rm GW}^{(0)}(f) = \frac{4 \pi^2}{3 H_0^2} \, f^3 \, S_h(f) \, .
\label{OmandSh}
\ee
Inserting $H_0$, in terms of the dimensionless amplitude $h_c = \sqrt{2 f\,S_h}$, we have
\bea
\label{Omandhc}
S_h(f) = 7.98 \times 10^{-37} \, \left(\frac{\mathrm{Hz}}{f}\right)^3 \, h^2\,\Omega_{\rm GW}(f)\,\frac{1}{\rm Hz}\,, \\
h_c(f) = 1.26 \times 10^{-18} \, \left(\frac{\mathrm{Hz}}{f}\right) \, \sqrt{h^2\,\Omega_{\rm GW}(f)} \, .
\eea
 
It is important to notice that, when performing a GW direct detection experiment, the expansion of the universe is completely negligible on the time scales of interest (with the exception of very particular cases, see e.g. \cite{Bonvin:2016qxr}). This allows us to make the connection between the quantities introduced above, in particular expansion \eqref{hrketa}, and the expansion of $h_{ij}(\mathbf{x},t)$ often used in the literature, for example in Refs.~\cite{MaggioreBook,Maggiore:1999vm,Romano:2016dpx}:
\be
\label{Exphij}
h_{ij}(\mathbf{x},t) = \sum_{r = + , \times} \int_{-\infty}^{+\infty} df\,\int d^2\mathbf{\hat{k}}~\bar{h}_{r}(f, \mathbf{\hat{k}})\,e^{i\,2\pi\,f (t - \mathbf{\hat{k}} \cdot \mathbf{x})}\,
e_{ij}^{r}(\mathbf{\hat{k}})\,,
\ee
where the integration over negative frequencies is obtained via the definition $\bar{h}_{r}(-f, \mathbf{\hat{k}})\equiv \bar{h}_{r}^*(f, \mathbf{\hat{k}})$ that is necessary for $h_{ij}(\mathbf{x},t)$ to be real.
To connect the above equation to expansion \eqref{hrketa}, the starting point is \eqref{eq:plane_wave_A} for sub-Hubble GWs, those of interest for detectors. The scale factor must be fixed to the scale factor today, while neglecting the expansion of the universe means that it is possible to perform a time Fourier Transform:
the time in the exponentials of \eqref{eq:plane_wave_A} becomes therefore the Fourier conjugate variable to the frequency $f$. One can then rewrite \Eqref{eq:plane_wave_A} in terms of frequency via the change of variable $d^3\mathbf{k}=(2\pi\,a_0)^3\,f^2\,df\,d\mathbf{\hat k}$, and extend to negative frequencies by identifying ${A}_{r}(-f, \mathbf{\hat{k}})\equiv {A}_{r}^*(f, \mathbf{\hat{k}})$, to get
\be
\label{ExphijA}
h_{ij}(\mathbf{x},t) = a_0^2 \sum_{r = + , \times} \int_{-\infty}^{+\infty} f^2\, df\,\int d^2\mathbf{\hat{k}}\,
A_{r}(f, \mathbf{\hat{k}})\,e^{i\,2\pi\,f (t - \mathbf{\hat{k}} \cdot \mathbf{x})}\,
e_{ij}^{r}(\mathbf{\hat{k}})\,.
\ee
If the expansion of the universe can be neglected, \Eqref{eq:plane_wave_A} becomes therefore equivalent to \Eqref{Exphij}, provided one identifies $\bar{h}_{r}(f, \mathbf{\hat{k}})=a_0^2\,f^2\,A_{r}(f, \mathbf{\hat{k}})$ (the same can be done with $B_{r}(f, \mathbf{\hat{k}})$).
 
At this point, the usual expression for the power spectrum of the Fourier amplitudes $\bar{h}_{r}(f, \mathbf{\hat{k}})$ given in Refs.~\cite{MaggioreBook,Maggiore:1999vm,Romano:2016dpx} and defining the power spectral density can be easily recovered. First of all, one needs to observe that \Eqref{eq:h_sub_hub_AB} can also be written as (by using \Eqref{eq:plane_wave_B})
\begin{equation}
           \langle h_r(\mathbf{k}, \eta) \, h^*_{p}(\mathbf{q}, \eta) \rangle = \frac{2}{a^2(\eta)}\langle A_r(\mathbf{k}) \, A^*_{p}(\mathbf{q})  \rangle\,.
\end{equation}
One obtains therefore
\begin{eqnarray}
\langle \bar{h}_{r}(f, \mathbf{\hat{k}}) \bar{h}_{p}^*(g, \mathbf{\hat{q}})\rangle &=& a_0^4\,f^2\,g^2 \langle A_r(\mathbf{k}) \, A^*_{p}(\mathbf{q})  \rangle = \nonumber\\
&=&\frac{1}{8\pi} \,
\delta(f - g) \, \delta^{(2)}(\mathbf{\hat{k}} - \mathbf{\hat{q}}) \, \delta_{r p} \, S_h(f)\,,
\end{eqnarray}
where the second equality has been obtained inserting \Eqref{powerspec}, changing variable to $f=k/(2\pi \,a_0)$ and using definition \eqref{Shf}. From the above equation, one can appreciate that the definition of the spectral density given in \cite{MaggioreBook,Maggiore:1999vm,Romano:2016dpx} is equivalent to the one of \Eqref{powerspec}.

\subsection{Propagation of gravitational waves through cosmic history}
\label{redshift}
 
As mentioned in section \ref{sec:intro}, the weakness of the gravitational interaction guarantees that GWs are decoupled from the rest of the universe since the Planck scale. Furthermore, because of the symmetries of FLRW space-time, for most practical purposes there is no need to go beyond linear order in perturbation theory (c.f. \Eqref{GWcosmo}). One can therefore neglect both interactions with ordinary matter and self-interactions, and assume that sub-Hubble GWs propagate freely once they have been produced (or once they re-enter the Hubble radius, in the case of an inflationary produced signal). Consequently Eq.~(\ref{subHub}) applies for sub-Hubble modes as soon as the GW source has stopped operating. As demonstrated in the above sub-section, in this case the GW energy density spectrum redshifts with the expansion of the universe, but it retains its initial shape. In fact, the GW energy density is diluted as radiation, $\rho_{\rm GW} \propto a^{-4}$, while the GW physical wave-numbers evolve simply as $k / a$. Normalizing the GW energy density at the time of production to the total energy density in the universe at that time $\rho_p$ (the subscript $p$ indicates that a quantity is evaluated at the time of GW production), the GW spectrum today is given by
\be
\label{spec0}
h^2\,\Omega_{\rm GW} (k)= \frac{h^2}{\rho_c}\,\left(\frac{a_p}{a_0}\right)^4\,\rho_p\,\left(\frac{1}{\rho}\,\frac{d \rho_{\rm GW}}{d \mathrm{log} k}\right)_p \hspace*{1cm}
\mbox{for } k \gg  \mathcal{H} \, .
\ee
Furthermore, normalizing the physical wave-number at the time of GW production to the Hubble rate at that time,
\be
\label{xk}
x_k = \frac{k/a_p}{H_p}
\ee
the corresponding GW frequency today is given by
\be
\label{f0}
f = \frac{1}{2 \pi}\,\frac{k}{a_0} = \frac{x_k}{2 \pi}\,\frac{a_p}{a_0}\,H_p\,.
\ee
As discussed in sub-section \ref{sec:stochback1}, the GW signal from a source operating at some time $t_p$ in the early universe cannot be correlated on length/time scales larger than $H_p^{-1}$. In general, the typical correlation scale shows up as a peak (or a feature) in the power spectrum: consequently, we expect the GW energy density spectrum to be peaked at a characteristic wave-number smaller than the Hubble radius at the time of production of the GWs, i.e. we expect its characteristic wave-number to be $k/a_p\geq H_p$. Thus, although the value $x_k$ in Eq.~(\ref{xk}) for the peak of the spectrum depends on the particular source under consideration, we can conclude that it certainly satisfies $x_k \geq 1$. Examples of such short-lasting sources include first-order phase transitions (discussed in section \ref{sec:FOPT}) and preheating after inflation (discussed in section~\ref{sec:PreheatingAndOthers}). Other sources like e.g.~cosmic strings (discussed in section \ref{sec:CosmicDefects}), on the other hand, produce GWs continually during a long period of time. In this case the source is active for a large range of values of $a_p\,H_p$ in Eq.~(\ref{f0}), and the resulting GW spectrum today covers a wide frequency range. Similarly, the GW spectrum produced by inflation is very broad because $a_p \, H_p$ varies exponentially during ({\it de Sitter}) inflation. Note also that, if a source produces GWs continually during the radiation era with an energy density $\rho_{\rm GW}$ that remains a constant fraction of the total energy density $\rho$ at the time of production, then the resulting spectrum (\ref{spec0}) is approximately flat because $a_p^4\,\rho_p$ is approximatively constant during the radiation era (up to variations of the number of relativistic species discussed in the previous sub-section). This is in fact a typical situation for long-lasting sources (including inflation for modes that re-enter the Hubble radius during the radiation era, as we will see in section \ref{sec:inf}).
 
For GWs produced during the radiation era, Eqs.~(\ref{spec0}) and (\ref{f0}) can be rewritten with the help of (\ref{a0ap}) and \eqref{rhorad} to give the present-day GW amplitude and GW frequency in terms of the temperature $T_p$ at the time of production:
\be \label{eq:Om_today}
h^2 \, \Omega_{\rm GW} = 1.6 \times 10^{-5} \, \left(\frac{100}{g_*(T_p)}\right)^{1/3} \, \left(\frac{1}{\rho}\,\frac{d \rho_{\rm GW}}{d \mathrm{log} k}\right)_p
\ee
and
\be
\label{fTp}
f = 2.6 \times 10^{-8}  \, \mathrm{Hz}\,\,\, x_k \, \left(\frac{g_*(T_p)}{100}\right)^{1/6} \, \frac{T_p}{\mathrm{GeV}}
\ee
where we have used $H^2 = 8 \pi G \rho / 3$ and again $g_*(T_p) = g_S(T_p)$ for $T_p \gsim$ MeV \cite{Kolb:1990vq}.
 
\begin{figure}[t]
\centering
\includegraphics[width=12cm]{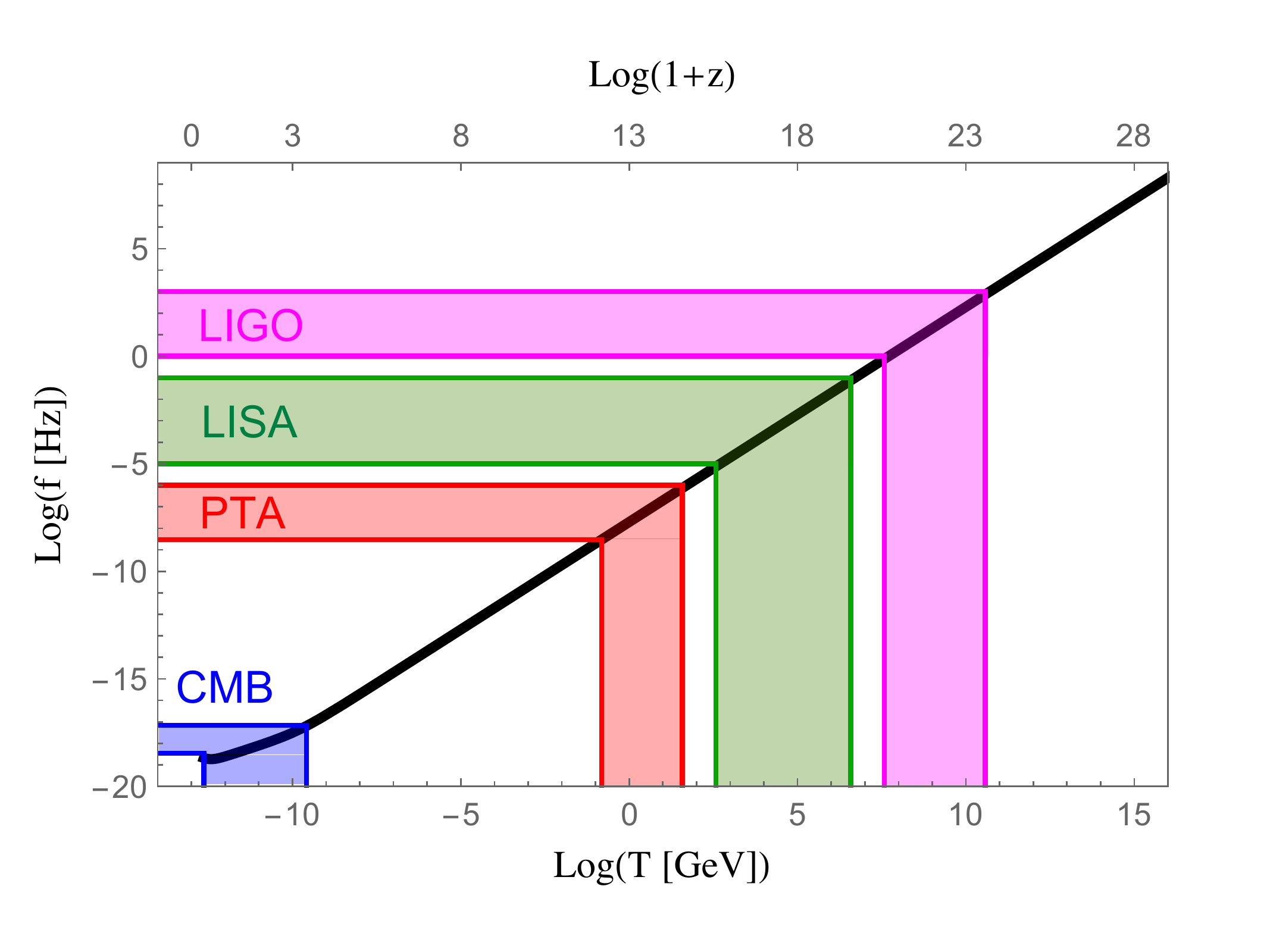}
\vspace*{8pt}
\caption{Black line:~the characteristic GW frequency of Eq.~(\ref{f0}) as a function of temperature (the corresponding redshift is shown above). Shaded regions: the frequency ranges detectable by several GW experiments, from right to left respectively $1\,{\rm Hz}\lesssim f\lesssim 10^3\,{\rm Hz}$ for ground-based interferometers, $10^{-5}\,{\rm Hz}\lesssim f\lesssim 0.1\,{\rm Hz}$ for LISA, $3\times 10^{-9}\,{\rm Hz}\lesssim f\lesssim 10^{-6}\,{\rm Hz}$ for Pulsar Timing Arrays, and $3.4\times 10^{-19}\,{\rm Hz}\lesssim f\lesssim 7\times 10^{-18}\,{\rm Hz}$ for the CMB.
\label{evol}}
\end{figure}

Eq.~(\ref{f0}), and its analogue in the radiation era Eq.~\eqref{fTp}, provide an interesting connection between the frequency of the GW today and the epoch in the early universe when a GW source was operating. The precise value of $x_k$ can only be determined within a specific GW generation process; however, since $x_k\geq 1$, one can still find, through these equations, the lowest possible frequency emitted by a process operating at a given time in the universe parametrised by $T_p$. Therefore, it is possible to associate to a given GW detection experiment, operating in a given frequency range, the epochs in the early universe during which a source should have been active to produce GWs detectable by that experiment. This is shown in Fig.~\ref{evol} for several GW detectors (for details, see section \ref{sec:interferometers}): $1\,{\rm Hz}\lesssim f\lesssim 10^3\,{\rm Hz}$ for ground-based interferometers, $10^{-5}\,{\rm Hz}\lesssim f\lesssim 0.1\,{\rm Hz}$ for LISA, $3\times 10^{-9}\,{\rm Hz}\lesssim f\lesssim 10^{-6}\,{\rm Hz}$ for Pulsar Timing Arrays, and $3.4\times 10^{-19}\,{\rm Hz}\lesssim f\lesssim 7\times 10^{-18}\,{\rm Hz}$ for the CMB. In this last case, the observable frequency window corresponds respectively to the Horizon today and at the epoch of photon decoupling, c.f. section \ref{sec:CMB}: $H_0/(2\pi)\leq f\leq H_{\rm dec}\,(a_0/a_{\rm dec})/(2\pi)$, with $T_{\rm dec}\simeq 0.26$ eV. Note that Eq.~(\ref{fTp}) does not hold in the case of the CMB, which extends beyond the radiation era, while Eq.~(\ref{f0}) is generically valid. Fig.~\ref{evol} illustrates how GW experiments have the potential to probe well separated energy scales and cosmological epochs that are not directly accessible by any other mean, since the universe was opaque to photons at that time. Other cosmological probes like e.g. Large Scale Structures, the CMB and Big Bang Nucleosynthesis (BBN) can probe temperatures $T_p\lesssim 1$ MeV (or the inflationary epoch for what concerns CMB anisotropies and polarisation), while GW experiments have in principle access to a wide range of energy scales beyond 1 MeV.

\subsection{Gravitational wave spectrum by a generic stochastic source}
\label{sec:spectrum_generic}
 
In section \ref{gweom} we have presented the free solutions for the GW Fourier modes $h_r(\mathbf{k},\eta)$ at sub- and super-horizon scales. Here we derive the amplitude of the Fourier modes and the energy density power spectrum Eq.~\eqref{drhodlogk} in the presence of a generic stochastic source of GW acting during the radiation or matter dominated eras\footnote{The solution for a source that acts continuously through the radiation-matter transition requires numerical integration and goes beyond the illustrative purposes of this sub-section.}.
 
The tensor anisotropic stress sourcing the GWs can be decomposed, analogously to $h_{ij}$, in two polarisation states (here and in the following we omit the superscript TT for brevity):
\be
\label{eq:PIr}
\Pi_{ij}(\mathbf{x}, t) = \sum_{r = + , \times} \, \int \frac{d^3 \mathbf{k}}{(2 \pi)^{3}} \, \Pi_r(\mathbf{k},t) \,
e^{- i \mathbf{k} \cdot \mathbf{x}} \, e_{ij}^{r}(\mathbf{\hat{k}})\,.
\ee
For the reasons put forward in section \ref{sec:stochback1}, it is enough to describe the source stochastically. Though there are exceptions (as we will see in the following of this review), here we assume that the properties of statistical homogeneity and isotropy, gaussianity and the absence of a preferred polarisation, apply for the GW source as well, as assumed for the GW spectrum (c.f. section \ref{sec:stochback1}). The power spectrum of the Fourier components of the tensor anisotropic stress can therefore be written as
\begin{equation}\label{eq:PIspec}
                      \langle \Pi_r(\mathbf{k}, \eta) \, \Pi_{p}^*(\mathbf{q}, \zeta) \rangle = \frac{(2\pi)^3}{4} \,
\delta^{(3)}(\mathbf{k} - \mathbf{q}) \, \delta_{r p} \, \Pi(k, \eta, \zeta)\,.
\end{equation}
Note the chosen the normalisation and that we have written the correlator at unequal times for future convenience.
 
In terms of the Fourier amplitudes of $H_{ij}=a\,h_{ij}$, the evolution equation \eqref{gweq2} reads
\be
\label{gweq2Fou}
H_{r}''(\mathbf{k}, \eta) + \left(k^2 - \frac{a''}{a}\right)\,H_{r}(\mathbf{k}, \eta) =
16\pi G \, a^3\,\Pi_{r}(\mathbf{k}, \eta)\,.
\ee
In the radiation dominated era with $a(\eta)=a_*\eta$, and in terms of the dimensionless variable $x=k\eta$, the above equation becomes simply
\begin{equation}
           \frac{d^2H_r^{\rm rad}(\mathbf{k},x)}{dx^2}+H_r^{\rm rad}(\mathbf{k},x)=\frac{16\pi G a_*^3}{k^5} \, x^3 \,\Pi_r(\mathbf{k},x)\,,
           \label{eq:GWeomRDwithSources}
\end{equation}
which has $\lbrace \sin x,\cos x\rbrace$ as homogeneous solutions and $\mathcal{G}(x, y)=\sin(x-y)$ as the Green function associated to the differential operator in the $lhs$ of Eq.~(\ref{eq:GWeomRDwithSources}). Let us suppose that the source starts operating at a time $x_{\rm in}=k\eta_{\rm in}$ and stops operating at $x_{\rm fin}=k\eta_{\rm fin}$, with $\eta_{\rm fin}$ still well into the radiation dominated era. While the source is active $x< x_{\rm fin}$, one has therefore (assuming vanishing initial conditions)
\begin{equation}\label{eq:Hradsource}
           H_r^{\rm rad}(\mathbf{k},x<x_{\rm fin})=\frac{16\pi G a_*^3}{k^5}\int_{x_{\rm in}}^x dy\, y^3\,\sin(x-y)\, \Pi_r(\mathbf{k},y)\,.
\end{equation}
We are interested in the GW spectrum today: we need therefore the solution at $x_0=k\eta_0\gg x_{\rm fin}$. For this, one has to match \eqref{eq:Hradsource} with the homogeneous solution
\be
H_r^{\rm rad}(\mathbf{k},x>x_{\rm fin})=A_r^{\rm rad}(\mathbf{k})\cos x+B_r^{\rm rad}(\mathbf{k})\sin x
\ee and its first derivative, to find the matching coefficients $A_r^{\rm rad},B_r^{\rm rad}$. This procedure leads to
\begin{eqnarray}
           A_r^{\rm rad}(\mathbf{k})&=& \frac{16\pi G a_*^3}{k^5}\int_{x_{\rm in}}^{x_{\rm fin}} dy\, y^3\,\sin(-y)\, \Pi_r(\mathbf{k},y), \\
           B_r^{\rm rad}(\mathbf{k})&=& \frac{16\pi G a_*^3}{k^5}\int_{x_{\rm in}}^{x_{\rm fin}} dy\, y^3\,\cos(y)\, \Pi_r(\mathbf{k},y).
\end{eqnarray}
One can now apply Eq.~(\ref{eq:h_sub_hub_AB}) together with \Eqref{powerspec} and \Eqref{eq:PIspec} to find the power spectrum amplitude today due to a GW source acting in the radiation era:
\begin{equation}
           h_c^2(k, \eta_0)|_{\rm rad}=64\,\frac{G^2}{a_0^2}\,\frac{a_*^6}{k^7}\int_{x_{\rm in}}^{x_{\rm fin}} dy\, y^3\int_{x_{\rm in}}^{x_{\rm fin}} dz\, z^3\,\cos(y-z)\, \Pi(k,y,z)\,.
\end{equation}
It is now straightforward to get the energy density power spectrum by applying Eq.~\eqref{drhodlogk}:
\begin{eqnarray}\label{eq:rhogwPiRad}
\lefteqn{\left. \frac{d \rho_{\rm GW}}{d \mathrm{log} k}(k,\eta_0) \right|_{\rm rad}=}\nonumber\\
&&\hspace{1cm}=\frac{4}{\pi} \,\frac{G}{a_0^4}\, k^3 \int_{\eta_{\rm in}}^{\eta_{\rm fin}} d\eta\, a^3(\eta)\int_{\eta_{\rm in}}^{\eta_{\rm fin}} d\zeta\, a^3(\zeta)\,\cos[k(\eta-\zeta)]\, \Pi(k,\eta,\zeta)\,,
\end{eqnarray}
where we have rewritten the integrals in terms of conformal time. The above equation represents the energy density power spectrum for a generic stochastic source operating during the radiation dominated era.
 
If the source is operating during the matter dominated era with $a(\eta)=a_*\eta^2$, the solution of the GW equation
\begin{equation}
           \frac{d^2H_r^{\rm mat}(\mathbf{k},x)}{dx^2}+\left(1-\frac{2}{x^2}\right)H_r^{\rm mat}(\mathbf{k},x)=\frac{16\pi G a_*^3}{k^8} \, x^6 \,\Pi_r(\mathbf{k},x)\,,
\end{equation}
after the source has ceased acting becomes
\begin{equation}\label{eq:Hrsolmat}
           H_r^{\rm mat}(\mathbf{k},x>x_{\rm fin})=A_r^{\rm mat}(\mathbf{k})\left(\frac{\cos x}{x}+\sin x\right)+B_r^{\rm mat}(\mathbf{k})\left(\frac{\sin x}{x}-\cos x\right)\,,
\end{equation}
with matching coefficients
\begin{eqnarray}
           A_r^{\rm mat}(\mathbf{k})&=&  \frac{16\pi G a_*^3}{k^8}\int_{x_{\rm in}}^{x_{\rm fin}} dy\, y^5\,[y\cos y - \sin(y)]\, \Pi_r(\mathbf{k},y), \\
           B_r^{\rm mat}(\mathbf{k})&= & \frac{16\pi G a_*^3}{k^8}\int_{x_{\rm in}}^{x_{\rm fin}} dy\, y^5\,[\cos y+y\sin y]\, \Pi_r(\mathbf{k},y).
\end{eqnarray}
Note that, since we are interested in wave-numbers satisfying $x_0\gg 1$, the terms proportional to $x_0^{-1}$ are sub-dominant and can be dropped in Eq.~\eqref{eq:Hrsolmat}, which then takes the same form as the sub-Hubble solution \Eqref{subHub}. We can adopt the same procedure as for the radiation case and find the GW energy density power spectrum for a source operating during the matter era:
\begin{eqnarray}
&&\hspace{-1.2cm}\left. \frac{d \rho_{\rm GW}}{d \mathrm{log} k}(k,\eta_0) \right|_{\rm mat}=\frac{4}{\pi} \,\frac{G}{a_0^4}\, k \,a_* \int_{\eta_{\rm in}}^{\eta_{\rm fin}} d\eta\, a^{5/2}(\eta)\int_{\eta_{\rm in}}^{\eta_{\rm fin}} d\zeta\, a^{5/2}(\zeta)\,\Pi(k,\eta,\zeta)\nonumber\\
&&\hspace{2.2cm}\times [(1+k^2\eta\zeta)\cos(k(\eta-\zeta))+(k\eta-k\zeta)\sin(k(\eta-\zeta))]\,.
\end{eqnarray}


\section{Bounds and detectors}
\label{sec:bounds}

In this section we review the present observational constraints on stochastic backgrounds of GWs (Sects.~\ref{sec:BBN}-\ref{sec:PTA}), and describe the basic features of current and future GW direct detection experiments (Sect.~\ref{sec:interferometers}). In Section~\ref{sec:BBN} we discuss how  BBN (Big Bang Nucleosynthesis) and the CMB can be used to set upper bounds on the total energy density of a cosmological background of GWs. In section \ref{sec:CMB} we discuss the imprint on the CMB temperature and polarisation anisotropies from a stochastic GW background. We present the current upper bound on the spectrum of a stochastic GW background at the CMB scales, as inferred from current CMB measurements. In section \ref{sec:PTA} we review the ability of pulsar timing arrays (PTA) to probe a stochastic GW background, and discuss the present and future upper bound(s) that can be placed using this technique. In section~\ref{sec:interferometers} we turn to the direct detection of a stochastic GW background via interferometers. We review the basic principles for the detection of a stochastic GW background using interferometry, and briefly survey the present and future Earth- and space-based interferometric detectors. 

\subsection{Constraints on the gravitational wave background energy density}
\label{sec:BBN}

As discussed in Section~\ref{sec:general}, the energy density of a GW background decays with the expansion of the universe as relativistic degrees of freedom, i.e.~$\rho_{\rm GW}\propto a^{-4}$. This means that a GW background acts as an additional radiation field in the universe\footnote{Here we implicitly assume that the characteristic wavelengths of the GW background are well inside the horizon.}, contributing to the background expansion rate as 
\begin{equation}\label{eq:HBBN}
	H^2(a)=H_0^2~\left[\left(\frac{\rho_{\rm GW}^{0}}{\rho_c^{0}} + \Omega^{0}_{\rm rad}\right) \left(\frac{a_0}{a}\right)^4 +\Omega^{0}_{\rm mat} \left(\frac{a_0}{a}\right)^3 + \Omega^{0}_\Lambda \right]\,.
\end{equation}
Any observable capable of probing the background evolution of the universe (and hence its energy content), has therefore the potential ability to constrain the integrated GW energy density $\rho_{\rm GW}/{\rho_c} = (\rho_{\rm GW}^{0}/\rho_c^{0})(H_o/H(a))^2 ({a_0}/{a})^4$ present in that moment. In particular, two events that yield a very precise measurement of the expansion rate of the universe are Big Bang Nucleosynthesis (BBN), and the process of photon decoupling leading to the CMB. An upper bound on the energy density of a GW background present at the time of BBN and CMB decoupling, can therefore be derived from the constraint on the amount of radiation tolerated at those cosmic epochs (when the Universe had a temperature of $T_{\rm BBN} \sim 0.1$ MeV and $T_{\rm CMB} \sim 0.3$ eV, respectively). Using Eq.~\eqref{a0ap}, and Eq.~\eqref{rhorad} for the photon component $\rho_\gamma(T)=(\pi^2/15)\,T^4$, we can write
\begin{equation}\label{eq:OmgwforBBN}
\left(\frac{h^2\rho_{\rm GW}}{\rho_c}\right)_0 = h^2\Omega_{\gamma}^{0}\left(\frac{g_S(T_0)}{g_S(T)}\right)^{4/3}\frac{\rho_{_{\rm GW}}(T)}{\rho_\gamma(T)}\,,
\end{equation}
 where $h^2\Omega_{\gamma}^0=2.47\times 10^{-5}$ is the density parameter of photons today. The GW energy density $\rho_{{\rm GW}}(T)$ must not exceed the limits on the abundance of radiation during BBN and CMB decoupling. A constraint on the presence of `extra' radiation is usually expressed in terms of an effective number of neutrinos species $N_{\nu}$, as follows. The radiation energy density in the universe is given by Eq.~(\ref{rhorad})
\begin{equation}
\label{eq:rhorad_compl}
	\rho_{\rm rad} = \frac{\pi^2}{30}g_{*}(T)T^4 \equiv \frac{\pi^2}{30}\left[{\sum_b}~g_{*b}\left(\frac{T_b}{T}\right)^4 +{7\over8}\,{\sum_f}~g_{*f}\left(\frac{T_f}{T}\right)^4 \right]\,T^4\,,
\end{equation}
where we have made explicit the contributions to $g_*(T)$: the sum $\sum_b$ is over bosonic species and the sum $\sum_f$ over fermionic species. Before electron-positron annihilation, at $T \sim$ MeV, the total number of relativistic degrees of freedom was
\be\label{eq:gstarMeV}
g_*(T = \mathrm{MeV}) = 2 + \frac{7}{8}\,\left(4 + 2 \, N_{\nu}\right)
\ee
where the first term is due to the photons (with two helicity states), the second one is due to the electrons and positrons (with two helicity states each), and the last term is due to the $N_{\nu}$ species of neutrinos and anti-neutrinos (with one helicity state each). In the Standard Model, $N_{\nu} = 3$. From Eqs.~(\ref{eq:rhorad_compl}) and \eqref{eq:gstarMeV}, we see that an extra amount of radiation can be put in the form of $\Delta N_{\nu}$ extra neutrino species as
\begin{equation}
 	\Delta \rho_\mathrm{rad} =  \frac{\pi^2}{30}\, \frac{7}{4}\, \Delta N_{\nu} \, T^4\,. 
 \end{equation} 
Thus, an upper bound on any extra radiation component in addition to those of the Standard Model, can be seen as an upper bound on $\Delta N_{\nu}$. Since the energy density in GW must satisfy 
$\rho_{\rm GW}(T) \leq \Delta \rho_\mathrm{rad}(T)$, one has
\be
\label{rhogwrhogammaBBN}
\left(\frac{\rho_\mathrm{GW}}{\rho_{\gamma}}\right)_{T = \mathrm{MeV}} \leq \frac{7}{8}\,\Delta N_{\nu} \, .
\ee 
Inserting the above equation into \eqref{eq:OmgwforBBN}, one finds a constraint on the GW energy density redshifted to today in terms of the number of extra neutrino species,
\begin{eqnarray}
\label{ConsRhoBBN}
\left(\frac{h^2\rho_{\rm GW}}{\rho_c}\right)_0 \leq  h^2\Omega_{\gamma}^{0}\left(\frac{g_S(T_0)}{g_S(T)}\right)^{4/3} \frac{7}{8}\,\Delta N_{\nu}= 5.6 \times 10^{-6} 
\Delta N_{\nu}\,,
\end{eqnarray}
where we have inserted $h^2\Omega_{\gamma}^{0}=2.47\times 10^{-5}$, $g_S(T = \mathrm{MeV}) = 10.75$ and $g_S(T_0) \simeq 3.91$~\cite{Kolb:1990vq}, see discussion below Eq.~(\ref{a0ap}). 

Before proceeding to discuss how BBN and CMB can limit $\Delta N_\nu$, let us note that the bound in Eq.~\eqref{ConsRhoBBN} is often quoted in terms of $N_{\rm eff}$, the effective number of neutrino species present in the thermal bath after $e^+e^-$ annihilation. In the Standard Model $N_{\rm eff}=3.046$, differing slightly from $N_\nu = 3$ simply due to the fact that neutrinos are not yet completely decoupled when $e^+e^-$ annihilation takes place, and hence they are partially 'reheated' along with the rest of the plasma~\cite{Mangano:2005cc}. Instead of normalising~\eqref{rhogwrhogammaBBN} at $T=$MeV we can choose a temperature below $e^+e^-$ annihilation. In this case we have to use
\be\label{eq:gstarepem}
g_*(T < T_{e^+e^-}) = 2 + \frac{7}{4}\,N_{\rm eff} \left(\frac{4}{11}\right)^{4/3}\,,
\ee
since, after the photons have been reheated by $e^+e^-$ annihilation, the temperature of the neutrinos satisfies $T_\nu=(4/11)^{1/3}\,T$. Carrying on the same procedure as before, leads to 
\begin{equation}\label{ConsRhoBBNII}
	\left(\frac{h^2\rho_{\rm GW}}{\rho_c}\right)_0 \leq h^2\Omega_{\gamma}^0 ~\frac{7}{8}\,\left(\frac{4}{11}\right)^{4/3}\Delta N_{\rm eff} = 5.6 \times 10^{-6} \, 
\Delta N_{\rm eff} \,,
\end{equation}
equivalent to Eq.~\eqref{ConsRhoBBN}. 

The above bound applies on the integrated energy density, defined in Eq. (\ref{Omegagw}), as
\begin{equation}
	\left(\frac{h^2\rho_{\rm GW}}{\rho_c}\right)_0 = \int \frac{df}{f} \, h^2 \, \Omega_{\rm GW}(f)\,.
\end{equation}
Except in the extreme case of a GW spectrum with a very narrow peak of width $\Delta f \ll f$, the above bound can be interpreted as a bound on the amplitude of a GW spectrum, $h^2 \, \Omega_{\rm GW}(f) \lsim 5.6 \times 10^{-6} \Delta N_\nu$, over a wide frequency range. This of course only applies to GWs with characteristic wavelength well inside the horizon at the time (or slightly before) when the constraint on $\Delta N_\nu$ is established; otherwise, if the wavelength is super-horizon, the tensor mode does not propagate as a wave and hence it cannot affect the expansion rate of the universe. As shown in section \ref{sec:general}, the comoving scale entering the horizon at time $t_p$ is $k=H_{p}\,a_{p}$, corresponding to a frequency $f=(H_{p}/2\pi)(a_p/a_0)$. 
Furthermore, the bounds in Eq.~\eqref{ConsRhoBBN} (equivalently \eqref{ConsRhoBBNII}) obviously applies only to GW backgrounds produced before the physical mechanism (BBN or CMB decoupling) considered to infer the constraint on $N_\nu$ ($N_{\rm eff}$), takes place. 

Let us now turn to actual constraints on $N_\nu$ ($N_{\rm eff}$). We first discuss BBN, which can be used to place an upper limit on $N_\nu$ through the predictions of the primordial abundances of light elements. For a review, see e.g.~Ref.~\cite{Iocco:2008va}. BBN successfully predicts the primordial abundances of $^2\mathrm{H}$ (Deuterium), $^3\mathrm{He}$, $^4\mathrm{He}$, in very good agreement with measurements from the CMB and astrophysical observations\footnote{The predictions for $^6\mathrm{Li}$ and $^7\mathrm{Li}$ are in contrast with observations \cite{Anders:2014ija,Cyburt:2008kw}, and this remains an open problem today.}~\cite{Cyburt:2004yc}. In particular, the abundance of $^4\mathrm{He}$ is very sensitive to the expansion rate $H(T)$ at the beginning of BBN ($T \sim$ MeV), and therefore to the total amount of radiation at that time, including a possible GW background (Eq.~\ref{eq:HBBN}). 
The expansion rate in fact controls the relative abundance of neutrons and protons $n_n / n_p$. At $T \gsim$ MeV, neutrons and protons are kept in thermal equilibrium by weak processes like 
$p + e^- \leftrightarrow n + \nu_e$, simply thanks to their interaction rate $\Gamma_{\mathrm{weak}} \sim G_F^2\,T^5$, which is larger than the expansion rate $H(T)$. However, as $T$ decreases, the interaction rate falls below $H(T)$, and the interactions freeze-out at the temperature $T_f\simeq 0.7$ MeV, for which $\Gamma_{\mathrm{weak}}(T_f) = H(T_f)$. The neutron-to-proton ratio then freezes at the value $n_n/n_p \sim e^{-Q / T_f}$, where $Q = m_n - m_p$ is the difference between the neutron and proton masses, whereas the exponential (Boltzmann) suppression comes from the fact that the protons and neutrons are non-relativistic. Until the actual onset of BBN, $n_n$ is reduced only by neutron decays. Even though $T_f$ is smaller than the Deuterium biding energy (2.2 MeV), the huge amount of photons present in the early universe delays the Deuterium synthesis by photo-dissociation, until $T_N\simeq 0.1$ MeV, the temperature at which BBN truly starts: the Deuterium formation in fact initiates a chain of nuclear processes leading to heavy nuclei production, especially $^4\mathrm{He}$. Since practically all the available neutrons eventually form $^4\mathrm{He}$, its abundance depends directly on $e^{-Q / T_f}$, and on the baryon-to-photon ratio $\eta_B = n_B/n_\gamma=6\times 10^{-10}$ \cite{Ade:2015xua}. An extra radiation component parametrised by $N_\nu$ increases the Hubble rate, leading to a larger freeze-out temperature $T_f$, and therefore to more neutrons and hence to a larger abundance of $^4\mathrm{He}$. 

The latest constraints on $N_\nu$ by BBN can be found in Ref.~\cite{Cyburt:2004yc}. From $^4\mathrm{He}$ measurements only, the constraint is $N_\nu<4$, quite loose because of the strong degeneracy with the baryon to photon ratio $\eta_B$. This improves considerably if considered in combination with the Deuterium abundance, and quite strong bounds can be put on both $N_\nu$ and $\eta_B$. However, the best determination of both parameters is obtained adding the CMB data, as the latter yield a very good measurement of $\eta_B$. In particular Ref.~\cite{Cyburt:2004yc} finds $N_\nu<3.2$ at 95\% confidence level. Eq.~(\ref{ConsRhoBBN}) then gives 
$(h^2 \, \rho_{\rm GW} / \rho_c)_0 < 1.12 \times 10^{-6}$. As mentioned above, this constraint applies only to GW backgrounds produced before BBN, and not to stochastic backgrounds e.g.~from astrophysical sources.
Furthermore, it applies only to GWs that were inside the Hubble radius at the time of BBN, which corresponds to present-day frequencies $f \geq 1.5 \times 10^{-12}$ Hz (we have used Eq.~(\ref{fTp}) with $x_k = 1$ and $T = 0.1 \,{\rm MeV}$). 

The CMB constitutes as well a very precise measurement of the radiation energy density, and can be used therefore to infer an upper bound on extra radiation components parametrised by $N_\nu$, and in turn on the presence of a GW background. The effects that an extra radiation component can have on the CMB are multiple, see e.g.~\cite{Rossi:2014nea}. At the background level, it alters the redshift of matter-radiation equality and of photon decoupling: this leads to a change of all angular scales, shifting the position and amplitude of the CMB acoustic peaks, as well as of the baryon acoustic oscillations observed in galaxy catalogues; it also leads to less growth of the perturbations inside the Hubble radius, affecting the matter power spectrum. For a representation of the effects on the CMB, see Fig.~1 of \cite{Sendra:2012wh}. 

The implications of this for GWs have been first analysed in Ref.~\cite{Smith:2006nka}, using a combined dataset including WMAP, ACBAR, CBI, VSA and BOOMERnaG for the CMB, and 2dF and SLOAN galaxy surveys and Lyman-$\alpha$ forest for the matter structure. Two cases were identified, depending on the initial conditions. In the first case, labelled as the `adiabatic initial condition', the GW background is assumed to be alike a gas of free-streaming neutrinos, so that there are perturbations imprinted on its energy density, following the same distribution as all other components in the universe. The constraint in this case is $ (h^2 \, \rho_{\rm GW} / \rho_c)_0  \leq 3.9 \times 10^{-5}$. In the second case, labelled as the `homogeneous initial condition', the GW background is not perturbed and the curvature perturbation is the one of the standard adiabatic case (therefore it would vanish in the limit of a universe made exclusively by GWs). We view this second option for initial conditions as more justified, since it applies to all known mechanisms of generation of a GW background\footnote{In most circumstances the energy density of GWs is not perturbed at first order in cosmological perturbation theory, though there are exceptions as in certain cases of preheating, see~\cite{Bethke:2013aba,Bethke:2013vca} for a discussion on this. Furthermore, GWs are not expected to be produced by the decay of the inflation field in the same manner as all other matter/radiation fields in the universe, so there is really no clear justification to assume they have the same adiabatic initial conditions that apply to neutrinos.} (from the irreducible GW background due to quantum fluctuations during inflation, to all active sources operating during or after inflation in the early Universe, see sections \ref{sec:inf}-\ref{sec:CosmicDefects}). In the case of homogeneous initial conditions, Ref.~\cite{Smith:2006nka} finds a stronger bound as $(h^2 \, \rho_{\rm GW} / \rho_c)_0   \leq 6.9 \times 10^{-6}$, since all degeneracy with the neutrino parameters (number of species, sum of the masses) is broken. 

Note that this bound extends over a wider range of frequencies than the analogous one from the BBN one, as the frequency corresponding to the comoving scale entering the horizon during decoupling is $f=(H_{\rm dec}/2\pi)(a_{\rm dec}/a_0)\simeq 7\times 10^{-18}$ Hz. More realistically, Ref.~\cite{Smith:2006nka} considers that the the tensor modes will have to oscillate for a while, once inside the horizon, before they can be fully considered as GWs (and hence as a radiation component), in order to provide the effect under analysis. Therefore Ref.~\cite{Smith:2006nka} tentatively sets the lowest frequency for which the bound applies to $f> 10^{-15}$ Hz. 

The analysis of \cite{Smith:2006nka} has been redone more recently by \cite{Sendra:2012wh,Pagano:2015hma}. In particular, the latests analysis of Ref.~\cite{Pagano:2015hma} uses Planck data, together with CMB lensing, Baryon Acoustic Oscillations and also Deuterium abundances, and finds a constraint that goes down to $(h^2 \, \rho_{\rm GW} / \rho_c)_0  \leq 1.2 \times 10^{-6}$. Not surprisingly, this is comparable to what is obtained from the BBN analysis by~\cite{Cyburt:2004yc}, which also uses WMAP data to pin down the baryon to photon ratio $\eta_B$. However, Ref.~\cite{Pagano:2015hma} only analyses adiabatic initial conditions. From the results of Refs.~\cite{Smith:2006nka,Sendra:2012wh}, one can infer that the gain obtained imposing homogeneous initial conditions, due to the breaking of degeneracies with neutrino parameters, is of a factor of a few, of the order of $\sim 5$. Using this, we can tentatively estimate the constraints that could be put by Planck and other actual cosmological data in the hypothesis of GW with homogeneous initial conditions, as $(h^2 \, \rho_{\rm GW} / \rho_c)_0  \lesssim 2 \times 10^{-7}$. 

The bounds presented above apply to any GW background produced before BBN or before CMB decoupling, as those we will discuss in the following sections. They are not relevant however for the irreducible background of GWs expected from inflation, that we will introduce in section~\ref{sec:PrimTensInf}. The inflationary background of GWs is in fact constrained much better (down to many orders of magnitude lower in amplitude) by another kind of constraint, formerly know as the `COBE bound', which we describe in the next section\footnote{As we will see, the `COBE' bound attains only lower frequencies (corresponding to CMB scales), but assuming the standard almost-scale invariant form of the inflationary spectrum (c.f.~section \ref{sec:inf}), the corresponding bound on $h^2 \, \rho_{\rm GW} / \rho_c  $ at CMB scales can be extrapolated to much higher frequencies (corresponding to scales much smaller than the CMB scales).}.

\subsection{Constraints from Cosmic Microwave Background anisotropies}
\label{sec:CMB}

As the CMB is our best probe of the homogeneity and isotropy of the universe, it can also be used to place constraints on the amplitude of metric perturbations over the FLRW metric, in particular in the presence of a stochastic background of GWs\footnote{Let us recall here that tensor perturbations can only be properly interpreted as a background of GWs for modes well inside the horizon, see discussions in Sect.~\ref{sec:GWdef}. The imprint of tensor perturbations on the CMB, similarly as in the case of scalar perturbations, is however due to both super- and sub-horizon modes at the CMB time. Thus, we should be really speaking about stochastic tensor modes, rather than of a stochastic background of GWs. We will nonetheless allow ourselves such abuse of language, as it is customary when discussing the effect of tensor perturbations on the CMB.}. A stochastic background of GWs induces fluctuations in both the temperature and polarisation of the CMB, as the GWs affect the space-time (with its characteristic quadrupolar pattern) through which the photons propagate. The measurement of the CMB temperature and polarization angular power spectra at large angular scales, can be used therefore to constrain the amplitude and spectral index of a stochastic background of GWs at the very low frequencies $f=k/(2\pi a_0)$, comprised between the Hubble parameter today and the Hubble parameter at matter-radiation equality, i.e. $H_0a_0< k < H_{\rm eq}a_{\rm eq}$ (for a universe dominated by radiation and matter only, $H_{\rm eq}a_{\rm eq}=\sqrt{2}\,a_0\,H_0\,\Omega_{\rm mat}/\sqrt{\Omega_{\rm rad}}$, c.f. Eq.~\eqref{eq:keq}). 

The constraint that can be inferred from the measurement of CMB fluctuations holds, in principle, irrespective of the origin of the stochastic GW background. It is however only relevant, in practice, for mechanisms capable of producing a GW spectrum with a non-negligible amplitude on scales around the Hubble scale at the epoch of CMB decoupling (both sub- and super-horizon scales). These mechanisms include an early epoch of accelerated expansion, inflation (sect.~\ref{sec:PrimTensInf}), or its alternatives like pre-big bang and ekpyrotic scenarios (sect.~\ref{sec:InfAltern}). They also include active seeds like cosmic defects, which continuously source metric perturbations around the horizon scale at every moment of cosmic evolution (sect.~\ref{sec:IrreduclibleGWCD}). On the contrary, typical GW generation mechanisms from the early Universe, operating for a short amount of time and within the causal horizon (e.g.~a phase transition), cannot be probed well by the CMB fluctuations. As we will see for example in section \ref{sec:FOPT}, short-time mechanisms lead generically to a SGWB energy density spectrum scaling as $\Omega_{\rm GW}(k) \propto k^3$, on scales $k<H_{\rm in}a_{\rm in}$ larger than the size of the horizon at the time $\eta_{\rm in}$ of production. These mechanisms are typically operative after inflation but well before Nucleosynthesis. Their characteristic time and length scale, of the order of $1/(H_{\rm in}a_{\rm in})$ at most, is therefore much smaller than CMB scales, which span around the size of the horizon at photon decoupling $c_s(\eta_{\rm dec}) /(H_{\rm dec}a_{\rm dec})$. 

A stochastic background of GW induces temperature fluctuations in the CMB through the Sachs Wolfe effect \cite{Sachs:1967er,Starobinsky:1985ww} as
\begin{equation}
\frac{\Delta T}{T}=-\int_i^f {h}'_{jl}({\bf x},\eta) n^j n^l d\lambda\,,
\label{SW_COBE}
\end{equation}
where $\lambda$ is an affine parameter along the photon geodesic with tangent vector ${\bf n}$, ${h}'_{jl}({\bf x},\eta)$ is the time derivative of the tensor metric perturbation and $i$ and $f$ denote the initial (e.g.~photon decoupling) and final (e.g. today) times. Note that the above temperature fluctuation, due to presence of the tensor mode, has a quadrupolar pattern. Around the recombination time, photons perform Thomson scattering with the electrons of the primordial plasma. If the incident radiation on the electron is not isotropic, the resulting scattered light becomes polarised. Consequently, the quadrupole anisotropy in the photon distribution, due to the presence of tensor modes, induces a net polarisation in the Thomson-scattered photons. The polarization pattern is then maintained until today, as the photons travel freely since they performed their last scattering~\cite{Kosowsky:1994cy,Bond:1984fp, Bond:1987ub}. 

This polarisation signal in the CMB is generated only at last scattering, contrary to the temperature anisotropies sourced continuously up to today as in Eq.~\eqref{SW_COBE}. Customarily, CMB polarisation is decomposed into two polarisation patterns that have the advantage of being independent of the reference frame: E and B polarisation modes. The E-mode is sourced by all scalar, vector and tensor metric perturbations, while the B-mode is only sourced by vector and tensor perturbations, but not by scalar perturbations. As in the dominant paradigm of the early Universe, inflation, vector perturbations died away during the accelerated expansion, a B-mode polarization represents a unique imprint of the presence of primordial inflationary GWs. Unfortunately, more important sources of B-polarisation than the primordial tensor perturbations are gravitational lensing and galactic foregrounds, such as galactic dust and synchrotron emission. This poses a serious challenge for CMB detectors aiming at measuring the tensor primordial spectrum. While lensing can be distinguished in the angular CMB spectrum by the characteristic shape it produces, and by the fact that it becomes important only at relatively high multipoles, distinguishing galactic foregrounds is more challenging\footnote{In 2014, BICEP2 announced the detection of B-mode polarization of primordial origin \cite{Ade:2014xna}, before a joint analysis using Planck data proved that it was in reality polarised emission from galactic dust~\cite{Ade:2015tva}, see footnote in Sect.~\ref{sec:PrimTensInf}.}. Envisaged solutions are to measure several CMB frequencies (in order to distinguish the CMB black-body spectrum in frequency from the foreground ones), observations (from the ground) of regions of the sky poorly contaminated by foreground emissions, and full-sky observations with space satellites. 

The contribution of tensor modes to the CMB temperature and polarisation anisotropy angular power spectra today can be written in a very compact form using the formalism of~\cite{Hu:1997hp},
\begin{eqnarray}\label{eq:CMBspec}
	\hspace*{-1cm} C_{\ell, T}^{XY}= \frac{2}{\pi}\frac{1}{(2\ell+1)^2} \int dk\,k^2 [X^2_\ell(k,\eta_0)Y^{2*}_\ell(k,\eta_0)+X^{-2}_\ell(k,\eta_0)Y^{-2*}_\ell(k,\eta_0)]
\end{eqnarray}
where $T$ stands for tensor, and $X$ and $Y$ can be, respectively, the temperature $\Theta^{\pm 2}_\ell$, or the polarization $E^{\pm 2}_\ell$ or $B^{\pm 2}_\ell$, anisotropies today. These are here expanded using the normal modes defined in Eqs.~(10-11) of \cite{Hu:1997hp}, which combine the angular dependence from the spherical harmonic decomposition (adopted since temperature and polarisation transform, respectively, as a scalar and a spin two field, on the surface of the sky) with the one of a plane wave decomposition (in terms of which the spatial dependence of the temperature and polarisation fields is expanded, and which accounts for the photon propagation after CMB decoupling). Note that, in the formalism of~\cite{Hu:1997hp}, tensors are expanded in the helicity basis (see Eq.~(34) of~\cite{Hu:1997hp}, from which we borrow the normalisation, and also Sect.~\ref{sec:general})
\begin{equation}
	e^{\pm 2}_{ij}=-\sqrt\frac{3}{8}\,(\hat m\pm i \,\hat n)_i\times (\hat m\pm i \,\hat n)_j
\end{equation}
where $\hat m$ and $\hat n$ are the unit vectors of Eq.~\eqref{eij}. In terms of the usual polarisation tensors defined in section~\ref{sec:general} one has $e_{ij}^{\pm 2}=-\sqrt{3/8}\,(e_{ij}^{+}\pm i\,e_{ij}^{\times})$. The coefficients of the tensor-induced temperature $\Theta^{\pm 2}_\ell$ and $E^{\pm 2}_\ell$, $B^{\pm 2}_\ell$ polarisation anisotropies today, are given by time integrals of the sources as~\cite{Hu:1997hp}
\begin{eqnarray}
\frac{\Theta^{\pm 2}_\ell(k,\eta_0)}{2\ell+1}&=& -\frac{2}{3}\int_0^{\eta_0} d\eta\,e^{-\tau}\, (h^{\pm 2}(k,\eta))' j_\ell^{(2\,2)}(k(\eta_0-\eta))\,,\label{eq:Thetaell}\\
\frac{E^{\pm 2}_\ell(k,\eta_0)}{2\ell+1}&=& -\frac{\sqrt{6}}{10}\int_0^{\eta_0} d\eta\, \tau' e^{-\tau}\,[\Theta_2^{\pm 2}-\sqrt{6}E_2^{\pm 2}] \,\epsilon_\ell^{\pm 2}(k(\eta_0-\eta))\,,\label{eq:Eell}\\
\frac{B^{\pm 2}_\ell(k,\eta_0)}{2\ell+1}&=& -\frac{\sqrt{6}}{10}\int_0^{\eta_0} d\eta\, \tau' e^{-\tau}\,[\Theta_2^{\pm 2}-\sqrt{6}E_2^{\pm 2}] \,\beta_\ell^{\pm 2}(k(\eta_0-\eta))\,,\label{eq:Bell}
\end{eqnarray}
where $\tau(\eta)$ is the optical depth between time $\eta$ and today, whereas the radial functions $j_\ell^{(2\,2)}$, $\epsilon_\ell^{\pm 2}$ and $\beta_\ell^{\pm 2}$ represent how the total angular power is transferred into the $\ell$ modes, and can be found in Eqs.~(15), (17), (18) of \cite{Hu:1997hp}. Eq.~\eqref{eq:Thetaell} follows directly from \eqref{SW_COBE} and shows that the source of temperature anisotropy is the time derivative of the tensor perturbations. Moreover, the factor $e^{-\tau(\eta)}$ becomes non-zero only from the time of photon decoupling onwards, naturally selecting the period since the last scattering surface till today, as the time window during which the source of temperature anisotropy is active. This is to be compared with Eqs.~(\ref{eq:Eell}, \ref{eq:Bell}), where on the contrary $\tau' e^{-\tau}$ is strongly peaked around decoupling time, indicating that CMB polarisation is generated only around this time. Moreover, Eqs.~(\ref{eq:Eell}, \ref{eq:Bell}) show that only the quadrupole $\ell=2$ in the temperature anisotropy $\Theta_2^{m}$ and in the polarisation $E_2^{m}$ can source the polarisation signal. 

In the following we derive the temperature anisotropy angular spectrum, as a worked out example of the effect of tensor modes on the CMB fluctuations. For the initial tensor spectrum leading to the signal we want to evaluate, we consider the case of a super-horizon spectrum with constant amplitude $A_T$ and spectral index $n_T$, c.f.~Eq.~\eqref{eq:Pheta0}. This leads to equations which are sufficiently simple to be tackled analytically, at least if one is looking for an approximate (but instructive) result. Furthermore, it is directly applicable to the case of inflation (c.f. section \ref{sec:inf}), even though it is in principle valid for any generation mechanism that produces such a spectrum on super-horizon scales. 

In order to find the CMB temperature spectrum, one has to combine Eqs.~\eqref{eq:Thetaell} and \eqref{eq:CMBspec}. Using $j_\ell^{(2\,2)}(x)=\sqrt{3(\ell+2)!/(8(\ell-2)!)}\,\,j_\ell(x)/x^2$, this gives  
\begin{eqnarray}\label{eq:CMBspecInf}
\hspace*{-1cm}C_{\ell, T}^{\Theta\Theta} &\simeq& \frac{1}{3\pi}\frac{(\ell+2)!}{(\ell-2)!} \int dk\,k^2 \int_{\eta_{\rm dec}}^{\eta_0} d\zeta \,\frac{j_\ell(k(\eta_0-\zeta))}{k^2(\eta_0-\zeta)^2} \int_{\eta_{\rm dec}}^{\eta_0} d\xi\,\frac{j_\ell(k(\eta_0-\xi))}{k^2(\eta_0-\xi)^2} \nonumber \\
	& & \times [(h^2(k,\zeta))'(h^{2*}(k,\xi))'+(h^{-2}(k,\zeta))'(h^{-2*}(k,\xi))']\,,
\end{eqnarray}
where $\eta_{\rm dec}$ denotes the photon decoupling time, inserted as a lower bound of integration in order to account for the effect of the optical depth. The tensor perturbation $h^{\pm 2}(k,\eta)$ can be inferred from the non-decaying solution of the wave equation (\ref{gweq3}). Tensor perturbations contribute to the present CMB temperature fluctuations roughly from the decoupling time onwards. If the GW source operated for a finite period of time in the early universe and ceased to act before decoupling time, we have seen in section \ref{sec:general} that the homogeneous solution $h_r(k,\eta>\eta_{\rm dec})$ of equation (\ref{gweq3}) is a constant fixed by the initial conditions on super-horizon scales, while it oscillates and decays as $a^{-1}$ once the mode has entered the horizon. We will re-derive and confirm this result in the context of slow roll inflation in section \ref{sec:PrimTensInf}. As a consequence, the dominant contribution to the time integrals of Eq.~\eqref{eq:CMBspecInf} comes from modes that entered the horizon during the matter dominated era, since the amplitude of those that entered before has been further suppressed by the sub-horizon decay, giving a negligible contribution. As we will derive in section \ref{sec:EvolTensInf}, in the matter era the solution is particularly simple, c.f. Eq.~\eqref{eq:solmatIC}:
 \begin{eqnarray}
  && h^{\pm 2}(x) = h_{\rm in}^{\pm 2}(k)\, \frac{3\,j_1(x)}{x}\\
  && (h^{\pm 2}(x))' = h_{\rm in}^{\pm 2}(k)\, \frac{-3\,k\,j_2(x)}{x} \label{eq:hmatpm2}
 \end{eqnarray} 
where $x=k\eta$ and $h_{\rm in}^{\pm 2}(k)$ denotes the initial amplitude at super-horizon scales. To make contact with definitions in other sections (e.g.~sections \ref{sec:general} and \ref{sec:inf}), we go from the helicity basis to the usual tensor basis, and define the initial GW spectrum (see e.g.~equation \eqref{eq:defDeltah2}):
\begin{eqnarray}
	\left\langle h^{+2}_{\rm in}(k)h^{+2*}_{\rm in}(k)+h^{-2}_{\rm in}(k) h^{-2*}_{\rm in}(k) \right\rangle &=&\frac{4}{3}\left\langle[h^+_{\rm in}(k)h^{+*}_{\rm in}(k)+h^{\times}_{\rm in}(k) h^{\times*}_{\rm in}(k)] \right\rangle \nonumber \\
	&=&\frac{4}{3}\,\pi^2\frac{\mathcal{P}_h(k)}{k^3}\,.
\end{eqnarray}
Because of the presence of the time integrals in \eqref{eq:CMBspecInf}, it is useful to normalise $\mathcal P_h(k)$ at the scale corresponding to the inverse comoving time today, $k_0=1/\eta_0$ (note however that this normalisation is completely arbitrary). Allowing for a generic spectral index, we write 
\begin{equation}\label{eq:Pheta0}
 	\mathcal P_h(k)=A_T(k_0)\,(k\eta_0)^{n_T}\,.
 \end{equation} 
In the inflationary case, the initial tensor spectrum is due to the amplification of vacuum metric fluctuations and it is given in Eq.~\eqref{eq:FullGWpowerInf} (see also Eq.~\eqref{eq:nT2e}) at first order in slow roll. Going back to Eq.~\eqref{eq:CMBspecInf}, and using Eq.~\eqref{eq:hmatpm2}, the CMB temperature spectrum becomes
\begin{eqnarray}
	C_{\ell, T}^{\Theta\Theta} &\simeq & 4\pi \frac{(\ell+2)!}{(\ell-2)!} \int \frac{dk}{k}\,\mathcal P_h(k) \left[ \int_{x_{\rm dec}}^{x_0} dx \,\frac{j_\ell(x_0-x)}{(x_0-x)^2}\frac{j_2(x)}{x}\right]^2 \nonumber \\
	&\simeq & \frac{\sqrt{\pi}}{3} \,A_T(k_0)\, \frac{\ell (\ell+2)!}{(\ell-2)!} \frac{\Gamma\big[  \frac{7-n_T}{2}\big]\Gamma\big[  \ell +\frac{n_T}{2}\big]}{\Gamma\big[  4-\frac{n_T}{2}\big] \Gamma\big[  \ell +7-\frac{n_T}{2}\big]} \label{Cell_COBE3} \\
	&\propto&   \,\,  \ell^{n_T-2} ~~~~{\rm for}~~\ell\gg 1\,, \nonumber
\end{eqnarray}
where the approximated result in the second line has been evaluated using Eq.~(B.5) of Ref.~\cite{Caprini:2003vc} for the integral in $x$, and then integrating exactly over $k$ using (6.574.2) of Ref.~\cite{TablesBook}. The spectrum in Eq.~(\ref{Cell_COBE3}) has been derived also e.g.~in~\cite{DurrerBook}. In the limit of big multipoles (small angular scales) $\ell\gg 1$, one recovers the usual shape for the CMB tensor angular spectrum, where $\ell(\ell+1)C_\ell \propto const.$ for a scale invariant tensor spectrum $n_T=0$~\cite{DurrerBook,PeterUzanBook}. However, in general, there is a dependence of the CMB spectrum on $n_T$ as $\ell^{n_T-2}$. Note that the latter result is valid only for $1 \ll \ell \lesssim 60$, as it is a quite crude estimation which does not take into account the decay of GW for modes that enter the horizon before matter-radiation equality, corresponding roughly to an angular scale of $\ell\simeq 60$~\cite{DurrerBook}. 

From Eq.~\eqref{Cell_COBE3} it appears that the measurement of the CMB spectrum at low multipoles can be used to infer a combined bound on the amplitude $A_T$ and the spectral index $n_T$ of a GW background (generated before decoupling) at super-horizon scales. Historically, in the GW literature, this constraint has been referred to as the `COBE bound', since it was first established using the measurement of the CMB quadrupole $\ell=2$ by the COBE satellite~\cite{Wright:1992tf,Allen:1994xz,Smoot:1992td,Krauss:1992ke,Koranda:1994bj}. The bound was derived setting $n_T=0$ and considering that the CMB temperature anisotropy at large angular scales was 
entirely due to the tensor mode. Today we know that this is not the case, and that the main contribution to the CMB anisotropy comes from scalar perturbations. In fact, the difference in the shape of the scalar and tensor temperature CMB spectra as a function of multipole $\ell$ allows to distinguish the two contributions. For what concerns the latest results from the Planck satellite, Ref.~\cite{Ade:2015xua} states that the strongest constraint on tensor modes still comes from the CMB temperature spectrum at $\ell<100$ (the tensor mode contribution decays at higher multipoles), so that the addition of E-polarisation does not change the result significantly. Therefore, the precision of this constraint is limited by cosmic variance, and can only be improved by adding a direct measurement of B-polarisation. This has been done in~\cite{Array:2015xqh}, where the data of the BICEP2 and Keck Array B-polarisation detectors have been combined with the Planck data, yielding the strongest constraint to date on the amplitude of tensor modes, coming from CMB only. The bound is given in terms of the tensor to scalar ratio (see section \ref{sec:PrimTensInf} and in particular Eq.~\eqref{eq:T2Sratio} for a definition) and reads (for the pivot scale $k_*=0.05\,{\rm Mpc}^{-1}$)~\cite{Ade:2015xua} 
\begin{equation}\label{eq:rbound}
	r_{0.05} = \frac{\mathcal{P}_h}{\mathcal{P}_{\mathcal{R}}}\leq 0.07~~~~~{\rm at}~~~95\%~{\rm c.l.}
\end{equation}
where $\mathcal{P}_{\mathcal{R}} \simeq 2\cdot 10^{-9}$ denotes the primordial curvature power spectrum amplitude at the pivot scale $k_*=0.05\,{\rm Mpc}^{-1}$, see Eq.~\eqref{eq:DeltaR2}. 

Note that the above constraint is derived assuming a $\Lambda$CDM model, fixing the tensor spectral index to zero, $n_T=0$. Current CMB measurements do not have the ability to constrain $n_T$, since the measurement of $r$ is still compatible with zero, and for low enough $r$, practically any value of $n_T$ is acceptable. For this reason, the constraints on $n_T$ depend on the chosen prior on $r$, as pointed our for example in \cite{Cabass:2015jwe}. This situation will change if, in the future, a positive detection of a non-zero tensor amplitude is obtained from primordial B-modes (see~\cite{Cabass:2015jwe} for forecasts concerning a COrE-like mission~\cite{Bouchet:2011ck} exclusively dedicated to B-mode polarisation). In the meanwhile, other observations like those from Pulsar Timing Arrays (PTA) (c.f. section \ref{sec:PTA}), or from direct GW detection ground-based interferometers (c.f. section \ref{sec:interferometers}), can be used to place upper bounds on $n_T$ in the case of blue tilted SGWB spectra, see e.g.~Refs.~\cite{Lasky:2015lej,Cabass:2015jwe,Meerburg:2015zua,Bartolo:2016ami}.

\begin{figure}
\begin{center}
\includegraphics[width=0.7\textwidth]{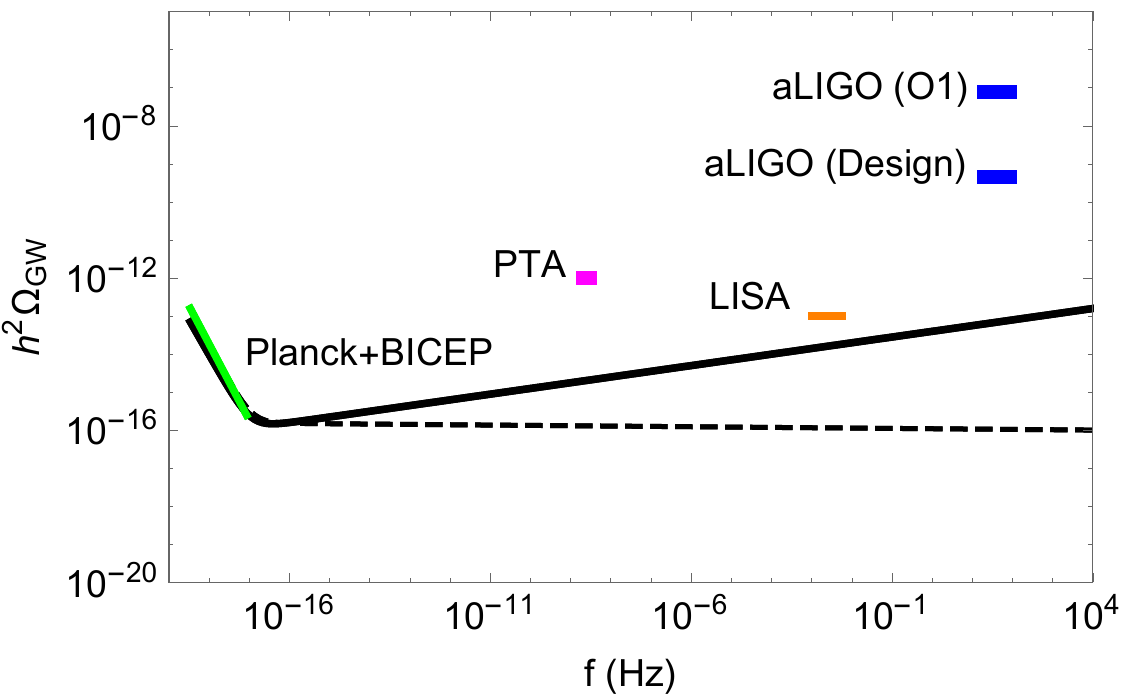}
\caption{\label{fig:BlueSpec} The power spectrum of Eq.~\eqref{eq:omegaKnt} for $r=0.07$ and $n_T=-r/8$ (black dashed line), and $n_T=0.2$ (black solid line), together with the constraint given by Planck+BICEP2+Keck Array data given in Eq.~\eqref{eq:rbound}, i.e.~setting $r=0.07$ and $n_T=0$ (green solid line), and the reach of current and future GW detectors: PTA (magenta, solid), advanced LIGO at the first run and at design sensitivity (blue, solid) and LISA (orange, solid).}
\end{center}
\end{figure}

The GW energy density fraction today can be expressed in terms of the tensor to scalar ratio $r$ and a generic spectral index $n_T$ using Eq.~\eqref{eq:omegaK} (c.f. section \ref{sec:EvolTensInf} and the discussion therein),
\begin{equation}\label{eq:omegaKnt}
	\Omega_{\rm GW}(f)=\frac{3}{128}\,\Omega_{\rm rad}\,r\,{\mathcal{P}_{\mathcal{R}}}\left(\frac{f}{f_{*}}\right)^{n_T}\,\left[\frac{1}{2}\left(\frac{f_{\rm eq}}{f}\right)^2+\frac{16}{9}\right]\,, 
\end{equation}
where $f_*=k_*/(2\pi a_0)$ is the pivot frequency at which the primordial scalar amplitude is normalised, $k_*=0.05\,{\rm Mpc}^{-1}$, and $f_{\rm eq}=H_0\,\Omega_{\rm mat}/(\pi\sqrt{2\,\Omega_{\rm rad}})$ is the frequency entering the horizon at matter-radiation equality, c.f. Eq.~\eqref{eq:keq}. The energy density spectrum of Eq.~\eqref{eq:omegaKnt} is shown in Fig.~\ref{fig:BlueSpec} for $r$ saturating the bound in Eq.~\eqref{eq:rbound} and two values of $n_T$: the red-tilted value predicted by slow roll inflation, corresponding to the consistency relation $n_T=-r/8$ (c.f.~sec.~\ref{sec:PrimTensInf} and in particular Eq.~\eqref{consistency}), and a blue-tilted case with large index, $n_T=0.15$. In this figure we also show the sensitivity of current and future GW detectors (which we will describe in Sect.~\ref{sec:interferometers}) and the CMB bound derived in Eq.~\eqref{eq:rbound}, which applies for $n_T=0$. Note that, as discussed before, tensor modes decay once they enter the horizon; this determines the regime of validity of the CMB bound, namely frequencies which were outside the horizon at matter-radiation equality but are inside the horizon today: $3.4 \cdot 10^{-19}< f < 2.1 \cdot 10^{-17}$ Hz. For forecasts on the constraints that can be derived on the couple of parameters $(n_T,r)$ by the combination of data from CMB temperature anisotropies and B-polarization, with direct bounds established by advanced LIGO/Virgo and PTA, and with indirect bounds described in section \ref{sec:BBN}, we refer the reader for example to Fig. 2 of \cite{Lasky:2015lej}, see also \cite{Cabass:2015jwe,Meerburg:2015zua}. 

At last, it is worth mentioning another way to probe the presence of primordial tensor modes via the CMB, namely spectral distortions of the CMB black-body spectrum \cite{Ota:2014hha,Chluba:2014qia}. These are tiny deviations of the CMB monopole from a perfect black-body, produced at redshift $z\lesssim 10^6$, when thermalisation processes like Compton scattering and Bremsstrahlung start to be less efficient. Spectral distortions can be due to several processes that cause energy injection in the photon distribution, both in the context of the standard cosmological model (recombination, the dissipation of primordial fluctuations, reionisation, structure formation...) and of its extensions (annihilation or decay of particles, primordial magnetic fields, primordial black holes, cosmic strings...). See~\cite{Chluba:2011hw,Chluba:2014sma} for recent analyses, and references therein. The CMB spectral distortion due to the presence of primordial tensor modes is produced via Thomson scattering in a similar way to the one due to the dissipation of scalar perturbations, but the tensor anisotropic stress (shear viscosity) in the photon distribution, necessary to activate the dissipation process, is directly induced by the tensor modes instead of by photon free-streaming \cite{Ota:2014hha}. As a consequence, the dissipation occurs over a wider range of scales than those probed by Silk damping, as in the case of scalar modes. For the almost scale invariant tensor spectrum produced in slow roll inflationary scenarios, it has been shown that the amplitude of the distortion is very small, providing only a tiny correction to the signal expected from the dissipation of scalar perturbations~\cite{Ota:2014hha,Chluba:2014qia}. A more important deviation from the CMB black-body can be induced if the tensor spectrum is very blue, say with $n_T\simeq 1$. In this case, the level of the distortion can become comparable to the one induced by scalar modes in the standard inflationary scenario \cite{Chluba:2014qia}.

\subsection{Pulsar timing (arrays)}
\label{sec:PTA}

Pulsars are highly magnetized, rotating neutron stars, that emit a beam of electromagnetic radiation in the direction of their magnetic axis (which spins along with the rotation of the neutron star). It is the magnetic axis of the pulsar, which does not necessarily coincide with the rotational axis, that determines the direction of the electromagnetic beam. If the beam sweeps the line of sight between Earth and the pulsar, a regular train of radiation pulses is observed, similarly as when we observe a lighthouse from the distance. Pulsars are thus colloquially referred to as `cosmic lighthouses'. The arrival times of the pulses are extremely regular and can be predicted very accurately over long times. Pulsars represent therefore very stable `clocks' scattered around the sky, which allow for a variety of precise astronomical measurements. For instance, the first extra-solar planet was discovered around a pulsar~\cite{Wolszczan:1992zg}, and the decrease of the orbital period of a binary system with a pulsar -- the {\it Hulse-Taylor binary} -- provided the first indirect evidence for the existence of GWs~\cite{Hulse:1974eb,Taylor:1989sw}. As we will discuss in the following, pulsars can be used also as a direct probe of stochastic GW backgrounds, since the arrival times of the electromagnetic pulses experience a frequency shift in the presence of a GW background.

The use of pulsars as a precise clock, rely on a technique known as `pulsar timing', see e.g.~Ref.~\cite{Lorimer:2008se} for a review. This basically involves two main steps: first, the beam profiles are analysed so that a `time of arrival' (TOA) is assigned to each pulse. Secondly, the measured TOA's are compared to a theoretical modelling that incorporates many circumstances possibly affecting the signal, from the evolution of the pulsar's rotation, to the relative motion between the pulsar and Earth, the propagation of the pulses through the interstellar medium, etc. Parameters such as the pulsar's spin period, spin-down rate, proper motion, and others, are obtained from fits to the data. The difference between the best-fit model predictions and the measured TOA, for a given pulsar, is called 'timing residual'. Among the different pulsars, those with rotation periods of the order of a millisecond, have very small timing irregularities, and their TOA's can be measured with very high precision. Millisecond pulsars are therefore particularly suited for the detection of GW backgrounds, as timing residuals varying by no more than few micro seconds over several years can be obtained for them, see e.g.~Fig.~5 of Ref.~\cite{Kaspi:1994hp}.

The use of pulsar timing observations as a way to detects/constrain GWs was first studied by Sazhin~\cite{Sazhin:1977tq} and Detweiler~\cite{Detweiler:1979wn} and further developed in Refs.~\cite{Romani:1983fq,Hellings:1983fr, Bertotti:1980pg,Stinebring:1990px}. The very small timing residuals observed in various pulsars can be used to set upper bounds on the presence of a GW background between a given pulsar and Earth. The maximum sensitivity can be achieved for a GW background with power at frequencies of the order of the inverse total observation time $T$, typically\footnote{The fitting procedure to a pulsar model described before removes the signal effect from GWs with frequencies much smaller than $\sim 1/T$.} at $f \sim 10^{-9} - 10^{-8}$ Hz for $T \sim \mathrm{few} \, \mathrm{years}$. For a timing residual varying by an amount of $\delta t$, a constraint on the characteristic amplitude of a stochastic GW background, typically of the order of $h_c \lsim \delta t / T$ [see Eqs.~(\ref{hcketa},\ref{hcf})], can be obtained. Using this fact and e.g.~$\delta t \sim \mathrm{few} \times 10^{-6}$ sec and $T \sim$ few years, Eq.~(\ref{Omandhc}) leads to a constraint $h^2\,\Omega_{\rm GW} \lsim \mathrm{few} \times 10^{-8}$ at $f \sim 10^{-8}$ Hz. More precisely, using long-term observations of stable millisecond pulsars, Kaspi et al~\cite{Kaspi:1994hp} obtained, for a flat spectrum of GW, an upper bound of $h^2\,\Omega_{\rm GW} < 6 \times 10^{-8}$ (at $95 \%$ confidence level) at $f \approx 4.5 \times 10^{-9}$ Hz. This was further improved to $h^2\,\Omega_{\rm GW} < 10^{-8}$ at $f \approx 4.5 \times 10^{-9}$ Hz in Ref.~\cite{Thorsett:1996dr}, and to $h^2\,\Omega_{\rm GW} < 2 \times 10^{-9}$ at $f \approx 2 \times 10^{-9}$ Hz in Ref.~\cite{Lommen:2002je}. Note however that the statistical method used in the latter works has been criticized in the literature, see Refs.~\cite{McHugh:1996hd}, \cite{Jenet:2006sv}. 

The previous technique allows to establish strong upper bounds on the presence of a stochastic GW background in the Universe. However, since it is not possible to determine the exact origin of a given timing residual, the observation of an individual pulsar does not allow by itself, to actually detect a GW stochastic background. The timing residuals might be due to a number of reasons, not always under control, such as irregularities in the pulsar's rotation or in the terrestrial time standards. The effect of a stochastic GW background on the timing residuals of various pulsars, however, may be distinguished and isolated by looking for correlations among the different pulsars' residuals. This can be done by the so called `pulsar timing arrays' (PTA), where timing observations are performed for many pulsars distributed over the sky. The presence of an isotropic stochastic background of GWs will induce a correlation between the timing residuals from different pulsars, depending only on the angular separation between the location of the pulsars in the sky. The GW induced frequency shifts in the arrival times of a train of pulses (from a given pulsar) depends on the GW amplitude along the path of the pulse,
\begin{eqnarray}
{\Delta \nu\over \nu} = -{1\over2}\int_{\lambda_e}^{\lambda_r}d\lambda\, {h}'_{ij}(x(\lambda))\hat{\zeta}_i\hat{\zeta}_j\,,
\end{eqnarray}
where $\hat{{\zeta}}_i$ is the unit vector in the pulsar-Earth direction, and $\lambda$ is the affine parameter parametrizing the spatial trajectory of the pulse [c.f.~\Eqref{SW_COBE}]. The correlation between the frequency shifts of two pulsars comes from the Earth-contribution to each pulsar. Using the decomposition of a GW background given by Eq.~(\ref{hrketa}), with $e_{ij}^r$ the polarization tensors Eq.~(\ref{eij}), the correlation can be written as~\cite{Hellings:1983fr}
\begin{eqnarray}
\left\langle \frac{\Delta \nu}{\nu}(\hat{\boldsymbol{\zeta}}) \, \frac{\Delta \nu}{\nu}(\hat{\boldsymbol{\xi}}) \right\rangle &\propto& \sum_{r = + , \times} \int d^2\mathbf{\hat{k}}\,
\frac{e_{ij}^{r}(\mathbf{\hat{k}}) \, e_{kl}^{r}(\mathbf{\hat{k}}) \, \hat{\zeta}_i \, \hat{\zeta}_j \, \hat{\xi}_k \, \hat{\xi}_l}{(1 + \hat{k}_m \, \hat{\zeta}_m) \, (1 + \hat{k}_n \, \hat{\xi}_n)}\\
&\propto& x \, \mathrm{log} x - \frac{x}{6} + \frac{1}{3}
\hspace*{0.5cm} \mbox{with} \hspace*{0.5cm} 
\,x = \frac{1 - \mathrm{cos} \theta}{2}\,,
\end{eqnarray}
where in the second line we have used the polarization tensor completeness relation Eq.~(\ref{completeness}), and defined the angle $\mathrm{cos} \theta = \hat{\zeta}_i \, \hat{\xi}_i$ between the unit vectors in the pulsar-Earth direction for each pulsar. The specific dependence on the angle $\theta$ exhibited by this correlation between frequency shifts, represents a characteristic and unique signature of the presence of a stochastic GW background in PTA observations. 
  
There are various active PTA collaborations searching for GW backgrounds: the {\it Parkes Pulsar Timing Array}~\cite{Manchester:2012za} (PPTA), the {\it European Pulsar Timing Array}~\cite{Ferdman:2010xq} (EPTA) and the {\it North American Nanohertz Observatory for Gravitational Waves}~\cite{Jenet:2009hk} (NANOGrav). They have recently joint forces forming the {\it International Pulsar Timing Array}~\cite{Hobbs:2009yy,IPTA):2013lea} (IPTA). The current upper bounds on GW backgrounds obtained by these collaborations have been reported in Refs.~\cite{Jenet:2006sv}, \cite{Demorest:2012bv}, \cite{Lentati:2015qwp}, and \cite{Verbiest:2016vem}. In these works, a GW background is characterized by its amplitude $h_c(f)$ (see Eq.~(\ref{Omandhc}) for the relation to $\Omega_{\rm GW}(f)$), parametrized like
\be   
h_c(f) = A_{\alpha} \, \left(\frac{f}{\mathrm{year}^{-1}}\right)^{\alpha} \,.
\ee
The upper limits obtained on $A_{\alpha}$ depend, in principle, on each chosen value for $\alpha$. In particular $\alpha = -2/3$  (i.e. $\Omega_{\rm GW}(f) \propto f^{2/3}$) is expected in the case of the GW background from unresolved supermassive black hole binaries, whereas $\alpha = -1$ corresponds to a flat spectrum $\Omega_{\rm GW}(f) \propto~const.$ (see section \ref{sec:inf}), and $\alpha = -7/6$ (i.e.~$\Omega_{\rm GW}(f) \propto f^{-1/3}$) is a typical value expected for the GW background from cosmic strings (see section \ref{sec:CosmicDefects}). The upper bounds at the frequency where the sensitivity is best achieved, $f \sim 1 / T$ ($T$ here is the effective total observation time of the multiple pulsars), are approximately independent of the slope of the GW spectrum. The $\alpha$-dependence of the upper limits on $A_{\alpha}$ comes about mostly because the chosen reference frequency $f = 1 / \mathrm{year}$ differs from $f \sim 1 / T$, see Refs.~\cite{Jenet:2006sv}, \cite{Demorest:2012bv}. For instance, Jenet et al~\cite{Jenet:2006sv} obtained, for a range of values of $\alpha$, an upper bound ($95 \%$ confidence) $h^2\,\Omega_{\rm GW} < 2 \times 10^{-8}$ at $f = (8 \, \mathrm{years})^{-1} \simeq 4 \times 10^{-9}$ Hz, whereas Lentati et al~\cite{Lentati:2015qwp} obtained $h^2\,\Omega_{\rm GW} < 1.2 \times 10^{-9}$, at $f = (18 \, \mathrm{years})^{-1} \simeq 2.4 \times 10^{-9}$ Hz. For $\alpha = -2/3$, the results yield $h^2\,\Omega_{\rm GW} < 3.1 \times 10^{-8}$ at $f \sim 3 \times 10^{-8}$ Hz (Demorest et al~\cite{Demorest:2012bv}) and $h^2\,\Omega_{\rm GW} < 1.1 \times 10^{-9}$ at $f \sim 2.4 \times 10^{-9}$ Hz (Lentati et al~\cite{Lentati:2015qwp}). As we will discuss in Section~\ref{sec:CosmicDefects}, the PTA bounds on GW backgrounds are particularly constraining for cosmic strings. Furthermore, contrary to the BBN and CMB bounds discussed in Sect.~\ref{sec:BBN}, the PTA bound can be applied as well to stochastic backgrounds produced (way after BBN and photon decoupling) by unresolved astrophysical sources. 

The sensitivity of PTA to stochastic GW backgrounds is expected to improve in the near future. There are good prospects for improving the timing precision, the total observation time, the number of observations and the number of millisecond pulsars that will be observed. The IPTA~\cite{Hobbs:2009yy} is expected to reach a sensitivity of the order of $h^2\,\Omega_{\rm GW} \sim 10^{-11}$. In the longer term, the {\it Square Kilometer Array} (SKA) is expected to reach sensitivities, in the nHz frequency range, down to $h^2\,\Omega_{\rm GW} \sim 10^{-15}$ assuming 50 pulsars monitored every two weeks for 20 years \cite{Moore:2014lga}. Let us note however, that one of the most significant stochastic backgrounds of GWs in the PTA frequency range, is produced by super-massive black hole (SMBH) binaries that coalesce when galaxies merge~\cite{Jaffe:2002rt,Wyithe:2002ep,Enoki:2004ew,Sesana:2008mz}. Even though this astrophysical background has a characteristic frequency-dependence as $\Omega_{\rm GW} \propto f^{2/3}$ that distinguishes it from other backgrounds, it will 'hide' other cosmological backgrounds with smaller amplitude. The amplitude of the SMBH background depends on the galaxy merger rates during the cosmological evolution and on the typical SMBH masses. Given the current uncertainties on the astrophysical parameters characterizing these mergers, a conservative lower limit on this background can be obtained using the results from Refs.~\cite{Sesana:2008mz}, \cite{Sesana:2012ak},
\be
\label{SMBH}
h^2 \, \Omega^{SMBH}_{gw} \gsim 10^{-11} \, \left(\frac{f}{10^{-8} \, \mathrm{Hz}}\right)^{2/3} \hspace*{0.5cm} 
\mbox{ for } \hspace*{0.5cm} f \lsim 10^{-8} \, \mathrm{Hz} \, .
\ee
A cosmological background with a smaller amplitude than in Eq.~(\ref{SMBH}) will thus be unlikely to be observable by PTA. Note also that the signal amplitude Eq.~(\ref{SMBH}) is well within the reach of an experiment like SKA.

\subsection{Gravitational wave interferometers}
\label{sec:interferometers}	


\subsubsection{Principles of the detection of a stochastic background.}	
\label{sec:principledet}

We present here a basic description of the principles of SGWB detection by GW interferometers, based mainly on \cite{Maggiore:1999vm}. We point the reader interested in a closer examination of this topic to the exhaustive treatments of Refs.~\cite{MaggioreBook,Romano:2016dpx}.

The detection of GWs consists in measuring the quadrupolar deformation of spacetime that they induce. This can be done through a resonating mechanical system (the first GW detectors were resonant massive bars, for an overview see \cite{MaggioreBook}), but a much more efficient method is via a Michelson interferometer operating between freely suspended masses. In the idealized case of a linearly polarized wave hitting perpendicularly the interferometric system, with its polarization axes aligned with the arms in the proper detector frame as defined in \cite{MaggioreBook}, one arm contracts while the other expands. Therefore, the laser beams returning from each arm pick up a phase difference which changes in time following the passage of the wave, influencing the interferometric patterns: this is the principle of the GW detection via interferometers\footnote{As analysed in detail in \cite{MaggioreBook}, the interpretation of the physical effect a GW has on the experimental device differs for different coordinate choices: in the TT gauge for example, since the coordinates expand with the metric perturbation, the GWs affect the propagation of photons along the geodesic instead of affecting the masses position.} (for a review, see e.g.~\cite{Riles:2012yw}). 

In the case of one interferometer with arms of equal length $L$ in the $x$ and $y$ directions, the output of the detector {due} to the passing of a GW is the strain $h(t)=(\Delta L_x(t)-\Delta L_y(t))/L$. This scalar signal is related to the GW in the TT gauge as $h(t)=F_+ h_+(t)+F_\times h_\times(t)$, where $F_{+,\times}(\hat\Omega, \psi)$ are the detector pattern functions, which depend on the geometry of the system \cite{Maggiore:1999vm,Schutz:2011tw}: the direction of arrival of the wave $\hat{\Omega}$, and the choice of axes with respect to which one defines the polarizations, represented by a rotation angle $\psi$ in the plane orthogonal to $\hat\Omega$. The Fourier transform of the signal becomes then \cite{Maggiore:1999vm}
\begin{equation}
\tilde h(f)=\int d\hat\Omega \, [F_+(\hat\Omega, \psi) h_+(f,\hat\Omega)+F_\times (\hat\Omega, \psi) h_\times(f,\hat\Omega)]\,.
\end{equation}
For an unpolarized stochastic background, there is no privileged choice of axes to define the polarizations, and the angle $\psi$ effectively cancels from the final result. Furthermore, the average of the signal vanishes, while the second moment of the signal distribution can be written simply as \cite{Maggiore:1999vm}
\begin{equation}
\langle h^2(t)\rangle = F \int_0^\infty df \, S_h(f)\,, ~~~~~~~~~ F=\int \frac{d\hat\Omega}{4\pi}\,[F_+^2(\hat\Omega, \psi)+F_\times^2(\hat\Omega, \psi)] \,,
\end{equation}
where $S_h(f)$ is the spectral density defined in Eq.~(\ref{Shf}), and the factor $F$ is the average of the detector pattern functions over all directions of  arrival of the waves. This factor quantifies the loss of sensitivity due to the fact that the GW background hits the detector isotropically, with respect to the sensitivity of the detector in the optimal direction: if $\alpha$ is the angle between the two arms, one has \cite{Maggiore:1999vm} $F=2/5\sin^2\alpha$. 

\begin{figure}
\begin{center}
\includegraphics[width=0.7\textwidth]{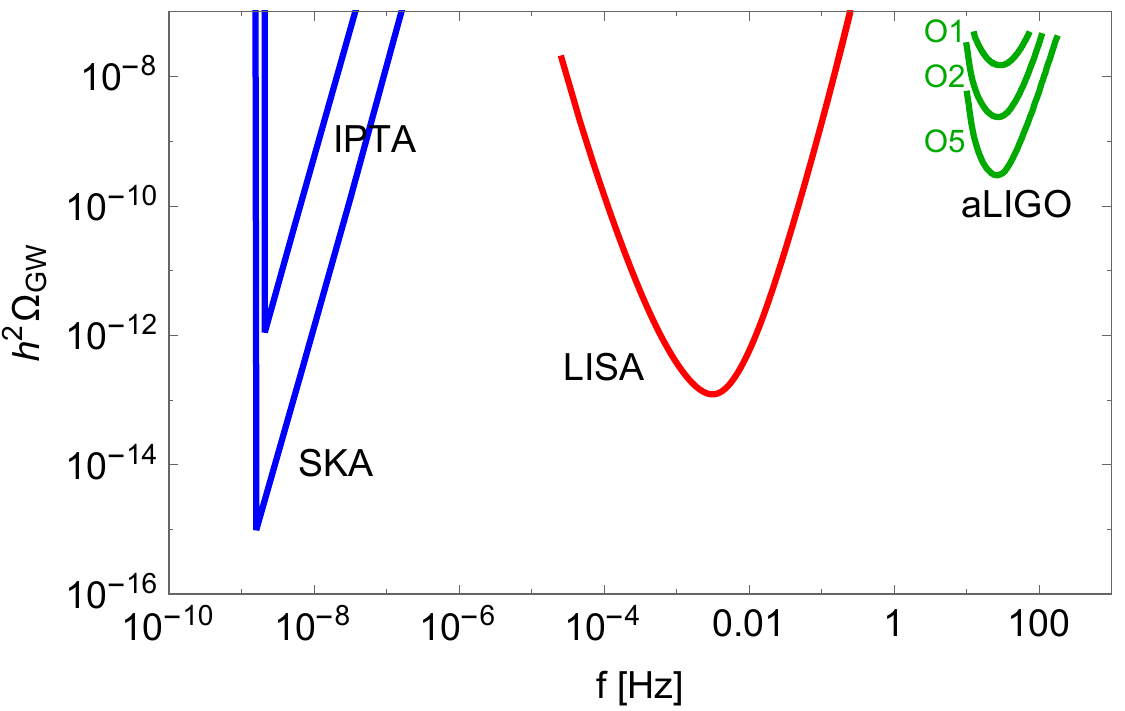}
\caption{\label{fig:detectors} The Power Law-Integrated Curves of current and future GW detectors. From left to right: for PTA (blue) we show the predictions for the International Pulsar Timing array and for a network monitored by the SKA, taken from \cite{Moore:2014lga}; for LISA (red), we show the expected power law-integrated curve adapted from Ref.~\cite{Audley:2017drz}; for Advanced LIGO/Virgo (green) we show the sensitivities given in \cite{TheLIGOScientific:2016wyq} for the first (O1) and second (O2) runs, and at design sensitivity (O5).}
\end{center}
\end{figure}

The total output of a detector is given by the GW signal plus the noise, $S(t)= h(t)+n(t)$. In the same way as done for the signal $h(t)$, one can define the noise spectral density $S_n(f)$ from the averaged squared noise: 
\begin{equation}
\langle \tilde n^*(f) \tilde n(f') \rangle =\delta(f-f') \frac{S_n(f)}{2}\,, ~~~~~~~~~ \langle n^2(t)\rangle =  \int_0^\infty df \, S_n(f)\,. 
\end{equation}
The level of noise in the detector is given by the strain sensitivity $h_f=\sqrt{S_n(f)}$. 
In a single detector the SGWB manifests itself simply as an extra source of noise. This makes impossible to isolate the detector from the signal, in order to measure the actual detector noise, subtract it, and thereby distinguish the two components. Therefore, when a single detector is operating, a stochastic background can be measured if it overcomes the detector noise level (this is what happened e.g.~with the measurement of the CMB black-body spectrum) or it is of comparable amplitude. Methods to distinguish the signal from the noise are, for example, a substantial difference between their spectral shapes, or the distinctive time modulation of an anisotropic signal due to the detector motion (as expected, for instance, in the case of the SGWB from galactic binaries visible by the space-based interferometer LISA \cite{Audley:2017drz}). Furthermore, a good knowledge of the noise can be achieved using null channels: particular combinations of the interferometer outputs to which the contribution of the signal is strongly suppressed \cite{Romano:2016dpx}. Null channels allow to measure the noise spectral density directly from the detector output; the knowledge of the noise is then used to assess the presence of a SGWB in the standard interferometric channels. However, the interferometer structure must be such as to allow the construction of a null channel: in the case of the triangular LISA configuration, for example, the null channel is the so called {\it symmetrised Sagnac} combination~\cite{Hogan:2001jn,Tinto:2001ii}. Once the noise has been evaluated, the technique to dig out a signal buried in it is optimal Wiener filtering, often using a parametrized model for the signal. This has been demonstrated e.g.~in the third Mock LISA Data Challenge \cite{Babak:2009cj}: a relatively loud isotropic stochastic background, with amplitude slightly below the LISA instrument noise (c.f. Fig.~\ref{fig:detectors}), had been buried in the simulated detector output, and the teams participating in the data analysis challenge were able to recover the presence of this signal and its spectral characteristics. 

The minimal condition to observe the SGWB with a single interferometer can be defined then as $S_h(f)>S_n(f)/F$. In terms of the GW energy density parameter $\Omega_{\rm GW} =(4\pi^2/3H_0^2)\,f^3\,S_h(f)$ (c.f. Eq.~(\ref{OmandSh})), this condition can be translated into a minimum value for detection assuming a signal to noise ratio of one, at a frequency $f$, and given the interferometer strain sensitivity $h_f(f)$ \cite{MaggioreBook}:
\begin{equation}
h^2\Omega_{\rm GW}(f) \gtrsim \frac{10^{-2}}{F} \left( \frac{f}{100\,{\rm Hz}} \right)^3 \left( \frac{h_f}{10^{-22}\,{\rm Hz}^{-1/2}} \right)^2 \,.
\label{det_OmGW1d}
\end{equation}
From the above equation it appears that, for the same noise level $h_f$, detectors with lower detection frequency window are more sensitive to a stochastic background.

Concerning currently operating detectors on Earth, for which Eq.~(\ref{det_OmGW1d}) has been normalised, the predicted amplitude of both astrophysical and cosmological stochastic backgrounds is still below their sensitivity. On the other hand, as several detectors are present, this allows to adopt a more elaborated detection strategy, which consists in cross-correlating the output of two (or more) different detectors, exploiting the fact that different detectors have independent noise. We present a brief description of this strategy, based on Refs.~\cite{Maggiore:1999vm,Allen:1996vm}. The cross-correlation signal $S_{12}$ is constructed multiplying the two detector outputs $S_{1}(t)$ and $S_{2}(t)$, and integrating over the observation time $T$ with a filter function $Q$,
\begin{equation}
S_{12}=\int_{-T/2}^{T/2} dt \int_{-T/2}^{T/2} dt'  S_1(t)S_2(t') Q(t-t')\,.
\label{dec_S12}
\end{equation}
As we will see later, the filter function must be chosen to maximize the signal to noise ratio: it depends on the position and orientation of the detectors, on the features of their characteristic noise, as well as on the spectrum of the stochastic background. However, since one considers that both the signal and the noise are stationary in time, it is reasonable to assume that the best choice for $Q$ depends only on the time difference $t-t'$. Furthermore, $Q(t-t')$ is different from zero only for $|t-t'| \ll T$: it must fall off rapidly for time intervals which are larger than the light-travel distance between the two detectors, which is of the order of $10^{-2}$ sec if they are both on Earth. On the other hand, the integration time is typically of the order of one year: one can therefore Fourier transform Eq.~(\ref{dec_S12}) in the limit of large integration time $T$. Moreover, because the noise in each detector is uncorrelated with the other one, and with the GW strain, it turns out that the expected value of the cross-correlation of the detector outputs $S_i(t)= h_i(t)+n_i(t)$, obtained by taking the expectation value of Eq.~(\ref{dec_S12}), depends only on the GW stochastic background. Using Eq.~(\ref{Shf}) together with the response of the detector through the pattern functions, one finds finally \cite{Maggiore:1999vm} 
\begin{equation}
\langle S_{12} \rangle = \frac{T}{2} \int_{-\infty}^{\infty} df  S_h(f)\Gamma(f) \tilde{Q}(f)  \,,
\end{equation}
where $\tilde{Q}(f)$ denotes the Fourier transform of the filter function, and $\Gamma(f)$ is the overlap function. This latter characterizes the reduction in sensitivity to the GW background arising from the relative positions and orientations of the pair of detectors: 
\begin{equation}
\Gamma(f) = \frac{1}{4\pi} \int d\hat\Omega \, [F^+_1 F^+_2 + F^\times_1 F^\times_2] \exp\left[ 2\pi {\rm i} \hat\Omega \cdot \Delta {\bf x} \right]  \,,
\end{equation}
where $\Delta {\bf x}$ denotes the separation between the two detectors. In the case of two coincident interferometers, $\Delta {\bf x}=0$ and $\Gamma(f)=F=2/5$. The exponential phase factor in the definition of $\Gamma$ is the phase shift arising from the time delay between the two detectors for radiation arriving along the direction $\hat\Omega$. As an example, the overlap function between the LIGO interferometers at Hanford and Livingston is shown in Ref.~\cite{Thrane:2013oya}. From the definition of $\langle S_{12} \rangle$, it appears that the expected signal grows linearly with the observation time. We will see that the r.m.s noise, on the other hand, grows only as $\sqrt{T}$: in principle, with enough observation time one can detect a GW stochastic background buried in any level of detector noise.

The form of the optimal filter function $\tilde Q(f)$ for a stochastic background search is the one that maximizes the signal to noise ratio. The noise in the detector is given by the variation of $S_{12}$ away from its mean value: $N=S_{12}-\langle S_{12}\rangle $. We are dealing with the case in which the noise is much higher than the gravitational strain signal, $n_i\gg h_i$, so that one can approximate $S_i\simeq n_i$. With the further condition that the noise is Gaussian and uncorrelated among the two detectors, one gets for the squared r.m.s. value of the expected noise \cite{Maggiore:1999vm} 
\begin{equation} 
\langle N^2 \rangle = \langle S_{12}^2 \rangle - \langle S_{12} \rangle^2 = \frac{T}{4} \int_{-\infty}^{\infty} df  \, S_n^{(1)}(f)\,S_n^{(2)}(f) \, |\tilde{Q}(f)|^2  \,,
\end{equation}
where $S_n^{(i)}(f)$ denote the spectral noise densities in each detector. As anticipated, one sees that the r.m.s.~noise grows as $\sqrt{T}$. The value of $\tilde Q(f)$ that maximizes the signal to noise ratio ${\rm SNR}=\langle S_{12} \rangle / \sqrt{\langle N^2 \rangle }$ is found by solving a variational problem and turns out to be \cite{Maggiore:1999vm}
\begin{equation}
\tilde Q(f)=c\,\frac{\Gamma(f)\,S_h(f)}{S^{(1)}_n(f) S^{(2)}_n(f)}\,, 
\label{dec_Q}
\end{equation}
with $c$ a normalization constant. The optimal filter function depends then on the GW spectral density: in order to search for the signal, one has to perform the data analysis using several forms for the filter function. $\tilde Q(f)$ can be chosen either so as to match the theoretical predictions for the GW stochastic backgrounds, or assuming a simple power law behavior $S_h(f)\propto f^\alpha$, so as to generically model the frequency dependence of a stochastic background over the (often narrow) frequency range a detector. 
The value of the signal to noise ratio SNR using the optimal filter function for two interferometers is then given by
\begin{equation}
{\rm SNR}=  \frac{\langle S_{12} \rangle}{\sqrt{\langle N^2 \rangle }}=\left[ 2\,T \int_0^\infty df\,\frac{\Gamma^2(f)\, S_h^2(f)}{S^{(1)}_n(f) S^{(2)}_n(f)} \right]^{1/2}\,. \label{eq:SNR}
\end{equation}
Starting from this expression for the signal to noise ratio, it is possible to estimate the gain in sensitivity due to the cross-correlation of two detectors with respect to the single detector case. Let us suppose that the integrand in the above equation is approximately constant over the frequency range of sensitivity of the detector, say $\Delta f\simeq 100$ Hz for a typical Earth-based interferometer. Setting ${\rm SNR} >1$, the GW spectral density must satisfy
\begin{equation}
S_h(f)>\frac{\sqrt{S^{(1)}_n(f) S^{(2)}_n(f)}}{\Gamma(f)\sqrt{2\,T\,\Delta f}}\,.
\end{equation}
We assume the same noise spectral density in the two detectors, so that $\sqrt{S^{(1)}_n(f) S^{(2)}_n(f)}=S_n(f)=h^2_f$, where $h_f$ is the strain sensitivity. Moreover, we assume that the two detectors are coincident, and as a plausible observation time, we take $T\simeq 1$ year. Through the relationship between $S_h$ and $\Omega_{\rm GW}$, we can then evaluate the minimum detectable value of $\Omega_{\rm GW}$ from an ideal two detector correlation $\Omega_{\rm GW}^{\rm 2d}(f)$, and compare it with the single detector case given in Eq.~(\ref{det_OmGW1d}) labelled now as $\Omega_{\rm GW}^{\rm 1d}(f)$:
\begin{equation}
h^2\Omega_{\rm GW}^{\rm 2d}(f) \gtrsim  10^{-5} \,\, [h^2\Omega_{\rm GW}^{\rm 1d}(f)]\,\,  \sqrt{\frac{1\,{\rm year}}{T}} \, \sqrt{\frac{100\,{\rm Hz}}{\Delta f}}\,.
\end{equation}
The cross-correlation between two coincident detectors, with the same noise characteristics but fully uncorrelated noise, helps in the detection of a GW stochastic background by about five orders of magnitude. The minimum detectable value of $\Omega_{\rm GW}(f)$ for 1 year of integration time and assuming that everything can be treated as constant over a frequency range of about $\Delta f\simeq 100$ Hz becomes then \cite{Maggiore:1999vm}
\begin{equation}
h^2\Omega_{\rm GW}^{\rm 2d}(f) \simeq  \frac{10^{-7}}{F} \left( \frac{f}{100\,{\rm Hz}} \right)^3 \left( \frac{h_f}{10^{-22}\,{\rm Hz}^{-1/2}} \right)^2 \,.
\end{equation}

From the expression of the signal to noise ratio \eqref{eq:SNR}, it appears that the detectability of a SGWB improves with the integration time $T$ and with the frequency span of the signal. In order to visualise this effect when plotting the sensitivity of GW detectors as a function of frequency, Ref.~\cite{Thrane:2013oya} proposes a method that holds for signals having a power law frequency dependence in the frequency band corresponding to the detector sensitivity. This consists in plotting the so-called Power Law-Integrated Curve, composed by the envelope of a set of power laws,
\begin{equation}
	\Omega_{\rm PI}=\stackunder{\max}{\scriptstyle\beta}\,\,\Omega_\beta\left(\frac{f}{f_*}\right)^\beta\,,
\end{equation}
where $f_*$ is a reference frequency and the amplitudes $\Omega_\beta$ are chosen such that they provide a given value for the SNR: converting the noise spectral density into an energy density using $S_n(f)=(3H_0^2/4\pi^2)\Omega_n(f)/f^3$, one defines
\begin{equation}\label{eq:SNROm}
	\Omega_\beta=\frac{\rm SNR}{\sqrt{2T}}\left[\int_{f_{\rm min}}^{f_{\rm max}}df \,\,\frac{(f/f_*)^{2\beta}}{\Omega_{n}^2(f)}\right]^{-1/2}\,,
\end{equation}
where $f_{\rm min},~f_{\rm max}$ denote the detector bandwidth. The meaning of the Power Law-Integrated Curve is that, a power-law background lying above it, is detectable with signal to noise ratio larger than the actual value of ${\rm SNR}$ chosen in Eq.~\eqref{eq:SNROm}.

Note that, since the bandwidths of current GW detectors are relatively small, it seems justified to construct a sensitivity curve which is meaningful for single power-law SGWBs. On the other hand, this method is going to become inadequate in the future, especially with space-based detectors such as LISA or with third generation Earth-based detectors such as the Einstein Telescope (c.f.~next subsections). As we will see in the following of this review, depending on the generation model, there are several predictions for the spectral shape of a cosmological SGWB, which go beyond simple power laws. Often, this constitutes the only handle one has to possibly distinguish the origin of the SGWB. Nevertheless, the Power Law-Integrated Curve remains the best currently available visualisation of the sensitivity of detectors to SGWB, and this is what we plot in Fig.~\ref{fig:detectors} for PTA, LISA and several observation runs of advanced LIGO/Virgo.

\subsubsection{Earth-based detectors.} 

The first generation of earth-based GW interferometers included four detectors: LIGO, Virgo, GEO 600 and TAMA. 
The {\bf GEO 600} interferometer \cite{geowebsite} is a 600-meter interferometer located near Hannover (Germany), with best sensitivity of $h_f\sim  3 \cdot 10^{-22}~{\rm Hz}^{-1/2}$ at about $700$ Hz. The {\bf TAMA} interferometer \cite{tamawebsite} has 300-meter arm, is located near Toyko (Japan), and was the first of the network to be operational: it reached its best sensitivity of $h_f\sim 1.5 \cdot 10^{-21}~{\rm Hz}^{-1/2}$ at about 1 kHz. Both these interferometers have been important for testing several technical innovations needed for the second generation experiments. {\bf LIGO (Laser Interferometer Gravitational-wave Observatory)} \cite{Harry:2010zz} consisted of three multi-kilometers interferometers, at two locations: H1 with 4 km arms and H2 with 2 km arms located at Hanford (Washington State), and L1 also with 4 km arm located in Livingston Parish (Louisiana). They have been operational since 2002. The {\bf Virgo} interferometer \cite{Acernese:2015gua}, located near Pisa (Italy), has a 3km arm. 

In general, for ground based detectors, the most important limitations to sensitivity result from the effects of seismic noise in the low frequency range (up to a few Hz), thermal noise associated with the test masses and their suspensions (up to a few hundreds Hz), and finally laser shot noise (from a few hundreds to $10^4$ Hz). All these aspects have been improved in the upgrade that brought the initial devices to the {\bf the second generation of GW detectors}, {\bf Advanced LIGO} \cite{ALwebsite} and {\bf Advanced Virgo} \cite{AVwebsite}. At the level of sensitivity reached by the first generation of Earth-based interferometers, there was no guaranteed source of detection. In 2015, the planned upgrade of LIGO into advanced LIGO was ready, with an improvement in sensitivity by about a factor of ten. This enhanced considerably the physics reach: the distance searched for detectable sources improves linearly with the strain sensitivity, so an order of magnitude in strain sensitivity means the ability to probe a thousand times more volume in the sky. This has guaranteed the first direct detection of GWs, announced in February 2016: the GW emission from the coalescence of two black holes with masses around $30\,M_\odot$ \cite{Abbott:2016blz}. Since, the detection of other three black hole binary coalescences have been announced, performed by the Advanced LIGO interferometers only \cite{Abbott:2016nmj,Abbott:2017vtc,Abbott:2017gyy}. In the spring of 2017, Advanced Virgo was ready to join the search, and not much later the network of three detectors performed the first the common detection of another black hole binary coalescence \cite{Abbott:2017oio}, followed by the first detection of the merger of two neutron stars that, most remarkably, has been detected at the same time in various electromagnetic bands \cite{TheLIGOScientific:2017qsa,GBM:2017lvd}. 

Concerning stochastic backgrounds, so far there is no detection but only upper bounds. The most recent one has been obtained with the data of the first Advanced LIGO run (without Advanced Virgo) in \cite{TheLIGOScientific:2016dpb}. We have seen in Eq.~(\ref{dec_Q}) that the ideal filter function $\tilde Q(f)$ depends on the shape of the GW spectrum: Ref.~\cite{TheLIGOScientific:2016dpb} assumes then a power law template GW spectrum of the form $\Omega_{\rm GW} (f) = \Omega_\alpha (f/25\,{\rm Hz})^\alpha$. This procedure yields the 95\% confidence upper limit $\Omega_{\alpha=0} < 1.7\cdot 10^{-7}$. For other values of the power index $\alpha$ in the range between -5 and 5, the 95\% upper limit varies between about $2\cdot 10^{-7}$ and about $4 \cdot 10^{-10}$ (c.f. Figure 2 of Ref.~\cite{TheLIGOScientific:2016dpb}). These constraints are stronger than both the BBN (c.f. Section \ref{sec:BBN}) and integrated CMB (c.f. Section \ref{sec:CMB}) constraints in the above mentioned frequency range. 

The next technological step is to use cryogenic mirrors, which is now being explored by the KAGRA collaboration. {\bf KAGRA} \cite{KAGRAwebsite} is a 3 km arm interferometer currently under construction, which besides using cooled mirrors to reduce the thermal vibrations, is placed underground in the Kamioka mine (Japan) to suppress also the seismic noise. KAGRA should join the network in the next years, and can be considered a path-finder for the third generation GW detectors to be placed underground (c.f. next paragraph). Furthermore, there is a planned advanced detector to be located in India (IndIGO) \cite{indigowebsite}. 

Since the second generation of Earth-based GW interferometers is expected to get to the lowest possible sensitivity given their technical structure, a conceptual design study for a {\bf third generation detector, the Einstein Telescope (E.T.)}, has been funded by the European Framework Programme FP7 \cite{Sathyaprakash:2012jk}. The third generation detector should be sensitive to a great variety of sources, at much larger distances and with higher signal to noise ratio than Advanced LIGO and Virgo. Via the implementation of new technologies, the aim is to provide a strain sensitivity about ten times better than second generation detectors (corresponding to scanning a thousand times larger volume of the universe), and to shift the minimal detectable frequency to approximately 1 Hz. Consequently, E.T. will allow to probe the stochastic background down to a level of $\Omega_{\rm GW} \sim 10^{-12} $\cite{Sathyaprakash:2011bh})=. A proposed configuration \cite{Freise:2008dk,Hild:2008ng} is that of a 10-km arm triangular set of three Michelson interferometers, situated underground to significantly reduce seismic noise and allow for very long low frequency suspensions, with 500 W lasers using squeezed light to beat down quantum noise, and cryogenic test masses of 120 Kg kept at 20 K. The arms of the triangle, which has an opening angle of 60 degrees, are each used twice to form three co-located interferometers, which allows to measure both GW polarization states, and to use time delay interferometry: a technique developed to suppress the noise in LISA (c.f. section \ref{sec:space}), which consists in constructing virtual output signals  by time shifting and linearly combining the actual interferometer output signals. As an example, one can construct the Sagnac observable \cite{Hogan:2001jn}, a combination of interferometric output signals that is insensitive to GW and can be used to firmly identify the instrument noise, improving greatly the performance of the instrument. Another proposed configuration for E.T. is the xylophone design \cite{Hild:2009ns,Hild:2010id}: this is composed by a high-power, high-frequency interferometer (ET-HF), which employs powerful lasers to suppress the high-frequency photon shot noise, and a cryogenic low-power, low- frequency interferometer (ET-LF), with less powerful lasers reducing the thermal noise which would dominate at low frequency. The two interferometers would be co-located and with the same orientation, but only ET-LF would be situated underground. Compared to the single-interferometer design, the xylophone configuration improves the sensitivity by a factor of 2-10 in the frequency range 6-10 Hz.

\subsubsection{Space-based detectors.}  
\label{sec:space}

Earth-based interferometers, even situated underground, are limited by seismic noise and cannot probe frequencies smaller than about 1 Hz. It is possible to reach much lower frequencies placing interferometers in space: the principle is to put drag-free spacecrafts into orbit, and compare the distances between test masses in the spacecrafts using laser interferometry. 

The most advanced and long-studied space-based project is {\bf LISA (Laser Interferometer Space Antenna)}: it consists in an array of three drag-free spacecrafts at the vertices of an equilateral triangle of side-length $2.5 \cdot 10^9$ m, orbiting at a distance of 1 AU from the Sun on a Earth-like orbit, but  20 degrees behind the Earth and inclined at 60 degrees with respect to the ecliptic \cite{Audley:2017drz}. This configuration allows to probe the frequency band between 0.1 mHz and 0.1 Hz, which is expected to be richly populated by signals from sources such galactic binaries of white dwarfs and neutron stars, stellar-origin black hole binaries, coalescing massive black hole binaries in the mass range $10^4-10^7~M_\odot$, binaries formed by a massive black hole and a stellar-mass compact object, and a stochastic background generated by cosmological sources \cite{Audley:2017drz}. 

The key technology for LISA, namely the ability to keep the test masses in free-fall with extremely low residual acceleration, was successfully tested by the {\bf LISA Pathfinder} mission in 2016 \cite{Armano:2016bkm}. This mission, consisting in one LISA arm reduced in one spaceship, has measured a differential acceleration among the two test masses that fully meets (and even overcomes, depending on the frequency range) the requirements for LISA. This has led the European Space Angency (ESA) to issue a call for mission concerning a GW observatory in space at the end of 2016, and in response to that call the LISA proposal has been accepted by ESA in spring 2017, in the configuration proposed in \cite{Audley:2017drz}. Currently, the planned launch date is 2034. 

In the LISA configuration, the direct reflection of laser light like in standard Michelson interferometers, is not feasible due to the large distance between the spacecrafts. Therefore, each arm is composed of two laser beams: the first one is sent out from a spacecraft, and received by the other spacecraft; there, instead of begin reflected, it is phase locked to the second laser beam, which is then send back to the first spacecraft. The combination of the emitted and sent back lasers gives the information about the arm length. The same procedure is repeated in the adjacent arms, and with the information about the length of the three arms one then constructs the interferometry signal. LISA has therefore six masses and six laser links joining the three satellites (two in opposite directions for each side of the triangle): this three-interferometer configuration was chosen first of all since it provides redundancy against component failure. Moreover, it allows to use time delay interferometry, i.e.~the use of virtual interferometric observables to effectively reduce the laser noise level \cite{Estabrook:2000ef}, and it improves the capability to disentangle an isotropic cosmological (or astrophysical) background from the instrumental noise through Sagnac calibration \cite{Hogan:2001jn}. As mentioned for the E.T., the symmetrized Sagnac observable is a combination of six interferometric signals that is much less sensitive to GWs at low frequencies than other combinations, and thus can be used to determine the instrumental noise level. Furthermore, concerning stochastic backgrounds, there is a considerable gain in moving to lower frequencies: the factor $f^3$ in Eq.~(\ref{det_OmGW1d}) leads to an improvement in the minimum detectable value of $h^2\Omega_{\rm GW}$ at  1 mHz by fifteen orders of magnitude with respect to the minimum detectable value at 100 Hz, for the same instrument strain sensitivity $h_f$. Therefore, whereas going to space prevents the possibility of increasing the detection performance through cross-correlation with other detectors (as in the case of Earth-based interferometers), on the other hand it provides a mean to retrieve a very high sensitivity to a stochastic background just by the opportunity of reaching a lower frequency range in the absence of seismic noise. LISA is expected to probe a SGWB down to a level of $h^2\Omega_{\rm GW}\sim 10^{-13}$ \cite{Audley:2017drz}. 

There are also concepts for two other space-based projects using optical interferometry, {\bf DECIGO (DECi-hertz Interferometer Gravitational Wave Observatory)} \cite{Seto:2001qf,Kawamura:2006up} and {\bf BBO (Big Bang Observer)} \cite{Crowder:2005nr,Harry:2006fi}. The proposed configuration for these two missions is quite similar: they consist of four LISA-like interferometers orbiting the Sun at 1 AU, two of which are co-located in a `star of David' shape, while the other two are ahead and behind by an angle of $2\pi/3$ respectively, on the same orbit. The reason for this design is that it allows to measure with high-precision the stochastic background by cross-correlating the outputs of the two overlapping constellations; while the other two constellations are there to improve the angular resolution, which is useful for characterizing and removing the compact binary `foreground' \cite{Cutler:2005qq}. The angular position of the source is determined by exploiting the differences in the arrival times of the GW at the different constellations. Both missions are designed to probe the 0.1-10 Hz frequency band, where the stochastic background from white dwarf binaries is absent: the primary goal of these missions is in fact to reach a sensitivity of about \cite{Cutler:2009qv,Kudoh:2005as} $\Omega_{\rm GW}\gtrsim 10^{-17}$, in order to detect the primordial stochastic background from inflation (see Section~\ref{sec:inf}). The dominant astrophysical foreground in this band is due to compact binaries of neutron stars and stellar-origin black holes; however, the concept missions are planned to be sufficiently sensitive to individually detect and subtract out every merging compact binary out to high redshift, thereby uncovering the primordial GW background. Note that this `foreground removal', which in practice consists in the detection of hundreds of thousands of merging binaries, allows to use this kind of detectors also for doing precision cosmology \cite{Cutler:2009qv} and tests of general relativity \cite{Yagi:2013du}. The BBO mission is a follow-up of LISA with the previously described constellation, with shorter arms of $5\cdot 10^7$ m, and exploiting very powerful lasers of about 300W. On the other hand, the DECIGO mission, though sharing the same constellation, would have shorter arms of $10^6$ m which form Fabry-Perot cavities, i.e.~the lasers (of 10 W in this case) are reflected among the arms, and would be 2 to 3 times less sensitive than BBO. 

There are also proposals for space-based missions using atom interferometry instead of optical interferometry, like {\bf AGIS (Atomic Gravitational-wave Interferometric Sensor)} and the more recent proposal AGIS-LEO \cite{Dimopoulos:2008sv,Hogan:2010fz}. The principle of such a GW measurement is to combine the use of the atom interferometry with lasers traveling a long distance. The atom wave function is first split coherently by a pulse of light, so that the atom follows a superposition of two spatially separated free-fall paths. Subsequent lights pulses are then used to redirect and finally recombine the atom trajectories. When the atom wave function is recombined, the resulting interference pattern depends on the relative phase accumulated along the two paths, which contains both a contribution due to the free evolution of the wave function and a contribution due to the local phase of the laser. The GW strain changes the light travel time between the atom and the laser, contributing then to the total phase shift. To remove the laser phase noise and vibration noise, the experiment is constructed such that it compares the phase shifts of two separate atom interferometers that are manipulated by the same laser, and that are therefore subject to the same laser noise. The differential phase shift is still sensitive to the GW strain, while the laser noise is suppressed as a common mode. Typically, the separation between the two interferometers is of the order of 1000 km, therefore probing the frequency band $10^{-2}-10$ Hz, with a strain sensitivity of the order of \cite{Hohensee:2010mw} $h\sim 10^{-18}~{\rm Hz}^{-1/2}$.

\section{Inflationary period, part I: irreducible GW background}
\label{sec:inf}

The inflationary period, defined as an early phase of accelerated expansion, provides a natural solution to the shortcomings of the hot Big Bang framework~\cite{Guth:1980zm, Linde:1981mu, Albrecht:1982wi}, namely the horizon and flatness problems; see also Refs.~\cite{Brout:1977ix,Starobinsky:1980te,Kazanas:1980tx,Sato:1980yn} for early works. The major success of inflation is to provide an explanation for the physical origin of the primordial density perturbations, required to start the process of structure formation in the Universe. 
Inflation leads naturally to the stretching of quantum fluctuations~\cite{Mukhanov:1981xt, Guth:1982ec, Starobinsky:1982ee, Hawking:1982cz, Bardeen:1983qw}, which result parametrically amplified into classical density perturbations~\cite{Polarski:1995jg,Kiefer:1998qe,Kiefer:1998pb}. Later on, during the decelerated evolution of the universe following after the inflationary period, the perturbations re-enter the Hubble radius, providing the required `seed' to trigger, via gravitational collapse, the formation of structures in the universe. The perturbations leave at the same time an imprint in the CMB, in the form of temperature and polarization anisotropies. The simplest models of inflation lead to a homogeneous and isotropic spatially flat universe, with adiabatic, Gaussian, and approximately scale-invariant density perturbations. These predictions have been spectacularly confirmed over the years by increasingly accurate observations of the CMB and of the large-scale structure in the universe.

During inflation any light field with a mass smaller than the the inflationary Hubble rate $m^2 \ll H^2$, experiences quantum fluctuations. Due to the accelerated expansion, initially small fluctuations with wavelength smaller than the inflationary Hubble radius, $k > a H$, result amplified and stretched to super-Hubble scales, $k < a H$. This applies, in particular, to tensor metric perturbations~\cite{Grishchuk:1974ny,Starobinsky:1979ty, Rubakov:1982df,Fabbri:1983us}, as these correspond to massless fields. As we will show in detail, the resulting spectrum of tensor modes is quasi scale-invariant, spanning over a wide range of scales (from the Hubble scale at the end of inflation, to at least the Hubble scale today). When the tensor modes re-enter the Hubble radius during the post-inflationary era, they turn into a proper classical (yet stochastic) background of GWs, with a quasi scale-invariant spectrum. This background constitutes an irreducible emission of GWs expected from any inflationary model.

The irreducible background of GWs from inflation is expected to create a pattern of B-modes in the polarization of the CMB~\cite{Kamionkowski:1996zd,Seljak:1996gy,Zaldarriaga:1996xe,Kamionkowski:1996ks}. This major prediction from inflation remains however unverified, as to date (May~2018) this effect has not been observed\footnote{Let us recall that on March 2014, the BICEP2 collaboration announced the detection of B-modes due to the irreducible background of GWs from inflation~\cite{Ade:2014xna}. Even though the detection of B-modes was very real (later on confirmed by other experiments), unfortunately the interpretation of the signal as due to inflationary tensors was mistaken. An underestimation of the contribution to polarized light from dust in the interstelar medium~\cite{Flauger:2014qra,Mortonson:2014bja,Adam:2014bub}, led the BICEP2 team to consider such dust contribution negligible as compared to the measured signal. It turned out that the signal was however only due to (or at least mostly dominated by) the dust contamination~\cite{Ade:2015lrj,Ade:2015tva}.}. If B-modes due to primordial tensors are eventually detected in the CMB, this will constitute a very strong evidence in favor of inflation\footnote{This will not be, however, a definitive proof of inflation. It has been shown nonetheless, that primordial sources of B-modes (other than astrophysical ones) such as primordial magnetic fields or topological defects, do not produce a B-mode angular spectrum resembling close enough the inflationary one, unless parameters in these scenarios are highly (and unnaturally) fine-tuned~\cite{Bonvin:2014xia,Lizarraga:2014eaa,Moss:2014cra,Durrer:2014raa,Lizarraga:2014xza}. Thus, in principle, in the event of a future detection of a primordial signal (assuming astrophysical contaminants have been properly removed), the shape of the B-mode angular power spectrum could be a good discriminant by itself, to differentiate whether a signal is due to the inflationary irreducible background of GWs, or rather due to other primordial sources.}. Besides, a detection of primordial B-modes will provide a powerful tool to discriminate between the (currently) many different inflationary models compatible with the data. 

Multiple CMB polarization B-mode experiments are currently ongoing or under construction, aiming to detect or further constrain in the near future, the irreducible GW background from inflation. In light of present CMB constraints [recall Eqs.~(\ref{eq:rbound}), (\ref{eq:omegaKnt})], we know however that the amplitude of this background is, unfortunately, too small [see e.g.~Eq.~(\ref{eq:maxHubbleInf})] to expect detecting it with current or planned direct detection GW observatories like aLIGO/Virgo, LISA, ET and others. If the energy scale of inflation is sufficiently high, futuristic GW detectors such as BBO may perhaps have a chance to detect this primordial background. As we will see in Section~\ref{sec:infII}, the irreducible contribution may not be, however, the only GW background expected from inflation. Under some circumstances, if new species or symmetries are at play during inflation, GWs with a high amplitude and a significant deviation from scale-invariance, can also be produced. Contrary to the irreducible contribution, these backgrounds are model dependent. However, whenever produced, they are expected to have a much larger amplitude than the irreducible background, particularly at the frequencies accessible to direct detection experiments. In light of this possibility, GWs from inflation remain a relevant target for the upcoming ground- and space-based interferometers. 

In this section, we first describe in Sect.~\ref{sec:PrimTensInf} the irreducible background expected from any inflationary model, due to the amplification of initial quantum fluctuations of the gravitational field. In Sect.~\ref{sec:EvolTensInf} we discuss the evolution of this primordial background until the present, including post-inflationary effects that may affect its present-day amplitude. We postpone for Section~\ref{sec:infII} the discussion of production mechanisms during inflation different than the standard paradigm of quantum fluctuations, leading to GW backgrounds beyond the irreducible contribution, raising up in frequency (i.e.~`blue tilted').

\subsection{Irreducible GW background: amplification of vacuum fluctuations}
\label{sec:PrimTensInf}

The simplest inflationary models involve a single inflaton scalar field $\phi$ slowly rolling down its potential $V(\phi)$ during inflation, minimally coupled to gravity, and with a canonical kinetic term. We will refer to these models as canonically normalized single-field slow-roll (SFSR) scenarios. These are characterized by an action
\be
\label{eq:Sinf}
S = \int d^4 x \, \sqrt{-g} \, \left[\frac{\mP^2}{2}\,R - 
\frac{1}{2} \, \partial^\mu \phi \, \partial_\mu \phi - V(\phi)\right]\,,
\ee 
where $R$ is the Ricci scalar and $\mP = 1 / \sqrt{8 \pi G}$ is the reduced Planck mass. In the slow-roll regime the kinetic energy of the scalar field is negligible compared to its potential energy, ${1\over2}\dot{\phi}^2 \ll V(\phi)$. This is a requisite to inflate the universe. To sustain this regime sufficiently long, it is also necessary that the acceleration of the field is suppressed compared to the field velocity per Hubble time, i.e.~$|\ddot{\phi}| \ll |\dot\phi|/H^{-1}$. These two conditions allow to simplify both the Friedmann equation for the homogeneous and isotropic background, and the equation of motion of the homogeneous inflaton, 
\bea
\label{eq:slowroll1}
3\mP^2 H^2 &=& V(\phi)\left(1+\epsilon_{\phi}/3\right) \simeq V(\phi)\\
\label{eq:slowroll2}
-V'(\phi) &=& 3H\dot{\phi}\left(1-\eta_{\phi}/3\right) \simeq 3H\dot{\phi}
\eea
where 
we have defined the two slow-roll parameters 
\bea
\epsilon_\phi \equiv {3\over 2}{ {\dot\phi}^2\over V}\,, \hspace*{1cm}
\eta_\phi \equiv -{\ddot \phi\over H\dot\phi}~~\,. 
\label{eq:slowrollKin}
\eea
The approximations in $rhs$ of Eqs.~(\ref{eq:slowroll1}), (\ref{eq:slowroll2}) are consistent only as long as both slow-roll parameters are sufficiently small, $\epsilon_{\phi} \ll 1$ and $\eta_{\phi} \ll 1$. It is then useful also to define the potential slow-roll parameters 
\bea
\epsilon_V \equiv \frac{\mP^2}{2} \left(\frac{V'}{V}\right)^2 \,,\hspace*{1cm} \eta_V \equiv \mP^2 \, \frac{V''}{V}~\,,
\label{eq:slowroll}
\eea
related to the former parameters by $\epsilon_{\phi} \simeq \epsilon_V$ and $\eta_{\phi} \simeq \eta_V - \epsilon_V$. Demanding $\epsilon_V \ll 1$ and $\eta_V \ll 1$ represents therefore a sufficient (though not necessary) condition, for obtaining and sustaining an inflationary slow-roll regime. The $\epsilon$ slow-roll parameter controls the deviation of the equation of state from pure {\it de Sitter}, $w \equiv ({1\over2}{\dot\phi}^2 + V)/({1\over2}{\dot\phi}^2 - V) \simeq -1 + {2\over3}\epsilon_{\phi}$, hence determining as well the rate of change of the inflationary Hubble rate,
\begin{equation}\label{eq:eH}
\epsilon_{H} \equiv -{\dot H\over H^2} \equiv {3\over 2}(1+w) \simeq \epsilon_{\phi}\,.
\end{equation}
Note that even though we have argued that $\epsilon_{\phi} \simeq \epsilon_{V} \simeq \epsilon_{H}$, this double equivalence does not necessarily hold in scenarios beyond the SFSR paradigm. Nevertheless from now on, for simplicity we will write without distinction $\epsilon$, unless we explicitly go beyond SFSR.

Once inflation starts (say at some value $\epsilon_V \ll 1$), if the inflaton potential is sufficiently flat (i.e.~$\eta_V \ll 1$) over a wide range of scalar field values, this ensures that the universe will inflate during a sufficiently long period, so that the initial condition problems of the Hot Big Bang model are solved. In the slow-roll regime $\epsilon_V, \eta_V \ll 1$, the equation of state $w$ is close to $-1$, producing a quasi-exponential expansion $a(t) \sim e^{H \, t}$, with a Hubble rate $H$ approximately constant. In reality, according to Eq.~(\ref{eq:eH}), the Hubble rate decreases with time, but the decreasing rate is `slow-roll suppressed' as $\Delta H / H \simeq \epsilon \Delta N$, where $N$ is the number of e-folds $dN=d\log \,a\simeq H dt$.

Let us note that although the anisotropic stress of a scalar field $\sim \partial\delta\phi\partial\delta\phi$, can act as a source term in the equation of motion of tensor perturbations, it is intrinsically of second order in the field fluctuations. Therefore, to linear order in field fluctuations, there is typically no active source of GWs during inflation\footnote{GWs produced by a non-zero anisotropic stress at second order in the scalar perturbations will be discussed in Sect.~\ref{sec:SecGWScal}. Other sources of a non-zero anisotropic stress during inflation, like gauge fields excited through a topological term or scalar fields with a non-standard kinetic term, will be discussed in Sect.~\ref{sec:PartProdInf}.}. However, unavoidable quantum fluctuations of $h_{ij}$ are parametrically amplified by the quasi-exponential expansion of the universe. To describe this phenomenon, we need to quantize the tensor modes of the metric, considered as perturbations over the homogeneous and isotropic inflationary background.

The first step to quantize the system is to identify the canonical degrees of freedom. This can be done by expanding the pure gravitational part of action (\ref{eq:Sinf}) with the metric Eq.~(\ref{GWconf}), at second order in $h_{ij}$ and expressed in conformal time $\eta$~\cite{Mukhanov:1990me}
\bea\label{eq:Sg2}
S_g^{(2)} &=& - \, \frac{\mP^2}{8} \, \int d\eta \, d^3\mathbf{x} \, a^2(\eta) \, 
\eta^{\mu\nu} \, \partial_\mu h_{ij} \, \partial_\nu h_{ij} \nonumber \\
&=& \frac{\mP^2}{4} \, \sum_{r = + , \times} \, \int d\eta \, {d^3\mathbf{k}\over(2\pi)^3} \, a^2(\eta) \, 
\left[|h'_r(\mathbf{k}, \eta)|^2 - k^2 \, |h_r(\mathbf{k}, \eta)|^2\right] \\
&=& \frac{1}{2} \, \sum_{r = + , \times} \, \int d\eta \, {d^3\mathbf{k}\over(2\pi)^3} \, 
\left[|v'_r|^2 - k^2 \, |v_r|^2 + \frac{a''}{a} \, |v_r|^2\right] \, ,
\eea
where we have used the decomposition Eq.~(\ref{hrketa}) and the orthonormal condition (\ref{orthonormal}) in the second equality, whereas for the third equality we have introduced the variables
\be
\label{eq:vrhr}
v_r(\mathbf{k}, \eta) = \frac{\mP}{\sqrt{2}} \, a(\eta) \, h_r(\mathbf{k}, \eta) \, .
\ee
Action Eq.~(\ref{eq:Sg2}) is equivalent to the action of two real scalar fields $v_r(\mathbf{x}, \eta)$ in Minkowski spacetime, with canonically conjugate momenta $\pi_r(\mathbf{x}, \eta) \equiv v'_r(\mathbf{x}, \eta)$, and time-dependent frequency $\omega_k^2(\eta) = k^2 - \frac{a''}{a}$. 
The quantization proceeds by promoting $v_r$ and $\pi_r$ to operators $\hat{v}_r$ and $\hat{\pi}_r$ that satisfy the standard commutation relations on hypersurfaces of constant time $\eta$,
\bea
\label{eq:ETCR}
\left[\hat{v}_r(\mathbf{x}, \eta) \, , \, \hat{\pi}_{r'}(\mathbf{x'}, \eta) \right] &=& 
i \, \delta_{r r'} \, \delta^{(3)}(\mathbf{x} - \mathbf{x'})\,, \\
\left[\hat{v}_r(\mathbf{x}, \eta) \, , \, \hat{v}_{r'}(\mathbf{x'}, \eta) \right] &=& 
\left[\hat{\pi}_r(\mathbf{x}, \eta) \, , \, \hat{\pi}_{r'}(\mathbf{x'}, \eta) \right] = 0 \, . 
\eea 
The fields can be decomposed on the basis of the solutions of the dynamical equations derived from action Eq.~(\ref{eq:Sg2}). Since the background is spatially isotropic, we can write
\be
\label{eq:vhat}
\hat{v}_r(\mathbf{x}, \eta) = \int \frac{d^3\mathbf{k}}{(2 \pi)^{3}} \, 
\left[ v_k(\eta) \, e^{-i \mathbf{k} \mathbf{x}} \, \hat{a}_{\mathbf{k} r} + 
v_k^*(\eta) \, e^{+i \mathbf{k} \mathbf{x}} \, \hat{a}^+_{\mathbf{k} r} \right]\,,
\ee
with $\hat{a}^+_{\mathbf{k} r}$ and $\hat{a}_{\mathbf{k} r}$ creation and annihilation operators satisfying the usual commutation relations
\bea
\left[\hat{a}_{\mathbf{k} r} \, , \, \hat{a}^+_{\mathbf{k'} r'} \right] = 
(2 \pi)^{3}\delta_{r r'} \, \delta^{(3)}(\mathbf{k} - \mathbf{k'}) \,,\\
\left[\hat{a}_{\mathbf{k} r} \, , \, \hat{a}_{\mathbf{k'} r'} \right] = 
\left[\hat{a}^+_{\mathbf{k} r} \, , \, \hat{a}^+_{\mathbf{k'} r'} \right] = 0\,,
\label{eq:aa+} 
\eea
while the mode functions $v_k(\eta)$ satisfy the equation of motion
\be
\label{eq:eomv}
v''_k + \omega_k^2(\eta) \, v_k = 0\,,~~~~~~{\rm with} ~~~\omega_k(\eta)^2 \equiv k^2 - {a''\over a}~~.
\ee
Consistency between the commutation relations Eq.~(\ref{eq:ETCR}) and Eq.~(\ref{eq:aa+}) requires the normalization condition
\be
\label{eq:normalv}
v_k \, v_k^{\prime *} - v_k^* \, v_k^{\prime} = i \, .
\ee

Eq.~(\ref{eq:eomv}) describes an harmonic oscillator with a time-dependent frequency varying from $\omega_k^2 \simeq k^2$, when the modes are sub-Hubble $a H \ll k$, to $\omega_k^2 \simeq a''/a$ when the modes become super-Hubble $a H \gg k$. For sub-Hubble modes, Eq.~(\ref{eq:eomv}) reads $v_k''+ k^2 \simeq 0$, which has two linearly independent solutions, $v_k = c_{k,+}v_k^{(+)} + c_{k,-}v_k^{(-)}$, with $v_k^{(\pm)} \equiv e^{\mp ik\eta}$, and $c_k^{(\pm)}$ constants. Defining a vacuum state $|0\rangle$ as $\hat{a}_{\mathbf{k} r} \, |0\rangle = 0$, we can associate the annihilation operators $\hat{a}_{\mathbf{k} r}$ in Eq.~(\ref{eq:vhat}) to the `positive frequency modes'\footnote{They are referred to as `positive' because they correspond to the eigenfunctions of the energy operator $\hat H = i \partial_{\eta}$ with positive eigenvalues, $\hat H v_k^{(+)} = +k v_k^{(+)}$.} $v_k^{(+)}$. This standard prescription corresponds to the so-called Bunch-Davies vacuum. A discussion of the consequences of other vacuum prescriptions for the GW background can be found in Ref.~\cite{Hui:2001ce}. The initial condition is then set with $c_{k,-} = 0$, so that initially $v_k(\eta) \propto v_+ \propto e^{-ik\eta}$. The value of $c_{k,+} \neq 0$ is determined by the normalization condition Eq.~(\ref{eq:normalv}), so that the physical solution of Eq.~(\ref{eq:eomv}) for sub-Hubble modes, finally reads
\be
\label{eq:vsubH}
v_k \simeq \frac{e^{-i k \eta}}{\sqrt{2 k}} \hspace*{0.5cm},~~ \mbox{ for } \hspace*{0.5cm} k \gg a H \, .
\ee
The fluctuations with deep sub-Hubble wavelengths have therefore an amplitude exactly like in flat spacetime. This should not come as a surprise, as in the ultraviolet regime $k \gg a H$, the oscillations of any field fluctuations are dominated by the momenta, so one expects flat spacetime to be a good approximation. 
   
After a mode leaves the Hubble radius during inflation ($a H \gg k$), Eq.~(\ref{eq:eomv}) reads $v_k''/v_k \simeq a''/a$, which is satisfied by $v_k \simeq C_k \, a(\eta)$ (with $C_k$ a constant), up to a sub-leading term that becomes quickly negligible [see Eq.~(\ref{supHub})]. In the slow-roll regime, we can determine the constant $C_k$ by simply matching the super-Hubble solution with the sub-Hubble solution (\ref{eq:vsubH}) at $a H = k$. This gives
\be
\label{eq:vsupH}
C_ka_k = {1\over\sqrt{2k}} ~~~~~\Rightarrow ~~~~ |v_k(\eta)| \simeq \frac{H_k}{\sqrt{2 \, k^3}} \, a(\eta) \hspace*{0.5cm} 
\mbox{ for } \hspace*{0.5cm} k \ll a H\,, 
\ee
where a subscript $_k$ indicates, from now on, that the quantity is evaluated when the mode is crossing the Hubble radius during inflation, $a_k \, H_k = k$. As $h_k \propto v_k/a$, Eq.~(\ref{eq:vsupH}) indicates that the amplitude of the tensor fluctuation at super-Hubble scales is constant in time. Although Eq.~(\ref{eq:vsupH}) has been derived following a rather imprecise method, it provides nonetheless a very good approximation to the solution in the slow-roll regime. 

We can actually do much better, as Eq.~(\ref{eq:eomv}) admits analytic solutions, for constant slow-roll parameters. In particular, let us notice that if we consider $\epsilon_H \equiv - d\log H/d N$ constant, then we can write the Hubble rate during inflation as $H(N+\Delta N) \simeq H(N) e^{-\epsilon \, \Delta N}$. This implies that the conformal time is $\eta \equiv \int {dN\over aH}\simeq -{1\over(1-\epsilon)\mathcal{H}}$, with $\mathcal{H} \equiv aH$. Taking derivatives (with respect to conformal time) at both sides of the last expression, we obtain $\mathcal{H}' \simeq (1-\epsilon)\mathcal{H}^2$. Hence, the term involving the scale factor in Eq.~(\ref{eq:eomv}) is $a''/a \equiv \mathcal{H}' + \mathcal{H}^2 \simeq (2-\epsilon)\mathcal{H}^2 \simeq {(2-\epsilon)\over(1-\epsilon)^2\eta^2} \simeq {1\over\eta^2}(2+3\epsilon)$, so that Eq.~(\ref{eq:eomv}) can be written as
\begin{equation}\label{eq:eomInfNuEpilon}
v_k'' + \left[k^2 - {1\over\eta^2}\left(\nu^2-{1\over4}\right)\right]v_k = 0\,,~~~~~\nu \equiv {3\over2} + \epsilon ~~.
\end{equation}
A slow-roll parameter $\epsilon \neq 0$ represents, therefore, a linear deviation from the exact {\it de Sitter} value $\nu \equiv {3\over 2}$ corresponding to $\epsilon = 0$. The general solution to Eq.~(\ref{eq:eomInfNuEpilon}), for constant $\epsilon$, is
\begin{equation}
v_k = (-\eta)^{1/2}\left[c_1(k)H_\nu^{(1)}(-k\eta) +  c_2(k)H_\nu^{(2)}(-k\eta)\right]\,,
\end{equation}
where $H_\nu^{(1)}(-k\eta)$, $H_\nu^{(2)}(-k\eta)$ are Hankel functions of the first and second kind. In the deep ultraviolet regime $(-k\eta) \rightarrow \infty$, this general solution must match the plane-wave solution Eq.~(\ref{eq:vsubH}). Hence, using the large argument expansion of the Hankel functions, 
$H_\nu^{(1)}(x \gg 1)\simeq \sqrt{2\over\pi x} e^{i(x-\nu-\pi/4)}$, $H_\nu^{(2)}(x \gg 1)\simeq \sqrt{2\over\pi x} e^{-i(x-\nu-\pi/4)}$, we conclude that $c_2(k) = 0$ and $c_1(k) = {\sqrt{\pi}\over 2}e^{{i\over2}\left(\nu+{1\over2}\right)}$. The exact solution then becomes
\begin{equation}\label{eq:vkExact}
v_k = {\sqrt{\pi}\over 2}e^{{i\over2}\left(\nu+{1\over2}\right)}\sqrt{-\eta}\,H_\nu^{(1)}(-k\eta)\,,~~\forall\,k, \eta ~~.
\end{equation}
Depending on the scale, this solution reduces to 
\begin{eqnarray}\label{eq:vkExpandedsubH}
v_k & ~~~\simeq ~~~ & \frac{e^{-i k \eta}}{\sqrt{2 k}}  \hspace*{5.65cm}\,,~~~{\rm for}~-k\eta \gg 1\vspace*{0.2cm}\\
\label{eq:vkExpandedsupH}
v_k & ~~~\simeq ~~~ & e^{i{\pi\over2}\left(\nu-{1\over2}\right)}2^{\left(\nu-{3\over2}\right)}{\Gamma(\nu)\over\Gamma(3/2)}{1\over\sqrt{2k}}(-k\eta)^{{1\over2}-\nu}\,~,~~~{\rm for}~-k\eta \ll 1\,,
\end{eqnarray}
where in the second expression we have used the small argument expansion $H_{\nu}^{(1)}(x \ll 1)$ $\simeq \sqrt{2\over\pi}e^{-i{\pi\over 2}}2^{\nu-{3\over2}}{\Gamma(\nu)\over\Gamma(3/2)}{1\over x^\nu}$. 
The exact solution Eq.~(\ref{eq:vkExact}) reduces correctly at sub-Hubble scales, to Eq.~(\ref{eq:vsubH}), as it should. At super-Hubble scales  and in the limit $\epsilon \rightarrow 0$, the amplitude of Eq.~(\ref{eq:vkExact}) reduces to $|v_k| \simeq -{1 \over \eta\sqrt{2k^3}}$, which thanks to $aH_k\eta = -1$, matches exactly Eq.~(\ref{eq:vsupH}) (which was derived for exact {\it de Sitter}). In reality, as one typically expects $\epsilon \neq 0$ (i.e.~the inflationary space-time is typically quasi-{\it de Sitter}), we see from the $-k\eta \ll 1$ limit in Eq.~(\ref{eq:vkExpandedsupH}), that $|v_k|$ has a tilt at super-Hubble scales; something we will discuss shortly.

Let us remark that Eq.~(\ref{eq:eomv}) describes an harmonic oscillator with a time-dependent frequency, which varies from $\omega_k^2 \simeq k^2$ to $\omega_k^2 \simeq a''/a$ ($\simeq -2 a^2 H^2$), when the initially sub-Hubble modes $a H \ll k$ eventually turn super-Hubble $a H \gg k$, due to the quasi-exponential expansion. When $\omega_k^2(\eta)$ varies only adiabatically in time, i.e.~$\omega_k' \ll \omega_k^2$, as long as $\omega_k^2(\eta)$ is positive, we can associate an occupation number $n_k$ to each mode $k$, so that $|\Delta \mathbf{k}|^3n_k$ represents the number density of gravitons with momenta $[\mathbf{k},\mathbf{k}+\Delta\mathbf{k}]$. This is given by the energy $E_k \equiv {1\over2}\left(|v'_k|^2 + \omega_k^2 |v_k|^2\right)$ per mode divided by the energy $\omega_k$ per particle,
\be
\label{eq:nk}
E_k = \left(n_k+{1\over2}\right)\omega_k ~~~~~\Rightarrow~~~~~ n_k+{1\over2} \equiv \frac{1}{2 \omega_k} \, \left(|v'_k|^2 + \omega_k^2 \, |v_k|^2\right)
\ee 
where the ${1\over2}$ term corresponds to the usual quantum vacuum contribution. Inserting the solution at deep sub-Hubble scales Eq.~(\ref{eq:vsubH}) into Eq.~(\ref{eq:nk}), gives $n_k = 0$, as it should for vacuum in flat space-time. The occupation number $n_k$ is an `adiabatic invariant' when $\omega_k^2(\eta)$ is positive and varies slowly as $|\omega_k'| \ll \omega_k^2$. However, the stretching of modes during the inflationary expansion violates both conditions, resulting in an abundant production of gravitons as the modes leave the Hubble radius, turning the initial $n_k = 0$ into $n_k \gg 1$. We can check that the solution for super-Hubble modes Eq.~(\ref{eq:vkExpandedsupH}) [Eq.~(\ref{eq:vsupH}) for exact {\it de Sitter}], corresponds in fact to a very large number of gravitons. Strictly speaking, the occupation number Eq.~(\ref{eq:nk}) is not well-defined during inflation, as $n_k$ is not adiabatically conserved during the inflationary period. Let us therefore evaluate it just after inflation, assuming an instantaneous transition into a power law expansion era $a(\eta) \propto \eta^p$ after inflation ($p \geq 1/2$), established at some time $\eta = \eta_e$. Using for simplicity the result in exact {\it de Sitter}, we can plug Eq.~(\ref{eq:vsupH}) into Eq.~(\ref{eq:nk}), and find that for super-Hubble modes $a_e \, H_e \gg k$, $n_k \sim (H_k / H_e)^2 \, (a_e \, H_e / k)^3$ if $p \neq 1$ [$n_k \sim (H_k / H_e)^2 \, (a_e \, H_e / k)^4$ for a RD background with $p = 1$]. Thus, super-Hubble modes exhibit a very large occupancy $n_k \gg 1$, as it corresponds to a large ensemble of gravitons. The originally quantum nature of the tensor perturbations (e.g.~non-commutation of variables) is lost (due to the time evolution leading to squeezing), but reflected in the stochastic nature of the emerging effectively classical field distribution. The quantum-to-classical transition, which occurs basically when the modes leave the Hubble radius, is studied in detail in Ref.~\cite{Polarski:1995jg,Kiefer:1998qe,Kiefer:1998pb}.  

In terms of the original GW field $h_{ij}$, Eqs.~(\ref{hrketa}), (\ref{eq:vrhr}) and (\ref{eq:vhat}), we can write
\be	\label{eq:hijPhysical}
\hat{h}_{ij}(\mathbf{x}, \eta) = \sum_{r = + , \times} \, \int \frac{d^3\mathbf{k}}{(2 \pi)^{3/2}} \, 
\left( h_k(\eta) \, e^{i \mathbf{k} \mathbf{x}} \, \hat{a}_{\mathbf{k} r} + 
h_k^*(\eta) \, e^{-i \mathbf{k} \mathbf{x}} \, \hat{a}^+_{\mathbf{k} r} \right) 
\, e_{ij}^{r}(\mathbf{\hat{k}})\,,
\ee  
with the amplitude of the physical tensor modes $h_k$ at super-Hubble scales determined by Eqs.~(\ref{eq:vrhr}) and Eq.~(\ref{eq:vkExpandedsupH}), as 
\be
\label{eq:hksupHarbitraryEpsilon}
|h_k(\eta)| \simeq \frac{H}{\mP \, k^{3/2}}\,f(\epsilon)\left({k\over aH}\right)^{-\epsilon}\,,\hspace*{0.5cm} \mbox{ for } \hspace*{0.5cm} k \ll a H
\ee
with $f(\epsilon) \equiv 2^{\epsilon}(1-\epsilon)^{1+\epsilon}{(\Gamma\left({3\over2}+\epsilon\right)/\Gamma\left({3\over2}\right))}$ $\simeq 1 - \left(1-\ln(2) - \psi_0\left({3\over2}\right)\right)\epsilon$ $\simeq 1 - 0.27\epsilon$, $\psi_0(x)$ the Digamma function. In the limit of exact {\it de Sitter} $\epsilon \rightarrow 0$, we obtain $\dot H \rightarrow 0$, $f(\epsilon \rightarrow 0) \rightarrow 1$ and $\left({k/aH}\right)^{-\epsilon} \rightarrow 1$. Hence the amplitude reduces to\footnote{Alternatively we could have deduced Eq.~(\ref{eq:hksupH}) by using Eq.~(\ref{eq:vsupH}) valid for exact {\it de Sitter}, instead of Eq.~(\ref{eq:vkExpandedsupH}).}
\be
\label{eq:hksupH}
\left|h_k(\eta)\right| 
~~~\longrightarrow~~~ \frac{H}{\mP \, k^{3/2}}\,,\hspace*{1cm} \mbox{ for } ~~\epsilon \rightarrow 0\,,~k \ll aH\,.
\ee
As discussed below Eq.~(\ref{supHub}), $h_k(\eta)$ remains constant in time after the modes leave the Hubble radius during inflation, until they re-enter the Hubble radius during the post-inflationary evolution. Eq.~(\ref{eq:hksupHarbitraryEpsilon}) evaluated at Hubble radius crossing $k = a_kH_k$ (which is exactly equivalent to Eq.~(\ref{eq:hksupH}) if we approximate $f(\epsilon_k) \simeq 1$), will thus provide the initial condition for the evolution of the modes once they re-enter the Hubble radius, to be discussed in the next sub-section. 

It is convenient to define a dimensionless {\it tensor power spectrum} $\mathcal{P}_h(k)$ as
\begin{eqnarray}\label{eq:Ph2}
\langle 0 | \hat{h}_{ij}(\mathbf{k},\eta)\hat{h}^*_{ij}(\mathbf{k}',\eta) | 0 \rangle = 
{2\pi^2\over k^3}\mathcal{P}_h(k)\delta^{(3)}(\mathbf{k}-\mathbf{k}')\,,
\end{eqnarray} 
so that 
\bea
\label{eq:defDeltah2}
\langle 0 | \hat{h}_{ij}(\mathbf{x}, \eta) \hat{h}_{ij}(\mathbf{x},\eta) | 0 \rangle = 
\int {{\rm d} k \over k}~\mathcal{P}_h(k)\,.
\eea
Using Eq.~(\ref{eq:hijPhysical}) evaluated at the super-Hubble solution Eq.~(\ref{eq:hksupHarbitraryEpsilon}), the orthonormal relation (\ref{orthonormal}) for the polarization tensor, and the commutation relations Eq.~(\ref{eq:aa+}) for the creation and annihilation operators, we obtain
\be
\label{eq:FullGWpowerInf}
\mathcal{P}_h(k) \simeq \frac{2}{\pi^2} \, \frac{H^2}{\mP^2} \, f^2(\epsilon)\left({k\over aH}\right)^{-2\epsilon} 
,\hspace*{0.5cm} \mbox{ for } \hspace*{0.5cm} k \ll a H\,.
\ee
At horizon-crossing $k = a_k H_k$, this expression reduces simply to (taking $f(\epsilon) \simeq 1$)
\be
\label{eq:Deltah2Prim}
\mathcal{P}_h(k) 
\simeq ~\frac{2}{\pi^2} \, \frac{H_k^2}{\mP^2}\,~, 
\hspace*{0.5cm} \mbox{ for } \hspace*{0.5cm} k = a_k H_k\,.
\ee
Since we saw above that the super-Hubble modes behave as a classical random field, the vacuum expectation value in Eq.~(\ref{eq:Ph2}) can be interpreted as a classical ensemble average over a stochastic field variable. The tensor power spectrum interpreted this way is then related to the characteristic GW amplitude introduced in Eqs.~(\ref{powerspec}), (\ref{hcketa}) by $\mathcal{P}_h(k) = 2h_c^2$.  

A small difference between the result in exact {\it de Sitter} Eq.~(\ref{eq:Deltah2Prim}) and quasi-{\it de Sitter} Eq.~(\ref{eq:FullGWpowerInf}), is the factor $f^2(\epsilon) \simeq 1 - 0.54\epsilon$, which simply amounts for a tiny correction in amplitude of $\sim 0.5(\epsilon/0.01) \%$. A more notable difference arises however due to the fact that the spectrum becomes slightly red-tilted in the quasi-{\it de Sitter} case, i.e.~$\mathcal{P}_h(k) \propto k^{n_T}$, with $n_{T} < 0$ but $|n_{T}| \ll 1$. More precisely, the spectral tilt $n_T(k)$ can be defined as\footnote{With this definition we also encompass the possibility that $n_T$ is a function of the scale $k$, even if this is not the case in canonical SFSR scenarios.}
\be\label{eq:ntDef}
n_T(k) = \frac{d \mathrm{log} \mathcal{P}_h(k)}{d \mathrm{log} k} \, .
\ee 
Applying this formula to Eq.~(\ref{eq:FullGWpowerInf}) leads immediately to the result for canonical SFSR inflation scenarios (where $\epsilon$ is small but non-vanishing),
\be
\label{eq:nT2e}
n_T \simeq - 2 \epsilon \, .  
\ee
Of course we could have directly read out this tilt from Eq.~(\ref{eq:FullGWpowerInf}) from the explicit $\mathcal{P}_h(k) \propto k^{n_T}$ behavior.

Let us emphasize that the redness of the spectrum $n_T < 0$, is a direct consequence of the fact that the amplitude of the tensor spectrum at horizon crossing 
Eq.~(\ref{eq:Deltah2Prim}), is directly proportional to the (inflaton potential) energy density $H_k^2 \propto V_k(1+\epsilon/3) \simeq V_k$. Because the Hubble rate decreases slowly during inflation, like $\dot{H} = -\epsilon_{H} H^2$ [recall Eq.~(\ref{eq:eH})], the amplitude of the spectrum Eq.~(\ref{eq:Deltah2Prim}) at different moments of horizon crossing, changes accordingly to the change of $H_k^2$ as time goes by. Applying therefore Eq.~(\ref{eq:ntDef}) over the spectrum Eq.~(\ref{eq:Deltah2Prim}) at horizon crossing\footnote{Note that this is different from what we did previously, when we applied Eq.~(\ref{eq:ntDef}) to the general spectrum Eq.~(\ref{eq:FullGWpowerInf}) at arbitrary super-Hubble scales $-k\eta \ll 1$.}, gives
\be
\label{eq:nTepsilon}
n_T(k) \simeq \frac{d \mathrm{log} V_k}{d \mathrm{log} a_k} \simeq 
\frac{\dot{\phi}_k}{H_k} \, \frac{V'_k}{V_k} \equiv - (2\epsilon_{\phi})^{1/2}(2\epsilon_{V})^{1/2}\left(1-\epsilon_\phi/3\right)^{-1/2} \simeq -2 \epsilon \,. 
\ee
As expected, this alternative computation leads consistently to the same result as in Eq.~(\ref{eq:nT2e}).

The primordial scalar perturbations generated from inflation can be studied in a similar way as we did for GWs. In single field inflationary models, the so-called comoving curvature perturbation $\mathcal{R}$ is conserved on super-Hubble scales~\cite{Liddle:2000cg}. In slow-roll models with a canonical kinetic term, its (dimensionless) spectrum, 
\begin{eqnarray}\label{eq:eq:DeltaR2Spectrum}
\langle 0 | \mathcal{R}(\mathbf{k},\eta)\mathcal{R}^*(\mathbf{k}',\eta) | 0 \rangle = 
{2\pi^2\over k^3}\mathcal{P}_{\mathcal R}(k)\delta^{(3)}(\mathbf{k}-\mathbf{k}')\,,
\end{eqnarray} 
takes a value at horizon crossing as
\be
\label{eq:DeltaR2}
\mathcal{P}_{\mathcal{R}} \simeq \frac{1}{4 \pi^2} \frac{H_k^4}{\dot{\phi}_k^2} \simeq {H_k^2\over 8\pi^2\epsilon m_{\rm Pl}^2}
\hspace*{0.5cm} \mbox{ for } \hspace*{0.5cm} k = a_k H_k \, .
\ee
The relative contribution of GW is often indicated by the tensor-to-scalar ratio $r$, defined as
\be\label{eq:T2Sratio}
r = \frac{\mathcal{P}_h(k)}{\mathcal{P}_{\mathcal{R}}(k)} \, .
\ee
Using Eqs.~(\ref{eq:Deltah2Prim}), (\ref{eq:DeltaR2}) for the spectra and the slow-roll equations (\ref{eq:slowroll1}, \ref{eq:slowroll2}), we find that, at horizon crossing $k = a_kH_k$, 
\be
r_k = 16 \, \epsilon \, .
\ee
Together with Eq.~(\ref{eq:nTepsilon}), this gives the so-called consistency relation for SFSR inflationary models (at the lowest order in the slow-roll parameters),
\be
\label{consistency}
n_T(k) = - \frac{r_k}{8} \, .
\ee
Remarkably, this relation is independent of the micro-physical details of the potential that is responsible for inflation: it only involves quantities that are in principle observable. The observational verification of this relation would provide a spectacular confirmation of the simplest models of inflation, and would certainly constitute a definite proof for inflation. As we will see, this relation can be modified however in inflationary models, e.g.~if several scalar fields are involved~\cite{Polarski:1995zn, GarciaBellido:1995qq} or if vacua prescriptions differ from the standard Bunch-Davies vacuum~\cite{Hui:2001ce}. 

To conclude this part, let us translate the CMB constraint on $r$ into a bound for $\Omega_{\rm GW}$. In particular, using the second term in the $rhs$ of Eq.~(\ref{eq:omegaKnt}) (corresponding to the modes crossing the horizon during RD), the CMB bound $r_{0.05} \leq 0.07$ in Eq.~\eqref{eq:rbound} (obtained assuming $n_T = 0$) can actually be re-written as an upper bound for the sensitivity that any direct detection experiment should achieve in order to detect the inflationary background (here we ignore changes in the number of relativistic species), 
\begin{eqnarray}\label{eq:maxHubbleInf}
\Omega_{\rm GW} \simeq 5 \cdot 10^{-16}\left({H_{k}\over H_{\rm max}}\right)^2\,,
\end{eqnarray}
where $H_{k}$ is the inflationary Hubble rate (at the time when the CMB scales left the horizon), bounded by $H_{\rm max} \simeq 9\cdot 10^{13}$ GeV, the maximum Hubble rate tolerated by the CMB (obtained via the Friedmann equation converting the maximum tensor-to-scalar ratio $r_{0.05} = 0.07$ into a Hubble rate). Thus, even in the most favorable case, assuming almost an exact scale invariant tensor spectrum all the way from CMB scales to the shorter scales probed by GW interferometric detectors, today's amplitude of the inflationary background of GWs is below the sensitivity of all planned interferometers, except perhaps for BBO and only if $H_{k}$ is somehow smaller but still of the order of magnitude of $H_{\rm max}$. In reality, even if the energy scale of inflation is not significantly lower than its current upper bound, the expected red-tilt of the background (which becomes more severe the smaller the scales we want to probe, as inflation progresses) will most likely make its detection impossible by interferometers, unless new detection technologies with much better sensitivity than a BBO-like experiment are eventually developed. Let us finally point out that the amplitude in Eq.~\eqref{eq:maxHubbleInf} is further reduced during the evolution of the universe by the change in the number of relativistic species, c.f. section \ref{sec:EvolTensInf} and especially Eq.~\eqref{eq:rhoGWrelspecies}.


\subsection{Evolution of the inflationary background after inflation}
\label{sec:EvolTensInf}
In section \ref{sec:general} we have derived the solution of the GW evolution equation at super-Hubble scales and found that its dominant part is constant in time, see Eq.~\eqref{supHub}. In the previous section~\ref{sec:PrimTensInf}, we have demonstrated that this behavior is indeed confirmed for the quantum tensor fluctuations produced during inflation, after they have become super-Hubble. We have specified solution \eqref{supHub} to the case of inflation, by choosing the correct behaviour at sub-Hubble scales (Bunch-Davies vacuum): this gives Eqs.~\eqref{eq:vkExpandedsubH}-\eqref{eq:vkExpandedsupH}, and in turn solution \eqref{eq:hksupHarbitraryEpsilon} [\eqref{eq:hksupH}) in the limit of exact {\it de Sitter}] at super-Hubble scales . 

Eq.~\eqref{eq:hksupHarbitraryEpsilon} [or Eq.~\eqref{eq:hksupH}] provides therefore an initial condition for the evolution of the tensor modes produced during inflation. When the modes re-enter the horizon\footnote{In the radiation and matter dominated phases, the horizon evolves in time like the Hubble scale, contrary to the inflationary period when it diverges, see e.g. \cite{DurrerBook}.} during the subsequent phases of the evolution of the universe, they acquire a time dependence, in particular they start oscillating and decaying like $1/a(\eta)$ (c.f. section \ref{sec:general}). The full solution for a generic power-law expansion factor is given in Eq.~\eqref{hsol_for_n}, for which one has to choose the right initial conditions for $k\eta\ll 1$, i.e.~a constant amplitude given by Eq.~\eqref{eq:hksupH}. For a universe that is radiation dominated, once the mode has re-entered the horizon, one finds 
\begin{equation}\label{eq:solradIC}
h_r^{\rm RD}({\bf k},\eta)= h^{\rm inf}(k) \,j_0(k\eta)\,,~~~~~~r = +, \times
\end{equation}
where
\begin{equation}
	h^{\rm inf}(k) \simeq \frac{H}{m_{\rm Pl}k^{3/2}}
\end{equation}
is the tensor amplitude set by inflation in the limit of exact {\it de~Sitter}. Note that the same expression \eqref{eq:solradIC} is valid for both polarisations. For a matter dominated universe, on the other hand, the relevant solution is 
\begin{equation}\label{eq:solmatIC}
h_r^{\rm MD}({\bf k},\eta)= h^{\rm inf}(k) \,\frac{3\,j_1(k\eta)}{k\eta}\,,~~~~~~r = +, \times~.
\end{equation}


To get the amplitude today of a tensor mode produced during inflation, one has to distinguish the modes that entered the horizon during the matter dominated era and those that entered the horizon during the radiation dominated era. In the first case, the relevant solution is simply Eq.~\eqref{eq:solmatIC}. In the second case, however, neither Eq.~\eqref{eq:solradIC} nor Eq.~\eqref{eq:solmatIC} apply. 
To find the result at present time, for a mode that crossed the horizon during radiation domination, one can match solution \eqref{eq:solradIC} at the time of radiation-matter equality with the full solution valid in the matter era with free coefficients (see Eq.~\eqref{hsol_for_n})
\begin{equation}\label{eq:solmatICfull}
	h_r^{\rm MD, full}({\bf k},\eta)= \bar A({\bf k})\frac{j_1(k\eta)}{k\eta}+\bar B({\bf k})\frac{y_1(k\eta)}{k\eta}\,.
\end{equation}
This procedure holds under the assumption that the radiation-matter transition is instantaneous. We therefore expect it to fail for modes which are comparable to the inverse duration of the transition \cite{Pritchard:2004qp}. A solution valid for these modes as well\footnote{For modes that enter the horizon after the universe starts accelerating at late-time, c.f.~the discussion around Eq.~\eqref{eq:TMDTLambda}.} can be obtained by integrating numerically the GW evolution equation~\eqref{gweq3}, accounting for the fact that the background evolves smoothly from radiation to matter-domination, i.e.~using the scale factor $a(\eta)=H_0^2\,\Omega_{\rm mat}\,a_0^3\,\eta^2/4+H_0\,\sqrt{\Omega_{\rm rad}}\,a_0^2\,\eta$ instead of the single power-law behaviours considered in section \ref{gweom}. Still, we present here the matching procedure in order to provide some analytical insight into the numerical solution. 

\begin{figure}
\begin{center}
\includegraphics[width=0.49\textwidth]{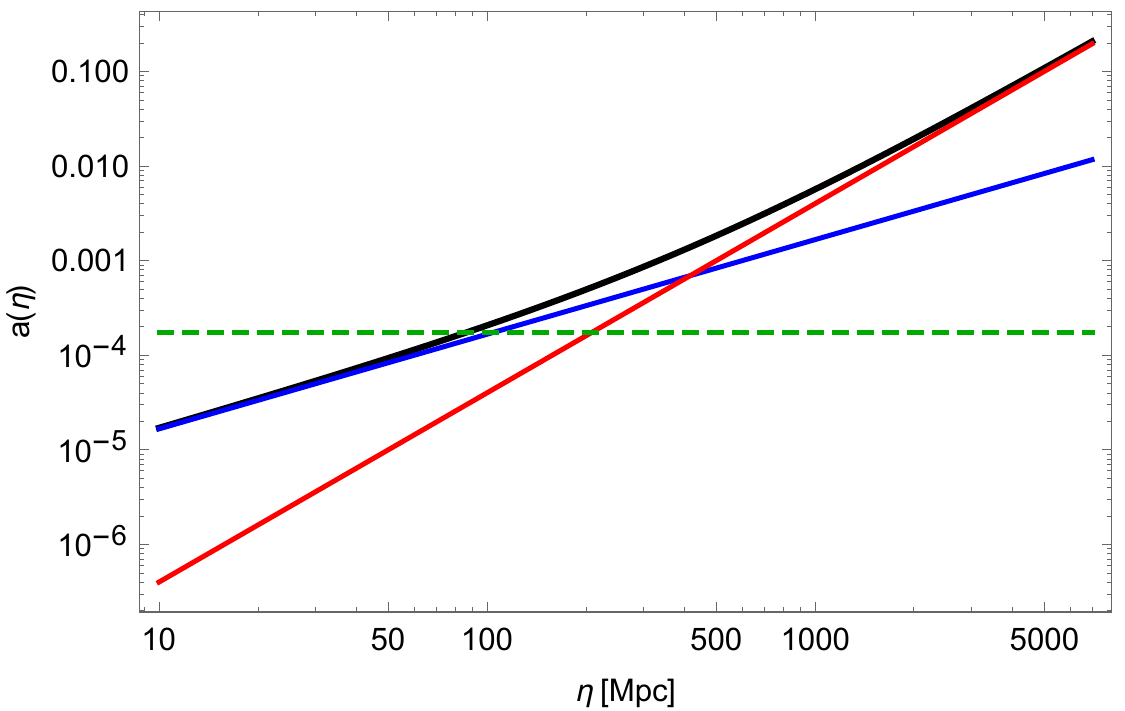}
\includegraphics[width=0.49\textwidth]{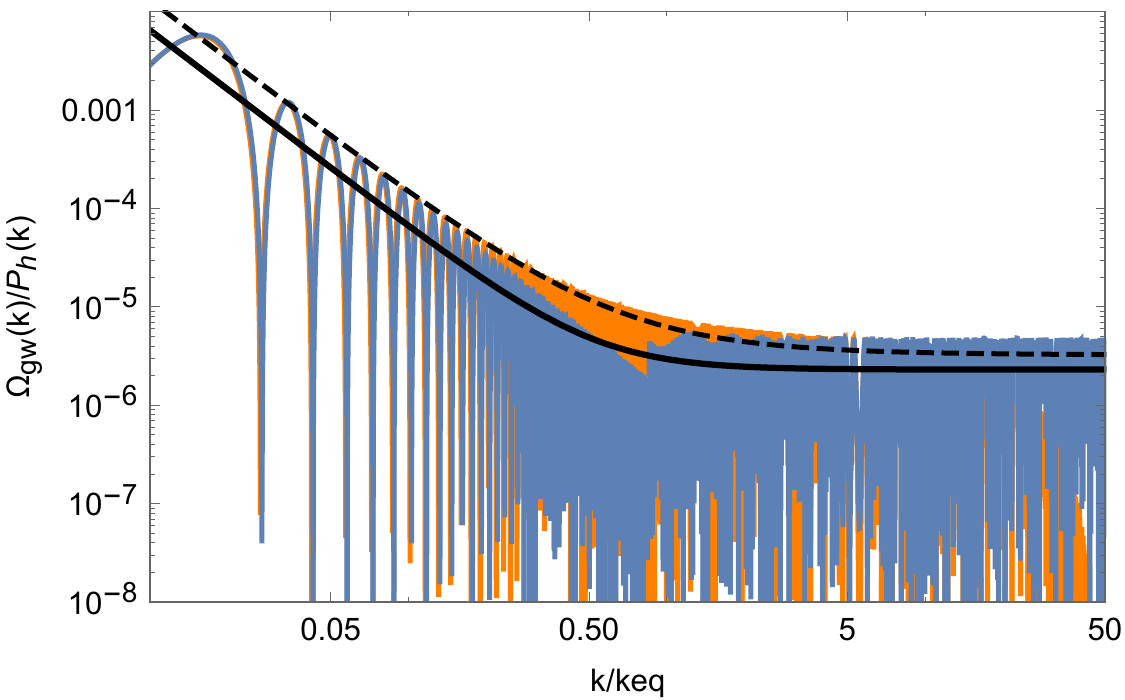}
\caption{\label{fig:TF} Left panel: the scale factor for a radiation-dominated universe (blue curve), a matter-dominated one (red curve), and a universe performing the transition from radiation to matter domination (black curve). The green, dashed line represents the scale factor at equality $a_{\rm eq}= a_0 (\Omega_{\rm rad}/\Omega_{\rm mat})$. Right panel: the GW energy density power spectrum, normalised to the primordial inflationary spectrum $\mathcal{P}_h(k)$, as a function of normalised wave-number $k/k_{\rm eq}$. The aim is to show the effect of the transfer function when modes enter the horizon (hence the normalisation). The orange curve represents Eq.~\eqref{eq:rhogwT}, where we have inserted the correct transfer function, found by numerically integrating the GW evolution equation through the radiation-matter transition. The blue curve is the analytical approximation, i.e. Eq.~\eqref{eq:rhogwT} inserting Eq.~\eqref{eq:Tfunct}. The black curve is approximation \eqref{eq:omegaK}, where we have accounted for oscillations by inserting a factor $1/2$. The black, dashed curve shows the result of Ref.~\cite{Turner:1993vb}, i.e. using Eq.~\eqref{eq:TWL} for the transfer function.}
\end{center}
\end{figure}

In order to proceed with the analytical approach, one must choose the time $\eta_*$, representing radiation-matter equality, at which to match the radiation solution \eqref{eq:solradIC} with the full solution \eqref{eq:solmatICfull}. One possibility is to use (note that the numerical values below are found using the cosmological parameters of \cite{Ade:2015xua} and setting $a_0=1$)
\begin{equation}
\eta_{\rm eq}=\frac{2(\sqrt{2}-1)\sqrt{\Omega_{\rm rad}}}{a_0 \,H_0\,\Omega_{\rm mat}}\simeq 86 \,{\rm Mpc}\,,\end{equation}
derived by solving $a_{\rm eq}= a_0 (\Omega_{\rm rad}/\Omega_{\rm mat}) = H_0^2\,\Omega_{\rm mat}\,a_0^3\,\eta_{\rm eq}^2/4+H_0\,\sqrt{\Omega_{\rm rad}}\,a_0^2\,\eta_{\rm eq}$. The corresponding equality scale $k_{\rm eq}$ (the wave-number entering the horizon at $\eta_{\rm eq}$) is 
\begin{equation}\label{eq:keq}
k_{\rm eq}=\frac{\sqrt{2}\,a_0\,H_0\,\Omega_{\rm mat}}{\sqrt{\Omega_{\rm rad}}}	\simeq 1.3\cdot 10^{-2}\,{\rm Mpc}^{-1}\,.
\end{equation}
However, from the left panel of Fig.~\ref{fig:TF} one can infer that a better choice for the matching time is the time at which the radiation $a_{\rm rad}(\eta)=H_0\sqrt{\Omega_{\rm rad}}\,a_0^2\,\eta$ and matter $a_{\rm mat}(\eta)=H_0^2\,\Omega_{\rm mat}\,a_0^3\,\eta^2/4$ solutions cross, i.e. $\eta_* = 4\sqrt{\Omega_{\rm rad}}/(a_0 H_0\Omega_{\rm mat})\simeq 417$ Mpc. Matching at $\eta_*$ better approximates the true background evolution, and therefore provides an analytical solution to the GW evolution equation that better reproduces the numerical one (as can be inferred from the right panel of Fig.~\ref{fig:TF}). The corresponding scale $k_*$ can be evaluated using the radiation solution $a_{\rm rad}(\eta)$: the wave-number entering the horizon at $\eta_*$ is then $k_*=1/\eta_*$.  

The free coefficients of Eq.~\eqref{eq:solmatICfull} are found via matching $h_r^{\rm MD, full}({\bf k},\eta)$ and its derivative $\partial_\eta h_r^{\rm MD, full}({\bf k},\eta)$, with $h_r^{\rm RD}({\bf k},\eta)$ and $\partial_\eta h_r^{\rm RD}({\bf k},\eta)$, respectively, at $\eta_*$. With $x_*=k\eta_*$, they read \cite{Watanabe:2006qe}:
\begin{eqnarray}
	\bar A(k)&=& h^{\rm inf}(k) \left[\frac{3}{2}-\frac{\cos(2x_*)}{2}+\frac{\sin(2x_*)}{x_*}\right]\,,\\
	\bar B(k)&=& h^{\rm inf}(k) \left[\frac{1}{x_*}-x_*-\frac{\sin(2x_*)}{2}-\frac{\cos(2x_*)}{x_*}\right]\,.
\end{eqnarray}
The full solution today can therefore be written as \cite{Watanabe:2006qe}
\begin{equation}\label{eq:h_inf_transfer}
	h_r({\bf k},\eta_0)= h^{\rm inf}(k) T(k,\eta_0)\,,
\end{equation}
where $T(k,\eta_0)$ is the {\it transfer function}:
\begin{eqnarray}\label{eq:Tfunct}
	T(k,\eta_0)=\left\{\begin{array}{lr}\vspace{0.2cm}
	\frac{3\,j_1(k\eta_0)}{k\eta_0}\,, & k<k_*\\
\frac{\bar{A}(k)}{h^{\rm inf}(k)}\frac{j_1(k\eta_0)}{k\eta_0} + \frac{\bar{B}(k)}{h^{\rm inf}(k)} \frac{y_1(k\eta_0)}{k\eta_0}\,, & k>k_*
	\end{array}\right.\,.
\end{eqnarray}

The GW energy density today from tensor modes produced during inflation becomes (c.f Eq.~\eqref{rhogw})
\begin{eqnarray}\label{eq:rhogwT}
	\rho_{\rm GW}(\eta_0)&=&\frac{\langle h'_{ij}({\bf x},\eta_0)h'_{ij}({\bf x},\eta_0)\rangle}{32\pi G a_0^2} \nonumber \\
	&=&\frac{1}{64\pi^3 G a_0^2}\int_0^\infty dk\,k^2 \,[T'(k,\eta_0)]^2\,|h^{\rm inf}(k)|^2\,,
\end{eqnarray}
where in the last equality we have used the decomposition in Eq.~\eqref{eq:hijPhysical} and relations \eqref{eq:aa+}. Using the above relation and definition \eqref{eq:defDeltah2}, the GW energy density parameter today can be written as
\begin{equation}\label{eq:omegaT}
	\Omega_{\rm GW}(k)=\frac{1}{12\,H_0^2\,a_0^2}\,[T'(k,\eta_0)]^2 \mathcal{P}_h(k)\,,
\end{equation}
with the inflationary tensor power spectrum given by
\begin{equation}
	\mathcal{P}_h(k)\simeq \frac{2}{\pi^2}\frac{H^2}{m_{\rm Pl}^2}\,,
\end{equation}
in the limit of exact de Sitter as we have seen in the previous section. Note that in general one is interested in Eq.~\eqref{eq:omegaT} at sub-horizon scales. Hence, it is customary to approximate $[T'(k,\eta_0)]^2\simeq k^2 \, T^2(k,\eta_0)$, given the oscillatory behaviour of the tensor modes inside the horizon (c.f. section \ref{gweom}). 

As described in~\cite{Boyle:2005se}, the in-phase oscillation of all modes with a given wave-number $k$ which re-enter the horizon at the same epoch, apparent in Eq.~\eqref{eq:Tfunct}, is a physical effect due to the common origin (inflation) of the modes. This effect is captured by the oscillating transfer function. However, from the observational point of view, as these modes correspond to a stochastic background of GWs, it is appropriate to average the transfer function over several oscillations. At sub-horizon scales $k\eta_0\gg 1$, and performing an oscillation-averaging procedure, one obtains 
\begin{eqnarray}\label{eq:Taverage}
[T'(k,\eta_0)]^2 \stackunder{\longrightarrow}{k\eta_0\gg 1}\left\{\begin{array}{lr}\vspace{0.2cm}
\eta_*^2/(2\eta_0^4)\,, & k>k_*\\
9/(2\eta_0^4\,k^2)\,, & k<k_*
\end{array}\right.\,.
\end{eqnarray}
By substituting the above equation in \eqref{eq:omegaT}, it appears that the energy density spectrum today of the tensor modes generated during inflation is flat in $k$ (assuming exact de Sitter) for modes that entered the horizon during the radiation era, and scaling as $k^{-2}$ for modes that entered the horizon during the matter era. 

The right panel of Fig.~\ref{fig:TF} shows the GW energy density power spectrum, calculated both by numerically integrating the GW evolution equation through the radiation-matter transition, and using the analytical transfer function \eqref{eq:Tfunct}. It appears that the analytical solution performs well, a part for modes around $k_{\rm eq}$, that enter the horizon during the radiation-matter transition. 

The GW energy density power spectrum can be approximated as 
\begin{equation}\label{eq:omegaK}
\Omega_{\rm GW}(k)=\frac{3}{128}\,\Omega_{\rm rad}\,\mathcal{P}_h(k)\,\left[\frac{1}{2}\left(\frac{k_{\rm eq}}{k}\right)^2+\frac{16}{9}\right]\,, 
\end{equation}
where we have used $\eta_0=2/[a_0H_0(\sqrt{\Omega_{\rm rad}}+\sqrt{\Omega_{\rm rad}+\Omega_{\rm mat}})]$, and we have interpolated oscillations simply by inserting a factor $1/2$. The pre-factor $3/128$ and the pivot wave-number $k_{\rm eq}$ have been chosen in order to recover Eq.~(4) of \cite{Lasky:2015lej}. Fig.~\ref{fig:TF} also shows the transfer function computed in Ref.~\cite{Turner:1993vb}, where it was derived for the first time (note that in this case oscillations have not been accounted for by introducing a factor $1/2$):
\begin{equation}\label{eq:TWL}
	[T'(k,\eta_0)]^2_{\rm TWL}= \left[\left(\frac{3\,j_1(k\eta_0)}{k\eta_0}\right)'\right]^2 \left(1 + 1.34\frac{k}{k_{\rm eq}} + 2.5\frac{k}{k_{\rm eq}}\right)\,.
\end{equation}


\vspace*{1cm}

The above analysis is valid only under several assumptions, some of which we are going to review now in some detail. Note that we concentrate only on the effects caused in the context of the standard model of particle physics and cosmology: for more exotic effects related to super-symmetry, the presence of dark fluids and dark interactions, exotic phase transitions or reheating models, see e.g. \cite{Watanabe:2006qe,Liu:2015psa,Seto:2003kc,Nakayama:2008wy,Boyle:2005se,Kuroyanagi:2008ye,Ghosh:2017jdy}. 

First of all, we have assumed that the transition between the radiation and the matter era is instantaneous. Consequently, solution \eqref{eq:h_inf_transfer} is valid only for modes with wavelength much larger than the duration of this transition. For solutions without this restriction, either evaluated under the WKB approximation or fully numerical, see \cite{Pritchard:2004qp,Turner:1993vb}. 

Second, we have assumed that the universe evolves from a phase of radiation domination with $a(\eta)\propto \eta$ to a phase of matter radiation with $a(\eta)\propto \eta^2$. This is quite simplistic since the evolution of the universe can be characterised by other phases, as for example the phase of late accelerated expansion following the matter era. In order to investigate how a change in the laws of the evolution of the universe affects the above analysis and the SGWB energy density spectrum, the simplest way is to use the solution given in Ref.~\cite{Boyle:2005se}. In this work, the behaviour of the GW amplitude at sub-horizon scale is derived by pushing the validity of the sub-horizon solution (e.g.~Eq.~\eqref{subHub}) up to horizon crossing, and matching it there with the constant inflationary solution. This gives
\begin{equation}\label{eq:hr_boyle}
	h_r\left(k>\frac{1}{\eta}, \,\eta\right) \simeq h^{\rm inf}(k) \cos[k(\eta-\eta_k)+\phi_k] \,\frac{a_k}{a(\eta)}\,,
\end{equation}
 where $a_k=k/H_k$ and $\eta_k$ denote respectively the scale factor and the conformal time at horizon crossing, and $\phi_k$ is a phase that does not interest us at this point. This solution is approximate, but for $k/\mathcal{H}\gg 1$ it recovers asymptotically the behaviour of Eq.~\eqref{eq:h_inf_transfer}-\eqref{eq:Tfunct}, as well as \eqref{eq:solradIC} and \eqref{eq:solmatIC}. The transfer function today becomes, using this solution, $T(k,\eta_0)\simeq \cos[k(\eta_0-\eta_k)+\phi_k] \,(a_k/a_0)$.  

Let us start with the radiation dominated era. In the above derivation (as well as in section \ref{sec:general}) we have assumed that the scale factor always evolves as $a(\eta)\propto \eta\propto 1/T$. However, this neglects the fact that particles content of the primordial thermal bath changes its nature as the temperature decreases, because particles become non-relativistic and/or get out of thermal equilibrium at different times. In the standard model this occurs for example at $e^+e^-$ annihilation, neutrino decoupling, the QCD phase transition, the EW phase transition, etc. When a given particle species gets out of thermal equilibrium, the effective number of relativistic species contributing to the entropy, $g_S(T)$, decreases, causing the scale factor to increase faster than $1/T$ during this phase (c.f. section \ref{sec:general}). Note that here we distinguish $g_S$, the effective number of relativistic species contributing to the entropy, from $g_*$, those contributing to the energy density: $g_S=g_*$ for $T> 0.1$ MeV (before neutrinos decouple), while they differ later (c.f. section \ref{sec:general}). The fast increase of $a(\eta)$ leads to an extra suppression of the tensor modes that entered the horizon before the time at which $g_S(T)$ changes. The extra suppression can be evaluated without modelling the time behaviour of the effective number of relativistic species in any detail, by using Eq.~\eqref{eq:hr_boyle} (c.f.~\cite{Watanabe:2006qe} for a more refined derivation). Let us suppose that $g_S(T)$ and $g_*(T)$ change at a time $\bar\eta$, going from $\bar g_S$, $\bar g_*$ to their value today $g_S^0$, $g_*^0$, and consider a mode that entered the horizon at $\eta_k<\bar\eta$. Using Eq.~\eqref{eq:hr_boyle} one can compare the amplitude of the tensor mode accounting for the change in $g_S(T)$, $g_*(T)$ with the one neglecting it, i.e.~assuming that $g_S$ and $g_*$ have always remained constant and equal to today's values, also in the past. Using $a(T)=(a_0\,T_0/T)(g_S^0/g_S(T))^{1/3}$, $\rho_{\rm rad}(T)=(\pi^2/30) g_*(T) T^4$, and identifying at horizon crossing $k=\mathcal{H}(T_k)=\sqrt{8\pi G/3}\, a_k \, \sqrt{\rho_{\rm rad}(T_k)}$, one obtains that the amplitude of the tensor mode with wave-number $k$ is suppressed as
\begin{eqnarray}
	\frac{h_r(k,\eta_0)|_{\bar g}}{h_r(k,\eta_0)|_{g^0}} \simeq \frac{(a_k/a_0)|_{\bar g}}{(a_k/a_0)|_{g^0}}= \frac{T_k|_{g^0}}{T_k|_{\bar g}}\left(\frac{g_S^0}{\bar g_S}\right)^{1/3}=\left(\frac{g_S^0}{\bar g_S}\right)^{2/3}\left(\frac{\bar g_*}{g_*^0}\right)^{1/2}\,,
\end{eqnarray}
where the notation is such that $|_{\bar g}$ means a quantity accounting for the change in the effective number of relativistic species, while $|_{g^0}$ neglecting it. For the energy density this leads to 
\begin{equation}
	\frac{\rho_{\rm GW}(k,\eta_0)|_{\bar g}}{\rho_{\rm GW}(k,\eta_0)|_{g^0}}\simeq \left(\frac{g_S^0}{\bar g_S}\right)^{4/3}\left(\frac{\bar g_*}{g_*^0}\right)\,. \label{eq:rhoGWrelspecies}
\end{equation}
Here we assumed a sudden decrease of $g_S$ and $g_*$ at $\bar T$, but actually $g_S(T)$ and $g_*(T)$ decrease continuously during the radiation dominated era for $T>0.1$ MeV. Therefore, the reduction in the tensor amplitude and energy density of a given mode $k$ is stronger, the earlier the mode enters the horizon. As a consequence, the energy density spectrum today is no longer scale invariant for modes that entered the horizon in the radiation era, as we have found above 
neglecting this effect (and assuming pure {\it de Sitter}, c.f. Eq.~\eqref{eq:omegaK}). If one accounts for the change in the effective number of relativistic degrees of freedom, the tensor energy spectrum today decreases with $k$, as shown for example in Fig.~(4) of \cite{Watanabe:2006qe}. 

A more exotic scenario that leads in fact to an enhancement of the GW energy density, is the presence of a stiff component in the evolution of the universe, i.e.~a component with equation of state parameter $w>1/3$ \cite{Giovannini:1999bh,Giovannini:1998bp,Boyle:2005se,Boyle:2007zx,Liu:2015psa,Figueroa:2016dsc}. If such a fluid is present in the early universe it would dominate over radiation at sufficiently early times, since the scale factor for a stiff component increases slower than radiation, $a(\eta)\propto \eta^{2/(3w+1)}$ with $2/(3w+1)<1$ for $w>1/3$ (and we know that at some point radiation must have dominated the universe). Clearly, a stiff component can only play a relevant role before BBN, when no constraint is present on the evolution of the universe. It can drive the expansion of the universe only for a finite amount of time, at some point between the end of inflation and the onset of BBN. During this phase, the amplitude of a GW background increases with respect to the standard radiation-dominated scenario. In fact, comparing the GW amplitudes in the two regimes as done previously, one obtains (note that $k\gg a_0 H_0$)
\begin{equation}
	\frac{h_r^w(k,\eta_0)}{h_r^{\rm rad}(k,\eta_0)} \simeq \frac{a_k^w}{a_k^{\rm rad}} \simeq \frac{\Omega_{\rm stiff}^{\frac{1}{3w+1}}}{\sqrt{\Omega_{\rm rad}}}\left(\frac{a_0 H_0}{k}\right)^{\frac{1-3w}{3w+1}} >1 ~~~~~~{\rm for}~w>\frac{1}{3}\,,
\end{equation}
where $\Omega_{\rm stiff}$ denotes the energy density parameter for the stiff fluid today, and we have used $a_k^{\rm rad}=H_0\sqrt{\Omega_{\rm rad}}\,a_0^2/k$ and 
\begin{equation}\label{eq:akw}
	a_k^w=H_0^{\frac{2}{3w+1}}\,a_0^{\frac{3(1+w)}{3w+1}}\,\Omega_{\rm stiff}^{\frac{1}{3w+1}}\, k^{-\frac{2}{3w+1}}\,.
\end{equation}
Note that $\eta_k=[2/(3w+1)k]^{2/(3w+1)}$. Furthermore, the GW energy density power spectrum becomes strongly blue-tilted, as opposed to the quasi-scale invariant case of the modes crossing during the radiation dominated era (c.f. Eq.~\eqref{eq:omegaK}). In order to evaluate the GW energy density spectrum today for the modes that crossed the horizon during the stiff era, let us first define $k_{\rm RD} = a_{\rm RD}H_{\rm RD}$, as the comoving horizon scale at the onset of radiation-domination, when the stiff fluid just became sub-dominant. The background of GWs is enhanced for the modes $k > k_{\rm RD}$ that crossed before the end of the stiff period. Substituting $T'(k,\eta_0)^2\simeq {k^2\over2} (a_k/a_0)^2$ derived from \eqref{eq:hr_boyle} in Eq.~\eqref{eq:omegaT} [the factor $1/2$ comes from averaging over the oscillatory transfer function], using Eq.~(\ref{eq:akw}) in $(a_k/a_0)^2 = (k_{\rm RD}/k)^{4/(1+3w)}(a_{\rm RD}/a_0)^2$, and using $(a_{\rm RD}/a_0)^4 = \Omega_{\rm rad}(H_0/H_{\rm RD})^2$, we obtain
\begin{equation}
	\Omega_{\rm GW}(k)= {\Omega_{\rm rad}\over 24}{\mathcal P}_h(k)\left(\frac{k}{k_{\rm RD}}\right)^{\frac{6w-2}{3w+1}}\,~~~~{\rm for}~ k > k_{\rm RD} \,.
\end{equation}
The blue-tilted scaling as $\Omega_{{\rm GW}} \propto (f/f_{\rm RD})^{(6w-2)(3w+1)}$ for $f > f_{\rm RD}$, represents a strong enhancement of the short-wave modes of the inflationary background, and opens up the possibility of direct detection of a SGWB of inflationary origin. The standard vacuum contribution from inflation for modes entering during the period of radiation-domination, $\Omega_{\rm GW}\big|_{\rm rad} = {\Omega_{\rm rad}\over 24}{\mathcal P}_h(k)$, is way below the sensitivity of present and future GW observatories as PTA, LISA, and advanced LIGO/Virgo, c.f. Fig.~\ref{fig:BlueSpec} and section \ref{sec:interferometers}. The enhancement $\Omega_{{\rm GW}}\big|_{\rm stiff}/\Omega_{{\rm GW}}\big|_{\rm rad} = (f/f_{\rm RD})^{(6w-2)(3w+1)}$ is larger the bigger the frequency, and the longer the phase of stiff-domination lasts (i.e.~the smaller the frequency $f_{RD}$ is). If the stiff period lasts until just before BBN, say with $f_{\rm RD} \sim 10^{-10}$ Hz, the enhancement at the LISA frequencies $f_{\rm LISA} \sim 10^{-3}$ Hz, can be quite significant for a super stiff phase with $w \simeq +1$, as $\Omega_{{\rm GW}}\big|_{\rm stiff}/\Omega_{{\rm GW}}\big|_{\rm rad} = (f_{\rm LISA}/f_{\rm RD}) \sim 10^{7}$. For an inflationary model with energy scale saturating the upper bound determined by CMB anisotropies, $H \leq H_{\rm max} \simeq 9\cdot 10^{13}$ GeV, $\Omega_{\rm GW}\big|_{\rm rad} = {\Omega_{\rm rad}\over 24}{\mathcal P}_h(k) \sim 5\cdot 10^{-16}$, and hence $\Omega_{\rm GW}\big|_{\rm stiff}(f \simeq f_{\rm LISA}) \sim 5\cdot 10^{-9}$, way above the sensitivity of LISA $\Omega_{{\rm GW}}(f \simeq f_{\rm LISA}) \sim 10^{-13}$.

As mentioned above, in our previous discussions we have neglected the presence of late-time acceleration (see e.g.~\cite{Turner:1993vb,Kuroyanagi:2009br}). Using solution \eqref{eq:hr_boyle}, we can approximately quantify the difference in the GW solution today accounting (quantities denoted with superscript ${\Lambda}$) and not accounting (quantities denoted with superscript $m$) for late-time acceleration. Without accounting for late-time acceleration, the scale factor at horizon crossing is given by $a_k^m=H_0^2\,\Omega_{\rm mat}^m\,a_0^3\,(\eta_k^m)^2/4$, with $\eta_k^m=2/k$: in this case, one must set $\Omega_{\rm mat}^m=1$ today if one assumes spatial flatness (note that we are neglecting the residual presence of radiation in the late time universe). When accounting for late-time acceleration, the solution for the scale factor does not have an explicit analytic form; however, before the effect of the cosmological constant becomes relevant, it is well approximated by the same solution as above, accounting in this case for the fact that $h^2\Omega_{\rm mat}=0.14$ today: $a_k^\Lambda \simeq H_0^2\,\Omega_{\rm mat}\,a_0^3\,(\eta_k^\Lambda)^2/4$ and again $\eta_k^\Lambda\simeq 2/k$. A numerical evaluation indicates that this solution is valid approximately until $\eta_{\Lambda}\simeq 8000$ Mpc, corresponding to $z\simeq 2.7$ and $k_{\Lambda}\simeq 2.5 \cdot 10^{-4} \,{\rm Mpc}^{-1}$. The ratio of the GW solution today accounting and not accounting for late-time acceleration, becomes then (using $a_k=k/H_k$, and normalising both solutions such that $a_0^{\Lambda}=a_0^{m}=1$)
\begin{eqnarray}\label{eq:TMDTLambda}
	\frac{h_r^{\Lambda}(k,\eta_0)}{h_r^{m}(k,\eta_0)}&\simeq& \frac{a_k^{\Lambda}}{a_k^{m}}=\frac{H_k^{m}(\eta_k^m)}{H_k^{\Lambda}(\eta_k^\Lambda)} = \frac{\sqrt{(1/a_k^{m})^{3}}}{\sqrt{\Omega_{\rm mat}/(a_k^\Lambda)^3+\Omega_\Lambda}} \\ 
	&\simeq & 
	\left\{ \begin{array}{lr}
    \Omega_{\rm mat} & {\rm if}~~~k>k_{\Lambda} \\
    \frac{1}{\sqrt{\Omega_\Lambda}}\left(\frac{k}{H_0}\right)^3 >1 & {\rm if}~~~k<k_{\Lambda}
    \end{array}\right. \nonumber
\end{eqnarray}
It appears that the GW solution for the modes that enter the horizon well in the matter era, i.e.~before the moment when dark energy becomes relevant, is suppressed by a factor $\Omega_{\rm mat}\simeq 0.3$, when accounting for late-time acceleration~\cite{Turner:1993vb,Kuroyanagi:2009br}. This is however a calculation artefact, since we are obliged to set $\Omega_{\rm mat}^m=1$ if we neglect the cosmological constant but still assume spatial flatness. On the other hand, the modes that enter the horizon after the onset of the accelerated phase, have tensor amplitudes enhanced with respect to the case neglecting late-time acceleration; furthermore, the energy density power spectrum is changed to $\Omega_{\rm GW}(k)\propto k^4$ instead of the usual $k^{-2}$ dependence given for example in Eq.~\eqref{eq:omegaK}. Note however that the approximation $\Omega_{\rm mat}/(a_k^\Lambda)^3+\Omega_\Lambda\simeq \Omega_\Lambda$ is really crude: a numerical solution shows that in reality the transition occurs very gradually and it is, in fact, still taking place today. The solution given in Eq.~\eqref{eq:TMDTLambda} for $k<k_\Lambda$, as well as the claim that $\Omega_{\rm GW}(k)\propto k^4$, must be therefore considered only as indicative. Furthermore, a numerical evaluation also shows that wave-numbers around the horizon today, with $k_0=2.2\cdot 10^{-4}$ Mpc, have already started exiting the horizon due to the onset of the accelerated expansion. For those, the above solution clearly does not apply\footnote{Note that also Ref.~\cite{Zhang:2005nw} tackles the problem of the effect of the cosmological constant on the GW transfer function, but assumes an instantaneous transition between the matter dominated phase and a {\it de Sitter} expansion.}. 

Let us note that in all our derivations so far, we have neglected the presence of free-streaming neutrinos. Their contribution however must be taken into account. After they decouple, neutrinos are no longer in thermal equilibrium with the rest of the universe, and start to stream freely. Consequently, they cannot be described as a perfect fluid, and hence they develop certain out-of-equilibrium terms in their energy momentum tensor, including a tensor anisotropic stress. It is the presence of tensor perturbations in the metric that sources a tensor anisotropic stress in the neutrino fluid, analogously to what happens for an imperfect fluid with shear viscosity $\nu$, which develops a tensor anisotropic stress as $\Pi_{ij}^{\rm TT}=-\nu \dot h_{ij}$. In the case of neutrinos, Ref.~\cite{Weinberg:2003ur} has worked out the expression for $\Pi_{ij}$ which, once inserted into the evolution equation for tensor modes \eqref{gweqx}, gives an integro-differential equation for the tensor perturbation,
\begin{eqnarray}\label{eq:gweq_nu}
	h_{ij}''({\bf k}, \eta)+ 2 \mathcal{H} h_{ij}'({\bf k}, \eta)&+& k^2 h_{ij}({\bf k}, \eta)= \\
	&-& 24 \frac{\rho_\nu}{\bar\rho}\mathcal{H}^2\int_{\eta_\nu}^{\eta} d\tau \left[\frac{j_2(k(\eta-\tau))}{k^2(\eta-\tau)^2}\right] h_{ij}'({\bf k}, \tau)\,, \nonumber
\end{eqnarray}
where $\rho_\nu$ and $\eta_\nu$ denote respectively the background neutrino energy density and the time of neutrino decoupling, and $\bar\rho$ is the background energy density. This equation can be solved numerically, as done e.g. in \cite{Weinberg:2003ur,Watanabe:2006qe,Pritchard:2004qp,Kuroyanagi:2008ye}, to infer the detailed effect of the free-streaming neutrinos on the tensor energy density power spectrum today: see e.g.~Fig.~2 in \cite{Kuroyanagi:2008ye}. However some features of the solution can be appreciated by directly looking at Eq.~\eqref{eq:gweq_nu}. First of all, the minus sign on the right hand side confirms that the overall effect is a damping of the amplitude of the inflationary tensor modes, and hence of the energy density power spectrum, see e.g.~Fig.~2 in \cite{Kuroyanagi:2008ye} or Fig.~4 in \cite{Watanabe:2006qe}. Secondly, free-streaming neutrinos only affect modes that are inside the horizon, as outside the horizon we expect $h_{ij}'=0$ for the inflationary GW background. In other words, the effect from free streaming neutrinos respects, as it should, causality. On the other hand, the source term on the r.h.s.~of \eqref{eq:gweq_nu} is proportional to $\mathcal{H}^2$, meaning that modes that are inside the horizon $k\gg \mathcal{H}$ at the onset of neutrino decoupling, are not altered by neutrino free-streaming. Furthermore, the source term is proportional to the ratio of the neutrino energy density to the one of the background universe. As long as the universe is radiation dominated this remains constant, but starts decaying after the onset of matter domination: consequently, the effect of neutrino viscosity is largely suppressed for modes that enter the horizon during the matter dominated era. To summarise, neutrino free-streaming leads to a damping in the GW energy density power spectrum of about $\sim$ 35\% \cite{Watanabe:2006qe}, for modes that enter the horizon between the time of neutrino decoupling and the time of matter-radiation equality, i.e.~$10^{-17}\,{\rm Hz} \lesssim f \lesssim 10^{-11}$ Hz.

Finally, let us also notice that in our previous derivations, the period of reheating has not been modeled in any detail: the transfer function in Eq.~\eqref{eq:Tfunct} assumes an instantaneous transition directly from the inflationary to a radiation dominated era. However, if the scale of inflation is sufficiently low, scales entering the horizon at reheating time could fall in the sensitivity range of Earth- or Space-based detection: in this case, a more refined modeling becomes necessary. This can go from the standard accounting for inflaton oscillations around the minimum of its potential (corresponding to a matter-dominated phase if the inflaton potential is quadratic at the minimum) before the universe gets thermalised~\cite{Turner:1993vb}, to more complicated scenarios where other fields, interacting or not with the inflaton, are present, see e.g.~\cite{Seto:2003kc,Nakayama:2008wy,Kuroyanagi:2008ye,Kuroyanagi:2014nba}.

\section{Inflationary period, part II: beyond the irreducible GW background}
\label{sec:infII}
As we will see in this section, the irreducible contribution to GWs from vacuum quantum fluctuations discussed in Sect.~\ref{sec:PrimTensInf}, may not be the only GW background expected from inflation. In some circumstances, if new species or symmetries are at play during the inflationary period, GWs with a large amplitude and a significant deviation from scale-invariance, can also be produced during inflation. Contrary to the irreducible contribution, these predictions depend however strongly on model dependent assumptions. 

The details of GW production during inflation can change significantly if, $i)$ additional fields present during inflation (other than the inflaton), have interactions leading to strong particle production, $ii)$ spectator fields present during inflation exhibit a (time-dependent) sub-luminal speed of propagation, $iii)$ new symmetry patterns in the inflationary sector lead to the breaking of space reparametrizations, allowing the graviton to acquire a mass, and $iv)$ alternative theories of gravity other than general relativity, underlie the inflationary period. The GWs produced whenever either of the circumstances $i), ii)$ or $iii)$ are met during inflation, can significantly overtake the irreducible GW signal due to quantum fluctuations. However, as the features of the inflationary quantum vacuum fluctuations reflect the underlying gravity theory, circumstance $iv)$ may also affect significantly the form of the irreducible background. 

We discuss case $i)$ in Section~\ref{sec:PartProdInf}, case $ii)$ in Section~\ref{sec:SpectFlds}, case $iii)$ in Section~\ref{sec:EFT}, and finally case $iv)$ in Section~\ref{sec:ModGrav}. In all these circumstances, the spectrum of GWs can be rather large and blue-tilted, or exhibit a large-amplitude bump at specific scales. Therefore, the perspective of detecting these inflation-related backgrounds with GW interferometers is very compelling. These scenarios represent a new source of GWs, providing an attractive target for the upcoming GW detectors like e.g.~LISA, which will have the ability to probe a significant fraction of the parameter space of some of these scenarios~\cite{Bartolo:2016ami}. 

Let us note that other circumstances, not encompassed by $i)$-$iv)$ but still related to inflation, may also lead to large backgrounds of GWs. In particular, in Sect.~\ref{sec:SecGWScal}, we consider the possibility that scalar perturbations may be enhanced at short scales during inflation, so that they act as a source of GWs to second-order in perturbation theory. This GW background is guaranteed to exist (within the inflationary framework) as scalar perturbations are present at every scale. If the scalar perturbations are large enough at small scales, they may lead to a deformation of the otherwise quasi-scale invariant GW spectrum at frequencies accessible to GW detectors. Furthermore, in Section~\ref{sec:PBH} we consider the possibility that the amplitude of the scalar spectrum is sufficiently large at small scales, so that when these perturbations eventually re-enter the horizon, they may collapse into primordial black holes. Upon later merging, such population of primordial black holes would lead to a large background of GWs. For completeness, we also consider in Section~\ref{sec:InfAltern} the GW background produced from alternative theories to inflation, namely Pre-Big-Bang, string gas and bounce cosmologies.

\subsection{Particle Production during Inflation}
\label{sec:PartProdInf}
Gravitational waves can be emitted classically during inflation if a tensor anisotropic stress is present during the inflationary stage. If such is the case, GWs produced inside the Hubble radius during inflation are diluted by the exponential expansion of the background. Only once a given wavelength crosses outside the Hubble radius, does the GW amplitude remains constant. Therefore, in order to minimize the amount of dilution, in order to provide a non-negligible GW signal, mechanisms of GW generation by a non-zero anisotropic stress during inflation, must operate sufficiently close to the Hubble scale. 

The emission of GWs by particle production during inflation pertains to this category of GW generation. Several models of particle production have been discussed in the literature. In general, particle production during inflation is possible because, as the inflaton rolls down its potential, it provides a time-dependent background that carries the energy necessary for the production of sufficiently light species~\cite{Chung:1999ve}. The energy momentum tensor of the produced species represents an anisotropic stress over the background energy-momentum tensor, hence sourcing GWs. In the following we consider two cases, differentiated by the transient and sustainable nature of the particle production mechanism.

\subsubsection{Transient particle production.}
\label{sec:TransientPartProdInf}

Let us consider either a scalar field $\chi$ or some fermion species $\psi$, coupled to the inflaton $\phi$ with Lagrangian $-\mathcal{L}_\chi = (\partial \chi)^2/2 + g^2(\phi-\phi_0)^2\chi^2/2$, and $-\mathcal{L}_\psi = \bar\psi\gamma^\mu\partial_\mu\psi + g(\phi-\phi_0)\bar\psi\psi$, respectively. Alternatively, we can also consider the dynamics of a gauge field $A_\mu$ following the Lagrangian $\mathcal{L} = -{1\over4}F_{\mu\nu}F^{\mu\nu}-|(\partial_\mu-gA_\mu)\Phi)|^2-V(\Phi^\dag\Phi)$~\cite{Finelli:2000sh,Cook:2011hg}, where $F_{\mu \nu} = \partial_\mu A_\nu - \partial_\nu A_\mu$ is the field strength, and $\Phi=\phi e^{i\theta}$ is a complex field. In this latter case, we do not identify $\phi$ with the inflaton\footnote{To simplify the discussion on the particle production of the three cases (scalar $\chi$, fermion $\psi$ and vector $A_\mu$ fields), we maintain the same notation $\phi$ for the field causing the particle production, independently of whether we identify this field with the inflaton or not.}. We assume however that $\Phi$ evolves during inflation in such a way that its amplitude vanishes at some point $\phi(t_0) \equiv \phi_0 = 0$. In either of the three scenarios, when $\phi$ crosses around $\phi_0$ ($\phi_0 \neq 0$ if $\phi$ is the inflaton, $\phi_0 = 0$ otherwise), the mass $m = g(\phi(t)-\phi_0)$ vanishes exactly at $t = t_0$, when $\phi(t_0) = \phi_0$. For a short period of time $\Delta t_{\rm na}$ around $t_0$, 
\begin{eqnarray}
\Delta t_{\rm na} \sim 1/\mu\,,~~~~~~\mu^2 \equiv g\dot\phi_0\,,
\end{eqnarray}
the mass changes non-adiabatically as $\dot m \gg m^2$, leading to an explosive production of particles\footnote{$\Delta t_{\rm na}$ must be shorter than a Hubble time in order for the particle production to be efficient, i.e.~$\Delta t_{\rm na} \ll 1/H$. This implies a coupling range $g^2 \gg H^2/|\dot\phi|$.}~\cite{Kofman:1997yn}. The occupation number of the quanta created is actually independent of the spin of the excited species (given the interactions considered), and it reads $n_k = {\rm Exp}\lbrace -\pi(k/\mu)^2\rbrace$~\cite{Barnaby:2012xt}. This shows clearly that only long wave modes $k \ll \mu$ are excited, as short modes evolve adiabatically around $t_0$. 

In all three cases (scalars, fermions, and vectors), GWs are generated by the anisotropic distribution of the created species. Since particle production happens around the precise time $t_0$ when $\phi(t_0) = \phi_0$, the spectrum of GWs shows a feature at the frequency today corresponding to that moment. This feature represents an additional contribution on top of the standard irreducible vacuum tensor spectrum. 

Notably, even though the field structure of the energy-momentum tensor sourcing the GWs depends on the spin of the excited species, {\it Barnaby et al}~\cite{Barnaby:2012xt} has shown that, due to some cancellations, the GW produced by the created particles is essentially independent of their spin, modulo normalization factors of order $\mathcal{O}(1)$. To review this, let us write the total power spectrum as $\mathcal{P}_h^{\rm (tot)}(k) = \mathcal{P}_h^{\rm (vac)}(k) + \mathcal{P}_h^{\rm (pp)}(k)$, with $\mathcal{P}_h^{\rm (vac)}(k) \equiv (2/\pi^2)(H/m_{\rm Pl})^2$ the vacuum contribution given by Eq.~(\ref{eq:Deltah2Prim}). Detailed calculations~\cite{Cook:2011hg,Carney:2012pk,Barnaby:2012xt} show that the contribution $\mathcal{P}_h^{\rm (pp)}(k)$ from the newly created particles, distorts the total tensor power spectrum like
\begin{eqnarray}\label{eq:GWtransient}
\hspace*{-1cm}{\Delta \mathcal{P}_h\over \mathcal{P}_h} \equiv {\mathcal{P}_h^{\rm (tot)}-\mathcal{P}_h^{\rm (vac)}\over \mathcal{P}_h^{\rm (vac)}} \equiv {\mathcal{P}_h^{\rm (pp)}\over \mathcal{P}_h^{\rm (vac)}} \sim  few\times\mathcal{O}(10^{-4}){H^2\over m_{\rm Pl}^2}W(k\tau_0)\left({\mu\over H}\right)^{3}\ln^2(\mu/H)\,,\nonumber\\
\end{eqnarray}
with $W(x) \equiv {(\sin(x)-x\cos(x))^2\over x^3}$, and where the exact amplitude depends on the spin. This corresponds to a scale-dependent distortion which reaches its biggest amplitude around the Hubble scale $x_0 = k\tau_0 \simeq 1$, with $W(x_0) \simeq 0.5$ (note that in this section the conformal time is denoted $\tau$). The maximum distortion of the vacuum tensor spectrum peaks therefore around the horizon scale at the moment of particle production $\tau_0$. When the excited species are either a scalar or a fermion field, since $\phi$ is the inflaton, we can use Eq.~(\ref{eq:DeltaR2}) to derive the relations $H^2/m_{\rm Pl}^2 \simeq 8\pi^2\epsilon \mathcal{P}_{\mathcal{R}}$ and $\mu/H = (g/2\pi)^{1/2}/\mathcal{P}_{\mathcal{R}}^{1/4}$. Plugging these into Eq.~(\ref{eq:GWtransient}), we find the maximum distortion as $\Delta\mathcal{P}_h/\mathcal{P}_h \sim \mathcal{O}(10^{-2})\times\epsilon({g\over2\pi})^{3/2}\mathcal{P}_{\mathcal{R}}^{1/4}\left[\ln(g/2\pi) - {1\over2}\ln(\mathcal{P}_{\mathcal{R}})\right]^2$. Using CMB measurements, we know that $\epsilon < 0.0068$ and $\mathcal{P}_{\mathcal{R}} \simeq 2.2\cdot 10^{-9}$ at the CMB scales. Plugging in these numbers, we obtain that the new contribution represents a negligible distortion $\Delta\mathcal{P}/\mathcal{P} \ll 1$ of the quasi scale-invariant quantum fluctuations background (at the CMB scales). Assuming that the inflaton is in slow-roll all the way from the exit of the CMB scales till the end of inflation, we conclude that this distortion is always negligible, even if the particle production takes place towards the end of inflation when $\epsilon$ approaches unity\footnote{It is interesting to note that if the scalar perturbations were much larger at smaller scales than at CMB scales, i.e.~$\mathcal{P}_{\mathcal{R}} \gg 10^{-9}$ at the scales leaving the Hubble radius towards the end of inflation, we might expect a non-negligible distortion $\Delta\mathcal{P}/\mathcal{P} \gtrsim \mathcal{O}(1)$, of the otherwise quasi scale-invariant inflationary irreducible background. However this requires a specific set up, somehow contrived, demanding not only particle production at the right moment towards the end of inflation, but also the appropriate feature in the inflaton potential creating large curvature perturbations only at small scales.}.

Note that if there were several occurrences of particle production, like for example in the model of Ref.~\cite{Green:2009ds}, the sourced GWs add up and the resulting GW spectrum can become scale invariant. However, even in this case, it is largely subdominant with respect to the vacuum inflationary tensor spectrum, and it cannot be detected in the CMB~\cite{Cook:2011hg}. It seems that in these type of scenarios, we cannot obtain observationally interesting signatures in the GW background from inflation. 
Ref.~\cite{Barnaby:2012xt} remarks that only a sufficiently high amplification of the GW signal in a model with multiple bursts of particle production, could produce a GW signal observable in the CMB; however, such a model is not identified yet.

To work out the tensor spectrum in the case of gauge field excitation, we must specify the nature of the field $\Phi$, and thus its velocity at the time $\tau_0$. Due to gauge invariance the potential $V(\Phi)$ must be a function of $|\Phi|^2$, and hence $\Phi = 0$ must be either a maximum or a minimum of $V(\Phi)$, so that $V'(\Phi = 0) = 0$. Hence, we conclude that $\Phi$ cannot be in a slow-roll regime, as otherwise the velocity $\dot\Phi \propto V'(\Phi)$ would vanish. The field $\Phi$ must be therefore fast rolling. A simple set up consist in assuming that $\Phi$ has a super-Hubble mass $m_{\Phi} \gg H$, and a large amplitude $\Phi_i$ (acquired at some early stage during inflation). Under these circumstances, the field $\Phi$ oscillates many times per Hubble time, and the (absolute value of the) velocity when crossing around zero is given by $\dot\phi_0 \sim m_{\Phi}\Phi_i$. Requiring that the energy of $\Phi$ is sub-dominant versus the inflaton energy, implies that $m_{\Phi}^2\Phi_i^2/6H^2m_{\rm Pl}^2 \ll 1$. Eq.~(\ref{eq:GWtransient}) implies that, up to a logarithmic correction, $\Delta\mathcal{P}_h/\mathcal{P}_h \sim \mathcal{O}(10^{-4})g^{3/2}(H/m_{Pl})^{1/2} \ll 1$, rendering again this signal insignificant.

\subsubsection{Sustained particle production.}
\label{sec:SustainedPartProdInf}
A more promising scenario of particle production, which can give rise to observable GWs, consist in introducing a derivative coupling between the inflaton and a gauge field~\cite{Barnaby:2010vf}. In this case the gauge field can remain massless as the inflaton rolls down its potential, and particle production can happen continuously throughout inflation. This occurs in particular when the inflaton $\phi$ couples to a gauge field through an interaction term of the form 
\begin{eqnarray}\label{eq:phiFF}
\Delta{\cal L}=-\frac{1}{4\,\Lambda}\,\phi\,F_{\mu\nu}\,\tilde{F}^{\mu\nu}\,,
\end{eqnarray}
with $F_{\mu \nu} = \partial_\mu A_\nu - \partial_\nu A_\mu$ the standard field strength and $\tilde{F}_{\mu \nu} \equiv {1\over 2}\epsilon_{\mu\nu\alpha\beta}F^{\alpha\beta}$ its dual, whereas $\Lambda$ is a dimension-full constant. Such an interaction is natural in models where the inflaton field is a pseudo-scalar. These models and generalizations have received a significant amount of interest in the literature (including when the pseudoscalar field is a spectator field), see e.g.~\cite{Freese:1990rb,Anber:2006xt,Watanabe:2010fh,Senatore:2011sp,Sorbo:2011rz,Soda:2012zm,Maleknejad:2012fw,Pajer:2013fsa,Ferreira:2014zia,Namba:2015gja,Ferreira:2015omg,Peloso:2016gqs,Domcke:2016bkh,Agrawal:2017awz}. These scenarios are very interesting mostly due to two properties, $i)$ their radiative stability, and $ii)$ the very varied phenomenology they can generate. Radiative stability plays in fact a crucial role in the general construction of inflationary models, helping to discriminate between technically natural potentials (for which quantum properties of the theory are under control), and models which require some fine tuning one way or another (for which the UV part of the theory is undetermined). Ensuring radiative stability can be easily done by assuming a (softly broken) shift symmetry, i.e.~an invariance under the transformation $\phi\to\phi+\phi_0$, with $\phi_0$ an arbitrary constant. One of the very few low dimension operators that respect such shift symmetry, is precisely the interaction term in Eq.~(\ref{eq:phiFF}). 

An interaction like Eq.~(\ref{eq:phiFF}) is expected to be generated in a large of class of technically natural models of inflation, and it is expected to give rise to a very rich phenomenology. The equation of motion of the (conformally transformed) $\pm 1$-helicity modes $A_\pm({\bf k},\,\tau)$ of the gauge field reads~\cite{Anber:2006xt} 
\begin{eqnarray}\label{eq:XiParam}
A_\pm''({\bf k},\,\tau)+\left[k^2\pm 2\,\xi\,\frac{k}{\tau}\right]A_\pm({\bf k},\,\tau)=0\,,~~~~
\xi\equiv\frac{\dot\phi}{2\,\Lambda\,H}\,,
\end{eqnarray}
with $H$ the Hubble parameter, and $'$ and $\dot{\,}$ denoting derivatives with respect conformal time $\tau$ and cosmic time $t$. The $\pm$ sign in Eq.~(\ref{eq:XiParam}) implies that for sufficiently long wavelengths $-k\,\tau<2\,\xi$, one (and only one) of the two helicity modes is exponentially amplified during inflation.  For constant and positive parameter $\xi > 0$, an exact solution of Eq.~(\ref{eq:XiParam}) can be found~\cite{Anber:2006xt}, showing that only $A_+$ is amplified\footnote{In slow-roll inflation, as long as the back reaction of the gauge field into the background dynamics is negligible, we can write $\xi \simeq \frac{m_{\rm Pl}}{\Lambda}\,\sqrt{\frac{\epsilon_\phi}{2}}$. Hence, unless $\epsilon_\phi \lll 1$, a value of $\Lambda$ within 1-2 orders of magnitude below $m_{\rm Pl}$, leads always to a value $\xi\gtrsim {\cal O}(1)$.} by a factor $\sim e^{\pi\,\xi}$ (for $\xi\gtrsim {\cal O}(1)$). The fact that only one of the photon helicities is amplified, reflects indeed the parity-violating nature of the operator Eq.~(\ref{eq:phiFF}). 

The exponentially excited gauge fields act as a powerful source of GWs (as well as of scalar perturbations). Since the energy-momentum tensor of the gauge field is quadratic in the field amplitude, the tensor (and scalar) perturbations sourced by the gauge fields obey necessarily a non-Gaussian statistical distribution. The scalar bispectrum (three-point expectation value), in particular, has an approximate equilateral shape, with a non-linear parameter of the order $f_{NL}^{\rm {equil}} \sim 10^6\,(H^6/|\dot\phi|^3)(e^{6\pi\xi}/\xi^9)$~\cite{Barnaby:2010vf}. The non-detection of scalar non-Gaussianities at the CMB scales by the Planck experiment, implies a strong constraint on this scenario, with $\xi\lesssim 2.5$ at $95\%$ C.L.~\cite{Ade:2015lrj, Ade:2015ava}. As we will see shortly, the sourced GWs for such small values of $\xi$, are unfortunately too weak to be observed.

The parameter $\xi$ is however a time dependent quantity. The spectrum of gauge fields is scale dependent, rendering automatically as scale dependent the spectrum of GWs created by the gauge fields. In general, as the inflationary stage progresses, $|\dot\phi|$ typically increases while $H$ decreases, so $\xi$ increases as we approach towards the last stages of inflation. Thus, we can consider a parameter $\xi \lesssim 2.5$ or so at the time when the CMB scales left the Hubble radius, but growing later on when shorter scales leave the Hubble radius. This way, the Planck constraints~\cite{Meerburg:2012id,Ade:2015lrj} on the power spectra and bispectra at the CMB scales can be fully satisfied, while at the same time a large GW signal can be generated at shorter scales.  The time dependence of $\xi$, and hence the scale dependence of the spectrum of GWs,  depends of course on the specific choice of inflaton potential\footnote{It also depends eventually on the back-reaction of the excited photons over the background, if the backreaction becomes large enough.}.

In these models, the standard GWs from vacuum fluctuations and the additional sourced GWs created by the excited gauge fields, are statistically uncorrelated. The total tensor power spectrum is therefore simply the sum of the power spectra of these two contributions. For $\xi\gtrsim {\cal O}(1)$, a detailed calculation of the spectrum of tensor modes~\cite{Barnaby:2010vf,Sorbo:2011rz} shows that
\begin{eqnarray}\label{eq:pGW_sourced}
\mathcal{P}_{h}^{(tot)} \left( k \right) &\equiv & \frac{k^3}{2 \pi^2} \sum_{a=\pm} \left \vert h_{a} \left( k \right) \right \vert^2 = \mathcal{P}_{h}^{\rm (vac)}(k) + \mathcal{P}_{h}^{\rm (pp)}(k)\\
&=& \mathcal{P}_{h}^{\rm (vac)}\left(1+\Delta\mathcal{P}_{h}\right) \simeq \frac{2\,H^2}{\pi^2\,m_{\rm Pl}^2}\left(1 +  4.3 \cdot 10^{-7} \frac{H^2}{m_{\rm Pl}^2}\frac{e^{4 \pi \xi}}{\xi^6}\right) \,. 
\end{eqnarray} 
where $pp$ stands for particle production. Let us note that even though it is not indicated explicitly, the last expression in the $rhs$ depends on $k$, as both $H$ and $\xi$ are evaluated at the time when a given mode $k$ leaves the horizon during inflation. 

\begin{figure}[t]
\centerline{
\includegraphics[width=12.0cm]{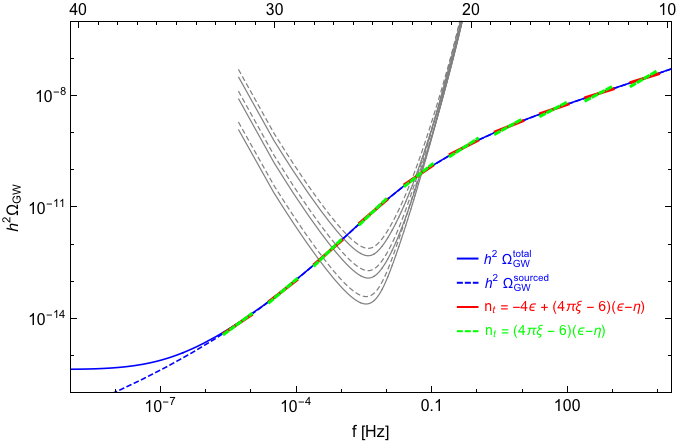}
}
\caption{Scenario of sustained particle production: numerical spectrum of GWs today $h^2\Omega_{\rm GW}$ for a model of quadratic inflaton potential $V(\phi) = {1\over2}m^2\phi^2$, with inflaton - gauge field coupling $\Lambda =  M_{Pl} / 35$ (continuous line) versus local parametrization $h^2\Omega_{\rm GW} \propto (f/f_*)^{n_T}$, evaluated at various pivot frequencies $f_*$ and with spectral tilt obtained from successive approximations to $n_T$. This figure is taken from Ref.~\cite{Bartolo:2016ami}, and also shows the Power Law-Integrated Curves of six LISA configurations that were still being considered at that time, c.f.~discussion in~\cite{Bartolo:2016ami}.}
\label{fig:local-nT}
\end{figure}

From the power spectrum $\mathcal{P}_{h}$ we can easily obtain the fractional GW energy density spectrum $h^2\Omega_{\rm GW}$. In Fig.~\ref{fig:local-nT} we plot the latter quantity as a function of frequency, for a quadratic inflaton potential and $\Lambda = m_{\rm Pl}/35$. We can notice three different regimes: {\it i)} dominance of the vacuum fluctuations contribution at large scales ($f \lesssim 10^{-5}$~Hz), {\it ii)} dominance of the sourced GWs contribution at intermediate scales ($10^{-5}$~Hz$ \lesssim f\lesssim 1$~Hz), but negligible back reaction of the gauge fields (so the evolution of $\dot\phi$ and $H$ is still determined by the standard slow-roll equations), and {\it iii)} dominance of the sourced GWs contribution at small scales ($f\gtrsim 1$~Hz) when the back reaction of the gauge fields cannot be neglected. Due to conservation of energy, the production of photons implies a reduction of the kinetic energy of the inflaton, so that the growth of $|\dot\phi|$ is slowed down, resulting into a flattening of $h^2\Omega_{\rm GW}$ at the smallest scales (highest frequencies). It is worth noticing that in the simplest slow-roll scenarios, as $|\dot\phi|$ and $H$ increase and decrease monotonically, respectively, the spectrum of the sourced GWs is always blue tilted. It is however possible to consider scenarios where $\xi$ has a transient behaviour, resulting into a localized bump in the GW spectrum, see~\cite{Namba:2015gja,Garcia-Bellido:2016dkw}. 

It is worth noting that in the above scenario, the amplitude of sourced GW background, i.e.~the power spectrum Eq.~(\ref{eq:pGW_sourced}), does not characterize completely this background. As a matter of fact, there are two very distinctive properties of the sourced GW background, namely its chirality and non-Gaussian statistics:

$\cdot$ {\em Chirality.} Eq.~(\ref{eq:pGW_sourced}) describes the total power in GWs, given by the sum of the individual contributions from left- and right-handed modes, $\mathcal{P}_{h}^{\rm (pp)} = \mathcal{P}_{h,+}^{\rm (pp)} + \mathcal{P}_{h,-}^{\rm (pp)}$. Each polarization is excited separately, with $\mathcal{P}_{h,+}^{\rm (pp)} \simeq 8.7\cdot 10^{-8} \frac{H^4}{m_{Pl}^4}\frac{e^{4 \pi \xi}}{\xi^6}$ representing the dominant contribution, and $\mathcal{P}_{h,-}^{\rm (pp)} \simeq 2.1\cdot 10^{-3} \mathcal{P}_{h,+}^{\rm (pp)}$ a sub-dominant part. The different amplitude between the left- and the right-handed tensor modes is a sign of the parity-violating nature of Eq.~(\ref{eq:phiFF}). Hence, a very distinctive signature of these scenarios is that the sourced GW background is highly chiral\footnote{For a discussion on the strategies for detecting a stochastic chiral backgrounds of GWs see~\cite{Crowder:2012ik,Smith:2016jqs}.}.

$\cdot$ {\em Non-Gaussianity.} Both the scalar and tensor perturbations sourced by the excited gauge field, obey non-Gaussian statistics. For an approximately constant $\xi$, the three point function of the GWs in this scenario has been computed in~\cite{Cook:2013xea}. The shape of the bispectrum is close to equilateral, with an amplitude (evaluated in the exact equilateral configuration $|{\bf{k}}_1|=|{\bf{k}}_2|=|{\bf{k}}_3|=k$) given by $\langle \hat{h}_+({\bf{k}}_1)\,\hat{h}_+({\bf{k}}_2)\,\hat{h}_+({\bf{k}}_3)\rangle_{\rm {equil}}= (2\pi)^2B(k,k,k)\,\delta^{(3)}({\bf{k}}_1+{\bf{k}}_2+{\bf{k}}_3)$ $\simeq 6.1\times 10^{-10} k^{-6}(H/M_{\rm Pl})^6(e^{6\pi\xi}/\xi^9)\,\delta^{(3)}({\bf{k}}_1+{\bf{k}}_2+{\bf{k}}_3)$. Hence, another distinctive signature in these scenarios, is the fact that there is a one-to-one correspondence of the power spectrum and the (equilateral) bispectrum, $k^6 B(k,k,k) \simeq 23 \mathcal{P}_h^{3/2}$. Constraints on the tensor bispectrum from CMB measurements~\cite{Ade:2015ava} yield a constraint on $\xi$ (on CMB scales) similar to that from the CMB scalar bispectrum\footnote{For a discussion on the detectability of non-Gaussian primordial signatures of GWs at interferometers, see~\cite{Thrane:2013kb}.} measurements $\xi \lesssim 2.5$. 

In summary, gauge field amplification during inflation from the interaction term (\ref{eq:phiFF}), leads naturally to a GW signal with a significant blue spectrum that can be probed by LISA and other upcoming experiments, hence allowing to probe the inflationary period well after CMB scales left the horizon. Furthermore, the resulting stochastic background of GWs has very distinctive properties that allow to distinguish it from other backgrounds, namely its high chirality and deviation from Gaussian statistics. Both of these properties enjoy specific predictions unique to this scenario, which therefore can be used to differentiate it unambiguously from other scenarios.

The sourced GW signal, which in terms of the energy fraction of GWs reads $h^2\Omega_{\rm GW}^{\rm (pp)} \simeq 1.5 \cdot 10^{-13} (H/m_{\rm Pl})^4(e^{4 \pi \xi}/{\xi^6})$ (for $\xi > 1$)~\cite{Bartolo:2016ami}, can be locally parametrized in the form $h^2\Omega_{\rm GW} \propto f^{n_T}$ at each frequency $f$. In order to figure out this parametrization, one just needs to evaluate $H$ and $\xi$ as a function of the e-folding $N$ corresponding the moment when a given wavenumber $k$ left the horizon. At any frequency we can define the spectral tilt as $n_T(f) \equiv \frac{d \ln \Omega_{\rm GW} h^2 }{d \ln f}$, which is equivalent to Eq.~(\ref{eq:ntDef}). Detailed calculations~\cite{Bartolo:2016ami} to first order in the slow-roll parameters, lead to 
\begin{eqnarray}
n_T \simeq  - 4 \epsilon + \left(4 \pi \xi - 6 \right) \left(  \epsilon - \eta \right) \;, 
\label{eq:nT}
\end{eqnarray}
with $\epsilon = \epsilon_H$ [Eq.~(\ref{eq:eH})] and $\eta = \eta_\phi$ [Eq.~(\ref{eq:slowroll})]. In the limit of negligible back reaction from the gauge fields, one can also use $\epsilon_H \rightarrow \epsilon_V$ and $\eta_\phi \rightarrow \eta_V - \epsilon_V$, with $\epsilon_V$ and $\eta_V$ related to the inflaton potential, see Eq.~(\ref{eq:slowroll}). 

For the range of $\xi$ that future detectors will be able to probe (e.g.~$\xi \gtrsim 3.5$ for LISA~\cite{Bartolo:2016ami}), the term $-4 \epsilon$ in the final expression of Eq.~(\ref{eq:nT}) is actually negligible compared to the other terms, so the tilt can be further approximated as $n_T \simeq  \left(4 \pi \xi - 6 \right) \left( \epsilon - \eta \right)$. The advantage of this simplified tilt is that it allows to reduce the number of independent variables in which the GW signal depends on, from $\lbrace H_N,\xi,\epsilon,\eta\rbrace$ to $\lbrace H_N,\xi,(\epsilon-\eta)\rbrace$. The local parametrization allows us this way to obtain a model-independent parameter estimation based on the sensitivity curves of a given GW detector. In particular, {\it Bartolo et al}~\cite{Bartolo:2016ami} have analyzed the ability of LISA to probe the parameter space of this scenario. They find that for sufficiently small slow-roll parameters, $(\epsilon-\eta)\ll 0.1$, the minimum value $\xi \geq \xi_{\rm min}$ required for a GW signal $h^2\Omega_{\rm GW}(f)$ to be above the sensitivity curve of the LISA configuration, is essentially independent of the spectral tilt $n_T$, and hence independent of the slow-roll parameters. The Hubble rate in chaotic inflation with quadratic potential at the e-fold $N_* \sim 25$ (corresponding to the frequency of maximum sensitivity of LISA) is $H_c \simeq 6.4 \cdot 10^{13}$ GeV. Taking this value as a reference, Ref.~\cite{Bartolo:2016ami} concluded that LISA cannot probe any Hubble rate smaller than $\sim \mathcal{O}(10^{-2})H_c$, as a too large $\xi_{\rm min}$ is in tension with perturbativity requirements~\cite{Peloso:2016gqs}. Taking $\xi_{\rm min} = 5.5$ as the maximum tolerated value at $N_* \simeq 25$, the minimum Hubble rate that can be probed by the (best) LISA configurations is $H_{\rm min} \simeq 6.3 \cdot 10^{11}$ GeV.

\subsection{Enhanced tensor perturbations at small scales}
\label{sec:BlueTilted}

\subsubsection{Spectator Fields.}
\label{sec:SpectFlds}
An inflationary scalar spectator field can be defined as a light scalar field which does not influence the inflationary background dynamics, but acquires perturbations due to quantum fluctuations. Several authors~\cite{Bartolo:2007vp,Biagetti:2013kwa,Biagetti:2014asa,Fujita:2014oba} have studied how the scalar perturbations of such a field, may act as a classical source of GWs during inflation. The total tensor power spectrum will then be the sum of two contributions, the standard vacuum part given by Eq.~(\ref{eq:Deltah2Prim}), plus the extra sourced contribution by the spectator field, that we will discuss next.

Let us anticipate that Refs.~\cite{Biagetti:2013kwa,Biagetti:2014asa,Fujita:2014oba} conclude that the contribution of the GWs produced by spectator fields, cannot create a large tensor-to-scalar ratio $r$ on CMB scales. As we will see, this is due to the fact that the scalar perturbations created by a spectator field, are determined by the same parameters controlling the tensor counterpart. Since scalar perturbations are well constrained by current CMB measurements~\cite{Ade:2015lrj}, the related GW production is constrained as well. These restrictions attain however only the amplitude of scalar and tensor perturbations at the CMB scales. In principle, under certain circumstances, the sourced contribution to the GWs may exhibit a blue-tilted spectrum. If sufficiently blue, the signal may become accessible to experiments at small scales, while keeping an acceptable amplitude at the CMB scales.

In order to understand all of this, let us consider the following Lagrangian
\begin{equation}\label{eq:LagSpectFlds}
\mathcal{L}= \frac{1}{2}m_{\rm Pl}^{2}R + P\left(X,\sigma\right) - \frac{1}{2}\partial_{\mu}\phi\partial^{\mu}\phi - V\left(\phi\right)\,,
\end{equation}
where $\phi$ is the inflaton, $\sigma$ is a spectator field, $X = -\frac{1}{2}\partial_{\mu}\sigma\partial^{\mu}\sigma$ and $P(X,\sigma)$ is a generic function of $X$ and $\sigma$. We consider that the inflaton is entirely responsible for the inflationary expansion. The spectator field acquires nonetheless quantum fluctuations, which in turn create curvature and tensor metric perturbations. In general, in order to have an interesting GW signal, the spectator field $\sigma$ must be characterized by a non-canonical Lagrangian, with a propagation speed of its perturbations different than the speed of light\footnote{In light of the Lagragian Eq.~(\ref{eq:LagSpectFlds}), the speed of sound is defined like $c_{s} \equiv P_{X}/(P_{X}+\dot{\sigma}_{0}^{2}P_{XX})$, where $\sigma_{0}$ is field background value, and $P_X \equiv \partial_X P$, $P_{XX} \equiv \partial^2_X P$.}, $c_s \neq 1$. As we will see, the sourced GW power spectrum scales as $\propto 1/c_s^3$, so that the smaller the speed of sound is, $c_s \ll 1$, the larger the amplitude of the sourced GWs will be. In general, $c_{s}$ may vary during inflation. In order to characterize its evolution, it is useful to introduce the (dimensionless) parameter
\begin{equation}\label{eq:esse}
	s \equiv {\dot{c}_{s}\over H c_{s}}\neq 0\,,
\end{equation}
requiring $\left|s\right| < 1$ for an adiabatic evolution of $c_s$.

The equation of motion of tensor modes is sourced by the anisotropic stress provided by the spectator field fluctuations~\cite{Fujita:2014oba}, which in conformal time reads
\begin{equation}\label{eq:GWeomSpectFld}
	h_{ij}''+2\mathcal{H}h_{ij}'-\nabla^2h_{ij} = \frac{2}{m_{\rm Pl}^{2}}\partial_{X}P(X,\sigma)\left\lbrace\partial_{i}\delta\sigma\partial_{j}\delta\sigma\right\rbrace^{\rm TT}\,.
\end{equation}
Detailed calculations developed in~\cite{Fujita:2014oba}, show that after solving Eq.~(\ref{eq:GWeomSpectFld}), the power spectrum of these classically generated GWs, is actually well described by a power law. Similarly, the sourced scalar power spectrum induced by the spectator field, is also well described by a power law. Ref.~\cite{Fujita:2014oba} finds
\begin{eqnarray}\label{eq:PSgwSpectFld}
\hspace*{-2cm}\mathcal{P}_{h}^{\rm (tot)}\left(k\right) = \mathcal{P}_{h}^{\rm (vac)}\left(k\right) + \mathcal{P}_{h}^{\rm (ext)}\left(k\right)
\simeq \frac{2H^{2}}{\pi^2 m_{\rm Pl}^{2}}\left(\frac{k}{k_{\ast}}\right)^{n_{T}^{\rm (vac)}} + \frac{8}{15\pi^3 c_{s}^{3}}\frac{H^{4}}{m_{\rm Pl}^{4}}\left(\frac{k}{k_{\ast}}\right)^{n_{T}^{\rm (ext)}} 
\hspace*{-0.2cm},\\
\label{eq:PScurvSpectFld}
\hspace*{-2cm}\mathcal{P}_{\mathcal{R}}^{\rm (tot)}\left(k\right) = \mathcal{P}_{\mathcal{R}}^{\rm (vac)}\left(k\right) + \mathcal{P}_{\mathcal{R}}^{\rm (ext)}\left(k\right) \simeq  \frac{H^{2}}{4\pi^2\epsilon m_{\rm Pl}^{2}}\left(\frac{k}{k_{\ast}}\right)^{n_{S}^{\rm (vac)}-1}+\frac{1}{32\pi^3 c_{s}^{7}}\frac{H^{4}}{m_{\rm Pl}^{4}}\left(\frac{k}{k_{\ast}}\right)^{n_{S}^{\rm (ext)}-1}\nonumber\\
\end{eqnarray}
with $H$ and $c_{s}$ evaluated at the pivot scale $k=k_{\ast}$. The spectral indexes of the sourced contributions (obtained from the scale dependence of $H$ and $c_{s}$), are given, at the lowest-order in $\epsilon$ and $s$, by
\begin{equation}\label{eq:nTspectFlds}
	n_{T}^{\rm (ext)}=-4\epsilon-3s\,,\hspace*{1cm}n_{S}^{\rm (ext)}-1=-4\epsilon-7s\,,
\end{equation}
with $\epsilon \equiv \epsilon_H$ [Eq.~(\ref{eq:eH})] and $s$ defined in Eq.~\eqref{eq:esse}. 

The total power spectra are thus the sum of two different power-laws. The smaller the speed of sound, $c_{s} \ll 1$, the more enhanced the sourced contribution is. If $s$ is negative with $-s > {4\over3}\epsilon$, the sourced GW background is actually blue tilted. This allows the sourced GWs to reach a large amplitude at small scales, while keeping a small amplitude at CMB scales. However, this enhances as well the amplitude of the sourced power spectrum of curvature perturbations, which scales as $\propto 1/c_s^7$. A minimum bound $c_s \geq c_s^{\rm (min)}$ is therefore required in order for the sourced scalar perturbations spectrum not to conflict with the current CMB measurement at a pivot scale $k_* = 0.05 {\rm Mpc}^{-1}$, $\mathcal{P}_{\cal R}^{[0.05]} \simeq 2.21\cdot 10^{-9}$. From $\mathcal{P}_{\cal R}^{\rm (ext)} \leq \mathcal{P}_{\cal R}^{[0.05]}$, one can derive that $c_s^{\rm (min)} \simeq 0.018(H_*/10^{14}{\rm [GeV]})^{4/7}$. 

Let us also note that a large negative value of $s$ implies, not only a blue tilted spectrum for the sourced GWs, but also a positive spectral index for the sourced curvature spectrum, see Eq.~\eqref{eq:nTspectFlds}. We require therefore the sourced scalar perturbations to be compatible with CMB measurements for the whole range of scales probed by the CMB, particularly keeping under control the amplitude of the curvature spectrum on the smallest scales measured at the CMB. An analysis along these lines~\cite{Bartolo:2016ami} turns out to add a more stringent (but $s$-dependent) lower bound on $c_{s}$, than the previous limit $c_s^{\rm (min)}$. Besides, an upper limit on the spectral index of the sourced scalar power spectrum, and hence on $\left|s\right|$, can also be found, for a fixed value of $H$ and for a given value of the sourced scalar amplitude. This constraint implies that CMB measurements do not admit a secondary contribution to the curvature power spectrum, with amplitude larger than $\sim 10\%$ of the main contribution (in agreement with e.g.~\cite{Kinney:2012ik}).

As both scalar and tensor perturbations are determined by the same parameters, significant constraints on the tensor power spectrum are then obtained thanks to current bounds on scalar perturbations. The computation of the sourced scalar perturbations, in particular of its bispectrum, remains however as an incomplete task. Several terms, {\it a priori} of the same order, appear in the the action expanded to third order in the scalar perturbations. It is not clear which one of them, if any, plays a main role in determining the scalar perturbations properties. Based on theoretical considerations, Ref.~\cite{Fujita:2014oba} selected the term $\delta \phi\left(\partial_{i}\delta\sigma\right)^{2}$ as the possible dominant one, and developed calculations assuming only such a contribution. The curvature power spectrum reported in Eq.~(\ref{eq:PScurvSpectFld}) is in fact derived under such assumption. There is however no general argument excluding the fact that other terms in the action may partially cancel the selected term. Therefore, it should be noted that, in the absence of a more elaborated analysis, any consideration on the sourced GWs derived from existing constraints on scalar perturbations, only represent the present (limiting) state of the art of this scenario. 

It is worth stressing that similarly to the GWs sourced by gauge fields in Section~\ref{sec:PartProdInf}, the sourced tensor perturbations created by a spectator field, must also follow a non-Gaussian statistical distribution. No particular quantification of this is available in the literature.

The parameter space of this model's power spectra is characterized by the inflationary Hubble rate $H$, the speed of sound $c_s$, and the time variation of the latter $s$. Considering the current limits from the CMB on slow-roll parameters~\cite{Ade:2015lrj}, $\epsilon<0.0068$ (at $95\%$ C.L.), and taking into account the measured amplitude of the scalar perturbations~\cite{Ade:2013zuv} $\mathcal{P}_{\cal R}^{[0.05]} \simeq 2.21\cdot 10^{-9}$ (at $68\%$ C.L), it is possible to explore whether the parameter space $s-c_s$ for a given value of $H$, can be probed by GW detectors. In first place, the upper bound on the integrated spectrum from BBN, or the null-detection of a stochastic background by aLIGO, must constrain this scenario. Each of these constraints provide an upper bound on the amplitude of the GW spectral energy density. However, the most stringent constraint comes from the fact that the sourced scalar power spectrum is expected to be notably enhanced at small scales, where one may run into trouble with PBH bounds. Unfortunately, these constraints are rather severe, resulting in a drastic reduction of the parameter space of the signal: only $c_s-s$ values well below the detection thresholds of aLIGO and LISA, are actually compatible with PBH constraints\footnote{An interesting perspective on how much PBH constraints limit the ability of observation of GWs by LISA can be found e.g.~in~\cite{Garcia-Bellido:2016dkw}}. See Section 4 of Ref.~\cite{Bartolo:2016ami} for further discussion on this. We conclude, therefore, that the sourced GW background from inflationary spectator fields, is inaccessible to direct detection GW experiments, unless one considers futuristic proposals like BBO or DECIGO. 

\subsubsection{Effective field theory of inflation.}
\label{sec:EFT}

Up to now in Section~\ref{sec:infII}, we have examined scenarios where the standard irreducible background of GWs from inflation, was surpassed at small scales, by tensor modes sourced by extra field species present during inflation, but not responsible for the exponential expansion. As long as the dynamics of such extra species does not lead to a significant back-reaction over the inflationary background, and do not affect significantly the properties of the scalar fluctuations (e.g.~creating too much non-Gaussianities), the presence of extra species can be made to be (choosing the right parameter space) absolutely compatible with all observational and theoretical constraints on inflation. An interesting question we may address now is whether we can build an inflationary model leading to similar enhanced small-scale GWs, but without invoking the presence of extra species beyond those responsible for driving inflation. 

As before, in order to obtain a GW background detectable by direct detection observatories, the tensor spectra must be enhanced at small scales, while keeping a small amplitude at large scales, compatible with CMB constraints. 
As we do not want to add extra species, a natural approach is to consider whether new symmetries within the inflationary sector can modify the vacuum tensor modes, so that these develop a blue tilt. Let us note that standard inflationary models involve scalar field(s) with a time-dependent homogeneous profile, thus breaking time-reparametrizations of the (quasi-){\it de Sitter} space during inflation. Space-reparametrizations are however typically preserved. This imposes a specific structure of the perturbed action describing the tensor fluctuations, with GWs represented by transverse-traceless modes, which are adiabatic, massless, and conserved at superhorizon scales. The tensor power spectrum is controlled, in general, by the value of the Hubble parameter $H$, but also by the tensor sound speed $c_T$ (naturally set to one in standard inflationary models). The latter can be different from unity, e.g.~in models with kinetic mixing between gravity and the inflaton like G-inflation~\cite{Kobayashi:2010cm,Kobayashi:2011nu}, resulting in a tensor power spectrum typically even more red than the standard tilt Eq.~(\ref{eq:nT2e}) in SFSR models. Therefore, inflationary scenarios breaking only time-reparametrization do not normally lead to an enhancement of the tensor modes at small scales (rather the opposite).

We clearly need some new ingredient, other than simply breaking time-reparametrization. A daring proposal is to break space-reparametrizations during inflation, considering inflaton fields with space-dependent profiles. Under these circumstances, tensor modes do not need to be massless any more, since no symmetry prevents them from acquiring a mass. Besides, if during inflation a fluctuating field posses a positive mass $m_h^2 > 0$, the spectrum due to quantum fluctuations can be blue, as long as $m_h$ is sufficiently large ($m_h > \sqrt{3\epsilon} H$), but yet small enough so that the field remains light ($2m_h < 3H$)~\cite{Riotto:2002yw}. If at the end of inflation, the inflaton field(s) arranges itself to recover the space-reparametrization symmetry, the graviton mass will then be switched off. As the tensor modes enter back into the horizon after inflation, they will behave as ordinary massless GWs, but preserving the blue-tilted scale-dependent spectrum imprinted during inflation, reflecting the inflationary tensor excitation when the graviton had a mass.

Various scenarios in the literature have implemented the possibility of breaking the space-reparametrization during inflation. Models of vector inflation~\cite{Golovnev:2008cf}, for instance, achieves this thanks to space-like vacuum expectation values acquired by a set of vector fields. In its original formulation, this scenario is however plagued by a ghost mode~\cite{Himmetoglu:2008zp}. Other scenarios like gauge or chromo-natural inflation~\cite{Maleknejad:2011jw,Adshead:2012kp,Dimastrogiovanni:2016fuu}, have been successfully constructed free from these instabilities. Models involving only scalar fields have been also proposed, like Solid~\cite{Endlich:2013jia} and Supersolid~\cite{Koh:2013msa,Cannone:2014uqa,Cannone:2015rra,Bartolo:2015qvr} inflation, where various scalar fields acquire time- and space-dependent vacuum expectation values, yet obeying internal symmetries which ensure homogeneity and isotropy of the spacetime background. The development of a blue tilt in the GW background generated in these models (as well as other distinctive properties of the tensor modes), has been studied e.g.~in Refs.~\cite{Endlich:2013jia,Bartolo:2015qvr,Cannone:2014uqa, Akhshik:2014gja,Akhshik:2014bla,Ricciardone:2016lym}.

A natural toolkit to study the GW production during inflation from scenarios breaking space-reparametrization, but from a model independent perspective, is the Effective Field Theory (EFT) of inflation~\cite{Cheung:2007st}. In the following, we will consider model independent features of tensor fluctuations in scenarios that break space-reparametrization during inflation, limiting ourselves to set-ups preserving isotropy and homogeneity of the inflationary space-time background. Under these assumptions, one can write the most general form for the second order action for tensor fluctuations as~\cite{Cannone:2014uqa}
\begin{equation} \label{eq:sol-qac}
S_{(2)}\,=\,\frac{m_{Pl}^2}{8}\int dt\,d^3 x \,a^3(t)\,n(t)\,\left[\dot h_{ij}^2 -\frac{c_T^2(t)}{a^2} \,\left( \partial_l h_{ij}\right)^2
-m_h^2(t)\,h_{ij}^2\right]\,,
\end{equation}
where $n$, $c_T$ and $m_h$ are functions of the time $t$, determined by the underlying (here unspecified) model. Let us emphasize that in writing Eq.~(\ref{eq:sol-qac}), we do not stick to any particular scenario, but simply include all terms allowed by the symmetries. The fact that we allow for a mass term for the graviton $m^2_h h_{ij}^2$, is a reflection of the breaking of space-reparametrizations. A tensor sound speed $c_T \neq 1$ normally emerges in scenarios with kinetic mixing among tensors and scalars, and in general in inflationary models described by Horndeski scalar-tensor theories with non-minimally coupled scalar fields~\cite{Horndeski:1974wa}. Scenarios with kinetic mixing also display typically an overall factor $n(t)$, `renormalizing' the Planck mass.

In principle, both positive or negative graviton square masses $m_h^2$ are compatible with the symmetries demanded. Let us recall that, when considering Lorentz invariant theories, the {\it Higuchi bound} states that tensor modes cannot have a mass in the interval $0\,<m_h^2\,\le \,2 H$ in a pure {\it de Sitter} background~\cite{Higuchi:1986py,Arkani-Hamed:2015bza}. This suggests that one may expect $m_h^2\,<\,0$ as a natural choice. However, as inflation does not correspond to an exact {\it de Sitter} background, a small positive mass $m_h^2 > 0$ is allowed in theories which preserve Lorentz symmetry, but spontaneously break the exact {\it de Sitter} symmetries. Besides, if one considers theories directly breaking Lorentz symmetry, e.g.~\cite{Blas:2009my} or Horava-Lifshitz scenarios~\cite{Horava:2009uw,Blas:2009qj,Blas:2009yd,Blas:2010hb}, the Higuchi bound does not necessarily hold. We will therefore allow also for a positive $m_h^2 > 0$. 

The power spectrum of tensor modes depends explicitly on the time-dependent functions entering in Eq.~(\ref{eq:sol-qac}). Analytical calculations can be carried out in a {\it de Sitter} or quasi-{\it de Sitter} space, only if $c_T$ and $m_h$ are functions changing adiabatically in time. For simplicity, we will model the inflationary period with an exact {\it de Sitter} background with constant Hubble rate $H=$ const. Taking then $c_T$ and $m_h$ also as constants, and fixing the overall factor to $n = 1$, we can consider the standard procedure for quantization of light degrees of freedom on a {\it de Sitter} background. Detailed calculations (similar to those displayed in Section~\ref{sec:PrimTensInf}) lead to
\begin{equation}  \label{eq:PSHankel}
{\cal P}_h = \frac{H^2}{4 \pi\,m_{Pl}^2} \,
\left(
\frac{k^3}{  \,k_{*}^3}\right)\,\Big| H^{(1)}_\nu\left( \frac{c_T\,k}{k_*}\right)\Big|^2\,,\hspace*{1cm} \nu\,=\,\frac32\,\sqrt{1-\frac{4 m_h^2}{9 H^2}}\,,
\end{equation}
with $H_\nu^{(1)}(x)$ the Hankel function of first kind, and $k_*$ an arbitrary reference scale. In the limit of small graviton mass $|m_h| \ll H$, the tensor power spectrum reduces at super-horizon scales [using the small argument expansion of $H_\nu^{(1)}$ similarly as in Eq.~(\ref{eq:vkExpandedsupH})], to
\begin{equation}\label{eq:PS_GW_EFT}
{\cal P}_h\,=\,\frac{H^2}{2\pi^2\,m_{Pl}^2\,c_T^3}\,\left( \frac{k}{k_*}\right)^{n_T}\,,\hspace*{1cm}n_T \equiv {2\over 3}\frac{m_h^2}{H^2}\,,
\end{equation}
where we recall that we have assumed an exact {\it de Sitter} background\footnote{Otherwise the spectral index would be also contributed by the standard red tilted slow-roll term given by Eq.~(\ref{eq:nT2e}), which here is considered negligible against the expression in Eq.~(\ref{eq:PS_GW_EFT}).}.

It is clear that in order to obtain a blue spectrum with $n_T > 0$, we need a positive mass $m_h^2 > 0$. This represents the most interesting case from the point of view of direct detection observatories, as a blue tilt will allow for a large tensor spectrum at the scales direct GW detectors, while keeping a small signal at CMB scales. 

\begin{figure}[t]
 \begin{center}
\includegraphics[width=11.5cm]{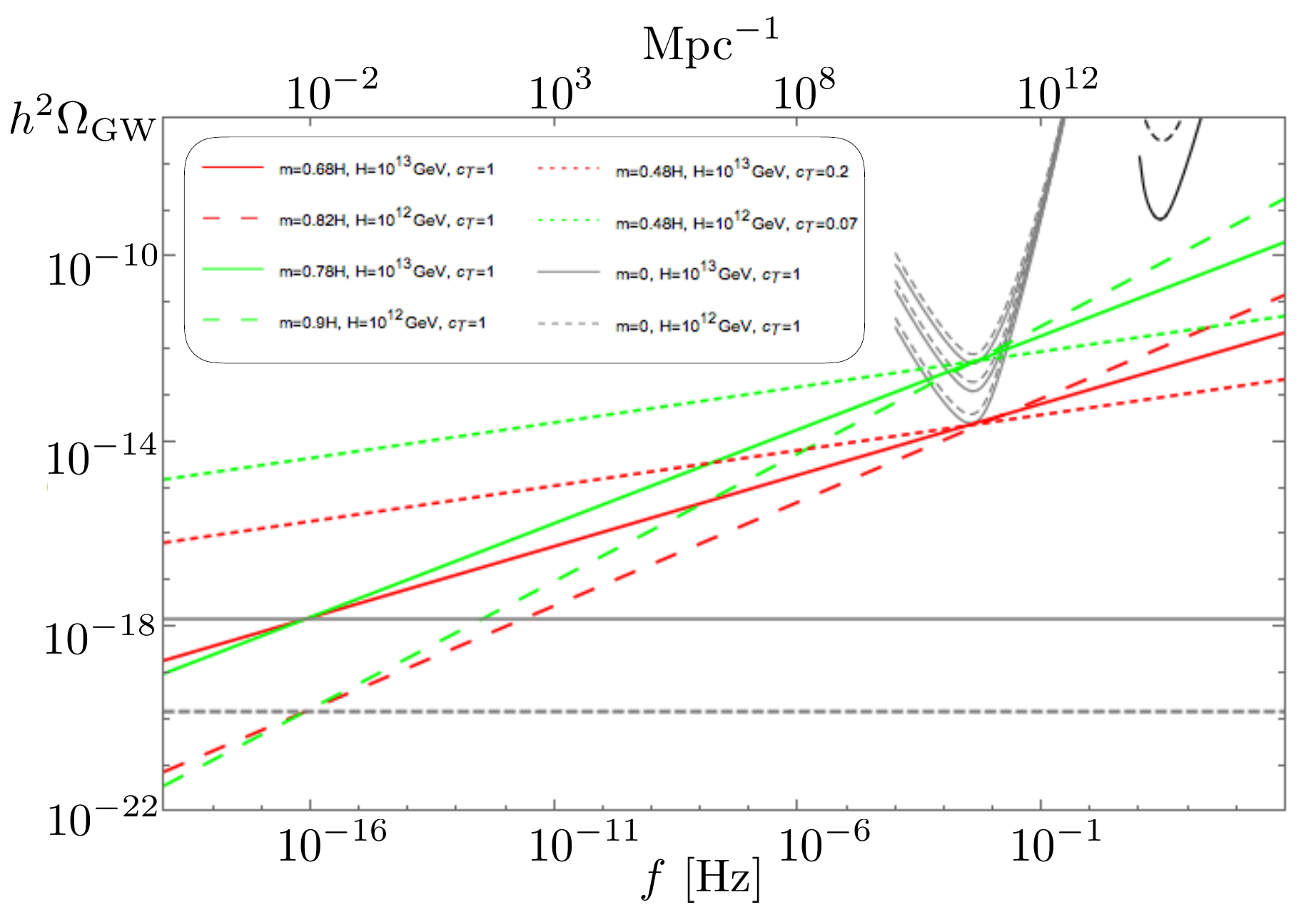}
\end{center}
    \caption{Effective field theory of inflation: spectrum of GWs energy density $h^{2}\Omega_{\rm GW}$ for different values of the effective mass of the graviton $m_{h}$, Hubble rate during inflation $H$, and tensor sound speed $c_{T}$, compared with sensitivities of LISA (grey curves, the six LISA configurations that were still being considered at that time, c.f.~discussion in~\cite{Bartolo:2016ami}) and aLIGO (black curves). The pivot scale used is $k_{*}=0.05~{\rm{Mpc^{-1}}}$. This plot is taken from Ref.~\cite{Bartolo:2016ami}.}
  \label{fig:EFT}
\end{figure}

Using Eq.~(\ref{eq:PS_GW_EFT}) and fixing the pivot scale at $k_{*}=0.05~{\rm{Mpc^{-1}}}$, Ref.~\cite{Bartolo:2016ami} has carried out a model-independent analysis of the ability of the LISA detector to probe scenarios with broken space-reparametrization during inflation. In the following, we summarize their analysis. The tensor power spectrum depends on three parameters: the inflationary Hubble rate measured in Planck mass units $H/m_{\rm Pl}$, the tensor speed of sound measured in speed of light units $0 \leq c_{T} \leq 1$, and the mass of the graviton measured in Hubble units, $m_{h}^2/H^2 > 0$, which we take positive so that the spectrum is blue tilted. While $H/m_{\rm Pl}$ and $c_T$ determine the amplitude of the spectrum, $m_h^2/H^2$ controls the spectral tilt. Let us stress that Ref.~\cite{Bartolo:2016ami} assumed for simplicity that all the parameters involved, $H^2$, $m_{h}^2$ and $c_T$, are time independent (for a different analysis assuming a certain time dependence, see ~\cite{Cai:2015yza,Cai:2016ldn}). In Fig.~\ref{fig:EFT}, we show  the GW energy density today versus frequency, for representative values of the $\lbrace m_H, H, c_T\rbrace$ parameters. The effect of the mass enhancing the tensor spectrum at high frequencies (small scales) is clearly appreciated, resulting in a bigger amplitude the larger the ratio $m_h/H$ is. This allow us to infer a minimum graviton mass in order to have this signal detectable by LISA, $m_{h} \geq m_h^{\rm (min)} \simeq 0.78\, H$, for an energy scale of inflation of $H=10^{13}{\rm{GeV}}$ and $c_{T}=1$. In general, the lower the energy scale of inflation the higher it needs to be the ratio $m_h/H$ in order to probe the signal at a given direct detection experiment. Fixing the value of the graviton mass to some representative value, one can do a different exercise, determining (for a given inflationary Hubble rate) the minimum graviton speed required for a detection at LISA. Ref.~\cite{Bartolo:2016ami} concludes that for a mass $m_{h} \simeq 8 \cdot 10^{12}\,{\rm{GeV}}$, LISA will be able to probe gravitons' speed of sound at the level of $\sim 20\%$ of the speed of light, while lowering the energy scale of inflation allows to reach a $\sim 1\%$ level. 

Ref.~\cite{Bartolo:2016ami} has characterized the ability of some GW detectors to probe the space of parameters $\lbrace m_H, H, c_T\rbrace$ of these scenarios. Their analysis conclude that probing this signal (either through its detection or simply by establishing constraints due to a non-detection) cannot be done with aLIGO O1, neither with the future aLIGO/aVirgo O5. It really requires a detector such as LISA to reach the sensitivity to probe this signal (or of course, the futuristic proposals BBO and DECIGO). Choosing e.g.~$n_{T} = 0.3$ (equivalently $2m_h^2 \simeq H^2$), which corresponds to a value that LISA can typically probe for sufficiently high inflationary energy scales, the best LISA configuration can probe down to inflationary energy scales of $\sim 10^{12} {\rm{GeV}}$. Of course a higher value of the spectral tilt, or a tensor sound speed lower than the speed of light, would allow to reach lower inflationary energy scales, but the graviton would then become more and more massive, challenging its light field condition.

Let us note that besides the small scale enhancement of tensor modes just discussed, other distinctive features are imprinted in the GW background created by scenarios breaking space-reparametrizations during inflation. These features make in fact these scenarios distinguishable from other scenarios with blue tilted GW backgrounds, like those discussed in Sections~\ref{sec:SustainedPartProdInf} and \ref{sec:SpectFlds}. These extra features, contrary to the short scale enhancement for a positive graviton mass, are however model-dependent properties of the GW background of each scenario. For instance, primordial non-Gaussianities can be enhanced without entering in conflict with current measurements. The bispectrum may have distinctive features~\cite{Endlich:2012pz,Endlich:2013jia,Bartolo:2015qvr}, potentially testable in the CMB~\cite{Ade:2013ydc,Ade:2015ava}. No dedicated template(s) have been however implemented so far for this, even though galaxy surveys can, in principle, offer the opportunity to test these features of the bispectra~\cite{Schmidt:2015xka,Chisari:2016xki}. Scalar-scalar-tensor three point functions can also be enhanced, so that this observable can be tested through B-modes searches of tensor non-Gaussianity~\cite{Cannone:2014uqa,Meerburg:2016ecv}. Similarly, the scalar four point function can be enhanced, leading to a peculiar angular dependence~\cite{Bordin:2016ruc}, although no particular constraints have been derived yet on this respect. Any of these additional effects could be used to place strong constraints on scenarios with broken space-reparametrizations during inflation, though a more concrete analysis of this remains to be done. In the same spirit of the effective field theory of inflation~\cite{Cheung:2007st}, constraints on the physical observables of the theory, could be automatically translated into operators appearing in the quadratic action for the tensors. 

\subsection{Modified gravity models}
\label{sec:ModGrav}
Modifications of gravity beyond General Relativity (GR) are motivated from both theoretical considerations, and cosmological observations. Quantization of gravity or the unification of fundamental interactions often lead to effective actions involving higher-order derivative terms, or non-minimal couplings to curvature invariants. The need to explain the current acceleration of the Universe has also motivated to consider adding extra gravitational degrees of freedom, including scalar, vector and further tensor fields. Besides, modified gravity scenarios offer as well the possibility to realize inflation within the gravitational sector only, without invoking the existence of an inflaton matter field (typically beyond the Standard Model).

In inflationary scenarios based on modified gravity theories, tensor modes are excited during inflation due to quantum mechanical effects, similarly as in the case of slow-roll single field scenarios discussed in Section~\ref{sec:PrimTensInf}. New features in the tensor sector may however appear, such as a non-vanishing graviton mass $m_T \neq 0$~\cite{Dubovsky:2009xk,Fasiello:2015csa,Malsawmtluangi:2016agy}, a sound speed different than the speed of light $c_T \neq 1$~\cite{Creminelli:2014wna,Amendola:2014wma,Pettorino:2014bka,Raveri:2014eea,Xu:2014uba,Saltas:2014dha,Noumi:2014zqa,Cai:2015dta,Cai:2016ldn,Burrage:2016myt}, and a modified dynamical propagation of the tensors~\cite{Pettorino:2014bka,Xu:2014uba}. From the recent detection of GWs from inspiriling binaries by the LIGO/VIRGO detectors~\cite{Abbott:2016blz,Abbott:2016nmj,Abbott:2017vtc}, constraints have been obtained for $c_{\rm T}$~\cite{Blas:2016qmn,Ellis:2016rrr,Collett:2016dey,Bicudo:2016pps,TheLIGOScientific:2017qsa} and $m_{\rm T}$~\cite{TheLIGOScientific:2016src}. Let us note that the first observation of GWs from a binary neutron star inspiral~\cite{TheLIGOScientific:2017qsa} in Sept.~2017, has placed a stringent constraint in the speed of propagation of GWs as $|c_T - 1| \leq 4.5\cdot 10^{-16}$. This has severe restrictive implications for a broad class of modified gravity models which can be used to explain dark energy~(see e.g.~\cite{Ezquiaga:2017ekz,Creminelli:2017sry,Baker:2017hug,Sakstein:2017xjx,Lombriser:2015sxa,Lombriser:2016yzn}).

Certain modified gravity theories with higher-order derivative terms in the Lagrangian, can provide second order field equations without gradient or ghost instabilities. The simplest approach consists in replacing the Hilbert-Einstein term, linear in the curvature $R$, by a function of it. These are correspondingly known as $f(R)$ theories. A remarkable property of the latter is that they can be mapped into another type of modified gravity scenarios, known as scalar-tensor theories, where a scalar field represents a new gravitational degree of freedom. One can then determine the structure of the most general scalar-tensor theory in a 3+1 dimensional space-time, leading to second order field equations (albeit starting from higher-order derivative terms in the Lagrangian) and avoiding gradient and ghost instabilities. These are known as Horndeski theories. For exhaustive reviews on modified gravity theories and their implications for cosmology see~\cite{Sotiriou:2008rp,DeFelice:2010aj,Clifton:2011jh}, and for constructions beyond Horndeski see~\cite{Gleyzes:2014dya,Zumalacarregui:2013pma,Gleyzes:2014qga}. In the following, we will restrict our attention to Horndeski theories, first proposed in the seminal paper~\cite{Horndeski:1974wa}. The Horndeski action can be written~\cite{Horndeski:1974wa,Deffayet:2010qz,Kobayashi:2010cm,Deffayet:2011gz} as~$S=\sum_{i=2}^{5}\int\mathrm{d}^{4}x\sqrt{-g}\mathcal{L}_{i}$, where
\begin{eqnarray}\label{eq:Horndeski}
	\hspace*{-2.1cm}\mathcal{L}_{2} &=& K\left(\Phi,X\right)\,,\\	
	\hspace*{-2.1cm}	\mathcal{L}_{3} &=& G_{3}\left(\Phi,X\right)\Box\Phi\,,\\	
	\hspace*{-2.1cm}	\mathcal{L}_{4} &=& G_{4}\left(\Phi,X\right)R+G_{4X}\left[\left(\Box\Phi\right)^{2}-\left(\nabla_{\mu}\nabla_{\nu}\Phi\right)^{2}\right]\,,\\ 
	\hspace*{-2.1cm}	\mathcal{L}_{5} &=& G_{5}\left(\Phi,X\right)G_{\mu\nu}\nabla^{\mu}\nabla^{\nu}\Phi-\frac{G_{5X}}{6}\left[\left(\Box\Phi\right)^{3}-3\left(\Box\Phi\right)\left(\nabla_{\mu}\nabla_{\nu}\Phi\right)^{2}+2\left(\nabla_{\mu}\nabla_{\nu}\Phi\right)^{3}\right],
\end{eqnarray}
with $G_{\mu\nu} \equiv R_{\mu\nu} - {1\over2}Rg_{\mu\nu}$ the Einstein tensor, $G_{iX}=\partial G_{i}/\partial X$, and $K,G_{i}$ are functions of $\Phi$ and $X\equiv -\partial_{\mu}\Phi\partial^{\mu}\Phi/2$. The system is described by four independent and arbitrary functions of $\Phi$ and $X$ (which can easily incorporate the Hilbert-Einstein term), which can be used to describe certain phenomena relying only on the gravitational sector, without adding matter fields beyond the SM. The action above encompasses, for instance, a large class of inflationary models, from single-field slow-roll inflation, k-inflation \cite{ArmendarizPicon:1999rj}, Higgs G-inflation \cite{Kamada:2010qe}, Galileon inflation \cite{Burrage:2010cu}, to Generalized G-inflation~\cite{Kobayashi:2011nu}. 

Let us focus on the tensor perturbations of the theory. Perturbing the Horndeski Lagrangian leads to a second order action for the tensors as~\cite{Kobayashi:2011nu}
\begin{eqnarray}\label{eq:HorndeskiTensors}
 S_{\rm T}^{\left(2\right)} = \frac{m_{\rm Pl}^{2}}{8}\int\mathrm{d}t\mathrm{d}^{3}x\,a^{3}\left(t\right)\left[\mathcal{G}_{\rm T}\dot{h}_{ij}\dot{h}_{ij}-\frac{\mathcal{F}_{\rm T}}{a^{2}}\left(\nabla h_{ij}\right)^{2}\right]\,,\\
	\mathcal{F}_{\rm T}\equiv\frac{2}{m_{\rm Pl}^{2}}\left[G_{4}-X\left(\ddot{\Phi}G_{5X}+G_{5\Phi}\right)\right]\,,\\
	\mathcal{G}_{\rm T}\equiv\frac{2}{m_{\rm Pl}^{2}}\left[G_{4}-2XG_{4X}-X\left(H\dot{\Phi}G_{5X}-G_{5\Phi}\right)\right]\,.
\end{eqnarray}
This action has the same structure as Eq.~(\ref{eq:Sg2}) in GR, i.e.~it corresponds to relativistic waves, but with non-standard coefficients. It implies that the speed of propagation of the tensors is $c_T^2 \equiv \mathcal{F}_{\rm T}/\mathcal{G}_{\rm T}$ (in general $\neq 1$), and that the friction term in the dynamical equation will not simply be proportional to the Hubble rate. To avoid ghosts and gradient instabilities, it is required that both $\mathcal{F}_{\rm T} > 0$ and $\mathcal{G}_{\rm T} > 0$. In order to compute the power spectrum of the tensor modes, we can define analogous Mukhanov variables and quantize the perturbations, by analogous procedure as we did in Sect.~\ref{sec:PrimTensInf}. Defining
\begin{equation}
\epsilon_H\equiv-\frac{\dot{H}}{H^{2}}\,,~~~~ f_{\rm T}\equiv\frac{\dot{\mathcal{F}_{\rm T}}}{H\mathcal{F}_{\rm T}}\,,~~~~ g_{\rm T}\equiv\frac{\dot{\mathcal{G}_{\rm T}}}{H\mathcal{G}_{\rm T}}\,,
\end{equation}
and assuming for simplicity that $\epsilon_H \simeq const.$, $f_{\rm T} \simeq const.$ and $g_{\rm T} \simeq const.$, the power spectrum of the tensor modes is obtained as a power law~\cite{Kobayashi:2011nu}
\begin{eqnarray}\label{eq:MGtensorSpectrum}
\mathcal{P}_{\rm T} = \frac{\mathcal{G}_{\rm T}^{1/2}}{\mathcal{F}_{\rm T}^{3/2}}\frac{8\gamma_{\rm T}}{m_{Pl}^{2}}\,\left(\frac{H}{2\pi}\right)^{2}\left({k\over c_TaH}\right)^{3-2\nu_{\rm T}},
\end{eqnarray}
with
\begin{eqnarray}\label{eq:MGspectralIndex}
\hspace*{-2cm}\nu_{\rm T}\equiv\frac{3-\epsilon_H+g_{\rm T}}{2-2\epsilon_H-f_{\rm T}+g_{\rm T}}\,,~~~\gamma_{\rm T}=2^{2\nu_{\rm T}-3}\left|\frac{\Gamma\left(\nu_{\rm T}\right)}{\Gamma\left(3/2\right)}\right|^{2}\left(1-\epsilon_H-\frac{f_{\rm T}}{2}+\frac{g_{\rm T}}{2}\right)\,,
\end{eqnarray}
Compared to the standard tensor power spectrum obtained for slow-roll single field inflation models Eq.~(\ref{eq:Deltah2Prim}), Eq.~(\ref{eq:MGtensorSpectrum}) exhibits two main differences: $i)$ the amplitude of the spectrum is not any more in one-to-one relation with the Hubble scale, as there is a new prefactor modulating the amplitude (depending on the aspects of the modified gravity model), and $ii)$ the spectral index depends now on the functions $f_{\rm T}$ and $g_{\rm t}$, so that contrary to what happens in GR, we can obtain a blue tilted spectrum if the condition $4\epsilon_H + 3f_{\rm T} < g_{\rm T}$ holds (while maintaining $\epsilon_H > 0$, so that $\dot H < 0$).

The results presented above, Eqs.~(\ref{eq:MGtensorSpectrum}), (\ref{eq:MGspectralIndex}), were derived for generalized G-inflation models, which have been shown to be equivalent to the Horndeski action~\cite{Kobayashi:2011nu}. The resulting tensor power spectrum depends on the details of the modified gravity scenario, i.e.~on the specifics of the functions $\mathcal{F}_{\rm T}, \mathcal{G}_{\rm T}, f_{\rm T}$ and $g_{\rm T}$. It is perhaps worth noticing that for a specific case of the Hordenski action, $S = \int\mathrm{d}^{4}x\sqrt{-g}\left[\frac{1}{2}m_{\rm Pl}^{2}R+\mathcal{L}_{\Phi}\right]$, with $\mathcal{L}_{\Phi} = X- V(\Phi) +g(\Phi)X\Box\Phi$ (which represents, for instance, the action of Higgs G-inflation~\cite{Kamada:2010qe}), the tensor power spectrum reduces to the standard expression $\mathcal{P}_{\rm T} = \frac{8}{m_{Pl}^{2}}\left(\frac{H}{2\pi}\right)^{2}\left({k\over{aH}}\right)^{-2\epsilon_H}$. This implies a degeneracy between this modified gravity scenario and the standard single field slow-roll in general relativity. The curvature power spectrum is however different than the standard scenario, allowing therefore to distinguish this modified gravity scenario via its specific consistency relation, which differs from the standard one $n_T = - \frac{r}{8}$. See~\cite{Kobayashi:2010cm} for more details.

\subsection{Enhanced scalar perturbations at small scales as a source of GWs}
\label{sec:ScalarEnhancement}

\subsubsection{Second-order scalar perturbation theory.}
\label{sec:SecGWScal}

One may assume that the amplitude of the primordial density perturbations, determined very precisely by the Planck satellite at cosmological scales, has a similar value at every scale. According to the standard single field slow-roll inflationary family of scenarios, this should be precisely the case\footnote{Of course, in reality, slow-roll models deviate from this simple power-law picture, as towards the end of inflation the inflaton accelerates more and more, resulting in a larger deviation with respect scale-invariance}. Let us take at face value the Planck measurement of the scalar spectral index $1-n_s \simeq 0.04$ and of the curvature amplitude $\mathcal{P}_{\mathcal R}(k_{\rm CMB}) \simeq 2\cdot 10^{-9}$ determined at a pivot scale of $k_{\rm CMB} = 0.05 {\rm Mpc}^{-1}$, and assume that the power-law behavior $\mathcal{P}_{\mathcal R}(k) \propto k^{n_s-1}$ is valid at arbitrary smaller scales. The amplitude of the curvature perturbation at some small scale, say $k \sim 10^X k_{\rm CMB}$, $X > 1$, should be of the order $\mathcal{P}_{\mathcal{R}}(k \sim 10^X k_{\rm CMB}) \sim  10^{-0.04\cdot X}\mathcal{P}_{\mathcal{R}}(k_{\rm CMB})$, with $0.1 \lesssim 10^{-0.04\cdot X} \lesssim 1$, i.e.~marginally smaller than at CMB scales. The previous estimation indicates that, if the spectral index has no significant running, the scalar perturbation amplitude would not change much between large and small scales. This is however a huge extrapolation, as we do not know the underlying model of inflation. It is therefore possible, that the spectrum of scalar perturbations deviates significantly from quasi-scale invariance at small scales. In the following we will assume, in fact, that such deviation occurs.

Various scenarios can accommodate easily the production of a large amplitude of scalar fluctuations at small scales, while respecting CMB constraints at cosmological scales, see e.g.~\cite{GarciaBellido:1996qt,Clesse:2015wea,Garcia-Bellido:2017mdw}. Very large density perturbations at small scales may lead, in fact, to the formation of primordial black holes, which potentially might explain the origin of (a fraction of) the dark matter, see e.g.~\cite{Carr:2016drx,Bird:2016dcv,Sasaki:2016jop,Clesse:2016vqa,Garcia-Bellido:2017mdw,Ballesteros:2017fsr,Ezquiaga:2017fvi,Carr:2017jsz,Kannike:2017bxn}. In this section, however, we will not be concerned with a particular scenario or assumption, about the origin of the scalar perturbation at small scales. We will rather focus on the fact that first-order scalar perturbations (independently of their amplitude and statistics) inevitably generate gravitational waves at second (and higher) order~\cite{Matarrese:1993zf,Matarrese:1996pp,Matarrese:1997ay,Noh:2004bc,Carbone:2004iv,Nakamura:2004rm,Ananda:2006af,Baumann:2007zm,Mangilli:2008bw,Malik:2008im}. In particular, second-order terms in the perturbed Einstein equations, built from first-order scalar perturbations, act as a source for tensor perturbations. The resulting stochastic background probes in fact the amplitude and statistical properties of the scalar perturbations at small scales.

In order to quantify this effect, let us consider the following perturbed FRW metric
\begin{equation}
\label{eq:LongitudinalPertMetric}
 ds^{2} = a^{2}(\eta)[-(1+2\Phi)d\eta^{2}+[(1-2\Psi)\delta_{ij}+h_{ij}]dx^{i}dx^{j}]
\end{equation}
with $\Phi$, $\Psi$ the scalar metric perturbations, and $h_{ij}$ the transverse-traceless (TT) tensor perturbation. Expanding to second-order in perturbations the Einstein field equations, one can recognize that the bi-linear terms made out of first-order scalar perturbations, act as a source term $S_{ij}^{\rm TT}$ for the second-order tensor perturbations. Passing all bi-linear terms made from first-order scalar perturbations, into the right-hand side of the transverse-traceless part of the perturbed Einstein equation, one can write
\begin{equation}
\label{eq:GWeom2ndOrder}
 h''_{ij}+2{\cal H}  h'_{ij}+k^{2}h_{ij} = S_{ij}^{\TT}
 \,,
\end{equation}
where the source term reads~\cite{Ananda:2006af,Baumann:2007zm}
\begin{eqnarray}
\label{eq:2ndOrderSource} 
\hspace*{-2.5cm}  S_{ij} =
 2\Phi\partial_{i}\partial_{j}\Phi - 2\Psi\partial_{i}\partial_{j}\Phi
 + 4\Psi\partial_{i}\partial_{j}\Psi + \partial_{i}\Phi\partial_{j}\Phi
 - \partial^{i}\Phi\partial_{j}\Psi - \partial^{i}\Psi\partial_{j}\Phi
 + 3\partial^{i}\Psi\partial_{j}\Psi \nonumber\\
\hspace*{-2cm} - \frac{4}{3(1+w){\cal H}^{2}}\partial_{i}(\Psi'+{\cal
H}\Phi)\partial_{j}(\Psi'+{\cal H}\Phi) -\frac{2c_{s}^{2}}{3w{\cal H }}\,[3{\cal H}({\cal
H}\Phi-\Psi')+\nabla^{2}\Psi]\,\partial_{i}\partial_{j}(\Phi-\Psi)\,,\nonumber\\
\end{eqnarray}
with $w=P/\rho$ the equation of state and $c_s^2=P'/\rho'$ the adiabatic sound speed. Notice that in the above Eqs.~(\ref{eq:GWeom2ndOrder}), (\ref{eq:2ndOrderSource}), first order tensor and vector perturbations are considered negligible, in comparison with first-order scalar perturbations. Hence $S_{ij}^\TT$ is only made by the transverse-traceless part of the quadratic terms in first-order scalar perturbations, which behave effectively as a source term for induced second-order GWs.

From an observational point of view, we are particularly interested in second-order induced GWs during the radiation era, as only modes entering the horizon during that early stage of the Universe, are probed by the frequency range of GW detectors. For the effect of second-order induced tensors (as well as vectors) on the CMB, see~\cite{Mollerach:2003nq}. There, it is found that second-order GWs dominate over first-order GWs only if the tensor-to-scalar ratio at CMB scales is $r < 10^{-6}$. For the rest of the section we rather focus on the amplitude of the second-order induced GW background at smaller scales than those of the CMB. 

Let us consider second-order induced GWs during the radiation-dominated era, so that $w = 1/3$. Due to the non-linear dynamics, the amplitude of the GW spectrum for a given wavenumber $k$, can in principle be contributed by scalar perturbations from a large range of scales. However, in practice, the second-order tensor modes are primarily sourced when first-order density perturbations on same scale $\sim k$, cross inside the Hubble scale during the radiation era~\cite{Ananda:2006af,Baumann:2007zm}. In the following computation, the anisotropic stress from neutrinos will be neglected, so that we can take~$\Phi = \Psi$~\cite{Mukhanov:1990me}. Neglecting the anisotropic stress from neutrinos implies a $\sim 10 \%$ error for both first-order tensor and scalar perturbations. Therefore we only expect a $\sim 1 \%$ error in the amplitude of the spectrum of the second-order induced GWs, which in fact has been proven explicitly by numerical integration in~\cite{Baumann:2007zm}. One last assumption is to consider that the scalar perturbations are mostly Gaussian and exhibit a smooth power-spectrum $\mathcal{P}_{\mathcal R}$ over increasingly smaller scales, so that $\mathcal{P}_{\mathcal R}$ can be well described by a power law $\propto (k/k_*)^{n_s-1}$ around any scale $k_*$, even though $n_s(k)$ may exhibit a significant running so that $\mathcal{P}_{\mathcal R}$ grows from large (small $k$) to small (large $k$) scales.

Under the circumstances just mentioned, the amplitude of the induced second-order GW background today, for the modes produced during the radiation-dominated era (i.e.~for frequencies $f > 10^{-17}$ Hz), is characterized by an energy density spectrum (normalized to the critical density at the present time) as~\cite{Ananda:2006af,Baumann:2007zm}
\begin{equation}
\label{eq:OmegaGW2ndOrder}
\Omega_{\rm GW}(k) = F_{\rm rad}\,\Omega_{\rm rad}^{(0)}\,\mathcal{P}_{\mathcal R}^2(k)\,.
\end{equation}
For modes well inside the horizon at the end of the radiation-dominated era, i.e.~$k\eta_{\rm eq} \gg 1$, the prefactor reads~\cite{Ananda:2006af}
\begin{equation}
\label{eq:PrefactorGW2ndOrder} 
F_{\rm rad} = \frac{8}{3} \left( \frac{216^{2}}{\pi^{3}}\right) 8.3 \times 10^{-3} f(n_s)\,,
\end{equation}
where $f(n_s)$ is a weakly-dependent function on the spectral tilt, with $f(1) \approx 1$, becoming slightly smaller than unity for a red spectrum with $n_s < 1$, e.g.~$f(0.9) \approx 0.97$, and larger than one for a blue spectrum with $n_s > 1$, e.g.~$f(1.1) \approx 1.05$. See Fig.~6 of~\cite{Ananda:2006af}. 

Using Eqs.~(\ref{eq:OmegaGW2ndOrder}), (\ref{eq:PrefactorGW2ndOrder}), Ref.~\cite{Assadullahi:2009jc} has placed bounds on the the primordial scalar perturbation by means of the constraints on a stochastic background of GWs. Their numerical estimates are normalized to $F_{\rm rad} \approx 30$, as this corresponds to an approximately scale-invariant spectrum of scalar perturbations, $n_s \approx 1$. This represents a conservative lower bound on $F_{\rm rad}$ as, in reality, a blue spectrum with $n_s>1$ (and hence $F_{\rm rad} > 30$) is required to produce a detectable background of GWs on scales much smaller than the CMB scales. Confronting Eq.~(\ref{eq:OmegaGW2ndOrder}) against existing bounds on a stochastic GW background (or against projected sensitivity curves from future experiments), translates into bounds on the primordial curvature perturbation $\mathcal{P}_{\cal R}(k)$, at the corresponding scales of the observation/detector. These constraints are listed in Table~\ref{tab:2ndOrderGWs}.

\begin{table}[t]
\begin{center}
\begin{tabular}{|c|c|c|c|}
\hline
{\rm Experiment} & {\rm GW bound} & {\rm Derived Scalar Bound} & {\rm Frequency}\\ \hline
{\rm BBN} & $\Omega_{\rm GW} \lesssim 2\cdot 10^{-6}$ & $\mathcal{P}_{\mathcal R} < 2.72\cdot10^{-2}(30/F_{\rm rad})^{1/2}$ & $f \gtrsim 10^{-10}$ {\rm Hz}\\ \hline
{\rm LIGO} & $\Omega_{\rm GW} \lesssim 1.1\cdot 10^{-7}$ & $\mathcal{P}_{\mathcal R} < 6.38\cdot10^{-3}(30/F_{\rm rad})^{1/2}$ & $f \sim 10^{2}$ {\rm Hz}\\ \hline
{\rm LISA} & $\Omega_{\rm GW} \lesssim 2\cdot 10^{-13}$ & $\mathcal{P}_{\mathcal R} < 8.61\cdot10^{-6}(30/F_{\rm rad})^{1/2}$ & $f \sim 10^{-3} {\rm Hz}$\\ \hline
{\rm BBO/DECIGO} & $\Omega_{\rm GW} \lesssim 10^{-17}$ & $\mathcal{P}_{\mathcal R} < 6.09\cdot10^{-8}(30/F_{\rm rad})^{1/2}$ & $f \sim 1~{\rm Hz}$ \\ \hline
\end{tabular}
\caption{\label{tab:2ndOrderGWs} Derived bounds on the primordial curvature power spectrum at different scales. We have updated some of these constraints obtained by~\cite{Assadullahi:2009jc}, based on the updated bounds and sensitivity curves of future experiments. We specify in the second column the amplitude in $\Omega_{\rm GW}$ that we have used to infer the constraint on $\mathcal{P}_{\mathcal R}(k)$.}
\end{center}
\end{table}

Let us note that using Eqs.~(\ref{eq:OmegaGW2ndOrder}), (\ref{eq:PrefactorGW2ndOrder}), one can also derive a constraint over an effective tilt $\overline{n}_s$, describing the averaged slope between the logarithm of the curvature power spectrum at a scale $k$, and at the CMB scale,
\begin{equation}
\label{eq:nsEffective}
\log[\mathcal{P}_{\mathcal R}(k)] = (\overline{n}_s-1)(\log f - \log f_{\rm CMB}) + \log[\mathcal{P}_{\mathcal R}(f_{\rm CMB})]\,,
\end{equation} 
with $f_{\rm CMB} \sim 10^{-18}$~Hz. The constraints (upper bounds) on $\mathcal{P}_{\mathcal R}(k)$ listed in Table~\ref{tab:2ndOrderGWs}, represent therefore also a constraint on the value of the effective spectral index $\overline{n}_s$, across a wide range of scales (from CMB scales to the shorter scales probed by a given experiment/constraint on stochastic GWs). These constraints are displayed in Table~\ref{tab:2ndOrderGWsNs}.

\begin{table}[t]
\begin{center}
\begin{tabular}{|c|c|c|}
\hline
{\rm Experiment} & {\rm GW bound} & {\rm Effective Spectral Index}\\ \hline
{\rm BBN} & $\Omega_{\rm GW} \lesssim 2\cdot 10^{-6}$ & $\overline{n}_s < 1.89-\frac{1}{37}\log_{10}
\left(F_{rad}/30\right)$ \\ \hline
{\rm LIGO} & $\Omega_{\rm GW} \lesssim 1.1\cdot 10^{-7}$ & $\overline{n}_s < 1.36 -\frac{1}{92}\log_{10}
\left(F_{rad}/30\right)$ \\ \hline
{\rm LISA} & $\Omega_{\rm GW} \lesssim 2\cdot10^{-13}$ & $\overline{n}_s < 1.24 -\frac{1}{69}\log_{10}
\left(F_{rad}/30\right)$ \\ \hline
{\rm BBO/DECIGO} & $\Omega_{\rm GW} \lesssim 10^{-17}$ & $\overline{n}_s < 1.08-\frac{1}{83}\log_{10}
\left(F_{rad}/30\right)$ \\ \hline
\end{tabular}
\caption{\label{tab:2ndOrderGWsNs} Derived bounds on the effective spectral index between CMB scales and smaller scales probed by GW bounds/detectors. As in table~\ref{tab:2ndOrderGWs}, we have updated some of these constraints obtained by~\cite{Assadullahi:2009jc}, based on new bounds and sensitivity curves of future experiments.}
\end{center}
\end{table}

Even though it is probably unrealistic that the power in the curvature perturbation can be extrapolated to small scales just as in Eq.~(\ref{eq:nsEffective}) (where $\overline{n}_s$ is constant), the constraints derived in this way in Table~\ref{tab:2ndOrderGWsNs}, can be very useful: they are informative about the amount of `averaged blue-tilting' required for a given experiment to see this second-order induced GW background (given the amplitude of the primordial curvature perturbation at CMB scales).

Finally, let us make the following important remark. If we do not assume Gaussianity, the second-order induced GW spectrum would not be proportional to the square of the scalar power-spectrum, as in Eq.~(\ref{eq:OmegaGW2ndOrder}). However, only a rather strong deviation from Gaussianity would prevent the term in the $r.h.s.$ of Eq.~(\ref{eq:OmegaGW2ndOrder}) from being the dominant contribution to the GW power-spectrum. For a discussion about how induced second-order GWs can probe the non-Gaussian statistics of the scalar perturbation, and the relation to the abundance of primordial black holes (interpreted as a fraction of the dark matter), see~\cite{Nakama:2016gzw,Garcia-Bellido:2017aan}. For analogous constraints on the primordial black hole abundance and their interpretation as dark matter, but based on scalar perturbation with Gaussian statistics, see~\cite{Saito:2008jc,Saito:2009jt}.

\subsubsection{Primordial black holes from inflation.}
\label{sec:PBH}

Large peaks in the curvature power spectrum at certain scales, can give rise to primordial black holes (PBHs), when those scales reenter the horizon during the standard evolution of the universe after inflation\footnote{Let us remark that PBHs produced in the early Universe can emit GWs via Hawking radiation, if certain conditions are met. We do not consider this circumstance in the remaining of the section, but we point the interested reader to~\cite{Dong:2015yjs}.}. This mechanism typically gives rise to a population of isolated PBHs that could act as cold dark matter~\cite{GarciaBellido:1996qt,Nakamura:1997sm}. In certain models, like in hybrid models with long-lasting waterfall regimes, the peak in the curvature perturbation spectrum not only can be large, but also very broad~\cite{Clesse:2015wea}. The production of PBHs in such scenarios occurs then in clusters, which subsequently merge in the matter era, thus creating a stochastic background of GWs. For PBH masses in the range $M_{\rm PBH} \sim 10^2 - 10^4 M_\odot$, the peak frequency of this background could lie, for instance, within the LISA sensitivity range~\cite{Clesse:2016ajp}. As PBHs can act as dark matter, and hence as seeds for early galaxy formation, the GWs from these merging PBHs represent a promising probe of structure formation. In fact, it has been pointed out that these PBHs could have already been detected by aLIGO~\cite{Bird:2016dcv,Clesse:2016vqa,Sasaki:2016jop,Cholis:2016kqi}. In the future, we may then expect to measure the broad PBH mass distribution with ground- and space-based interferometers~\cite{Clesse:2016vqa,Clesse:2016ajp,Raidal:2017mfl}. For a recent analysis of the full mass range constraints on PBHs as dark matter, see~\cite{Carr:2016drx}. For a recent criticism (based on microlensing) on the ability of PBHs to represent all the dark matter in the universe, see~\cite{Zumalacarregui:2017qqd}, and for an opposite view, see~\cite{Garcia-Bellido:2017imq}.

Here we will focus on variants of the Hybrid inflation scenario. In these set-ups inflation ends at some point of the inflaton $\phi$ slow-roll trajectory, due to the occurrence of a phase transition driven by a symmetry breaking field $\psi$, usually referred to as the `water fall' field. The water fall transition of $\psi$ is actually triggered when the inflaton crosses through a critical value $\phi_c$. The original hybrid potential~\cite{Linde:1993cn} predicts a blue-tilted scalar spectrum, which is clearly ruled out nowadays by CMB observations. Following~\cite{Clesse:2015wea,Clesse:2016ajp}, we consider therefore a more general form for the effective two-field potential, 
\begin{equation}\label{eq:HybridPotentialII}
V(\phi,\psi) = \Lambda \left[ \left(1-\frac{\psi^2}{v^2}\right)^2 + \frac{(\phi-\phi_c)}{\mu_1} 
-  \frac{(\phi-\phi_c)^2}{\mu_2^2}+ \frac{2 \phi^2 \psi^2}{\phi_c^2 v^2} \right] \,.
\end{equation}
This potential exhibits a negative quadratic term in $\phi$, so that it generates a red tilted spectrum at the CMB scales. It also has a linear term in $\phi$, which controls the duration of the waterfall phase. Initially, inflation takes place along the valley $\psi = 0$. Below the critical value, when $\phi < \phi_c$, the potential develops a tachyonic instability, forcing the field trajectories to reach a global minima located at $\phi = 0$, $|\psi| = v$. Apart from the additional quadratic and linear terms, the potential is identical to the one of the original hybrid model. Closely related models, like inverted hybrid, can also be constructed. Other well motivated scenarios, similar in spirit but different than hybrid variants, may also end inflation abruptly due to a field instability~\cite{Renaux-Petel:2015mga,Renaux-Petel:2017dia}. These scenarios can potentially generate a similar phenomenology to Hybrid models, but the details remain yet to be developed. Certain type of single field inflation has also been shown to create high peaks in the curvature power spectrum leading to PBH formation later on~\cite{Garcia-Bellido:2017mdw,Ballesteros:2017fsr,Kannike:2017bxn}. Therefore, even though the following discussion will be linked to the scenario Eq.~(\ref{eq:HybridPotentialII}), the details of the PBH merging history and associated GW production, can be associated to a broader group of scenarios.

In Eq.~(\ref{eq:HybridPotentialII}), $\mu_1$ and $\mu_2$ control the slope and curvature of the potential at the critical point, respectively. Taking $\mu_1$ sufficiently large compared to $\mu_2$, guarantees the slope along the valley to be constant over a sufficiently large range of scales, from CMB scales down to the scales exiting the Hubble radius when the critical instability point $\phi = \phi_c$ is reached. CMB observables fix in fact part of the parameters of the model. Namely, taking 2015 Planck's best fit for the scalar spectral index $n_s \simeq 0.967$, the amplitude of the curvature power spectrum (evaluated at a pivot scale $k_*= 0.05~ {\rm Mpc}^{-1}$) $\mathcal P_\zeta(k_*)  =  2.21\times 10^{-9}$ (in this section we denote $\mathcal P_\zeta$ what is $\mathcal P_\mathcal R$ in the rest of the review), and the upper bound on the tensor-to-scalar ratio $r \lesssim 0.07$, one arrives (assuming for simplicity instantaneous reheating) at $\mu_2 \simeq 10\,m_{\rm Pl}$, $\mu_1 \gtrsim 10\,m_{\rm Pl}$, and $\Lambda/m_{\rm Pl}^4 \simeq 2.6\cdot 10^{-7}m_{\rm Pl}^2/\mu_1^2$, see~\cite{Clesse:2016ajp,Bartolo:2016ami} for details.

The capability of observing (constraining in the absence of detection) the GW background associated with the peak(s) in the matter power spectrum at small scales well below the CMB scales, in turn constrains the model. The primordial curvature power spectrum in this scenario can be written as~\cite{Clesse:2015wea}
\be
{\cal P}_\zeta (k)  = \frac{\Pi^3}{\sqrt{2\pi}} \times \exp \left[-\frac{{4(N_c - N_k)}^2}{\Pi^2}  \right]\,, \hspace{0.5cm} \Pi \equiv {v \over m_{\rm Pl}}\sqrt{{\mu_1 \over m_{\rm Pl}}{\phi_c \over m_{\rm Pl}}}\,,
\label{eq:PShbrid}
\ee
which is maximal at the critical point of instability, $N_c$ e-folds before the end of inflation. A detail calculation leads to $N_c \simeq 2\Pi$. The mild-waterfall induces a broad peak in the scalar power spectrum for modes leaving the horizon just before the critical point, and just afterwards. Let us note that the geometrical combination of parameters entering in $\Pi$ controls completely the power spectrum shape. It is in fact difficult to modify independently the width, height and the position of the power spectrum peak, as these properties depend all directly on $\Pi$.

When a given mode re-enters the horizon during the matter era, it may lead to the formation of a black hole if the corresponding mode amplitude is sufficiently large. The amplitude of the fluctuation needs to be above a certain critical value of the order of $\zeta_c \simeq 0.03 - 0.3$, according to numerical and analytical estimations~\cite{Carr:2009jm}. Following~\cite{Bartolo:2016ami}, we will take $\zeta_c = 0.1$ as a fiducial value. Approximating the probability distribution of density perturbations to be Gaussian, the fraction $\beta$ of the Universe collapsing into primordial black holes of mass $M$ at the time of formation $t_M$, can be evaluated as
\bea
\beta^{\rm form}(M) &\equiv& \left. \frac{\rho_{\rm PBH} (M) }{\rho_{\rm tot}} \right|_{t=t_M} \nonumber\\
  &=&   \int_{\zeta_c}^{\infty} \frac{{\rm d} \zeta}{\sqrt{2 \pi} \sigma}\, e^{- \frac{\zeta^2}{2 \sigma^2}} = \frac{1}{2} \, {\rm erfc} \left( \frac{\zeta_c}{\sqrt 2 \sigma} \right) 
  \simeq \frac{ \sigma}{ \sqrt{2\pi}\, \zeta_c} \, e^{- \frac{\zeta_c^2}{2 \sigma^2 }}\,, \label{eq:betaapprox}
\eea
where the last expression is only valid in the limit $\sigma \ll \zeta_c$. Let us note that the variance $\sigma$ is related to the power spectrum simply as $\langle \zeta^2 \rangle = \sigma^2 = \mathcal P_\zeta (k_M) $, with $k_M$ the wavelength of the mode re-entering inside the Hubble radius at the time $t_M$.   

In the scenario of mild waterfall that we are considering, the peak in the power spectrum of scalar perturbations is broad, spanning over several orders of magnitude in wavenumber.  Hence, the PBHs are formed at different times during the expansion history of the Universe, and have a broad mass spectrum (as opposed to a `monochromatic' distribution of PBHs, which has been traditionally assumed in the past). As the relative contribution of PBHs to the total energy density in the radiation era grows like $\propto a(t)$, the contribution to the total energy density from PBHs with low masses, forming earlier\footnote{We neglect the cases where evaporation through Hawking radiation is relevant, since this only concerns PBHs with very low masses that were formed immediately after inflation. }, may weight more at the onset of the matter era, than the more massive PBHs forming later, for a given identical value of $\beta^{\rm form}$. The total density of PBHs at radiation-matter equality is obtained by integrating $\beta^{\rm eq}$ over masses $\Omega_{\rm PBH}(z_{\rm eq})  = \int_0^{M_{t_{\rm eq}}} \beta(M,N_{\rm eq} )  {\rm d} \ln M$, which must be equal to $\Omega_{\rm PBH}(t_{\rm eq}) \simeq 0.42$ if PBHs are to constitute the totality of the dark matter (the rest of the matter constituted by baryons). 

For every set of $\lbrace \mu_1,v,\phi_c \rbrace$ parameters one can determine the value of $\zeta_c$ necessary for the PBHs to represent the right amount of dark matter. To place the mass spectrum peak in the range where there is no solid observational constraint for PBHs, one can take $\Pi \sim \mathcal{O}(10^2)$. Taking combinations of parameters ranging as $\mu_1/m_{\rm Pl} \sim 10^2-10^8$, $v/m_{\rm Pl} \sim 0.01-1$, and $\phi_c/m_{\rm Pl} \sim 10^{-2}-1$, it is easy to find parameters for which $\Pi$ ranges between $\sim 150-375$, leading to critical values of the order of $\zeta_c \sim 0.1-10$. The mass range for PBHs is very broad, from $\sim 10^{-20} \msun$ to $\sim 10^{5} \msun$,  but for a given set of parameters, the mass spectrum typically covers 3-5 decades at matter-radiation equality. Given $\Pi^2$, PBHs can be made arbitrarily massive by increasing $\mu_1$ but reducing $v$ and $\phi_c$, hence lowering the energy scale of inflation, and increasing at the same time the PBH masses (this does not affect importantly the shape of the mass spectrum).  

Refs.~\cite{Clesse:2015wea,Clesse:2016ajp,Garcia-Bellido:2017imq} consider a broad lognormal distribution for the PBHs as
\begin{equation}
PDF(M) =  \frac{1}{M\sqrt{2\pi\sigma^2}} \exp\left(-
\frac{\log^2(M/\mu)}{2\sigma^2}\right)\,,
\label{eq:PDF}
\end{equation}
although the origin of this is not entirely clear. For the mild-waterfall hybrid model we have been discussing, $\sigma = \Pi$, whereas a detail calculation leads to $\mu \simeq 0.2 {4\pi m_{\rm Pl}^2\over H_{N_c}}e^{2{N_c}}$, with $N_c$ the number of e-folds till the end of inflation after the system reaches the critical point; typically $\mu \sim$ few $\times \mathcal{O}(10) \msun$.

The GW background from inspiraling BHs can be obtained from the GW emission of binary systems~\cite{Phinney:2001di},
\begin{equation}
\frac{d\,\rho_{\rm GW}}{d\ln f} = \int_0^\infty\frac{dz}{1+z}\,\frac{dn}{dz}\,
\frac{\pi^{2/3}}{3}\,{\cal M}_c^{5/3}\,(G\,f_r)^{2/3}\,,
\label{eq:rhoGW}
\end{equation}
where $dn/dz$ is the number density of black hole binaries per unit of redshift, 
${\cal M}_c^{5/3}=m_1m_2(m_1+m_2)^{-1/3}$ is the chirp mass, and $f_r = f(1+z)$ is the frequency at the source. Assuming a constant merger rate as a function of redshift, say $\tau_{\rm merger}\simeq 50$ events/yr/Gpc$^{3}$ as inferred from aLIGO detection events~\cite{Abbott:2016nhf}, after integrating over masses with a broad mass distribution like (\ref{eq:PDF}) for both $m_1$ and $m_2$ [with the same parameters $(\mu,\,\sigma)$], one arrives at~\cite{Clesse:2016vqa}
\begin{eqnarray}\label{eq:ratio}
\hspace*{-2cm} h^2\,\Omega_{\rm GW}(f) = \ 8.15\times10^{-15} \ \tau_{\rm merger} \,
\left(\frac{f}{{\rm Hz}}\right)^{2/3}
\left(\frac{\mu}{M_\odot}\right)^{5/3} R(\sigma) \,, \\
\hspace*{-2cm}R(\sigma) = \frac{e^{\frac{793}{882}\,\sigma^2}}{1245889}
\left(639009 + 583443\,e^{\frac{2}{21}\,\sigma^2} + 30429\,
e^{\frac{40}{21}\,\sigma^2} - 9177\,e^{\frac{122}{21}\,\sigma^2} 
+ 2185\,e^{\frac{82}{7}\,\sigma^2}\right)\,, \nonumber
\end{eqnarray}
which becomes $R(\sigma=0) = 1$ for a monochromatic spectrum with mass $M=\mu$. We see from this expression that the width of the mass spectrum can significantly enhance the GW amplitude. Depending on the parameters, this GW background may very well be well inside the sensitivity of GW direct detection experiments. From the point of view of the parameter space, there is a degeneracy between the merger rate $\tau_{\rm merger}$ and the mean mass $\mu$, which satisfies $\tau_{\rm merger}\times\mu^{5/3}(M_\odot)=$ const. Taking into account e.g.~the projected sensitivity of LISA, we notice that this particular GW background from inspiralling PBHs, can be detectable by this experiment for a very wide range of parameters, see e.g.~Fig.~18 in~\cite{Bartolo:2016ami}.

\subsection{Alternatives to inflation}
\label{sec:InfAltern}

Various scenarios of the early Universe have been proposed as a complementary approach, when not directly as an alternative, to an inflationary phase of accelerated expansion. Each of these scenarios present certain conceptual caveats, of one kind or another, that may challenge their viability. Given the observed properties of the cosmological perturbations, exhibiting Gaussian statistics, a scalar spectral index close to scale invariance but strictly less than unity (at $\sim 5\sigma$ level), and the absence of isocurvature modes at any detectable level, it is hard to imagine any mechanism responsible for these other than quantum vacuum fluctuations of an effective scalar degree of freedom. Thus, even though an intense debate about the pros and cons of alternative scenarios is still ongoing, inflation appears to be the most natural paradigm simultaneously solving the problems of the hot Big Bang, while explaining the origin of the cosmological perturbations. We have focused accordingly, so far, only on inflationary models. Alternative/complementary scenarios predict nonetheless, the emission of a stochastic background of GWs, with a spectrum rising up in frequencies. In some cases, such blue-tilted backgrounds can in fact evade bounds from CMB or PTA's, while still sustain a large background amplitude within the frequency range of GW direct detection experiments. In the following, we will discuss some of the basic features of the GW backgrounds predicted by these scenarios, and the potential constraints expected from GW detectors. We will simply focus on the properties of the GW backgrounds they predict, referring the reader interested on more elaborated discussions on the general properties of these scenarios, to the appropriate references.

\subsubsection{Pre-Big-Bang models.}

Underlying duality symmetries in string theory have led to suggest that a cosmological 'super-inflation' phase~\cite{Lucchin:1985wy} of growing space-time curvature and accelerated expansion, may arise in the early Universe. The resulting string cosmology set-up is called the {\it Pre-Big Bang} scenario~\cite{Gasperini:1992em}, and it is known that it can led to a stochastic background of GWs with a blue-tilted spectrum~\cite{Grishchuk:1991kk,Gasperini:1992pa}. This background can evade bounds from CMB or pulsar timing, while sustaining a relatively large amplitude at the frequency scale of GW interferometers~\cite{Gasperini:1992dp,Brustein:1995ah,Brustein:1996ut}. 

For a comprehensive review of the {Pre-Big Bang} scenario and the possible problems arising in the construction of a viable model, see~\cite{Gasperini:2002bn,Gasperini:2007zz}. A working scenario can be divided typically in three phases: first a low curvature isotropic super-inflationary evolution, characterized by a weak coupling regime controlled by a logarithmically growing dilaton. This is matched to a second 'string' phase, characterized by a large curvature regime, where the dilaton grows linearly. Finally this second phase is matched to a third and final RD era. A significant amount of work has been devoted to understanding how to make viable this type of scenarios, curing with string effects the apparent singularity that emerges\footnote{During the super-inflationary evolution the Hubble parameter $H = \dot a/a$ grows until it reaches the string length scale $\lambda_s$, hence requiring new corrections beyond the lowest order string action that is used in first place to describe the system.}, or dealing with matching the end of this 'pre-big bang' phase with the standard cosmological evolution, see~\cite{Gasperini:2007zz} and references therein. 

Here we simply focus on the basic properties of the GW expected in these set-ups. In the context of a minimal Pre-Big Bang model (characterized by the three phases mentioned above), the GW spectrum today $\Omega_{\rm GW}(f)$ depends on the redshift $z_s\equiv f_1/f_s$ of the string phase, where $f_s$ is just the inverse duration of the phase, and
\begin{eqnarray}
f_1\sim 4.3\times 10^{10} {\rm Hz}\, (H_s/(0.15 M_{Pl}))(t_1/\lambda_s)^{1/2}\,,
\end{eqnarray}
represents the cut-off frequency of the spectrum, with $H_s$ is the Hubble parameter during such phase, $\lambda_s$ the string length, and $t_1$ the time when the string phase ends. The spectrum also depends on the value $g_s$ of the string coupling at the onset of the high-curvature phase. These two parameters $\lbrace z_s, g_s\rbrace$ affect the frequency behaviour as well as the peak amplitude of the spectrum.
 
The low frequency end of the spectrum scales as a power law like $\Omega_{\rm GW}\propto f^3$. In the high frequency range ($f_s \ll f \ll f_1$) the spectrum scales as~\cite{Gasperini:1992dp,Brustein:1995ah}
\begin{eqnarray}
\Omega_{\rm GW} \sim {(2\pi f_1)^4\over H_0^2 M_{Pl}^2}(f/f_1)^{3-2\mu}\,, 
\end{eqnarray}
where $\mu$ is a dimensionless free parameter, which quantifies the growth of the dilaton during the string phase. Clearly, for $\mu=3/2$, the spectrum is flat at large frequencies. If $\mu$ departures from 3/2 towards the smaller values side, the spectral slope increases. 

For each point in the $(\mu, f_1)$ plane, one can evaluate the spectrum of GW as predicted by the Pre-Big bang scenario, and check whether it can be probed by interferometer experiments. Of course any viable scenario has to produce a spectrum of scalar perturbations compatible with CMB constraints. Using the sensitivity curves of LISA and/or LIGO, one can explore the parameter space that is compatible with both scalar perturbations at CMB scales, and tensor perturbations observable at interferometers in the form of a stochastic background of GWs. The most updated results in this respect have been presented recently in~\cite{Gasperini:2016gre}, where it was concluded that: $i)$ in order to have a GW background detectable by LISA, this must have been produced by a regime with a long `stringy' high curvature phase, with $\log_{10} z_s > 8$, and $ii)$ in order to make it detectable by the O5 stage of aLIGO ($\sim 2020$), the GW background must have been produced during a regime of small enough growth of the string coupling during the high-curvature phase, with $\log_{10} z_s \gtrsim -2$.

Let us finally note that given the definition above of $f_1$, a bound on this quantity can be used to get an exclusion curve in the $(t_1/\lambda_s, H_s/M_{Pl})$ plane. Hence, one could potentially use GW interferometers to constrain string related parameters. For more details, we refer the reader to~\cite{Gasperini:2016gre} and references therein. 

\subsubsection{String gas cosmology.}

String gas cosmology is also a scenario based on string theory. It was first proposed in Ref.~\cite{Brandenberger:1988aj}, in order to account for the observed dimensionality of spacetime, see also~\cite{Sakellariadou:1995vk, Easther:2004sd, Greene:2012sa}. It assumes that the universe consists of a gas of weakly interacting fundamental strings on a toroidal compactification. Independently of the 'decompactification' mechanism, the proposal also leads naturally to the causal generation of a slightly red-tilted spectrum of curvature perturbations, sourced by the thermal excitation of strings wrapped around the space-time~\cite{Nayeri:2005ck}. It also predicts the generation of tensor metric perturbations that can be identified, as usual, with GWs. The tilt for the tensor mode spectrum is however of opposite sign to the scalar perturbations [see Eq.~\eqref{eq:tiltrelation}], corresponding to a blue-tilted background of GWs~\cite{Brandenberger:2006xi}. Note that this is in contrast to the prediction we derived for the simplest models of inflation, see Section~\ref{sec:PrimTensInf}, where the tensor spectrum is expected to be slightly red-tilted, recall Eq.~(\ref{eq:nT2e}).

The curvature and tensor perturbations are both dictated by the scaling properties of the free energy of a weakly coupled gas of strings at high energies~\cite{Brandenberger:2014faa}. Making the scenario viable requires a `controlled' background to exist. This remains however as an open question. Leaving aside this difficulties, it can be shown that the tensor and scalar spectra are simple power-laws, with spectral indexes related as~\cite{Brandenberger:2014faa}
\begin{equation}
\label{eq:tiltrelation}
n_T \approx  (1 - n_s).
\end{equation}
As a general question, one can proceed to quantify how blue-tilted a tensor spectrum has to be, in order to be detectable by a given interferometric detector. Given a tensor to scalar ratio $r_*$ fixed at the pivot scale $k_* = 0.002$ Mpc$^{-1}$, the energy density per logarithmic interval of a stochastic background of GWs can be estimated as
\begin{equation}
h^2\Omega_{\rm GW}(k_*) \approx 10^{-14.3} r_* .
\end{equation}
Let us denote now the frequency of the peak sensitivity of a given detector as $f_{\rm exp}$, where a stochastic background is expected to be detected if its amplitude is higher than a given threshold $h^2\Omega_{\rm GW} \gtrsim 10^{-X_{\rm exp}}$. If a blue spectrum of GWs is to be detected by a given experiment, we then require
\begin{equation}
h^2\Omega_{\rm GW}(f_{\rm exp}) \approx 10^{-15} r_* \left(\frac{f_{\rm exp}}{f_*}\right)^{n_T} \gtrsim 10^{-X_{\rm exp}} \,,
\end{equation}
where the frequency $f_* \simeq 10^{-17}$ Hz corresponds to the CMB scale $k_* = 0.002$ Mpc$^{-1}$. We find that in order for an experiment to detect a blue-tilted GW background, we require
\begin{equation}
\label{eq:bluetilt}
n_T \gtrsim {14.3\over\log_{10}(f_{\rm exp}/f_*)} - {X_{\rm exp}\over\log_{10}(f_{\rm exp}/f_*)} - {\log_{10} r_*\over\log_{10}(f_{\rm exp}/f_*)}  
\end{equation}
If we take LISA as a reference, then $f_{\rm exp} \sim 10^{-3}-10^{-2}$ Hz, corresponding to frequencies approximately 15 decades above those associated to CMB scales. The LISA peak sensitivity corresponds to $h^2\Omega_{\rm GW} \geq 10^{-X_{\rm exp}} \sim 10^{-13.5}$. Thus, inserting $f_{\rm LISA}/f_* \sim 10^{15}$, $X_{\rm exp} \simeq 13.5$, and using the latest constraint on the tensor-to-scalar ratio $r_* < 0.07$, we obtain
\begin{equation}
\label{eq:bluetiltLISA}
n_T \gtrsim 0.95 - 0.90 + 0.08 \simeq 0.13\,,  
\end{equation}
in order for a signal to be detectable by LISA\footnote{Here we assumed that the blue-tilted power-law behavior holds exactly, spanning from the CMB scales all the way up to LISA frequencies.}. In light of Eq.~(\ref{eq:tiltrelation}), this implies that we need scalar fluctuations with a spectral index $n_s \lesssim 1 - 0.13 \simeq 0.87$, in clear incompatibility with the measured value $n_s \simeq 0.96$ at the CMB scales. Turning around Eq.~(\ref{eq:bluetilt}), we see that $n_s \simeq 0.96$ implies that $n_T \simeq 0.04$. From this we deduce that, in order to observe a blue-tilted background at a frequency say of $\sim 1$ Hz, we need a an experiment with a peak sensitivity of $h^2\Omega_{\rm GW} \geq 10^{-X_{\rm exp}} \sim 10^{-15}$. 

\subsubsection{Cyclic/Ekpyrotic Models.}

The {\it ekpyrotic} scenario proposes a contracting phase in the cosmic history, preceding the creation of the matter and radiation in the Universe, which potentially can solve the standard cosmological puzzles of the hot Big Bang theory~\cite{Khoury:2001wf,Khoury:2001bz,Khoury:2001zk}. The `Bang' is substituted by a bounce, which separates an initial era of contraction (before the bounce) from the standard era of expansion (after the bounce). A cyclic (ekpyrotic) cosmology incorporates this idea, proposing a cosmological history based on an eternal succession of bounces at regular intervals. 

In brief, the cyclic cosmology scenario can be described in terms of the evolution of a field $\phi$ in 3+1 dimensions, rolling back and forth in an appropriately `engineered' potential $V(\phi)$. This corresponds to the periodic collision of orbifold planes moving in an extra dimension, with the field theory description representing simply the long wavelength approximation to the brane picture, by which the potential represents the inter-brane interaction, whilst the modulus field determines the distance between branes. For a review on the Ekpyrotic/cyclic scenario(s) see~\cite{Lehners:2008vx}, and for a critical review on bouncing cosmologies see~\cite{Battefeld:2014uga}. Here we just focus on the prediction of the GW background expected from this scenario.

Under the assumption that the perturbations pass smoothly through the bounce (something unproven and controversial), it is found that a blue-tilted spectrum of tensor perturbations emerges during the contracting era. The spectral behavior of the background at large frequencies goes as~\cite{Boyle:2003km,Lehners:2008vx} 
\begin{equation}
\Omega_{\rm GW} \propto k^{n_T}, ~~~n_T \approx 3 - {|3-\epsilon|\over|\epsilon-1|}\,,
\end{equation}
with $\epsilon \equiv {3\over2}(1+w)$ the ``slow-roll" parameter, and $w$ the equation of state at the slowly contacting phase, during which the tensor modes are excited. In these scenarios $w \gg 1$, so the GW background corresponds therefore to a blue spectrum with tilt $n_T \simeq 2$. This may led us to think, erroneously, that if the amplitude of the tensor spectrum is sufficiently large at the CMB scales (yet below the current upper limit), we might detect this background at direct detection experiments. The strongest observational constraint on this scenario comes however from BBN, enforcing the amplitude of the signal (at the CMB scales) to be orders of magnitude smaller than the current CMB limit. This also implies that the signal is way below the sensitivities of aLIGO and LISA, at their corresponding frequency window. The detection of the imprint of a quasi-scale invariant tensor spectrum in the B-mode polarization of the CMB, or alternatively the direct detection of a stochastic background with a blue-tilted power-law spectrum by aLIGO or LISA, would rule out this cyclic model.

Let us also remark that other class of nonsingular bouncing cosmologies have been proposed, like the matter-bounce paradigm and its variants, which can be viewed as an extension of inflationary cosmology with a matter contraction before inflation, see e.g.~\cite{Finelli:2001sr,Cai:2012va,Cai:2014jla,deHaro:2015wda}. For a review on these scenarios and their confrontation to theoretical and observational constraints, see~\cite{Cai:2014bea,Brandenberger:2016vhg}.

\section{Preheating and other non-perturbative phenomena}
\label{sec:PreheatingAndOthers}

A simple consequence of inflation is that any particle number density that could be present during the inflationary period, is exponentially diluted away due to the exponential stretching of space. At the end of inflation, essentially only the energy density responsible for inflation is present. The hot Big Bang (hBB) framework describes however the early Universe as filled up with relativistic species in thermal equilibrium. How can the end of inflation be matched with the thermal era? It seems clear that the Universe must have undergone a period of particle creation after inflation. This is know as the {\it Reheating} stage, and we expect that (almost) all the matter in the Universe was created then. In a way, the reheating process represents the `Bang' of the hBB paradigm. 

The energy of the inflaton must be therefore converted somehow into the particle species that dominate the Universe energy budget during the thermal era. The idea is that by coupling the inflaton to other particle species, the inflaton will eventually decay into such fields. A specific realization of reheating consists of a given particle physics model, specifying the form of the couplings between the inflaton and other matter fields. Depending on the model, different mechanisms of particle creation may take place, from inflaton perturbative decays~\cite{Abbott:1982hn,Dolgov:1982th,Linde:2005ht} to non-perturbative effects via parametric resonance~\cite{Kofman:1994rk,Kofman:1997yn,Greene:1997fu,Traschen:1990sw,Shtanov:1994ce,Kaiser:1995fb,Kaiser:1997mp,Kaiser:1997hg} or spinodal instabilities~\cite{Felder:2000hj,Felder:2001kt,Copeland:2002ku,GarciaBellido:2002aj}. Whichever the mechanism of particle production, the created species are expected to interact among themselves, and eventually reach thermal equilibrium. The {\it Reheating temperature} $T_{\rm RH}$ is defined as the common temperature first reached by all (dominant) species produced during reheating. It characterizes the moment when the evolution of the Universe can be first ascribed to the hBB framework.

Whenever particle production during reheating takes place via non-perturbative effects, we refer to this stage of the evolution of the Universe as {\it preheating}. For a review on preheating scenarios see~\cite{Allahverdi:2010xz,Amin:2014eta}. Preheating mechanisms are actually particularly interesting from the point of view of GWs, as the non-perturbative effects help, not only to reheat efficiently the universe, but also to create a large amount of GWs. In the following we describe standard preheating mechanisms and their production of GWs. We consider standard parametric resonance of scalar fields in Sect.~\ref{sec:ParamRes}, fermionic parametric excitation in Sect.~\ref{sec:Fermionic}, and spinodal instabilites in Sect.~\ref{sec:Tachyonic}. We also consider similar non-perturbative phenomena, possibly occurring in the post-inflationary early universe, albeit not necessarily related to (p)reheating. In particular we study the decay of flat-directions (Sect.~\ref{sec:susyQ}) and the production of oscillons (Sect.~\ref{sec:Osci}), and how these phenomena may source as well significant backgrounds of GWs.

\subsection{Parametric resonance of Bosons.}
\label{sec:ParamRes}

If the potential of the inflaton has a monomial shape during the stages following the end of inflation, the inflaton field will start oscillating around the minimum of its potential, typically with a large amplitude. The inflaton oscillations can induce parametric resonance in those particle species coupled to the inflaton, depending on the strength of their coupling. This corresponds to the creation of particles in energetic bursts. In the case of bosonic species, the production of particles is resonant. The process of energy transfer into the created species is then exponentially fast, ending after few oscillations of the inflaton, see Refs.~\cite{Kofman:1994rk,Kofman:1997yn,Greene:1997fu}, \cite{Traschen:1990sw,Shtanov:1994ce}, \cite{Kaiser:1995fb,Kaiser:1997mp,Kaiser:1997hg}, or the more recent ones~\cite{Amin:2014eta,Figueroa:2016wxr}. In the case of fermions, Pauli blocking prevents resonance from developing. There is however a significant transfer of energy into the fermion species, as fermionic modes with successive shorter wavelength at each inflaton oscillation, result successively populated, see~\cite{Greene:1998nh,Greene:2000ew,Peloso:2000hy,Berges:2010zv}. We will focus first in the parametric excitation of bosons, and postpone the discussion on the fermionic case for Sect.~\ref{sec:Fermionic}. 

The violent excitation of bosonic species via parametric resonance is expected to produce a significant amount of GWs~\cite{Khlebnikov:1997di,Easther:2006gt,Easther:2006vd,GarciaBellido:2007af,Dufaux:2007pt,Dufaux:2008dn,Figueroa:2011ye,Bethke:2013vca,Bethke:2013aba}. In order to study the GWs created by the excitation of some field(s) undergoing parametric resonance, let us consider an (initially) homogeneous field $\phi$ oscillating around the minimum of its potential $V(\phi)$. Let us also consider that it is coupled to another scalar field $X$, with coupling $g^2\phi^2 X^2$. We will typically refer to $\phi$ and to $X$, as the mother and daughter fields, respectively. The classical equations of motion (EOM, describing the dynamics of this coupled system, are
\bea \label{eq:generic-eom}
\ddot \phi - \frac{1}{a^2} \nabla^2\phi + 3 H \dot \phi + g^2  X^2 \phi  + \frac{\partial V(\phi)}{\partial \phi} = 0 \ , \\
\label{eq:generic-eomII}
\ddot X -  \frac{1}{a^2} \nabla^2 X + 3 H \dot X + g^2 \phi^2 X = 0   \ ,\eea
where $H \equiv {\dot{a}/a}$ is the Hubble rate. Disregarding the terms $a^{-2}\nabla^2\phi$ and $g^2X^2\phi$ (i.e.~considering $\phi$ homogenous and assuming that the 'feedback' of $X$ into $\phi$ is initially negligible), the solution to Eq.~(\ref{eq:generic-eom}) admits, for a power-law potential $V(\phi) \propto \phi^n$, an oscillatory solution as~\cite{Turner:1983he} 
\begin{equation}\label{eq:ApproxVarPhi}
\phi(t) \approx \Phi(t)F(t)\,,
\end{equation}
with $F(t)$ an oscillating function,  and $\Phi(t) = \Phi_{\rm i}(t/t_{\rm i})^{-2/n}$ a damping amplitude, where $_{\rm i}$ labels the initial time at the onset of oscillations. For monomial potentials $V(\phi) = {1\over n}\lambda M^{4-n}\phi^n$, with $M$ some mass scale and $\lambda$ a dimensionless coefficient, $F(t)$ is actually not periodic, except for $n = 2$. Yet, the frequency of oscillations changes only relatively slowly in time, as $\Omega_{\rm osc} \equiv \sqrt{d^2V/d\phi^2} = \sqrt{\lambda} M^{2-n/2}\Phi^{(n/2-1)} \equiv \omega_{\rm i}(t/t_{\rm i})^{1-2/n}$, with $\omega_{\rm i} \equiv \sqrt{\lambda} M^{2-n/2}\Phi_{\rm i}^{(n/2-1)}$. 

In the most relevant situations where parametric resonance takes place in the early Universe, the oscillatory field $\phi$ is a classical homogeneous field, while the field $X$ is considered to be a quantum field, initially in vacuum. We therefore quantize the scalar field $X$ by a standard mode decomposition
\begin{equation}
\label{eq:ChiQuant}
X(\mathbf{x},t) = {1\over a(t)}\int\frac{d \mathbf{k}}{\left(2\pi\right)^{3}}e^{-i\mathbf{k\cdot x}}\left[\hat{a}_{\mathbf{k}}\chi_{\mathbf{k}}(t)+\hat{a}_{-\mathbf{k}}^{\dagger}\chi_{-\mathbf{k}}^{*}(t)\right],
\end{equation}
where the creation/annihilation operators satisfy the canonical commutation relations $
[\hat{a}_{\mathbf{k}},\hat{a}_{\mathbf{k'}}^{\dagger}]$ $= \left(2\pi\right)^{3}\delta^{(3)}(\mathbf{k-k'})$, with other commutators vanishing. The (initial) vacuum state is defined as usual, with $\hat{a}_{\mathbf{k}}|0\rangle = 0$. Introducing the field re-definitions
\bea
\phi ~~\longrightarrow~~ \varphi \equiv a(t){\phi\over\Phi_{\rm i}}\,,~~~~~~ X ~~\longrightarrow~~ \chi \equiv a(t)X\,,
\eea
the EOM of the sub-horizon modes of the $\chi$ field, read
\begin{equation}\label{eq:ParamResEqFourier}
{d^2\over dz^2}\chi_{k} + \left(\kappa^{2} + q_{\rm i} \varphi^2\right)\chi_{k} \simeq 0\,,~~~~~ \kappa \equiv {k\over \omega_{\rm i}}\,,~~~z \equiv \omega_{\rm i}\int_t {dt'\over a(t')}\,,
\end{equation}
where $q_{\rm i}$ is the so called {\it resonance parameter}
\be\label{eq:ResParam}
q_{\rm i} \equiv {g^2\Phi_{\rm i}^2\over \omega_{\rm i}^2}\,.
\ee
Depending on the potential $V(\phi)$, the resonance parameter may be written as $q_{\rm i} \propto g^2\Phi_{\rm i}^2/\omega_{\rm i}^2$, with some proportionality factor of order $\sim \mathcal{O}(1)$. For instance, for $V(\phi) = {1\over2}m_\phi^2\phi^2$, we have $\omega_{\rm i} = m_\phi$, so one defines $q_{\rm i} \equiv {g^2\Phi_{\rm i}^2\over 4m_\phi^2}$. Thus, the extra factor ${1/4}$ makes $q_{\rm i}$ to match the resonance parameter definition from the {\it Mathieu} equation, to which Eq.~(\ref{eq:ParamResEqFourier}) can be mapped to (in cosmic time), see~\cite{Kofman:1997yn}. Such prefactors are of course purely conventional, and what really controls the physics is the dimensionless ratio ${g^2\Phi_{\rm i}^2\over \omega_{\rm i}^2}$, given in Eq.~(\ref{eq:ResParam}).

Given the oscillatory behaviour of $\varphi$, Eq.~(\ref{eq:ParamResEqFourier}) can exhibit unstable solutions of the type $\chi_{\kappa} \sim e^{\mu_q({\kappa}) z}$, with $\mu_\kappa(q)$ a complex function known as the {\it Floquet} index. For a given value of the resonance parameter $q$, there are typically some 'resonance bands' of momenta, for which $\mathfrak{Re} [\mu_\kappa(q)] > 0$. The corresponding field mode amplitudes inside such bands, experience then an exponential growth. It is precisely this unstable behaviour (ultimately due to the oscillatory pattern of $\varphi$) that we call parametric resonance. For an elaborated description of the theory of parametric resonance in the early Universe see e.g.~\cite{Kofman:1997yn,Amin:2014eta}. For a numerical study and parameter  fit, see~\cite{Figueroa:2016wxr}. The exponential growth of the $\chi_k$ modes undergoing parametric resonance, is actually interpreted as an abrupt particle creation: the occupation number of unstable modes grows as $n_k \propto |\chi_\kappa|^2 \sim e^{2\mu(\kappa)z}$. This implies that the system develops large time-dependent inhomogeneities, which in turn source GWs, as we will see next. 

The energy density spectrum of a stochastic background of GWs at subhorizon scales, can be expressed, in terms of the comoving momenta $k$ and conformal time $\tau$, as\footnote{With respect to Eq.~(\ref{eq:rhogwPiRad}) given in Sect.~\ref{sec:spectrum_generic}, Eq.~\eqref{eq:GW_spectra(Pi)} has different factors, due to the definition of the source: here $\Pi_{ij}^{{\rm TT}}\rightarrow a^2 \Pi_{ij}^{{\rm TT}}$. Moreover, Eq.~\eqref{eq:UTC} has a factor $1/4$ less with respect to the definition given in Eq.~\eqref{eq:PIspec}, and the UTC is denoted as $\Pi^2(k,\tau',\tau'')$ instead of $\Pi(k,\tau',\tau'')$ as in Eq.~\eqref{eq:PIspec}.}
\begin{eqnarray}\label{eq:GW_spectra(Pi)}
\hspace*{-1cm}\frac{d\rho_{\mathrm{\rm GW}}}{d\log k}\left(k,\tau\right) = \frac{2}{\pi}{G k^3\over a^4(\tau)}\int_{\tau_{i}}^{\tau}d\tau'd\tau''\,a(\tau')\,a(\tau'')\cos[k(\tau'-\tau'')]\,\Pi^2(k,\tau',\tau''),
\end{eqnarray}
where the unequal-time-correlator (UTC) is defined as
\begin{eqnarray}\label{eq:UTC}
\langle \text{0}| \Pi_{ij}^{{\rm TT}}(\mathbf{k},\tau)\Pi_{ij}^{{\rm TT}^{\hspace*{0.2mm}*}}\hspace*{-0.7mm}(\mathbf{k'},\tau') |0\rangle \equiv (2\pi)^3 \Pi^2(k,\tau,\tau') \delta^{(3)}(\bk-\bk)\,,
\end{eqnarray}
and $\Pi_{ij}^{\rm TT}$ is the TT-part of the anisotropic stress of the system. In our case $\Pi_{ij}^{\rm TT} = {1\over a^2}\left\lbrace \partial_{i}\chi\,\partial_{j}\chi \right\rbrace^{{\rm TT}}$. In other words, it is the field gradients, developed due to the exponential growth of the modes undergoing parametric resonance, that are ultimately responsible for the generation of GWs. Using the quantized field amplitude Eq.~(\ref{eq:ChiQuant}), it is found~\cite{Figueroa:2017vfa}
\begin{equation}
\Pi^{2}(k,\tau,\tau') = \int \frac{dp \,d\theta \,p^{6}\sin^{5}\theta}{4\pi^{2}a^{2}(\tau)a^{2}(\tau')}\chi_{\mathbf{p}}(\tau)\chi_{\mathbf{\mathbf{k-p}}}(\tau)\chi_{\mathbf{k-p}}^{*}(\tau')\chi_{\mathbf{\mathbf{p}}}^{*}(\tau')\,.
\end{equation}
Plugging this into the spectrum Eq.~(\ref{eq:GW_spectra(Pi)}), one obtains~\cite{Figueroa:2017vfa}
\begin{eqnarray}\label{eq:GWspectrumParamRes}
\hspace*{-0.25cm}\frac{d\rho_{{\rm GW}}}{d\log k}\left(k,t\right) = \frac{Gk^{3}}{2\pi^{3}}\int dp\, d\theta\, p^{6}\sin^{5}\theta\,\left(\left|I_{(c)}(k,p,\theta,\tau)\right|^{2}+\,\left|I_{(s)}(k,p,\theta,\tau)\right|^{2}\right)\,\nonumber\\
\label{eq:IsIc}
\hspace*{-1.5cm}I_{(c)} \equiv \int_{\tau_{i}}^{\tau}\frac{d\tau'}{a(\tau')}\cos(k\tau')\chi_{\mathbf{\mathbf{k-p}}}(\tau')\chi_{\mathbf{p}}(\tau')\,,~~I_{(s)} \equiv \int_{\tau_{i}}^{\tau}\frac{d\tau'}{a(\tau')}\sin(k\tau')\chi_{\mathbf{\mathbf{k-p}}}(\tau')\chi_{\mathbf{p}}(\tau')\nonumber\\
\end{eqnarray}

The spectrum today of this GW background, normalized to the current critical energy density $\rho_c$, can be obtained as
\begin{eqnarray}\label{eq:GWtoday}
h^{2}\Omega_{\rm GW} = h^2\Omega_{\rm rad}\left(a_{{\rm f}}\over a_{_{\rm RD}}\right)^{1-3w}\left(g_{S,0}\over g_{S,_{\rm RD}}\right)^{4\over3}\left(g_{_{\rm RD}}^*\over g_{0}^*\right)\Omega_{{\rm GW}}^{({\rm f})}\,,
\end{eqnarray}
where labels $\,_{\rm i}$, $\,_{\rm f}$, $\,_{\rm RD}$ and $\,_{\rm o}$, indicate respectively the time scales for the onset of GW production, end of GW production, first moment when the Universe became radiation dominated (RD), and the present time. Here $w = p/\rho$ is the averaged equation of state (pressure-to-density ratio) of the Universe between $\tau_{\rm f}$ and $\tau_{_{\rm RD}}$, whereas $\Omega_{{\rm GW}}^{(f)}$ is the fraction of energy density in GW to the total energy at the end of GW production. The latter quantity can be expressed as~\cite{Figueroa:2017vfa}
\begin{eqnarray}\label{eq:AmpProd}
\Omega_{{\rm GW}}^{(f)} \equiv {1\over \rho_f}\left({d\rho_{_{\rm GW}}\over d\log k}\right)_f = \left(\omega_{\rm i}\over m_{Pl}\right)^2\left(a_{\rm i}\over a_{\rm f}\right)^{1-3w}{\kappa^3 \mathcal{F}_{\rm f}(\kappa)\over 8\pi^4 \tilde \rho_{\rm i}}\,,
~~~~~~~~\\
\label{eq:AmpProd2}
\mathcal{F}_{\rm f}(\kappa) \equiv \omega_{\rm i}^2\int d\tilde p\, d\theta\, {\tilde p}^{6}\sin^{5}\theta \,\left(\left|\tilde I_{(c)}(\kappa,\tilde p,\theta,z_{\rm f})\right|^{2} + \,\left|\tilde I_{(s)}(\kappa,\tilde p,\theta,z_{\rm f})\right|^{2}\right)\,,
\end{eqnarray}
where $\kappa \equiv k/\omega_{\rm i}$, $\tilde p \equiv p/\omega_{\rm i}$, $z_{\rm f} \equiv \omega_{\rm i} \tau_{\rm f}$, $\tilde \rho_{\rm i} \equiv \rho_{\rm i}/\omega_{\rm i}^4$, and $\tilde I_{(x)}$ are the same functions as in Eq.~(\ref{eq:IsIc}), but written in terms of the dimensionless variables $z, \tilde p$ and $\kappa$. 

The GW spectrum is expected to have a well defined peak at a scale determined essentially by the resonance parameter $q_{\rm i}$. The excitation of a scalar field undergoing broad resonance (i.e.~$q_{\rm i} > 1$) corresponds, in fact, to an exponential growth of field mode amplitudes inside a {\it Bose-sphere} of radius $\kappa \lesssim \kappa_{*} \sim q_{\rm i}^{1/4}$. Outside this sphere there is no mode excitation, and hence no GW production. Using this and scaling arguments, Ref.~\cite{Figueroa:2017vfa} determined the parameter-dependence of the GW peak amplitude at $t= t_{\rm f}$ as
\begin{eqnarray}\label{eq:PeakAmplitude}
\Omega_{{\rm GW}}^{({\rm f})}(\kappa_p) \simeq {C^2\over 8\pi^4}\frac{\omega_{\rm i}^{6}}{\rho_{\rm i} m_{Pl}^{2}}\,q_{\rm i}^{-\frac{1}{2}+\delta}\,,
\end{eqnarray}
where $\delta \equiv {d\log\Omega_{_{\rm GW}}\over d\log q_{\rm i}} + {1\over2}$ quantifies the deviation (due to non-linearities) of the otherwise simple theoretical scaling $\Omega_{\rm GW} \propto q_{\rm i}^{-{1\over 2}}$. Here $C$ is a dimensionless constant characterizing the amplitude of the field fluctuations inside the Bose-sphere, $k_*|X^{(c)}_k|^2 = C\,q_{\rm i}^{-1}\,\theta(k_*-k)$. 

Let us note that GWs are initially created during the linear regime, when the field fluctuations grow exponentially fast. When the energy transferred into the daughter field(s) becomes sufficiently large, the resonance reaches an end. The system dynamics becomes then non-linear, sourcing further GWs. Eventually the fields relax into a stationary state, and the GW production ceases. From that moment onward, the GWs decouple and travel freely, redshifting until now. The corresponding amplitude and frequency of the GW peak today, can be written as
\begin{eqnarray}\label{eq:PeakAmplitudeToday}
\hspace*{-1.35cm}h^{2}\Omega_{\GW}^{(p)} 
\simeq h^{2}\Omega_{\mathrm{rad}}\left(\frac{g_{0}^*}{g^*_{{\rm f}}}\right)^{1/3}\times \epsilon_{\rm i}\,{C^2\over 8\pi^4}\frac{\omega_{\rm i}^{6}}{\rho_{\rm i} m_{Pl}^{2}}\,q_{\rm i}^{-\frac{1}{2}+\delta} \,,\\
\label{eq:PeakFreqToday}
\hspace*{-1.35cm}f_{p} \sim 8\cdot10^{9}\epsilon_{\rm i}^{1/4}\left(\frac{1}{\tilde{\rho}_{\rm i}}\right)^{\frac{1}{4}}\kappa_*~{\rm Hz} \sim 8\cdot10^{9}\left(\frac{\omega_{\rm i}}{{\rho}_{\rm i}^{1/4}}\right)\,\epsilon_{\rm i}^{\frac{1}{4}}\,q_{\rm i}^{\frac{1}{4}+\eta}~{\rm Hz}\,,
\end{eqnarray}
where $\epsilon_{\rm i} \equiv (a_{\rm i}/a_{\rm RD})^{1-3w}$, and $\eta$ measures the deviation with respect to the simple (linear) scaling $\kappa_{*} \propto q^{1/4}$. Using that $h^{2}\Omega_{\mathrm{rad}} \sim \mathcal{O}(10^{-5})$ and $\left(g_{o}/g_{*}\right)^{1/3}\sim\mathcal{O}(0.1)$, we conclude that the prefactor in Eq.~(\ref{eq:PeakAmplitude}) is of order $\mathcal{O}(10^{-6})$. Let us note that if the GW production took place during a RD stage, then $\epsilon_{\rm i} = 1$. If the GW production took place in an expanding phase with $w \neq {1\over3}$ between $\tau_{\rm i}$ and $\tau_{_{\rm RD}}$, then $\epsilon_{\rm i} < 1$ for $w < {1\over3}$, or $\epsilon_{\rm i} > 1$ for a stiff equation of state $w > {1\over3}$ (recall the stiff transfer function we discussed at the end of Sect.~\ref{sec:EvolTensInf}).

A remarkable result is the fact that the peak amplitude of the GW background from parametric resonance, decreases with the resonance parameter $q_{\rm i}$. One may have naively expected the opposite, as the larger the value of $q_{\rm i}$, the broader the resonance is. However, for larger values of $q$, even though a wider daughter field spectrum is excited, the amplitude of such spectrum is also smaller. The two effects combine together, resulting in a spectrum of GWs that decreases in amplitude with increasingly larger $q_{\rm i}$'s.

Numerical simulations in Ref.~\cite{Figueroa:2017vfa} were used to quantify the goodness of this peak parametrization as a function of $q_{\rm i}$. The parameters $\lbrace \delta, \eta, C \rbrace$ were extracted from runs of different scenarios, sampling a wide range of $q_{\rm i}$ values. Calibration of the previous formulas against such numerical results, yields the following amplitude and position today of the dominant peak\footnote{In some scenarios of parametric resonance, like in ${\lambda\over4}\phi^4$ preheating, various peaks appear in the GW spectrum, as a result of the fields dynamics. In the formulas above we only quote the peak resulting from the initial instability within the Bose-sphere, after the system became non-linear and settled down into a stationary stage. See~\cite{Figueroa:2017vfa} for details.} of the GW spectra:
\begin{itemize}
\item \textbf{Preheating with inflationary potential $V (\phi) = \frac{1}{2} m^2 \phi^2$}:

Results are given in terms of the initial resonance parameter $q_{\rm i} \equiv \frac{g^2 \phi_{\rm i}^2}{4 m^2} \gg 1$. Only one peak appears in the GW spectrum, at a frequency and amplitude
\bea 
f_p =   \epsilon_{\rm f}^{1/4} \left( \frac{q_{\rm i}}{10^4} \right)^{0.67} \times 2.0 \cdot 10^{8} \ {\rm Hz}\,, \\
h^2 \Omega_{\rm GW} (f_p) =  \epsilon_{\rm f}  \left( \frac{q_{\rm i}}{10^4} \right)^{-0.43} \times 1.5 \cdot 10^{-11}   \hspace{0.3cm} ({\rm for}~q_{\rm i} \gtrsim 6 \cdot 10^3 )   \ , \label{eq:m2phi2-finalfitv2}
\eea
where the ratio $\epsilon_{\rm f} \equiv (a_{\rm f} / a_{_{\rm RD}})^{1-3w}$ quantifies the unknown period (with equation of state $w$) between the end of GW production and the onset of a RD universe.

\item \textbf{Preheating with inflationary potential $V (\phi) = \frac{1}{4} \lambda \phi^4$}:

Results are given as a function of the resonance parameter $q_{\rm i}\equiv \frac{g^2}{\lambda} > 1$. The dominant peak has a fixed location (for a given resonance parameter), but its amplitude exhibits an oscillatory pattern between some maximum and minimum values, depending on the value of $q_{\rm i}$,
\bea
f_{\rm hb} \approx   \left( \frac{q_{\rm i}}{100} \right)^{0.54} \times 5.3 \cdot 10^7 \ {\rm Hz}  \ , \hspace{0.4cm} ({\rm only\,\,for\,\,} q \gtrsim 1000) \ ,\\
 3.4 \cdot 10^{-12}  \left( \frac{q_{\rm i}}{100} \right)^{-0.68} \lesssim h^2 \Omega_{\rm GW} (f_{\rm hb}) \lesssim 1.6 \cdot 10^{-11}  \left( \frac{q_{\rm i}}{100} \right)^{-0.94}  \ . \hspace{0.3cm} 
\eea
The shape of the peak resembles to a 'humpback', and hence the label $_{\rm hb}$, see e.g.~Figs.~3 or 4 in~\cite{Figueroa:2017vfa}. 

\end{itemize}

In summary, preheating via parametric resonance in the early Universe, may generate a large background of GWs, as the non-equilibrium dynamics of the fields after inflation develop large energy gradients, which in turn source efficiently tensor metric perturbations. The (dominant) peak of the GW spectrum may exhibit a very large amplitude today, of the order $h^2 \Omega_{\rm GW} \approx \mathcal{O} (10^{-11} ) - \mathcal{O} (10^{-13})$, for canonical resonance parameters. However, the GW spectrum is always peaked at very high frequencies, of the order of $f_p \approx \mathcal{O} (10^7) - \mathcal{O} (10^8) \ {\rm Hz}$, out of the range of current or planned GW detectors. For larger resonance parameters than those considered numerically, the peak shifts to higher frequencies, and its amplitude decreases further, so it becomes even more undetectable. For smaller values of the resonance parameter, the system no longer develops a resonance of the broad type, and hence the GW production becomes inefficient. We discuss this aspect in Section~\ref{subsec:NarrowResonance}.

\subsubsection{Anisotropies in the gravitational wave background.}

In our previous modelling, we considered implicitly that the daughter or 'preheat' field $X$, was massive during inflation, with mass $m^2_X = g^2\phi^2 > H^2$. This assumption is normally guaranteed in chaotic inflationary models, where the inflaton field develops super-Planckian amplitudes. Thus, for the majority of coupling values $g^2$, $X$ is massive during inflation, and hence it does not develop super-Hubble fluctuations. However, for a narrow range of coupling values, it is possible that the daughter field becomes light towards the last e-foldings of inflation (when the inflaton rolls down its potential towards smaller values), while at the same we maintain the condition for parametric resonance with $q_{\rm i} > 1$ to be developed after inflation. As a consequence, after the inflationary period ends, the $X$ field possesses amplified perturbations on super-horizon scales, just as the inflaton. In this case, at the onset of preheating, the sub-horizon vacuum fluctuations that serve as an initial condition for parametric resonance, are super-imposed over almost homogeneous values $X_{\rm i}$ of the daughter field. Such initial values are actually constant over regions that extend beyond the Hubble radius, as they are generated by super-Hubble fluctuations\footnote{The super-Hubble scale at which $X_{\rm i}$ varies spatially depends on the modelling, and is determined essentially by the number of efolds towards the end of inflation, during which $X$ becomes light.}. If the process of preheating, i.e.~the parametric excitation of the sub-horizon fluctuations of $X$, depends on the initial value $X_{\rm i}$ within a given Hubble-patch, the GW energy-density spectrum will depend as well on the values of $X_{\rm i}$. Therefore, a different amount of GWs is expected from different angular directions of observation today, with an anisotropy pattern in the amplitude of the GW spectrum, modulated by the super-horizon variation of the $X_{\rm i}$ amplitudes at the time of GW production at preheating~\cite{Bethke:2013aba,Bethke:2013vca}.

A crucial point in order for the GW anisotropies to be developed, is the excitation of the daughter field super-Hubble modes. The values $X_{\rm i}$ are modulated at super-horizon scales at the time of preheating, according to a variance $\sigma_X^2 \sim {H_{\rm inf}^2\over 4\pi^2}\Delta N$, where $H_{\rm inf}$ is the inflationary Hubble scale, and $\Delta N$ the number of efolds during which $X$ is a light degree of freedom. Roughly speaking, initial quantum fluctuations of the daughter field will be superimposed over homogeneous values of the order $X_{\rm i} \sim H_{\rm inf}$, which only vary at super-horizon scales. When non linearities become important during the preheating process, there will be a feedback between the super- and sub-horizon modes. The dynamics of the unstable sub-horizon modes during preheating will be, correspondingly, influenced by the value of $X_{\rm i}$ within each given region. As a result, the spatial distribution of the field $X$, and hence the source of the GWs, will be affected by the $X_{\rm i}$ values at causally disconnected regions. A different amount of GWs will be produced at different super-horizon regions, in correspondence with the different values of $X_{\rm i}$. 

The GW anisotropies have been studied in the case of massless preheating with inflationary potential $V (\phi) = \frac{1}{4} \lambda \phi^4$, because of its computational convenience, due to its scale invariance. In this scenario, the super horizon modes $k \ll aH$ of the $X$ field may lie, in fact, well inside a resonant band, if appropriate values of the coupling $g^2$ are chosen. Ref.'s~\cite{Bethke:2013aba,Bethke:2013vca} have computed the anisotropies arising in the GW background from preheating in this model, and argued that these anisotropies are a common phenomenon, arising naturally in other preheating scenarios, as long as the super-horizon modulation of the daughter field is present. Hence, considering a scenario of parametric resonance with total potential $V = \frac{\lambda}{4}\varphi^{4}+\frac{1}{2}g^{2}\varphi^{2}X^{2}$, the lightness of $X$ during the last stages of inflation is guaranteed, if the coupling constant is taken e.g.~to be $g^{2}/\lambda=2$. This fixes automatically the resonant band for the parametric resonance as $0 \leq k \lesssim \sqrt{\lambda}\Phi_{\rm i}$~\cite{Figueroa:2016wxr,Figueroa:2017vfa}, where $\Phi_{\rm i}$ is the inflaton amplitude at the end of inflation. With this choice, a large amplification of the $X$ field long-wave modes is ensured during preheating. The $X$ field super-Hubble perturbations at the end of inflation, are the same as for any light degree of freedom, i.e.~a nearly scale-invariant power-spectrum $\mathcal{P}_{X_{\rm i}}\simeq H_{\rm inf}^{2}/4\pi^{2}$. The dynamics describing the preheating process proceeds identically as described in the previous section, but the initial conditions at the onset of parametric resonance are now different: at each super-horizon volume, there are different values $X_{\rm i}$ (whereas before, in the previous section, we had set this value set to zero). Employing the `separate Universe' approach, one can compare the peaks of the GW spectral energy-density of two different simulations, with different initial values of $X_{\rm i}$. While the GW backgrounds are peaked at the same frequency, determined by the resonance parameter, the peak amplitudes of the GW spectra may differ significantly. This means that actually $X_{\rm i}$ influences the sub-horizon gradients of $X$, and hence the production of GWs. More precisely $\Omega_{\rm GW}$ varies by as much as a factor $\sim 5$ between slightly different values of $X_{\rm i}$~\cite{Bethke:2013vca,Bethke:2013aba} (the non-linear dynamics is actually chaotic~\cite{Bond:2009xx} and hence small variation of $X_{\rm i}$ may produce a large variation of subhorizon dynamics of $X$, and thus of the GWs). The level of anisotropy produced in the resulting GW energy density spectrum, is characterized by the angular power spectrum $C_l^{GW}$ of the relative GW spectral energy-density fluctuations, as a function of the $X_{\rm i}$ values (see~\cite{Bethke:2013aba,Bethke:2013vca} for further explanation). A general formula applicable to all scenarios characterized by a light spectator field during inflation is~\cite{Bethke:2013aba,Bethke:2013vca}
\begin{equation}
l\left(l+1\right)C_{l}^{GW}=\frac{H_{\rm inf}^{2}}{8\pi}\frac{\langle \delta X_{\rm i}\,\Omega_{\rm GW}(X_{\rm i})\rangle^{2}}{\sigma_{X}^{4}\langle\Omega_{\rm GW}\rangle^{2}}\,,
\end{equation}
implying that the angular power spectrum of GW anisotropies is scale invariant, resulting in a {\it plateau} at small multipoles, $l\left(l+1\right)C_{l}^{GW} \propto const.$, analogous to the large angular scale Sachs-Wolfe {\it plateau} for the temperature anisotropies in the CMB. For massless preheating with $g^{2}/\lambda=2$, the relative amplitude of the GW energy density spectrum on large scales, due to the variation of the different $X_{\rm i}$ values, is of the order of the $\mathcal{O}(1)\,\%$ percent level~\cite{Bethke:2013aba,Bethke:2013vca}. For comparison, bear in mind that the relative amplitude of the CMB temperature fluctuations is of the order of $\mathcal{O}(0.001)\,\%$. The GW anisotropies obtained in this model are therefore really huge. The details of the GW anisotropy, if ever observed, could provide a powerful way to distinguish between different microscopic details of the inflationary and preheating sectors.

\subsubsection{What if there is narrow parametric resonance?}
\label{subsec:NarrowResonance}

The logic behind the derivation of Eqs.~(\ref{eq:PeakAmplitudeToday})-(\ref{eq:PeakFreqToday}), assumed implicitly that $q_{\rm i} > 1$. This corresponds to a regime of {\rm broad resonance}, where modes within a broad band are excited. The effective resonance parameter of a system, however, often decreases in time (the only exception being the case $V(\phi) \propto \phi^4$, as this theory exhibits conformal invariance). Thus, if eventually the resonance parameter becomes smaller than unity $q_{\rm i} < 1$ (or if it simply started as $q_{\rm i} < 1$), the structure of the excited momenta is much more complicated, than a simple sphere of excited modes below a cut-off radius $\kappa \sim q_{\rm i}^{1/4}$. Only specific narrow bands around a set of discrete modes $\tilde p_1 < \tilde p_2 < ... $ are excited~\cite{Kofman:1997yn}. Consequently, this is called the {\it narrow resonance} regime. The $\delta$-parametrization in Eq.~(\ref{eq:PeakAmplitudeToday}), quantifying the deviation with respect the $\Omega_{\rm GW} \propto q^{-1/2}$ behaviour, was however obtained only for broad resonance. Such parametrization looses therefore its meaning in the case of narrow resonance, as $\delta$ does not need to be a small correction to the exponent ${-1/2}$. Besides, numerical simulations cannot even capture the narrow resonance regime $q < 1$, so we cannot assess whether Eq.~(\ref{eq:PeakAmplitude}) is a good {\it antsaz}. The power exponent to the scaling law of the GW amplitude as $\Omega_{\rm GW} \propto q^\beta$ (assuming a power-law behaviour is correct), is simply not known. However, in physical situations where $q < 1$, as only narrow bands are excited, we do not expect large field gradients to be developed in real space. Therefore, contrary to broad resonance, there is no reason to expect a large signal in GWs. Besides, the rate of transfer of energy into the excited species is much slower in narrow resonance than in broad resonance. The most physically interesting cases of parametric resonance in the early Universe are therefore, arguably, those where broad resonance $q > 1$ occurs, for which Eqs.~(\ref{eq:PeakAmplitudeToday})-(\ref{eq:PeakFreqToday}) apply.

\subsubsection{What if the resonant fields are not scalar, but gauge fields?} 

The presence of gauge fields in inflationary set-ups has been considered in many scenarios. For instance, in those where the inflaton enjoys a shift-symmetry, and hence it has a topological coupling to gauge fields, see Section~\ref{sec:SustainedPartProdInf} and references therein. Other inflationary scenarios which naturally involve gauge fields are {\it Gauge-flation}~\cite{Maleknejad:2011jw,Maleknejad:2011sq}, where a configuration of non-Abelian gauge fields is capable of sustaining inflation. In Hybrid inflationary models, the presence and excitation of gauge fields has been also considered as a natural ingredient (we will discuss these scenarios in Section~\ref{sec:Tachyonic}). In the case of parametric resonance, typically one considers the excitation of a singlet scalar field coupled to another singlet (coherently oscillating) scalar field. Nothing prevents us however, to consider the oscillatory field to be charged under a gauge symmetry, so that it would be naturally coupled to the gauge bosons associated with the given symmetry. Such case has however not been considered very often in the literature. First of all, there is no particular `need', in principle, from the point of view of the realization of inflation, to `gauge' the inflationary sector. Secondly, radiative corrections of the inflaton potential from the gauge couplings, might spoil the conditions to sustain inflation. Assuming however that one can construct a viable working model, nothing is wrong {\it per se} about considering an inflaton charged under a gauge group, and hence coupled to some gauge field(s). In such a case, when the inflaton starts oscillating during the period following the end of inflation, the gauge bosons are expected to be parametrically excited. This was considered in details in Ref.~\cite{Lozanov:2016pac}, for both Abelian $U(1)$ and non-Abelian $SU(2)$ gauge groups. A natural realization of an inflationary set-up where the inflaton is charged under a gauge group is the Higgs-Inflation scenario~\cite{Bezrukov:2007ep,Bezrukov:2010jz}. There the inflaton field is played by the Standard Model (SM) Higgs. The electroweak gauge bosons of the SM are  naturally coupled to the Higgs, and thus they experience parametric resonance during the oscillations of the Higgs after inflation~\cite{Bezrukov:2008ut,GarciaBellido:2008ab,Figueroa:2009jw}. 

To our knowledge, there has been no computation to date on the production of GWs from the parametric excitation of gauge bosons after inflation, due to the oscillations of a (charged) inflaton. The corresponding computation of GWs has been carried out, however, in the case when a charged oscillatory field is not the inflaton, but rather an energetically sub-dominant field~\cite{Figueroa:2016ojl}. In this case, the final GW amplitude is much smaller than in the case where the charged oscillating field is the inflaton (we discuss the case of parametric resonance by sub-dominant fields in the next subsection). There is no reason, however, to expect that the scaling with the resonance parameter should be different than in the case of a charged oscillatory field identified as the inflaton. A remarkable difference arises in fact when the daughter fields experiencing the parametric excitation are gauge fields, and not scalars. In the case of gauge fields, it is found that the corresponding GW background scales as $\Omega_{\rm GW} \sim q^{3/2+\delta}$~\cite{Figueroa:2016ojl}, as opposed to Eq.~(\ref{eq:PeakAmplitude}) $\Omega_{\rm GW} \sim q^{-1/2+\delta}$ for scalars (in both cases $|\delta| \ll 1$ represents a small correction to the corresponding power-law index). This different dependence on the resonance parameter $q$, can actually be easily explained: even though gauge bosons experience the same dynamics as scalar fields when coupled to an oscillatory (homogeneous) field\footnote{This has been demonstrated explicitly in Ref.~\cite{Figueroa:2015rqa} for the case of Abelian gauge fields. In the case of Non-Abelian gauge fields, one may expect a different behaviour than in standard parametric resonance of scalar fields, as there are non-linearities due to the non-Abelian gauge interactions.}, the source of GWs (i.e.~their anisotropic stress) has a different structure than in the scalar field case. Analytical calculations based on a linear analysis of the gauge field excitation (similar to the calculations displayed above for the scalar field case), led theoretically to expect $\Omega_{\rm GW} \sim q^{3/2}$, as confirmed by the simulations in~\cite{Figueroa:2016ojl}. 

\subsubsection{What if the oscillatory field is energetically sub-dominant?}
\label{subsec:subdominantPR}

Inflationary preheating is actually not the only case where parametric resonance is naturally developed in the early Universe. If a light (scalar) spectator field is present during the inflationary stage, it naturally develops a large homogeneous amplitude (parametrically controlled by the inflationary Hubble rate), simply due to the stretching of its quantum fluctuations during the quasi-exponential expansion. At some point after inflation, once the Hubble rate falls bellow the mass of the spectator field, the field starts oscillating around the minimum of its potential. This is the case e.g.~of the curvaton scenario~\cite{Enqvist:2001zp,Lyth:2001nq,Moroi:2001ct,Mazumdar:2010sa}. The curvaton may decay abruptly after inflation, via parametric resonance, transferring exponentially fast its energy into the particle species coupled to it~\cite{Enqvist:2008be, Enqvist:2012tc, Enqvist:2013qba, Enqvist:2013gwf}. Another example of a spectator field, naturally decaying through parametric resonance after inflation, is the Standard Model Higgs field. If the Higgs is weakly coupled to the inflationary sector, the Higgs is expected to be excited either during inflation~\cite{Starobinsky:1994bd,Enqvist:2013kaa,DeSimone:2012qr}, or towards the end of it~\cite{Herranen:2015ima,Figueroa:2016dsc}. The Higgs decays then into the rest of the SM species at some point after inflation\footnote{Note that this differs from the Higgs-Inflation scenario~\cite{Bezrukov:2007ep,Bezrukov:2010jz}, where the SM Higgs also decays after inflation, via parametric resonance, into the SM species~\cite{Bezrukov:2008ut,GarciaBellido:2008ab,Figueroa:2009jw,Figueroa:2014aya}. The case of Higgs-Inflation belongs however to the category of preheating, as the Higgs plays the role of the inflaton, not of a spectator field. For the most recent development of preheating in Higgs-Inflation see~\cite{Ema:2016dny}.}, once it starts oscillating around zero, via parametric resonance~\cite{Enqvist:2013kaa,Enqvist:2014tta,Figueroa:2014aya,Kusenko:2014lra,Figueroa:2015rqa,Enqvist:2015sua,Figueroa:2016dsc}. 

The production of GWs during the parametric resonance of a field coupled to an inflationary spectator field, is driven by exactly the same mechanism as in the case of parametric resonance from an oscillatory inflaton. The main difference is that the energy stored in the spectator field is parametrically suppressed as $\sim (H/m_{Pl})^2 \ll 1$, compared to the inflaton energy (in large field inflationary models). This results in a suppression of the total amount of GWs produced in the case of a spectator field. In particular, the ratio of the GW energy density produced by parametric resonance due to the oscillations of a spectator field $\Omega_{\rm GW}^{\rm (sf)}$, to the GW energy density created (for the same daughter-mother coupling) by the oscillations of an inflaton $\Omega_{\rm GW}^{\rm (inf)}$, can be parametrically estimated as~\cite{Figueroa:2017vfa}
\begin{eqnarray}\label{eq:StoIcomparison}
{\Omega_{\rm GW}^{\rm (sf)} \over \Omega_{\rm GW}^{\rm (inf)}} \sim \left({H\over m_{Pl}}\right)^4 \ll 1~\,.
\end{eqnarray}
The total GW production from parametric resonance of a field coupled to a spectator field, is therefore much smaller than the GW production of the parametric resonance process from an inflaton field. 

This explains e.g.~the smallness of the amplitude of the GW background produced during the decay of the SM Higgs after inflation~\cite{Figueroa:2014aya,Figueroa:2016ojl}, when the latter is a spectator field. In the case of Ref.~\cite{Figueroa:2014aya} the GW production was rather due to the parametric excitation of the SM fermions coupled to the Higgs, so we postpone the discussion of this case for Section~\ref{sec:Fermionic}. In the case of Ref.~\cite{Figueroa:2016ojl}, the GWs were due to the parametric excitation of the electroweak gauge fields. The energy arguments provided above, remain however valid: the GW amplitude from the parametric excitation of daughter field(s) coupled to an oscillating spectator field, is very suppressed, independently of the daughter field(s) spin. 

As as example we quote below the results from the SM Higgs spectator scenario~\cite{Figueroa:2016ojl}, based on the resonance of the electroweak $Z$ and $W^\pm$ gauge bosons. For a radiation-domination (RD) background with EoS $w = 1/3$, or kinetion-domination (KD) background with $w = 1$, taking an initially large Higgs amplitude $|\Phi_{\rm i}|= 0.1H/\sqrt{\lambda}$, and normalizing the inflationary Hubble rate $H$ to its upper bound $H_{\rm max} \simeq 8.5\cdot 10^{13}$ GeV~\cite{Ade:2015xua}, Ref.~\cite{Figueroa:2016ojl} obtains for the spectral peak amplitude
\bea \label{eq:spectRD}
{\rm RD}: \,\,\,\,\, h^2 \Omega_{\rm GW} (f_p) &\lesssim & 10^{-29}(H/H_{\rm max})^4 \ , \,\,\,\,\,\,\,\,\, f_p \lesssim 3 \cdot 10^8 \,\,{\rm Hz} \ , \\
\label{eq:spectKDv2}
{\rm KD}: \,\,\,\,\, h^2 \Omega_{\rm GW} (f_p) &\lesssim & 10^{-16}(H/H_{\rm max})^4 \ , \,\,\,\,\,\,\,\,\, f_p \lesssim 3 \cdot 10^{11} \,\,{\rm Hz} \ .
\eea
The GW amplitude for matter-domination (MD) with $w \simeq 0$, results even smaller than in the RD case. In the case of KD, the expansion history until the universe becomes RD, boosts the signal by a factor $\epsilon_{\rm i} \equiv (a_{\rm i}/a_{\rm RD})^{1-3w} = (a_{\rm RD}/a_{\rm i})^2 \gg 1$, explaining the larger amplitude quoted in Eq.~(\ref{eq:spectKDv2}). However, a KD background also shifts the GW peak spectrum toward higher frequencies, as indicated also in Eq.~(\ref{eq:spectKDv2}). As the GW amplitude is suppressed as $\propto (H/H_{\rm max})^4$, for a small Hubble rate the final amplitude becomes tiny. As anticipated, the GW background resulting from parametric resonance due to the oscillations of a spectator field, is extremely small, out of the range and sensitivity of current or planned GW detectors. In the case of the SM Higgs as a spectator field, the GW signal will remain, most likely, as an (unobservable) curiosity of the SM.

\subsection{Parametric excitation of Fermions}
\label{sec:Fermionic}

Let us consider now, as a potential source of GWs, a spin-${1\over2}$ fermion field
\begin{eqnarray}
\psi(\mathbf{x},t) &=& \int\frac{d\mathbf{k}}{\left(2\pi\right)^{3}}\,e^{-i\mathbf{k\cdot x}}\left[\hat a_{\mathbf{k},r}{\tt u}_{\mathbf{k},r}(t)+\hat b_{-\mathbf{k},r}^{\dagger}{\tt v}_{\mathbf{k},r}(t)\right]\,,
\end{eqnarray}
where with $\hat a_{{\bf k},r}, \hat b_{{\bf k},r}$ the usual creation/annihilation operators satisfying the standard anti-commutation relations $\left\lbrace \hat a_{{\bf k},r}, \hat a_{{\bf q},s}^\dag\right\rbrace$ = $\left\lbrace \hat b_{{\bf k}r}, \hat b_{{\bf q},s}^\dag\right\rbrace$ = $ (2\pi)^3\delta_{rs}\delta_D({\bf k}-{\bf q})$, $\left\lbrace \hat a_{{\bf k},r}, \hat b_{{\bf q},s}^\dag\right\rbrace$ = 0, and where the 4-component spinors are expressed as
\begin{eqnarray}\label{eq:u,v_Appendix}
{\tt u}_{\mathbf{k},r}(t) \equiv 
\left(\begin{array}{c}
\vspace{0.4cm}u_{\mathbf{k},+}(t)\,S_{r}\\ u_{\mathbf{k},-}(t)\,S_{r}
\end{array}\right),\hspace*{1cm}
{\tt v}_{\mathbf{k},r}(t)\equiv
\left(\begin{array}{c}
\vspace{0.4cm}v_{\mathbf{k},+}(t)\,S_{-r}\\ v_{\mathbf{k},-}(t)\,S_{-r}
\end{array}\right),
\end{eqnarray}
with $\{S_{r}\}$ 2-component spinors given by\footnote{We choose eigenstates of the Pauli matrix $\sigma_{3}$ as a basis.}
\begin{eqnarray}\label{eq:Sr_Appendix}
S_{1}=-S_{-2}=\left(\begin{array}{c}
\vspace{0.2cm}1 \\ 0\end{array}\right),\hspace*{1cm}
S_{2}=S_{-1}=\left(\begin{array}{c}
\vspace{0.2cm}0 \\ 1\end{array}\right)\,.
\end{eqnarray}
The fermion energy-momentum tensor is given by~\cite{Birrell:1982ix}
\begin{eqnarray}
T_{\mn}(\bx,t) = {1\over2a(t)}\left(\bar\psi\gamma_{(\mu}\overrightarrow{D}_{\nu)}\psi - \bar\psi\overleftarrow{D}_{(\mu}\gamma_{\nu)}\psi \right),
\end{eqnarray}
with $D_\mu \equiv \partial_\mu + {1\over4}[\gamma_\alpha,\gamma_\beta]\omega^{\alpha\beta}_\mu$ the covariant derivative, $\gamma_\mu$ the standard Dirac matrices in flat-space, and $\omega^{\alpha\beta}$ the spin connection. The transverse-traceless part of $T_{ij}$ in Fourier space can be obtained by means of the orthogonal projector ${\mathcal P}_{ij} \equiv \delta_{ij} - \hat k_i \hat k_j$, as $\Pi_{ij}^\TT({\bf k},t) = {\mathcal P}_{il}T_{lm}{\mathcal P}_{mj} - {1\over2}{\mathcal P}_{ij}{\mathcal P}_{lm}T_{lm}$ (recall Sect.~\ref{gweom}). Using that $v_{{\bf p},\pm} = \pm u_{{\bf p},\mp}^*$, one can obtain the Unequal-Time-Correlator (UTC) as~\cite{Figueroa:2013vif}
\begin{eqnarray}\label{eq:UTCfermions}
\left\langle {\Pi}_{ij}^\TT({\bf k},t)\,{{\Pi}_{ij}^{\TT}}({\bf k}',t')\right\rangle \equiv (2\pi)^3\,{\Pi}^2(k,t,t')\,\delta^{(3)}({\bf k}-{\bf k}')\,,\\
\Pi^{2}(k,t,t')=\frac{1}{2\pi^{2}a^{2}(t)a^{2}(t')}\int dp\,d\theta\,p^{4}\sin^{3}\hspace*{-1mm}\theta\,W_{{\bf k},{\bf p}}(t)W_{{\bf k},{\bf p}}^*(t')\,,\\
W_{{\bf k},{\bf p}}^{}(t) \equiv u_{\mathbf{k-p},+}(t)u_{\mathbf{p},+}(t) - u_{\mathbf{k-p},-}(t)u_{\mathbf{p},-}(t).
\end{eqnarray}
Substituting the UTC into the formula of a stochastic background of GWs, we find that the spectrum of GW produced by fermions is given by~\cite{Enqvist:2012im,Figueroa:2013vif}
\begin{eqnarray}\label{eq:GWspectrumFermions}
\hspace*{-1cm}\frac{d\rho_{\GW}}{d\log k}(k,t) = \frac{Gk^{3}}{\pi^{3}a^{4}(t)}\int dp\,d\theta\, p^{4}\sin^{3}\hspace*{-1mm}\theta \,\left(\left|I_{(c)}(k,p,\theta,t)\right|^{2} + \,\left|I_{(s)}(k,p,\theta,t)\right|^{2}\right)\,,\nonumber\\
\hspace*{-2cm}I_{(c)}(k,p,\theta,t) \equiv \int_{t_{i}}^{t}\frac{dt'}{a(t')}\cos(kt')W_{{\bf k},{\bf p}}(t'), \hspace*{0.5cm} I_{(s)}(k,p,\theta,t) \equiv \int_{t_{i}}^{t}\frac{dt'}{a(t')}\sin(kt')W_{{\bf k},{\bf p}}(t')\nonumber\\
\end{eqnarray}
Let us note that the structure of the formulae in Eq.~(\ref{eq:GWspectrumFermions}) resembles that of scalar fields sourcing GWs, Eq.~(\ref{eq:GWspectrumParamRes}). In the bosonic case, however, there appears a power $p^6$ instead of $p^4$ in the integrand, and the fermionic mode functions $u_{{\bf k},\pm}(t)$ are of course replaced by the Klein-Gordon scalar modes $\phi_{{\bf k}}(t)$ (and correspondingly the sum over polarization indices $+,-$ is absent).

It is perhaps appropriate now to make a small digression, as vacuum expectation values like the UTC in Eq.~(\ref{eq:UTCfermions}), require regularization of ultraviolet divergences. This has not been much of an issue in the case of bosonic fields, as the large-momentum modes causing the divergence, are not captured in lattice simulations used to study the bosonic excitations. As the lattice spacing is finite, there is a maximum momentum that can be captured in any lattice simulation. Excited modes of boson fields develop typically a huge hierarchy in amplitudes, with the infrared (IR) modes of the spectrum exhibiting  a much larger amplitude than the ultraviolet (UV) modes. Because of this, even though a regularization procedure is formally required in order to obtain a finite UTC for bosons, there is no real need in practice. This is certainly the case for the preheating scenarios that we presented in Section~\ref{sec:ParamRes}, where the resulting UV-tails of the field mode distributions are exponentially suppressed. If we were to include all the infinite tower of bosonic modes, say all the way up to $p \rightarrow \infty$ (i.e.~well beyond the maximum UV scale captured in simulations), we would encounter a UV divergence. However, because of the mentioned IR/UV hierarchy of spectral amplitudes, it is sufficient to measure the decaying slope of the large-$p$ tail in the field spectra obtained in simulations, and extrapolate it all the way up to $p \rightarrow \infty$. In this way, the otherwise divergent contribution from the UV bosonic vacuum fluctuations (not excited in the GW production process), is automatically regularized. In the case of GWs produced from fermionic parametric excitation~\cite{Enqvist:2012im,Figueroa:2013vif,Figueroa:2014aya}, due to Pauli blocking, there is no similar hierarchy of amplitudes between IR and UV modes. Therefore, in the case of fermions, one must necessarily deal with regularization, otherwise integrals over fermion mode functions diverge in the UV. Applying a time-dependent normal-ordering procedure (which subtracts at every moment the contribution from fluctuations that remain in vacuum, but maintains otherwise the contribution from excited modes), leds to a regularized version of Eqs.~(\ref{eq:GWspectrumFermions}), (\ref{eq:UTCfermions}), as~\cite{Figueroa:2013vif}
\begin{eqnarray}\label{eq:GWspectrumFermionsReg}
\hspace*{-1cm}\frac{d\rho_{\GW}}{d\log k}(k,t) = \frac{Gk^{3}}{\pi^{3}a^{4}(t)}\int dp\,d\theta\, p^{4}\sin^{3}\hspace*{-1mm}\theta \,\left(\left|\tilde I_{(c)}(k,p,\theta,t)\right|^{2} + \,\left|\tilde I_{(s)}(k,p,\theta,t)\right|^{2}\right),
\nn\\
\hspace*{-2cm}\tilde I_{(c)}(k,p,\theta,t) \equiv \int_{t_{i}}^{t}\frac{dt'}{a(t')}\cos(kt')\tilde W_{{\bf k},{\bf p}}(t'), \hspace*{0.5cm} \tilde I_{(s)}(k,p,\theta,t) \equiv \int_{t_{i}}^{t}\frac{dt'}{a(t')}\sin(kt')\tilde W_{{\bf k},{\bf p}}(t')\,,\nonumber\\
\end{eqnarray}
where the regularized mode functions are given by
\begin{eqnarray}
\hspace*{-2cm}\tilde W_{{\bf k},{\bf p}}(t) \equiv 2\sqrt{n_{\mathbf{p}}(t)n_{\mathbf{k}-{\bf p}}(t)}{W}_{\mathbf{k},\mathbf{p}}(t)\,,~~~~ {W}_{\mathbf{k},\mathbf{p}}(t) \equiv u_{\mathbf{k-p},+}(t)u_{\mathbf{p},+}(t) - u_{\mathbf{k-p},-}(t)u_{\mathbf{p},-}(t)\,,\nonumber\\
\end{eqnarray}
and the occupation numbers are identified with the Bogoliubov coefficients $n_k\equiv |\beta_k|^2$. 
Let us assume that a given process only excites fermions up to a given UV scale $k_*$, and above this cut-off only fermion vacuum fluctuations are present. We expect therefore the fermion occupation number to vanish for modes $k > k_*$, i.e.~$n_{p > k_*} \equiv |\beta_{p > k_*}|^2 = 0$. Consequently $\tilde{W}^{\pm}_{\mathbf{k},\mathbf{p}}(t) = 2|\beta_{\mathbf{p}}(t)||\beta_{\mathbf{k}-{\bf p}}(t)|{W}_{\mathbf{k},\mathbf{p}}(t) \rightarrow 0$ when $p \gg k_*$ or $k \gg k_*$. The would be UV-divergence in Eq.~(\ref{eq:GWspectrumFermions}) is regularized by the suppression of the high-momentum tail of $\tilde{W}^{\pm}_{\mathbf{k},\mathbf{p}}(t)$. The convergence of the GW spectrum at large $k$ is then guaranteed\footnote{A new rigorous treatment, based on a generalization of adiabatic regularization applied to fermions, has been recently proposed in~\cite{Landete:2013axa,Landete:2013lpa,delRio:2014cha,delRio:2017iib}. The application of this technique to correlators of fermion operators evaluated at different times, like the UTC needed to compute the spectrum of GWs, is however not straight forward. Therefore, to date, there has been no attempt to apply this technique to the calculation of the GWs from fermions.}. 

Eq.~(\ref{eq:GWspectrumFermionsReg}) describes the GW spectrum (at subhorizon scales) generated by a fermionic field $\psi$ with eigenfunctions $u_{{\bf k},\pm}(t)$. The equation of motion for the latter follows from the Dirac equation, and reads
\begin{eqnarray}\label{eq:DiracSimple}
\hspace*{-1cm}{d^2\over dt^2}{u}_{\mathbf{k},\pm}(t)+\left(\omega_{\mathbf{k}}^{2}(t) \pm i{d\over dt}(am_\psi)\right) u_{\mathbf{k},\pm}(t)=0\,,\qquad
\omega_{\mathbf{k}}^{2}(t) =  k^{2}+a^{2}(t)m_{\psi}^{2}(t)\,.\nonumber\\
\end{eqnarray}
If there is a process in the early Universe where fermions are highly excited, we just need to plug in the solutions $u_{{\bf k},\pm}(t)$ into Eq.~(\ref{eq:GWspectrumFermionsReg}), in order to find the spectrum of GWs. For a Yukawa interaction with some scalar field, we obtain
\begin{eqnarray}\label{eq:Yukawa}
\mathcal{L}_{\rm int} = - h\varphi\bar\psi\psi\hspace*{0.5cm}\Rightarrow\hspace*{0.5cm} m_{\psi} \equiv h\varphi\,.
\end{eqnarray}
Several scenarios of the early Universe may create high-energy out-of-equilibrium fermions by non-perturbative processes. In particular, whenever a homogeneous scalar field oscillates around the minimum of its potential, fermions coupled to such field experience a parametric excitation of their modes. When the oscillatory field is the inflaton, this process is known as fermionic preheating~\cite{Greene:1998nh,Greene:2000ew,Peloso:2000hy,Berges:2010zv}. If the mother field is simply an energetically sub-dominant field oscillating after reheating, the fermion particle production mechanism is the same as in preheating, but the produced fermions only represent a sub-dominant component of the energy budget of the Universe. Either way, fermion parametric excitation is very similar to the excitation of bosons discussed in Section~\ref{sec:ParamRes}, except that there is no resonance (given that fermion occupation numbers are Pauli-blocked). Instead, every time the oscillatory field crosses around zero, higher and higher fermionic modes are excited, until the adiabaticity condition of the fermion modes is eventually not violated, and hence no more fermions can be excited. The non-equilibrium fermions created this way, correspond to a mode excitation up to a given finite momentum $k_*$, so that modes with higher momentum are simply not excited (i.e.~remain in vacuum). A fermion configuration like this possesses a nontrivial anisotropic-stress that may significantly source GWs. Using Eq.~(\ref{eq:GWspectrumFermionsReg}) and scaling arguments, Ref.~\cite{Figueroa:2013vif} has characterized the parameter dependence of the GW background, sourced when and oscillatory field $\varphi$ creates fermions via a Yukawa interaction $h\bar\psi\varphi\psi$. It is found that the peak amplitude scales like
\begin{eqnarray}\label{eq:PeakGWFermions}
\hspace*{-0.85cm}h^{2}\Omega_{\rm GW}\left(f_{p}\right)\simeq h^{2}\Omega_{\mathrm{rad}}\left(\frac{g_{0}}{g_{*}}\right)^{\frac{1}{3}}\times\epsilon\,C^{2}\frac{\omega_i^{6}}{\rho_{\rm i}M_{p}^{2}}\,q_{\rm i}^{\frac{3}{2}+\delta}\,,\\
f_{p} \sim 8\cdot10^{9}\left(\frac{\omega_{\rm i}}{{\rho}_{\rm i}^{1/4}}\right)\,\epsilon_i^{\frac{1}{4}}\,q_{\rm i}^{\frac{1}{4}}~{\rm Hz}\,,
\end{eqnarray}
where $\delta$ is a small corrections to the index power-law, and $C^2$ is a dimensionless numerical constant, that can only be obtained from the numerical evaluation of the GW spectrum Eq.~(\ref{eq:GWspectrumFermionsReg}). 

Let us recall that in the case of scalar daughter fields, the theoretical analysis predicts that $\Omega_{\rm GW} \sim q_{\rm i}^{-1/2}$, see Eq.~(\ref{eq:PeakAmplitudeToday}) in Section~\ref{sec:ParamRes}. In the case of fermions, the behaviour of the spectrum of GW with respect to the `resonance' parameter is then very different. This is due to the fact that fermion fluctuations are Pauli blocked. For larger values of $q_{\rm i}$ there is a larger range of fermion modes excited, but contrary to scalar fields, this does not imply a lowering of the power per mode, as the spectral amplitude is typically saturated by the exclusion principle. On top of this, the anisotropic stress from fermions has a different structure (number of derivatives) than that of scalar fields. All together, a scaling as $d\log \Omega_{\rm GW}/d\log q \sim {3/2}$ emerges, see~\cite{Figueroa:2013vif} for a detailed derivation.

For different physical cases, the parametrization in Eq.~(\ref{eq:PeakGWFermions}) was tested in Ref.~\cite{Figueroa:2013vif} against the numerical outcome. After solving Eq.~(\ref{eq:DiracSimple}) with initial zero-number fermion density, the mode function solutions were plugged in into Eq.~(\ref{eq:GWspectrumFermionsReg}), obtaining the following fits:
\begin{itemize}

\item \textbf{Massless Preheating ($V(\varphi) = {\lambda\over4}\varphi^4$, $q_{\rm i} \equiv {h^2\over\lambda} > 1$)}:
\begin{equation}
h^{2}\Omega_{\rm GW}\left(f_{p}\right)\simeq
1.2\cdot10^{-9}\lambda^{2}\left(\frac{\Phi_{_{\rm i}}}{M_{Pl}}\right)^{2}q_{\rm i}^{1.61},
\end{equation}
\begin{equation}
f_p \simeq 7\cdot10^{10}q_{\rm i}^{1\over4} \lambda^{1\over4}\,\mathrm{Hz} = 7\cdot10^{10}\,h^{1\over2}\,\mathrm{Hz}\,,\vspace*{1.0cm}
\end{equation}

\item \textbf{Massive Preheating ($V(\varphi) = {1\over2}m_\varphi^2\varphi^2$, $q_{\rm i} \equiv {h^2\Phi_{\rm i}^2\over 4m_{\varphi}^2} > 1$)}:
\begin{equation}
h^{2}\Omega_{\rm GW}\left(f_{p}\right)\simeq
2.5\cdot10^{-12}\left(\frac{m_{\varphi}^{2}}{\Phi_{\rm i}M_{Pl}}\right)^{2}\left(\frac{a_{*}}{a_{\rm i}}\right)^{0.78}q_{\rm i}^{1.78},
\end{equation} 
\begin{equation}
f_p \simeq 6\cdot10^{10}\left(\frac{m_{\varphi}}{\Phi_{\rm i}}\right)^{1\over2}q_{\rm i}^{1\over4}\,\mathrm{Hz} = 6\cdot10^{10}\,h^{1\over2}\,\mathrm{Hz} \,, \vspace*{1.0cm}
\end{equation}

\item \textbf{Massive Spectator in Thermal Era ($V(\varphi) = {1\over2}m_\varphi^2\varphi^2$, $q_{\rm i} \equiv {h^2\Phi_{\rm i}^2\over 4m_{\varphi}^2} >1$)}:
\begin{equation}
h^{2}\Omega_{\rm GW}\left(f_{p}\right)\simeq
2.7\cdot10^{-10}\left(\frac{m_{\varphi}}{M_{Pl}}\right)^{4}\left(\frac{a_{*}}{a_{\rm i}}\right)^{1.74}q_{\rm i}^{1.74},
\end{equation}
\begin{equation}
f_{p}\simeq5\cdot10^{10}\left(\frac{a_{*}}{a_{\rm i}}\right)^{\frac{1}{4}}\left(\frac{m_{\varphi}}{\rho_{\rm i}^{1/4}}\right)q_{\rm i}^{\frac{1}{4}}\,\mathrm{Hz}\simeq 10^{11}\,\left(\frac{a_{*}}{a_{\rm i}}\right)^{\frac{1}{4}}\left(\frac{m_{\varphi}}{M_{Pl}}\right)^{\frac{1}{2}}q_{\rm i}^{1\over4}\,\mathrm{Hz}\,\vspace*{1.0cm}
\end{equation}
\end{itemize}
The numerical results show some deviation with respect the expected $\Omega_{\rm GW} \propto q_{\rm i}^{3/2}$ behaviour, as the theoretical scaling is obtained ignoring the details of the UV tail of the fermion excitation above the cut-off scale $k_* \sim q_{\rm i}^{1/4}\omega_i$. More importantly, the signal is unfortunately peaked only at very high frequencies, and its amplitude is tiny. For instance, in chaotic inflation with $m_{\phi} \simeq 10^{-6}M_{Pl}$ and $\Phi_{\rm i} \sim 0.1M_{Pl}$, we obtain $\Omega_{\rm GW} \lesssim 10^{-34}q_{\rm i}^{1.78}$, which for e.g.~$q_{\rm i} \sim 10^{6}$ (corresponding to Yukawa coupling of $h \sim 0.01$), it represents still a very small amplitude $\Omega_{\rm GW} \sim 10^{-24}$. 

There are interesting physical cases where fermions will be parametrically excited due to the oscillations of a scalar field. In particular, both when the SM Higgs plays the role of the inflaton in Higgs-inflation, or when the Higgs is simply considered a spectator field, all the charged fermions of the SM (i.e.~charge leptons and quarks) are coupled to the Higgs directly, and hence will be excited after inflation. As the Yukawa coupling of the top quark is the largest in the SM, the peak of the GW will be dominated by the contribution from the top quark. Ref.~\cite{Figueroa:2014aya} estimates that in the case of the Higgs-inflation scenario, the GW peak from the top quark is peaked at $f_p \sim 10^{10}$ Hz, with an amplitude $\Omega_{\rm GW} \sim 10^{-14}$ (assuming that the Higgs self-coupling is $\lambda \simeq 0.01-0.1$). In the case of the Higgs-spectator scenario, it is obtained that the background is peaked at $f_p \sim 10^7$ Hz and, even assuming instant reheating into RD, the maximum Hubble rate compatible with CMB constraints, and the Higgs self-coupling as $\lambda \simeq 0.01-0.1$, the amplitude is still as small as $\Omega_{\rm GW} \sim 10^{-30}$. The gauge bosons are also excited through parametric resonance, as discussed in Section~\ref{sec:ParamRes}, and their contribution to the GW amplitude seems to be somewhat larger than that from the top quark, but still not large enough, see Eqs.~(\ref{eq:spectRD}), (\ref{eq:spectKDv2}). These signals are therefore too faint, and peaked at too high frequencies, to expect an eventual detection in any near future. Most likely these backgrounds will simply remain as a curiosity of the SM.

\subsection{Spontaneous symmetry breaking via tachyonic effects}
\label{sec:Tachyonic}

Let us now analyse the preheating stage 
in a class of inflationary models rooted in particle physics, where spontaneous symmetry breaking plays a central role. Let us consider an inflaton with potential $V(\phi)$, coupled to a symmetry breaking field $\varphi$. 
The total potential of the two-field system is
\begin{equation}\label{eq:HybridPotential}
V(\varphi,\phi) =  \frac{\lambda}{4}\left(|\varphi|^2 - 
v^2\right)^2 + \frac{1}{2}g^2\phi^2|\varphi|^2
+ V_{\mathrm{infl}}(\phi)\,,
\end{equation}
with $\lambda$ the self-coupling of the Higgs-like field, $v$ its vacuum expectation value (VEV) in the broken phase, and $g^2$ the strength of the coupling between the $\phi$ and $\varphi$. In these scenarios, known as {\it Hybrid inflation}~\cite{Linde:1993cn}, the inflationary period is sustained by the Higgs-like vacuum energy $V_{\rm inf} \simeq {\lambda\over4}v^4$, so long as $V_{\mathrm{infl}}(\phi) \ll \lambda v^4$. 
During inflation, the effective mass of the symmetry breaking field is positive define $m_{\rm eff}^2 = (g^2\phi^2 - \lambda v^2) > 0$. Eventually, the mass-squared of $\varphi$ changes from positive to negative, when the inflaton crosses around the {\it critical point} $\phi_c \equiv \sqrt{\lambda}v/g$. The end of inflation is triggered either when the slow-roll conditions are violated, or due to the rolling of the inflaton below the critical point, whichever occur first. In either case, when $\phi \leq \phi_c$, the system enters then into a {\it water-fall} phase, where the symmetry breaking field acquires a negative squared-mass (tachyonic mass), which increases in time. The details of the water-fall phase (say its length in e-folds) depend sensitively on the velocity of the inflaton across $\phi = \phi_c$. 

Due to the waterfall mechanism, these scenarios exhibit a very advantageous feature as compared to standard single field models. A small value of $\lambda$ is not required by the observed amplitude of the CMB anisotropies, and the scale of inflation can range from a GUT scale $\sim 10^{16}$ GeV, all the way down to $\gtrsim$ MeV. Depending on the model, the inflaton potential $V_{\mathrm{infl}}(\phi)$ can take different forms, so one does not need to stick to the canonical $V_{\mathrm{infl}}(\phi) \propto \phi^2$. However, the scalar spectrum of these scenarios is actually in tension with the latest (2015) Planck results, as they tend to produce a spectral index not as red as observed. One needs to consider some variant of the simplest Hybrid model described by Eq.~(\ref{eq:HybridPotential}), so that the potential has a negative curvature close to the critical point. The slope of the potential at the critical point must be sufficiently small so that the waterfall phase is mild enough, in the sense that it lasts for a sufficient number of efolds, typically of $\mathcal{O}(10)$. In mild water-fall Hybrid models, the second slow-roll parameter contributes the most to the scalar spectral index, so that the Planck constraints can be easily fulfilled (recall discussions at the beginning of Sect.~\ref{sec:PBH}). 

Even though the standard Hybrid models with a water-fall regime are in tension with observations, they still serve as a very good `arena' to understand the physics of inflationary models where preheating occurs through tachyonic effects. Therefore, in the following, we will discuss the GW signal produced by both fast and mild water-fall regimes, which essentially can be characterized by how fast the inflaton is crossing across the critical point. It is natural to define the inflaton dimensionless velocity around the critical point as
\begin{equation}
V_c \equiv {1\over \phi_c}{d\phi\over d(m_\varphi t)} = {g\dot {\phi}\over\lambda v^2}\,.
\end{equation}
If the velocity is sufficiently large, the waterfall regime will be short, as a wide range of modes of the field $\varphi$ will become tachyonic, leading the field to reach the true vacuum $\langle |\varphi| \rangle = v$ in less than one e-fold. If the velocity is low, the waterfall regime can be sustained by a sizeable number of efolds, as the tachyonic regime will be triggered only due to the onset of a quantum diffusion regime, which is not as efficient as in the large velocity regime. See e.g.~\cite{Clesse:2008pf,Clesse:2010iz,Kodama:2011vs,Clesse:2012dw,Clesse:2013jra} for more details. 

The production of gravitational waves from a spinodal instability like in tachyonic preheating was first
properly investigated in~\cite{GarciaBellido:2007dg,GarciaBellido:2007af}, for a specific region of the parameter space ($g^2 = 2\lambda = 0.25$) and a significantly large velocity $V_c = 0.05$. Due to the tachyonic instability for modes $k \lesssim k_* = \sqrt{\lambda}v$, unstable Higgs modes grow exponentially, breaking spontaneously the original symmetry, as the Higgs-like field grows towards its true vacuum with expectation value $\langle |\varphi| \rangle = v$. In real space this corresponds to the development and subsequent collision at relativistic speeds, of large concentrations of energy density in the form of bubble-like structures of the Higgs field, of typical size $l_* \sim 1/k_*$. This generates a significant fraction of energy in the form of a stochastic background of GWs, whose time evolution is determined by the successive stages of preheating: first, the Higgs tachyonic instability makes the amplitude of the GWs to grow exponentially fast, due to the exponential instability of the Higgs tachyonic modes. Second, bubble collisions of the Higgs configurations create a new burst of gravitational radiation, corresponding to when the system becomes non-linear. Third, a turbulent regime is finally established, which sets the end of GW production. The GW background redshifts from then onward, as a relativistic energy density component. The resulting GW spectrum can reach very large amplitudes ($h^2\Omega_{\rm GW}^* \sim 10^{-11}$) for GUT scale scenarios with $v \simeq 10^{16}$ GeV, but these are typocally peaked at too large frequencies ($f_* \sim 10^{8}$ Hz) to be observable, similarly as the GW background from parametric resonance after chaotic inflation models.

A systematic parameter and regime dependence was later presented in~\cite{Dufaux:2008dn}, exploring also the slow velocity $V_c \simeq 0$ and small coupling $g^2 \ll \lambda$ regimes. The resulting amplitude and frequency of the peak of the GW spectrum in each regime, can be completely characterized as a function of the parameters $\lbrace g^2/\lambda, v/M_{Pl}, V_c \rbrace$. Defining a critical threshold velocity as
\begin{equation}
V_c^* \equiv {1\over 2}C^3 g^2 \lambda v^2\,,~~~ C \simeq 10\,,
\end{equation}
numerical simulations in~\cite{Dufaux:2008dn} lead to the following parametrization:
\\

$\bullet~g^2 \lesssim \lambda\,,~ V_c > V_c^*$:
\begin{eqnarray}\label{eq:peakHyb1}
f_* \sim \lambda^{1/4}V_c^{1/3} \cdot 7\times10^{10}~{\rm Hz}\,,~~~h^2\Omega_{\rm GW}^* \sim 10^{-6} V_c^{-2/3}\left({v\over M_{Pl}}\right)^2\,,
\end{eqnarray}\\

$\bullet~g^2 \lesssim \lambda\,,~ V_c < V_c^*$:
\begin{eqnarray}\label{eq:peakHyb2}
f_* \sim C g \lambda^{1/4} \cdot 6\times10^{10}~{\rm Hz}\,,~~~h^2\Omega_{\rm GW}^* \sim 2\times 10^{-6} {1\over g^2C^{2}}\left({v\over M_{Pl}}\right)^2\,,
\end{eqnarray}\\

$\bullet~g^2 \ll \lambda\,,~ \forall~V_c$:
\begin{eqnarray}\label{eq:peakHyb3}
f_* \sim \lambda^{1/4}{g\over\sqrt{\lambda}} \cdot 2\times10^{10}~{\rm Hz}\,,~~~h^2\Omega_{\rm GW}^* \sim 3\times 10^{-6} \left({\lambda\over g^2}\right)^{1.16}\left({v\over M_{Pl}}\right)^2\,.\nonumber\\
\end{eqnarray}\\

Let us note that whereas $C \simeq 10$ was measured in simulations for $g^2 = 2\lambda = 2\cdot 10^{-5}$, numerically it was not possible to determine whether $C$ may exhibit a dependence on $\lambda$ for smaller values than $10^{-5}$, and thus one can only conjecture that this value will be roughly constant. 

The present day frequency and amplitude of the stochastic background of GWs from Hybrid preheating, may therefore cover a wide range of values, depending on the regime developed during the water-fall stage, which depends on the values of the parameters of the model $\lbrace g^2, \lambda, v, V_c \rbrace$. Imposing that $f_* < 10^{3}$ Hz, so that any of these backgrounds could be potentially observed by LIGO, leads however to discover that only unrealistically small couplings could led to a detection. For $g^2 \lesssim \lambda$ and a sizeable inflaton velocity $V_c \gg V_c^*$, it is required that $g^2\sim \lambda \sim 10^{-28}$, whereas for $g^2 \lesssim \lambda$ but negligible inflaton velocity $V_c \ll V_c^*$, the situation improves, but we still require that $g^2 \sim \lambda \lesssim 10^{-11}$. For $g^2 \ll \lambda$, independently of $V_c$, it is still required that $g^2 \lesssim 10^{-14}$ for the most interesting regime when $\lambda \sim 1$.

In summary, even though it is possible to conceive that such a background of GWs could be potentially detected by current or planned detectors, the required ingredients seem extremely unnatural from the point of view of particle physics, rendering this background (realistically speaking) undetectable. Only if high frequency technology in the range around the GHz was eventually developed, we might have a chance to detect a background like this.

\subsubsection{What if gauge fields are coupled to the water-fall field?} 

We have described so far the production of GWs during preheating after hybrid inflation, where the spontaneous symmetry breaking of a Higgs-like field is dynamically triggered by the rolling of the inflaton field amplitude. Nothing prevents us from considering the Higgs field to be charged under some gauge symmetry. The addition of gauge fields in the context of hybrid inflation, has been explored extensively in the literature with the purpose of realizing post-inflationary baryogenesis, magnetogenesis, or simply to realize preheating in a more realistic manner, see~\cite{Rajantie:2000nj,Copeland:2001qw,Smit:2002yg,GarciaBellido:2003wd,Tranberg:2003gi,Skullerud:2003ki,vanderMeulen:2005sp,DiazGil:2007dy,DiazGil:2008tf}. As we will see shortly, the addition of gauge also provides a new source of GWs during the tachyonic process~\cite{Dufaux:2010cf,Tranberg:2017lrx}. Like in the absence of gauge fields, when the inflaton crosses below the critical value, the mass squared of the Higgs field becomes negative, and drives the spinodal growth of long-wave modes of the Higgs. However, since the Higgs is now charged, its rapid growth induces
correspondingly also an excitation of the gauge fields. If the Higgs is charged under an Abelian group, the long-wave fluctuations of the Higgs become semi-classical during the tachyonic growth, inducing the generation of a topological winding number of the Higgs field. After the symmetry breaking, there is not enough energy to unwind the Higgs phase, and hence the system is left behind with Nielsen-Olesen string-like configurations~\cite{Dufaux:2010cf}. This corresponds to the formation of a network of cosmic strings, as we will discuss in Section~\ref{sec:CosmicDefects}. The presence of the strings play in fact a crucial role in the production of GWs at preheating. The complicated dynamics occurring at preheating in this abelian Higgs-inflaton scenario is the following: at the end of inflation, the tachyonic growth of the Higgs modes creates an inhomogeneous distribution of the Higgs field components, characterized by bubbles-like structures of the Higgs field that expand and collide. The Abelian gauge field develops structures that concentrate at the regions between bubbles, where the Higgs field has its lowest amplitude. The gauge fields form in  this way long flux tubes of magnetic fields, with a large concentration of energy. The elongated configurations correspond to Nielsen-Olesen vortices which are connected with each other in a cosmic string network formed by the magnetic flux tubes. 

The initial out-of-equilibrium dynamics of the system gives raise to a significant production of GWs, sourced by both the Higgs and the gauge fields. This background inherits therefore specific features from the string dynamics. For instance, in position space, it is observed that the distribution of the GWs follows very closely the spatial evolution of the string network. In Fourier space, one can observe the successive appearance of very distinct peaks in the GW spectra, as the dynamical process carries on. The position of each peak is directly related to the physical scales involved in the problem: the Higgs mass, the gauge field mass, and the typical momentum amplified by tachyonic 
preheating, which depends on the Higgs VeV in the broken phase, but also on the critical velocity and couplings. In general, depending on the parameters in the scalar sector, the same structure of peaks given by Eqs.~(\ref{eq:peakHyb1})-(\ref{eq:peakHyb3}), will emerge in the GW spectrum, like in the absence of gauge fields. However a new peak will emerge at high momenta, due to the presence of the gauge fields. Leaving aside the very IR peak that may arise when the critical velocity of the inflaton is very low, we expect in general two peaks in the UV, one due to the tachyonic scale of the Higgs (essentially the Higgs mass in the case of large critical velocity), and another due to the the mass of the gauge boson, dynamically acquired through the Higgs mechanism, as the Higgs develops its increasingly large non-vanishing expectation value during the tachyonic transition. The peak due to the gauge field is located at a frequency $f_A$, whereas the peak due to the Higgs is centred at another frequency $f_H$. The ratio of frequencies scales with the parameters as $f_A/f_H \sim (e/\lambda^{1/4})$, where $e$ is the gauge coupling~\cite{Dufaux:2010cf}. Therefore, if $e^2 \gg \lambda^{1/2}$, the two scales are very separated, and a two peak structure emerges in the spectrum of GWs (with the two amplitudes typically of the same order). If $e^2 \sim \sqrt{\lambda}$, the two peaks overlap with a similar amplitude, and a single peak structure emerges. If $e^2 \ll \sqrt{\lambda}$, the gauge peak not only is at lower frequencies than the Higgs peak, but also it is sub-dominant, and hence the presence of the gauge fields is irrelevant. Considering SU(2) gauge fields instead of an Abelian U(1) case, does not seem to change this~\cite{Tranberg:2017lrx}. As in the case with only scalar fields, only very small couplings can make these GWs fall into a frequency range accessible by interferometric experiments, see discussion above this subsection. Whether this is natural or not, depends on the underlying particle-physics models of hybrid inflation, see e.g.~\cite{Dufaux:2008dn} and references therein. 

\subsection{Flat-directions}
\label{sec:susyQ}

Supersymmetric extensions of the SM typically involve the existence of flat directions~\cite{Gherghetta:1995dv}, for a review see~\cite{Enqvist:2003gh}. These are directions in field space where the renormalizable part of the scalar potential is exactly flat. As supersymmetry must be broken, the flatness can be uplifted by SUSY-breaking and soft breaking terms, as well as by non-renormalizable terms~\cite{Dine:1995kz}. During inflation, field configurations may develop a large expectation value along these directions, due to quantum fluctuations. Such large amplitude condensates may have interesting cosmological consequences~\cite{Enqvist:2003gh}, like Baryogenesis via Affleck-Dine mechanism~\cite{Affleck:1984fy}.
Any scalar field condensate with soft mass $m$, will start oscillating in the post-inflationary era when the Hubble rate becomes $H \leq m$, with an initial amplitude that could be as high as the Planck scale~\cite{Dine:1995kz,Gaillard:1995az}. The oscillations may lead to an explosive decay of the field condensate due to non-perturbative resonant effects~\cite{Olive:2006uw,Basboll:2007vt,Gumrukcuoglu:2008fk}, similarly as in parametric resonance. Such violent decay can in fact generate a significant background of GWs~\cite{Dufaux:2009wn}.

The resulting GW background from the decay of a flat-direction has some advantages with respect to the analogous backgrounds from the preheating mechanisms discussed so far. Let us recall that in parametric resonance the peak frequency of the GWs is too large $f \gtrsim 10^7-10^8$ Hz to be observable, see Sections~\ref{sec:ParamRes}, \ref{sec:Fermionic}. In scenarios exhibiting a tachyonic instability of a symmetry breaking field, the GW signal may fall into lower (observable) frequencies, but only at the expenses of an extreme fine-tuning of the couplings. By contrast, GWs from the non-perturbative decay of a flat direction, may fall naturally into lower frequencies, while maintaining a relatively large amplitude if the scalar condensate has a sufficiently initial large VEV. 

Let us think of a flat direction as a complex field. Non-renormalizable terms are expected to generate a velocity phase when the oscillation condition $H \lesssim m$ is finally achieved~\cite{Affleck:1984fy,Dine:1995kz}. The oscillation dynamics describe correspondingly an elliptical motion in the complex plane, leading to a time-dependent (quasi-periodic) excitation of fields coupled to the flat direction. This results in a very efficient mechanism of particle production~\cite{Olive:2006uw,Gumrukcuoglu:2008fk}. The excited modes have a typical momentum $k_* \sim m$. This shifts the peak of the GWs to small frequencies, and makes its location independent of $\Phi_i$. For a typical value for the soft masses expected in SUSY theories, $m \sim$ TeV, the peak frequency is of the order of $\sim 10^3$ Hz.

The potential
\be
\label{eq:FlatDirection}
V = m^2\,|\phi|^2 + m_\chi^2\,|\chi|^2 + \frac{g^2}{2}\,\left(\phi\,\chi^* + \phi^*\,\chi\right)\,,
\ee
represents a simple model capturing the basic features of a flat direction $\phi$ coupled to a decay product $\chi$, where both fields are considered complex ($^*$ denotes complex conjugate). This modeling grasps well, for instance, the essence of more realistic scenarios studied e.g.~in \cite{Basboll:2007vt,Gumrukcuoglu:2008fk}. Ref.~\cite{Dufaux:2009wn} has performed lattice simulations of this model, determining numerically, at the same time, the GW spectrum from the field dynamics. The total amount of GWs produced can be parametrically estimated by standard arguments as $\rho_{\rm GW}/\rho_{\rm tot}$ $= (\rho_{\rm GW}/\rho_{\rm flat})(\rho_{\rm flat}/\rho_{\rm tot})^2$ $= \alpha (R_*H)^2(\rho_{\rm flat}/3m_{Pl}^2H^2)^2$, where $\alpha$ is some constant to be determined from the numerics, $R_* \sim a/k_*$ is the typical size of the spatial configurations of the `decay product' field acquires due to its parametric excitation, and $(\rho_{\rm flat}/3m_{Pl}^2H^2)^2$ represents the suppression discussed in Section~\ref{subsec:subdominantPR} for the GW production from a sub-dominant source. Taking $k_* \sim m$, $\rho_{\rm flat} = {1\over2}m^2\Phi_{\rm i}^2$ and $H = m$ (onset of flat-direction oscillation), one can parametrize the peak frequency and amplitude of the GW background as
\begin{eqnarray}
\label{eq:freqSUSY}
f_* \simeq \kappa_* \, \sqrt{\frac{m}{\mathrm{TeV}}} \, \epsilon_{\rm i}^{1/4} \, 5 \times 10^2 \, \mathrm{Hz}\,, \\
\label{eq:GW_SUSY}
h^2 \Omega_{\mathrm{GW}}^* = h^2\Omega_{\mathrm{rad}}\times \epsilon_{\rm i}{\alpha}\left(\frac{m}{k_*}\right)^2 \left(\frac{\Phi_i}{m_{Pl}}\right)^4 \,, 
\end{eqnarray}
with $\alpha$ some numerical constant to be determined by numerical simulations. Ref.~\cite{Dufaux:2009wn} has performed such lattice simulations of the model given by Eq.~(\ref{eq:FlatDirection}), and determined the parameter space for which this background of GWs could be observed. Namely, Ref.~\cite{Dufaux:2009wn} finds that for the extreme case of $\Phi_{\rm i} \sim m_{Pl} \sim 10^{18}$ GeV, it is possible to obtain a (peak) amplitude of the background today as large\footnote{This requires the flat-direction to start oscillating directly in a RD background, otherwise there is an extra suppression proportional to $\epsilon_i < 1$.} as $h^2 \Omega_{\mathrm{GW}}^* \lesssim 10^{-8}$. If the soft masses are of the order $m \sim 0.1-10$ TeV, this background is then peaked at frequencies around $10~{\rm Hz} \lesssim f_* \lesssim 5\times 10^3$ Hz, pretty much overlapping with the LIGO sensitivity range. 

The previous prediction makes, presumably, these scenarios very attractive, from the observational point of view: contrary to standard preheating mechanisms, they are capable to provide a large amplitude of the GW background at relatively small frequencies, without fine-tuning the couplings to extremely small values. However, having a flat direction with such a large amplitude as $\Phi_{\rm i} \sim 10^{18}$ GeV, is (probably) very unrealistic. First of all, such a large amplitude implies that when the flat direction starts oscillating, its energy density is as large as the background energy density. Hence it is not anymore a sub-dominant field, and this may lead to observational inconsistencies which need to be dealt in a case by case basis. Secondly, flat-directions are typically coupled to non-flat directions, so that in reality their flatness is uplifted during inflation, in such a way that their field amplitude is not capable to freely fluctuate (and grow) beyond a certain threshold during inflation. For the MSSM D-flat and F-flat directions, the usual expected behavior $\langle \phi^2 \rangle = {N\over4\pi}H^2$ 
results strongly violated after $N \sim 10^2-10^3$ efoldings~\cite{Enqvist:2011pt}. The flat direction reaches in fact an asymptotic amplitude of $\langle \phi^2 \rangle \sim \mathcal{O}(10)H^2$, and even smaller, depending on the case. As the upper bound on the inflationary scale is currently $H \lesssim 8.5\cdot 10^{13}$ GeV, this implies that, realistically, $\Phi_{\rm i} \lesssim 10^{14}-10^{15}$ GeV. The GW signal from the post-inflationary flat-direction dynamics scales as $\propto (\Phi_{\rm i}/m_{Pl})^4$, and hence this implies a suppression of the signal by a factor (at least) of $\sim 10^{-12}$, turning the final amplitude to $h^2\Omega_{\rm GW} < 10^{-20}$ (in the best possible case).

\subsection{Oscillons}
\label{sec:Osci}

Oscillons are long-lived spatially localized scalar field configurations, formed when a scalar field oscillates around a potential that becomes shallower than quadratic, away from the minimum~\cite{Amin:2013ika}. Oscillons can be produced during preheating after different models of inflation~\cite{Copeland:2002ku,Broadhead:2005hn,Amin:2011hj,Gleiser:2014ipa,Antusch:2015nla}, as well as in a variety of field theories~\cite{Gleiser:1993pt,Copeland:1995fq,Farhi:2005rz,Fodor:2006zs,Graham:2006vy,Gleiser:2007te,Achilleos:2013zpa,Bond:2015zfa}. If the spatial configuration of oscillons is not spherically symmetric, oscillons are naturally expected to emit GWs. The production of GWs from oscillons was first studied in~\cite{Zhou:2013tsa}, in the context of axion-monodromy models~\cite{Silverstein:2008sg}. In these scenarios oscillons are originally formed asymmetric, but they quickly tend towards a spherical shape. Ref.~\cite{Zhou:2013tsa} finds that oscillons source a GW background, for as long as the oscillons remain asymmetric after inflation. The resulting background has a well defined peak, and once the oscillons become spherically symmetric, the production of GWs stops, and the GW peak height saturates. Eventually the oscillons decay. 

The current upper bound on the tensor-to-scalar ratio $r < 0.07$ at CMB scales, and the red-tilted scalar spectral index $n_{\rm s} \simeq 0.96$, could be interpreted as a hint for inflationary potentials with negative curvature. In such scenarios, the universe inflates during the rolling of the inflaton along a 'plateau' with a large potential energy, i.e.~along a flat 'hilltop'~\cite{Linde:1981mu,Martin:2013nzq}. Hilltop models of inflaton are naturally embedded in particle physics scenarios, where a phase transition develops at high energies, see e.g.~\cite{Linde:1981mu,Antusch:2008gw}. Oscillons are in fact expected also in Hilltop models, even though Hilltop potentials are shallow only in one side of the potential (the potential is steeper than quadratic on the other side of the minimum). Studies of post-inflationary dynamics in Hilltop scenarios have shown the appearance of oscillons, demonstrating that they can live for several $e$-folds after the end of inflation~\cite{Antusch:2015ziz,Graham:2006xs}. In Ref.~\cite{Antusch:2016con} it was found that GW production from oscillons in asymmetric potentials is, in fact, enhanced, as the oscillons remain asymmetric configurations for a much longer time. As oscillons in asymmetric potentials converge less efficiently into a spherical configuration, GW production continues long after oscillons first formed. The GW amplitude is continuously sourced by the oscillons, until they finally decay. This yields to a pronounced peak in the GW spectrum, hence making very appealing the GW production from oscillons in e.g.~Hilltop type of potentials.

Ref.~\cite{Antusch:2016con} considered two type of Hilltop scenarios, described by potentials
\begin{eqnarray}\label{eq:hilltopPotI}
{\rm Scenario~I}:~~V(\phi) \,=\,V_0\,\left(1-\frac{\phi^p}{v^p}\right)^2\,,~~ p \geq 3\\
\label{eq:hilltopPotII}
{\rm Scenario~II}:~~V(\chi,\phi) \,=\,V_0\,\left(1-\frac{\phi^p}{v^p}\right)^2 + V_{\rm inf}(\chi,\phi)\,,~~ p\geq 2.
\end{eqnarray}
In Scenario I, $\phi$ is the inflaton, and $V_0$ is tied to the amplitude of the curvature perturbation $\mathcal{P}_{\mathcal R}$, once the VeV $v$ is fixed. For a large VeV, say of the order of GUT scale $v \sim 10^{16}$ GeV, one obtains a large potential amplitude $V_0 \sim (10^{13} ~{\rm GeV})^4$, leading to a GW background peaked at a high frequency today $f \sim 10^9-10^{10}\,{\rm Hz}$. Oscillons emerge as well in scenarios where a second field $\chi$ acts as the inflaton, i.e.~in hybrid-like inflationary models, like those described by Scenario II (for $p = 2$ this includes the case of Hybrid inflation). The main difference between models I and II is that, in the latter, $V_0$ and $v$ are essentially free parameters, independent from each other. This leads to the possibility of realizing a low-scale phase transition, say with $V_0 \sim {\cal O}({\rm 10^5~GeV})^4$, so that the frequency of the corresponding GW background can lie within the frequency range of GW detectors.

Three stages can be distinguished in the evolution of the spectrum of GWs produced by the post-inflationary dynamics in scenarios I and II. There is an initial linear stage chracterized by an abrupt growth of the GW energy density amplitude. This corresponds to the production of GWs during a first tachyonic preheating stage, characteristic of these scenarios. In a second stage, the scalar field dynamics becomes non-linear and it is then when oscillons form. This results in a widening of the GW spectrum. These two stages are also observed in the case of oscillon formation in symmetric potentials. In the case of asymmetric potentials, there is however a final third phase, where a well defined peak emerges in the GW spectrum, centered around the oscillons' oscillatory frequency. This peak continues growing until eventually the oscillons decay. It is during the third phase that the GW production from oscillons differs strongly from the case of symmetric potentials~\cite{Zhou:2013tsa}, where the growth ceases as the oscillons become spherically symmetric, suppressing the production of GWs. In the models like those described by potentials in Eqs.~(\ref{eq:hilltopPotI}),(\ref{eq:hilltopPotII}), the GW peak amplitude may presumably continue growing orders of magnitude above the amplitude set in the previous phases. 

In Ref.~\cite{Antusch:2016con} it was originally found that the corresponding amplitude of the GW background, 3-efolds after the formation of oscillons, would correspond to a signal today just above the sensitivity of aLIGO run O5 $h^2\Omega_{\rm GW} \sim 5 \cdot 10^{-10}$. In Ref.~\cite{Antusch:2017vga}, more refined lattice simulations of the same scenarios, with a lattice volume 64 times larger than in Ref.~\cite{Antusch:2016con}, found that reaching such a high amplitude today, would require instead 5-6 e-folds since the formation of the oscillons\footnote{This new result is however an extrapolation from the end time of the simulations presented in Ref.~\cite{Antusch:2017vga}}. 

Whichever the number of efolds required since the appearance of the oscillons, it is interesting that a GW background with such a high amplitude, can be reached naturally in these scenarios. Let us notice nonetheless, that these GW backgrounds are naturally peaked at large frequencies. Only if there is a extreme fine-tuned separation of scales between $V_0^{1/4}$ and $v$, we obtain a large amplitude (essentially controlled by $v$) at low frequencies (essentially controlled by $V_0^{1/4}/v$). To begin with, this can only be achieved in scenario II, not in scenario I, where $v$ and $V_o$ are not independent from each other. This is the same type of problem as in Hybrid preheating, where the amplitude of the GW background was controlled by the VeV $v$, whereas the peak position was controlled by the self-coupling of the water-fall field $\lambda \sim V_0^{1/4}/v$. For instance, for $v \sim \mathcal{O}(10^{16})$ GeV and $V_0^{1/4} \sim \mathcal{O}(10^5)$ GeV, so that $V_0^{1/4}/v \sim 10^{-11} \ll 1$, a large amplitude signal, peaked at $f \sim \mathcal{O}(10)$ Hz frequencies, can be obtained. This is however the same kind of fine-tuning discussed in Section~\ref{sec:Tachyonic} about Hybrid preheating scenarios. Only by an unnatural separation of the height ($V_0$) and width ($v$) of the potential, can we obtain a small peak frequency while sustaining a significant amplitude of the GW background.

Let us remark that other effects may change the final answer amplitude of the GW background from oscillons. For instance, if more realistic embedding of the scenarios with asymmetric potentials are considered, this may change the above conclusion about a long period sustaining a continuous increase of the GW background peak amplitude. If the oscillons were coupled to other fields (e.g.~in order to realize reheating in scenario I), oscillons might decay into such field species, way before the time needed to build up a sufficiently large amplitude of GWs. Another aspect that could be taken into account, at least on those cases where the GW amplitudes grows non stop, is the addition of back-reaction of the GWs into the oscillon dynamics. As simulations see a continuous growth in time of the GW peak amplitude during the third stage of long lasting oscillons, eventually the fractional energy density in GW may grow too large. At some point oscillons will not be able to act anymore as an unperturbed {\it reservoir}, continuously pumping energy into the GWs, without noticing their own loss of energy into GWs. This moment might however, never be reached in realistic scenarios.

For a study of the parameter dependence of the GW background from oscillons in asymmetric potentials, we point the reader to Ref.~\cite{Antusch:2017vga}, where the calculation of the GWs from oscillons was also extended to the post-inflationary dynamics in the KKLT scenario~\cite{Kachru:2003aw,Antusch:2017flz}.

\subsection{Thermal background}
\label{sec:ThermalBack}

So far in Sect.~\ref{sec:PreheatingAndOthers}, we have considered out-of-equilibrium GW sources, such as the non-perturbative parametric excitation of fields during preheating. There is however, no reason to expect that GWs cannot be produced from a plasma in thermal equilibrium. In thermal equilibrium particles scatter off from each other, as they continuously accelerate due to their interactions. For (physical) momenta $p > T$, with $T$ the temperature of the plasma, the GW production rate is expected to be suppressed as $\propto e^{-p/T}$, simply because the energy carried away by the gravitons must be extracted from thermal fluctuations. As in a thermal bath particle momenta are typically of order $\sim 3T$, while scattering rates are proportional to the coupling strengths of the interactions involved, it is natural to assume that the GW emission rate should scale as $\sim \alpha T^3 e^{-p/T} / M_{\rm Pl}^2$, with $\alpha = g^2/4\pi$ the fine-structure constant for a given interaction with coupling constant $g^2$. In weakly-coupled systems such as the Standard Model (SM) of particle physics, thermal GW emission is then expected to be parametrically small. 

A thermalized plasma experiences nonetheless, long-wavelength fluctuations, known as hydrodynamic fluctuations~\cite{Ghiglieri:2015nfa}. For the largest wavelengths, the (integration of the) unequal-time-correlator of the (transverse-traceless part) of the plasma energy momentum tensor, which characterizes the spectrum of the GW, is determined through a Kubo formula, as proportional to the shear viscosity $\eta$. The shear viscosity is a macroscopic property of the system that emerges from the undergoing microscopic collisions of the plasma species constituents. It is inversely proportional to a scattering cross section. Therefore, for a plasma with weakly interacting particles, $\eta$ is expected to be large. In the SM above the electroweak crossover, i.e.~at temperatures $T > 160$~GeV, right-handed leptons are the most weakly interacting species, exchanging momenta only via hypercharge gauge mediated interactions. Ignoring particle species that equilibrate faster than right-handed leptons, the shear viscosity in the SM can be obtained as~\cite{Arnold:2000dr,Arnold:2003zc}
\be
 \eta \simeq 400\, T^3\,,
 \label{Viscosity}
\ee
where $g_1 \sim 0.36$ has been used as the numerical value for the $U_Y(1)$ coupling constant. 

Ref.~\cite{Ghiglieri:2015nfa} has derived the corresponding spectrum of GWs for soft modes $p \lesssim \alpha^2T$, and estimated its amplitude today plugging in 
the above result Eq.~(\ref{Viscosity}). For completeness, they have also estimated (to leading log-order) the production of GWs by hard modes $p \gtrsim 3T$. See Ref.~\cite{Ghiglieri:2015nfa} for the details of these derivations. The final result, for flat space time, can be summarized as
\be
 \frac{{\rm d}\rho_{\rm GW}}{{\rm d}t\, {\rm d}\ln p} 
 \; = \; 
 \frac{16 p^3 \eta T}{\pi M_{Pl}^2 } \, \phi\Bigl(\frac{p}{T}\Bigr)
 \;, \label{graviton_rate_Flat}
\ee 
where the function $\phi$ is given by
\be
 \phi\left(\frac{p}{T}\right) \; \simeq \; 
 \left\{ 
  \begin{array}{ll}
    \displaystyle 1 & \;,\quad  p \lesssim \alpha^2 T \\
    \displaystyle \frac{  p n_B(p) }{8\pi\eta}
   \sum_{i=1}^3  d_i\, m_{Di}^2 
   \left( \ln \frac{5 T}{ m_{Di}} + ... \right)
    & \;, \quad p \gtrsim 3 T
  \end{array}
  \right. \,,
  \label{FunctionThermal}
\ee
with $n_B(p)$ the Bose-Einstein distribution, $d_1 \equiv 1$, $d_2 \equiv 3$, $d_3 \equiv 8$, and $m_{Di}$ the Debye masses $m^2_{D1} = 11 g_1^2 T^2/6$, $m^2_{D2} = 11 g_2^2 T^2/6$, and $m^2_{D3} = 2 g_3^2 T^2$, corresponding to the gauge groups U(1), SU(2) or SU(3), respectively. Because of the largest multiplicity, the result is dominated by the QCD contribution. Eq.~(\ref{graviton_rate_Flat}) is expected to be quantitatively correct at $k \lsim \alpha^2 T$, whereas at $k \gsim 3T$ it only represents the qualitative structure, as there could be substantial non-logarithmic corrections not considered.

The above result Eq.~(\ref{graviton_rate_Flat}) can be embedded in a cosmological setting, and subsequently be evaluated to obtain a numerical estimate. Ref.~\cite{Ghiglieri:2015nfa} takes as a reference temperature the corresponding electroweak crossover scale in the Standard Model, $T_{EW} \equiv 160$~GeV, and considers that the Universe has been in thermal equilibrium (and dominated by SM species) up to a temperature $T_{\rm max} \gg T_{EW}$. The general formulas for the GW energy density spectrum emitted by a thermal relativistic plasma while the universe expands dominated by radiation at temperatures $T > T_{EW}$, are given in Eqs.~(6.7), (6.8) of Ref.~\cite{Ghiglieri:2015nfa} (we do not reproduce them because they are not particularly illuminating). Plugging in Eq.~(\ref{Viscosity}), and using the fact that the number of SM degrees of freedom above the EW scale is $g_S \sim g_* \sim 100$, Ref.~\cite{Ghiglieri:2015nfa} concludes that, at the time of the EW crossover, the fraction of GW energy density to the total energy density, is given by
\be
  \Omega_{\rm GW}(p) \sim 3 \times 10^{-13}
 \,
 \times \frac{T_{\rm max}}{10^6\,\mbox{GeV}} 
 \times 
 \, \frac{ p^3}{T_{EW}^3}
 \, \phi\Bigl( \frac{p}{T_{EW}}  \Bigr)
  \;, \label{GWthermalEW}
\ee
where $p$ represents physical momentum at the time of the EW crossover. The low-frequency tail $\Omega_{\rm GW}(p) \propto p^3$ for $p \lesssim T_{EW}$, is thus extremely suppressed at the EW scale. Multiplying Eq.~(\ref{GWthermalEW}) by $h^2\Omega_{\rm rad}^{(0)} \simeq 4 \cdot 10^{-5}$ leads to the red-shifted amplitude today. For instance, at the LISA frequencies $f_{LISA} \sim 10^{-3}$ Hz (this corresponds to $p_{EW} \sim 10^3 H_{EW}$), the amplitude today of this thermal GW background is extremely suppressed, as $h^2\Omega_{\rm GW}^{(0)}(f_{LISA}) \sim 10^{-40}(T_{\rm max}/M_{Pl})$. The GW spectral amplitude however grows at higher frequencies, until the second expression in Eq.~(\ref{FunctionThermal}) kicks in for very hard modes. If we consider that the Universe has been in thermal equilibrium up to a maximum temperature $T_{\rm max}$, not much smaller than the Planck scale $M_{Pl} \sim 10^{19}$ GeV, the thermal GW background could be rather large. It is however peaked at very high frequencies, of the order of $\sim 10^9$ Hz~\cite{Ghiglieri:2015nfa}. In order to assess properly its final amplitude in the high-frequency domain, a computation of $\phi(p/T)$, beyond the leading-logarithmic terms captured in Eq.~(\ref{FunctionThermal}), must be obtained. Such computation represents however a challenge in thermal field theory, analogous e.g.~to that for the right-handed neutrino production rate from a SM plasma~\cite{Besak:2012qm,Ghisoiu:2014ena}. It is actually more challenging, as every particle species carries energy and momentum, and hence is involved in the computation. 

\section{First order phase transitions}
\label{sec:FOPT}

In the course of its adiabatic expansion, the universe might have undergone several phase transitions (PTs) driven by the temperature decrease. There are a variety of processes related to primordial PTs that can lead to the production of a SGWB. In fact, a relic SGWB is often the only observable remaining after the occurrence of a PT, which can bring us relevant information on the nature of the PT. In Section \ref{sec:CosmicDefects}, we present the SGWB signal produced by topological defects, stable configurations of the field(s) undergoing the PT that can be left over after a spontaneous symmetry breaking. In this Section, we concentrate specifically on SGWB generation by processes which are related to the occurrence of first order PTs. 

First order PTs are characterised by the appearance of a barrier in the potential of the order parameter that is driving the PT, separating the false, symmetric vacuum from the true, symmetry-breaking one, which becomes more energetically favourable as the temperature decreases. In order for the field to reach the true vacuum, the potential barrier must be overtaken by quantum tunnelling or thermal fluctuations. In real space this corresponds to the nucleation of bubbles of the true vacuum in the space-filling false vacuum. The bubbles then expand due to the pressure difference acting on their walls. As the bubbles expand, the free energy contained in the false vacuum is released. In the idealised case of a PT occurring in empty space, the released energy can only be converted into gradient energy of the bubble walls, which accelerate up to the speed of light. More realistically, since the PT is occurring in the early universe, space is filled with the primordial plasma; the greatest part of the released energy is then converted into thermal energy, raising the temperature of the surrounding plasma. Moreover, part of the energy still goes into gradient energy of the bubble walls. However, since the field driving the transition is very likely coupled to the other fields present in the plasma, part of the released energy is also transferred to bulk motion of the surrounding fluid. 

Both the field energy momentum tensor representing the gradient energy stored in the expanding bubble walls and the fluid energy momentum tensor representing the kinetic energy of the bulk plasma motions, have in general a non-zero anisotropic stress component in their space-space part $\Pi_{ij}$ (c.f.~Eq.~\eqref{Piij}). If the latter has a tensor component, it can act as a source of GWs [recall Eq.~\eqref{gweqx}]. 

Note, however, that spherically symmetric expanding bubbles cannot produce gravitational 
radiation since the transverse and traceless part of the energy
momentum tensor of a radial distribution of field gradients, or of velocity fields, is identically
zero (c.f.~e.g.~Appendix A of \cite{Caprini:2007xq}). GW production occurs since, towards the end of a first order PT, the true vacuum bubbles collide and convert the entire universe to the symmetry-broken phase. The collisions break the spherical symmetry of the bubble walls and of the bulk fluid velocity configuration surrounding them, generating a non-zero tensor anisotropic stress which actively sources GWs. 

The fact that a first-order PT occurring through the nucleation of broken phase bubbles can be a source of GWs, was first pointed out in the seminal works \cite{Witten:1984rs,Hogan:1986qda}. Here the GW signal was estimated using dimensional arguments and the quadrupole approximation; subsequent analyses performed numerical simulations of the collision of bubbles in vacuum, in order to give a more accurate prediction of the GW signal \cite{Kosowsky:1991ua,Kosowsky:1992vn}, and generalized the problem also to PTs happening in a thermal environment \cite{Kamionkowski:1993fg}. 

\subsection{Occurrence of first order phase transitions in the early Universe}

The nature of the primordial PTs depends on the particle theory model describing the universe at high energy. Our present knowledge of high energy physics indicates that there must have been at least two PTs in the early universe: the electroweak one and the QCD one. At temperatures higher than the energy scale probed by the Large Hadron Collider (LHC), corresponding to the electroweak symmetry breaking, there is no experimental guidance to indicate what is the most appropriate particle theory model, and the physical picture is open to several hypothesis.  

In the Standard Model (SM) of particle physics, the electroweak phase transition (EWPT) is a cross-over, and it is not expected to lead to any appreciable cosmological signal (see e.g.~\cite{Gurtler:1997hr,Aoki:1996cu,Kajantie:1995kf,Kajantie:1996mn,Laine:1998jb}). However, deviations from the SM in the Higgs sector or the introduction of additional fields (for example because of supersymmetry) can modify the order of the EWPT with respect to the SM scenario. The discovery of the Higgs boson at the LHC confirms the paradigm of a scalar field-driven symmetry breaking in the early universe \cite{Aad:2012tfa}. However, there is so far no solid indication from the LHC of new physics near the EW energy scale. On the other hand, the order of the EWPT is not constrained by LHC data: several models leading to a first order EWPT remain viable and complying with LHC bounds. Besides a SGWB, these extensions of the standard model can provide dark matter candidates, baryogenesis, and can also alleviate the hierarchy problem (see e.g. \cite{Fairbairn:2013uta,Dorsch:2013wja,Huang:2012wn,Laine:2012jy,AbdusSalam:2013eya,Davoudiasl:2012tu,Patel:2012pi}). 

In the Minimal Supersymmetric Standard Model (MSSM) with a light stop, large cubic terms to the finite temperature effective potential are thermally induced, but these models are now practically ruled out by LHC constraints~(see e.g.~\cite{Carena:2012np,Menon:2009mz,Cohen:2012zza,Curtin:2012aa}). More promising are singlet extensions of the MSSM~(see e.g.~\cite{Pietroni:1992in,Davies:1996qn,Apreda:2001us,Kozaczuk:2013fga, Huang:2014ifa, Kozaczuk:2014kva}): the presence of the scalar singlet can induce barriers at tree-level and strong first order PTs are predicted, in particular in the regime where the EWPT occurs in two steps (i.e. with the singlet acquiring a VEV first, and EW symmetry breaking occurring at a subsequent transition) \cite{Huber:2015znp}. One can also have non supersymmetric extensions of the Standard Model scalar sector. For example, the Higgs portal scenario with a real gauge singlet
scalar field~\cite{Chen:2017qcz,Espinosa:2011ax,Espinosa:2008kw}: in this case, the singlet can either acquire a VEV with a barrier separating it from the EW vacuum (where the singlet VEV is zero), realising a two-step PT one of which is of first order, or it can reduce the SM-like Higgs quartic coupling through loop effects, thereby changing the order of the EWPT. Similarly, the new scalar can be charged under the SM gauge group as in the two Higgs doublet model~\cite{Fromme:2006cm,Kakizaki:2015wua}: in this class of scenarios, a first order EWPT is driven by a decrease in the free-energy difference between the two EW vacua at zero temperature. One can also invoke the presence of unknown, high-energy new physics and study its effect on the Higgs PT using an effective, model-independent field theory approach, for instance by adding dimension-6 operators in the Higgs potential allowing for a negative quartic coupling \cite{Grojean:2004xa,Delaunay:2007wb,Huber:2007vva}: this can lead to very strongly first order EWPT. 

The QCDPT is also predicted to be a cross-over by lattice simulations run at zero baryon and charge chemical potentials (in the absence of lepton and baryon asymmetries)~\cite{Aoki:2006we}. However, the lepton asymmetry is very poorly constrained in the early universe, since it could be hidden in the neutrino sector. It has been claimed that `large' lepton asymmetry, still compatible with present constraints, might affect the dynamics of the QCDPT in a way to render it first order in the early universe \cite{Schwarz:2009ii}. A viable test of this hypothesis would be the detection of the GW thereby emitted (see e.g.~\cite{Caprini:2010xv,Anand:2017kar}).

Alternatively, there are various extensions of the SM that predict strong first-order cosmological
PTs not tied to the EW or QCD scales (or to baryogenesis). Models solving the hierarchy
problem via warped extra dimensions, as the Randall-Sundrum one, are a very promising example in what concerns the production of a detectable SGWB \cite{Randall:2006py,Nardini:2007me,Konstandin:2011dr}. These models feature a spontaneously broken (approximate) conformal symmetry to which the associated pseudo Nambu-Goldstone boson is the dilaton. The dilaton potential can be described by a scale invariant function modulated by a weakly scale-dependent one: it is therefore very shallow, and the position of the barrier and the minimum of the potential can be widely separated. As a result, a significant amount of supercooling and therefore a strong first-order PT, can be obtained without a substantial tuning of parameters. 

Other well motivated scenarios are those in which the dark matter is a stable bound state of a confining dark sector, often without interaction with the (beyond-)SM visible sector, except gravitationally. The DM candidate can be a dark baryon-like state of a $SU(N)$ dark sector with light dark quarks \cite{Bai:2013xga} or, in models without quarks, it can be dark glue-balls \cite{Boddy:2014yra}. The dark sector undergoes a confinement PT, at a scale which sets the mass of the DM candidate. In a large class of models the PT is first order, giving rise to a SGWB just as do PTs in scenarios interacting and/or extending the SM~\cite{Schwaller:2015tja}. However, unlike collider and direct DM searches, the GW signal is produced regardless of the interaction strength between the dark and visible sectors, and thus provides a very powerful way to test these models. 

There are also scenarios whereby inflation is ended through a PT, via tunnelling from a false to a true vacuum, or whereby first order PTs take place during inflation. The oldest proposals of the first category are ruled out nowadays by CMB anisotropy observations \cite{La:1989za,Turner:1990rc,Turner:1992tz,Steinhardt:1990zx}, but more recently, allowed models of this kind have been explored~\cite{Masina:2011un,Masina:2011aa,Masina:2012yd,DiMarco:2005zn}: for instance, Ref.~\cite{Biswas:2005vz} predicts a strong GW signal from the phase transition ending inflation. Models in the second cathegory can also lead to strong GW signals, depending on the details of the scenario considered: see e.g.~\cite{Chialva:2010jt,Wang:2018caj,Hebecker:2016vbl}. 

In general, the detection of a SWGB would provide a neat probe of the occurrence and the nature of cosmological first order PTs, bringing new information on the underlying high energy theory describing the primordial universe. We provide here a general description of the expected signal, and conclude with two well motivated examples relevant for the space-based interferometer LISA. 

\subsection{Relevant parameters entering the SGWB signal}

The GW signal from first order PTs only depends on a few parameters that determine the evolution of the broken phase bubbles (for example their size at collision and their wall speed) and the amount of energy which is available to source the GWs, i.e. the tensor anisotropic stresses (which depend on the strength of the PT and the coupling of the field undergoing the transition with the particles in the surrounding plasma). The values that these parameters can take depend on the details and the particle physics nature of the PT, but the GW signal can be described in terms of them in a phenomenological, practically model-independent way.   

Relevant for the GW production is $T_*$, the temperature of the thermal bath at the time $t_*$ when GWs are produced, i.e. towards the end of the PT when bubble collision occurs (from now on, a subscript $*$ denotes a quantity at the time of GW production). For PTs without significant supercooling and reheating, this is approximately equivalent to the nucleation temperature, $T_*\approx T_n$. The nucleation temperature is the one at which the probability of nucleating one bubble per horizon volume is of order one. This is determined by the nucleation rate (see e.g.~\cite{Grojean:2006bp})
\begin{equation}
\Gamma (t) =A(t)e^{-S(t)} \, ,	
\end{equation}
where $A$ is a pre-factor with unit of energy to the fourth power, and $S$ is the Euclidean action of a critical bubble~\cite{Quiros:1999jp,Laine:2016hma}: either $S_4$, given by the $O(4)$-symmetric solution for vacuum transitions, or $S_3/T$, given by the $O(3)$-symmetric bounce solution for transitions at finite temperature. One can define an approximate inverse time duration of the PT $\beta$, as the rate of variation of the bubble nucleation rate (accounting for the fact that most of the time variation of $\Gamma(t)$ is in $S(t)$) 
\begin{equation}
\beta \equiv - \left.\frac{dS}{dt}\right|_{t_*} \simeq   \left.\frac{\dot{\Gamma}}{{\Gamma}}\right|_{t_*}\,.
\end{equation}
The ratio of the PT inverse duration $\beta$ and the inverse characteristic rate of expansion of the universe at the PT time $H(T_*)$, is a fundamental parameter for the GW signal, as we will see:
\begin{equation}
\label{eq:boH_thermal}
\frac{\beta}{H_*}=T_* \left. \frac{dS}{dT}\right|_{T_*} \, .
\end{equation}
This parameter fixes $R_*$, the size of the bubbles towards the end of the PT: if $v_w$ is the bubble wall speed, in the rest frame of the fluid and far away from the bubble, one has simply $R_*\simeq v_w/\beta$. 

The strength of the PT is characterised by the ratio of the vacuum energy density released in the transition to the radiation energy density in the universe at the moment of the PT (this parameter is defined in the literature also in terms of the latent heat, instead of in terms of the vacuum energy)
\begin{equation}
	\label{eq:alpha_thermal}
\alpha=\frac{\rho_{\rm vac}}{\rho_{\rm rad}^*} \, .
\end{equation}
Note that, if the supercooling is large and the PT effectively occurs in vacuum, it must be characterised also by a consistent amount of reheating, in order to restore the universe in a thermal state after its completion. In this case, one expects $T_n\ll T_{\rm reh}\simeq T_*$, and the above definitions must be changed accordingly \cite{Caprini:2015zlo}:
\begin{equation}
\label{eq:boH_vacuum}
\frac{\beta}{H_*}=\frac{H(T_n)}{H_*}T_n \left. \frac{dS}{dT}\right|_{T_n}, \qquad \alpha = \frac{\rho_{\rm vac}}{\rho_{\rm rad}(T_n)}.	
\end{equation}
In this case, the nucleation temperature can be many orders of magnitude smaller than the energy scale corresponding to the VEV at the minimum of the potential. However, the relevant temperature for the GW generation remains $T_* \simeq T_{\rm reh}$; and if the reheating process is sufficiently fast, one also has $H(T_n)\simeq H_*$. 

As previously mentioned, the amplitude of the GW signal depends on the amount of energy which is available to source the GWs. The source can be in various forms, depending on the properties of the phase transition. In the most common cases, the PT occurs in a thermal environment, and the largest part of the free-energy liberated by the bubbles is converted into heat, which does not lead to any GW production. However, the PT proceeds through bubble nucleation and therefore some fraction of the free-energy also sets the bubble walls into motion. Towards the end of the phase transition, the bubbles collide and a non-zero tensor anisotropic stress is generated, which acts as a source of GWs. The anisotropic stress can be from the gradient energy of the bubble walls, or from the bulk motion that are set in the fluid by the bubbles sweeping through (provided the field performing the PT is coupled to the surrounding plasma). There are therefore two relevant parameters for the GW generation, the fraction of vacuum energy that gets converted into gradient energy of the Higgs-like field, and into bulk motion of the fluid, respectively: 
\begin{equation}\label{eq:kappas}
 \kappa_\phi=\frac{\rho_\phi}{\rho_{\rm vac}}\,,~~~~\kappa_v=\frac{\rho_{v}}{\rho_{\rm vac}}\,.
\end{equation}
Note that if the PT is characterised by a large amount of supercooling, and hence effectively happening in vacuum, the free-energy liberated by the bubbles is converted only into gradient energy of the bubble walls, and in this case one has simply $\kappa_\phi=1$. 

\subsection{General properties and frequency shape of the SGWB spectrum}

It is easy to obtain a rough estimate of the GW amplitude, which shows how it scales with the duration and the tensor anisotropic stress of the GW source (see e.g.~\cite{Caprini:2007xq}). Let us suppose that the process leading to the tensor anisotropic stresses has a typical duration corresponding to the PT duration $1/\beta$, and that this is less than one Hubble time: $\beta/H_*>1$. The usual equation for GW production, Eq~\eqref{gweqx}, can be rewritten under these hypotheses simply as $\beta^2 h \sim 16\pi G \,\Pi$, where $h$ denotes the amplitude of the tensor perturbation, $\Pi$ the tensor part of the energy momentum tensor of the source, and we inserted $1/\beta$ as the characteristic time on which the perturbation is evolving (we have dropped indices for simplicity). This suggests that $\dot h \sim 16\pi G \,\Pi/\beta$, and the GW energy density at the time of production can then be estimated as (c.f Eq.~\eqref{rhogw}) $\rho_{\rm GW}^* \sim \dot h^2/(32\pi G) \sim 8 \pi G\, \Pi^2/ \beta^2$. Dividing by the total energy density in the universe $\rho_{\rm tot}^* = 3 H_*^2 / (8 \pi G)$ at the time of GW production, one has 
\begin{equation}\label{eq:PTscaling}
\frac{\rho_{\rm GW}^*}{\rho_{\rm tot}^*} \sim \left(\frac{H_*}{\beta}\right)^2 \, \left(\frac{\Pi}{\rho_{\rm tot}^*}\right)^2	\,.
\end{equation}
The above equation shows that the GW energy density scales like the square of the ratio of the GW source duration and the Hubble time, and the square of the ratio of the energy density in the source and in the universe at the emission time. Using Eq.~\eqref{eq:Om_today}, the amplitude of the SGWB today becomes 
\begin{equation}
	h^2 \, \Omega_{\rm GW} \sim 1.6 \times 10^{-5} \, \left(\frac{100}{g_*(T_p)}\right)^{1/3} \, \left(\frac{H_*}{\beta}\right)^2  \left(\frac{\kappa \,\alpha}{1+\alpha}\right)^2\,,
\end{equation}
where, to rewrite $\Pi/\rho_{\rm tot}^*$, we have used $\rho_{\rm tot}^*=\rho_{\rm rad}^*+\rho_{\rm vac}$, the definition of $\alpha$ given in Eq.~\eqref{eq:alpha_thermal}, and we have set $\kappa\sim \Pi/\rho_{\rm vac} $, where $\kappa$ can be either of the parameters defined in \eqref{eq:kappas}. 
As a rule of thumb, a GW signal above the sensitivity of a future interferometric detector like LISA ($h^2\Omega_{\rm GW}\gtrsim 10^{-13}$) can be generated if 
$(\HH_*/\beta) (\Pi / \rho_{\rm tot})_* \gtrsim \mathcal{O}(10^{-4})$. Therefore, detectable signals arise from very energetic processes, which involve a sizeable fraction of the total energy density in the universe, and at the same time slow processes, which minimise the value of $\beta/\HH_*$. 
   
As mentioned above, the processes leading to the production of the SGWB operate towards the end of the PT, since they are related to the collision of bubbles. The characteristic wave-number $k_*$ of the SGWB generated by these processes, i.e.~the wave-number at which one expects the SGWB to peak, corresponds to the inverse typical time or length scale of the problem: in this case, either due to the duration of the PT or to the bubble size, $k_*/a_*\simeq 2\pi \,\beta$ or $k_*/a_*\simeq 2\pi/ R_*\simeq 2\pi\,\beta/v_w$, depending on the details of the source (note that we are equating the comoving wave number $k_*$ to physical quantities as $\beta$ and $R_*$, therefore we introduce the factor $a_*$). If the growth of the bubble proceeds at a highly relativistic speed, the two time/length-scales are equal. Setting e.g.~$k_*/a_*\simeq 2\pi \,\beta$, and using Eq.~\eqref{fTp} with $x_k=2\pi \,\beta/H_*$, one obtains the following order of magnitude estimate for the characteristic frequency today:
\begin{equation}
	f \sim 1.6 \times 10^{-5}\,{\rm Hz} \,\,\frac{\beta}{H_*}\, \left( \frac{g_*(T_*)}{100} \right)^{\frac{1}{6}} \,\,\frac{T_*}{100\,{\rm GeV}}\,.
\label{kstar}
\end{equation}
Since at the end of the PT one expects the entire universe to be converted to the broken phase, in general the PT must complete faster than a Hubble time, so that $\beta/H_*>1$. From Eq.~(\ref{kstar}) it appears that the characteristic frequency of GW emitted around the EW symmetry breaking at 100 GeV falls in the frequency range of LISA \cite{Audley:2017drz} for values $1\lesssim \beta/H_* \lesssim 10^5$. As another example, we see from the above formula that GW production at the QCDPT at $T_*\simeq 100$ MeV can fall into the frequency range of detection with pulsar timing array, since $f\geq 10^{-8}$Hz (see e.g. \cite{Caprini:2010xv} and references therein). The precise value of $\beta/H_*$ has to be determined in the context of a given model for the first order PT. 

The slope of the GW spectrum at wave-numbers smaller than the Hubble radius at the time of production, $k< a_* H_*$, can be determined on general grounds, valid for any transient stochastic source after inflation. It is a consequence of the fact that the causal process (the PT) generating the GW signal cannot operate on time/length-scales larger than $(H_*)^{-1}$. Therefore, the anisotropic stresses $\Pi_{ij}({\bf k}, t)$ sourcing the metric perturbations in Eq.~\eqref{gweqx}
are not correlated for $k<a_* H_*$, and the anisotropic stress power spectrum $\Pi(k,\eta,\zeta)$ (c.f. Eq.~\eqref{eq:PIspec}) is expected to be flat in $k$ (white noise) up to the wave-number $k_*$. Eq.~\eqref{eq:rhogwPiRad}, valid in the radiation era, then shows that the spectrum of GW energy density per logarithmic frequency interval must grow as $k^3$ at these large scales $k<a_*H_*\simeq 1/\eta_{\rm in}$. Thus, it is a general result, for SGWBs produced by a first order PT, that the infrared tail of the present-day GW spectrum behaves as $h^2 \Omega_{\rm GW} \propto f^3$ for scales that were super-Hubble at the time of production \cite{Caprini:2009fx}. 

Note that at sub-Hubble scales $H_*<k<k_*$, the SGWB spectrum may also continue to grow as $f^3$ until the characteristic frequency $f_*=k_*/(2\pi)$: the inverse typical time or length scale of the problem, determining $k_*$, can also play the role of a maximal correlation scale, so that the anisotropic stress power spectrum remains uncorrelated, white noise, for every $k<k_*$, and the above argument applies. However, the details of the time dependence of the anisotropic stress power spectrum can also play a role in this frequency range, modifying the expected $k^3$ slope, while this cannot happen for super-Hubble modes. While often the case, it therefore cannot be taken for granted that $h^2 \Omega_{\rm GW} \propto f^3$ for $f<f_*$, and there are exceptions (MHD turbulence being one of these, as we will see).

For $k>k_*$, the GW power spectrum decays with a slope that depends on the details of the process sourcing the SGWB, and no general consideration is possible, apart from the fact that the total $\rho_{\rm GW}$, c.f. Eqs.\eqref{rhogw} and \eqref{eq:rhogwPiRad}, must be finite when integrated on the interval $0<k<\infty$.  

As mentioned earlier, the anisotropic stresses acting as a source of GW can be due to the gradient energy of the bubble walls, or to the bulk motion in the surrounding fluid. We now proceed to present the details of all the SGWB sources acting during a first order PT.

\subsection{Contribution to the SGWB from the scalar field driving the PT: bubble wall collisions}
\label{sec:Bubbles}

The GW production due to the collision of the bubble walls is the easiest contribution to model because, since the seminal paper \cite{Kosowsky:1992vn}, it is estimated using the `envelope approximation'. This consists in numerically simulating the motion of the bubble walls as a propagation of spherical, infinitely thin shapes instead of using the Klein-Gordon equation to evolve the scalar field. In this approximation, the gravitational radiation is sourced only by the TT part of the energy-momentum tensor of the uncollided envelope of the spherical bubbles, ignoring the interaction region: this greatly simplifies the numerical simulation since it dispenses with the detailed dynamics of the scalar field and reduces the required computational power. 

The validity of the envelope approximation has been asserted in \cite{Kosowsky:1992vn,Kamionkowski:1993fg} for the case of strongly first order PTs happening both in vacuum and in a thermal environment, if they proceed through detonation (i.e. at supersonic speed \cite{Steinhardt:1981ct}). In these cases, the energy momentum tensor representing the propagation of the bulk fluid motions sourcing the GWs is effectively concentrated on a thin shell near the bubble wall. The latest numerical simulations using the envelope approximation with considerably improved numerical accuracy have been carried on in \cite{Huber:2008hg,Weir:2016tov}, providing a better determination of a larger portion of the GW spectrum and consequently a more careful analysis of the high frequency behaviour with respect to previous works. The resulting SGWB spectrum is\footnote{Note that, by the argument given in the previous subsection, causality should imply that at low frequency the SGWB grows as $f^3$. This must be the case at least for frequencies smaller than the inverse Hubble horizon at GW production; however, $f^{2.8}$ provides a better fit to the simulated result close to the peak of the spectrum \cite{Huber:2008hg}.}
\begin{eqnarray}\label{eq:Omenv}
h^2\Omega_{\phi}(f) &=& 1.67 \times 10^{-5}   \, \left( \frac{H_*}{\beta} \right)^2 \left( \frac{\kappa_\phi\, \alpha}{1+\alpha} \right)^2  
 \left( \frac{100}{g_*(T_*)} \right)^{\frac{1}{3}} \left(\frac{0.11\,v_w^3}{0.42+v_w^2}\right) \\
&\times& \frac{3.8 \,\,(f/f_\phi)^{2.8}}{1 + 2.8 \, (f/f_\phi)^{3.8}}\,, \nonumber
\end{eqnarray}
where the peak frequency $f_\phi$ corresponds roughly to the characteristic time-scale of the PT, i.e.~its duration $1/\beta$ \cite{Caprini:2009fx}. The simulations yield~\cite{Huber:2008hg,Weir:2016tov}
\begin{equation}
\frac{f_*}{\beta} = \frac{0.62}{1.8-0.1v_w+v_w^2}\,, 
\end{equation}
which becomes, once redshifted to today 
\begin{equation}
f_{\phi} = 1.65 \times 10^{-2} \, {\rm mHz} \, \left(\frac{f_*}{\beta}\right) \, \left(\frac{\beta}{H_*} \right) 
\left(\frac{T_*}{100\,{\rm GeV}}\right) \left( \frac{g_*(T_*)}{100}
\right)^{\frac{1}{6}}\,,
\label{eq:envPeak}	
\end{equation}
(c.f. Eq.~\eqref{kstar}). 
Besides numerical simulations \cite{Huber:2008hg,Weir:2016tov}, there have been also several works that have tried to model the SGWB spectrum from bubble collisions analytically, see \cite{Caprini:2007xq,Jinno:2016vai,Jinno:2017fby}. 
 
\subsection{Contribution to the SGWB from the bulk fluid motions: sound waves}
\label{sec:SoundWaves}

The characteristics of the bulk flow depend on the strength of the coupling of the field driving the PT to the fluid particles: this coupling strongly influences the bubble evolution, as demonstrated by many analyses (see e.g. \cite{Ignatius:1993qn,Moore:1995ua,Megevand:2009ut,Konstandin:2010dm,Megevand:2012rt,Megevand:2013yua}). In general, it can be assumed that the propagation quickly reaches stability and the bubble walls expand with constant velocity $v_w$, which can be either subsonic (deflagration) or supersonic (detonation) (see e.g. \cite{Espinosa:2010hh} and references therein)\footnote{Note that the possibility of having runaway solutions in the electroweak PT, first put forward in \cite{Bodeker:2009qy}, has been excluded by a recent analysis \cite{Bodeker:2017cim}.}. The bubble wall speed $v_w$ should be determined by a full analysis of the microscopic interactions, of the type of those carried out in e.g.~\cite {Moore:1995ua,Moore:1995si,Kozaczuk:2015owa}. However, from the point of view of GW production, the problem can be tackled by introducing a phenomenological parameter, the friction $\eta$, and by studying the bubble evolution as a function of this \cite{Espinosa:2010hh}. Both semi-analytical methods (see e.g. \cite{Huber:2013kj}) and numerical simulations (see e.g. \cite{Giblin:2013kea}) show that the friction modelling the interactions influences the bubble wall velocity and the transfer of kinetic energy of the scalar field to bulk kinetic energy of the fluid \cite{Espinosa:2010hh}, which are important parameters entering the GW production rate. 

The most recent and detailed numerical simulations of the full system of the scalar field performing the transition and the surrounding fluid coupled to it via a friction parameter $\eta$ have been performed in Refs.~\cite{Hindmarsh:2013xza,Hindmarsh:2015qta,Hindmarsh:2017gnf}. These have demonstrated that compressional modes, i.e.~sound waves, are induced in the surrounding fluid by the expansion of the bubbles, due to the coupling among the scalar field and the fluid. At bubble collisions, the sound waves give rise to a non-zero tensor anisotropic stress that is a powerful source of GWs. Simulations have furthermore found that the sound waves continue to act as a source of GWs well after the merging of the bubbles is completed and the scalar field has everywhere settled in the true vacuum. They remain present in the fluid until either they are damped by viscosity, or they generate shocks. The long-lasting nature of the sound waves in the primordial fluid enhances the GW signal by a factor $\beta/H_*$, rendering them the most relevant contribution to the SGWB spectrum in the case of PTs which are not very strongly first order and happen in a thermal environment \cite{Hindmarsh:2013xza}. Note that the fact that GW sources lasting long, more than one Hubble time, are amplified by an extra $\beta/H_*$ factor does not contradict the result given in Eq.~\eqref{eq:PTscaling}, and was  predicted on the basis of analytical arguments in Ref.~\cite{Caprini:2009yp}. 

The SGWB spectrum from sound waves, fitted from the numerical results of \cite{Hindmarsh:2015qta}, is given by
\begin{eqnarray}\label{eq:OmGWSound}
h^2\Omega_{\rm sw}(f) & = &  2.65 \times 10^{-6} \, \left( \frac{H_*}{\beta} \right) \left( \frac{\kappa_v \,\alpha}{1+\alpha} \right)^2  
 \left( \frac{100}{g_*(T_*)} \right)^{\frac{1}{3}} v_w \\
& \times &\left(\frac{f}{f_{\rm sw}}\right)^{3}\,\left(\frac{7}{4 + 3\,(f/f_{\rm sw})^{2}}  \right)^{7/2}\,, \nonumber
\end{eqnarray}
where the peak frequency is set by the characteristic size of the bubbles at the end of the transition, and it is approximatively given by $f_{\rm sw}\simeq (2/\sqrt{3})(\beta/v_w)$, which, after redshifting, becomes\footnote{Note that the most recent analysis \cite{Hindmarsh:2017gnf} finds a somewhat smaller peak frequency of the order of $f_{\rm sw}\simeq 0.3(\beta/v_w)$}
\begin{equation}
f_{\rm sw} = 1.9 \times 10^{-2} \, {\rm mHz} \, \frac{1}{v_w} \, \left(\frac{\beta}{H_*} \right) 
\left(\frac{T_*}{100\,{\rm GeV}}\right) \left( \frac{g_*(T_*)}{100} 
\right)^{\frac{1}{6}}\,.
\label{eq:SWPeak}
\end{equation}

\subsection{Contribution to the SGWB from the bulk fluid motions: MHD turbulence}
\label{sec:Turbulence}

Besides sound waves, the bubble merging could also induce vortical motions in the surrounding fluid, which would constitute an independent source of GWs. The primordial plasma is characterized by a very high Reynolds number (of the order of $10^{13}$ at 100 GeV and at the typical scale of the bubbles \cite{Caprini:2009yp}): therefore, the energy injection caused by the collision of the bubbles is expected to lead to the formation of magneto-hydrodynamic (MHD) turbulence, which generates GWs through the anisotropic stresses of the chaotic fluid motions and of the magnetic field. Note that the turbulence is expected to be accompanied by the presence of magnetic fields since the early universe plasma is fully ionised and has a very high conductivity \cite{Ahonen:1996nq,Baym:1997gq}. Turbulence can also lead to the amplification of small magnetic fields generated by charge separation at the bubble wall (see e.g. \cite{Stevens:2010ym}).

In the simulations of \cite{Hindmarsh:2013xza,Hindmarsh:2015qta,Hindmarsh:2017gnf}, the vortical component of the bulk fluid motions has been evaluated and was always largely sub-dominant with respect to the compressional one. However, after a characteristic time $\tau_{\rm sh} \sim (v_w/\sqrt{\kappa_v\alpha})\beta^{-1}$ (see e.g. \cite{Pen:2015qta}) one expects the formation of shocks, that will eventually convert the acoustic signal into a turbulent one. This happens of course only if $\tau_{\rm sh}\leq H_*^{-1}$, i.e.~if shocks can develop within one Hubble time. Up to now, the numerical simulations did not simulate strong enough PTs capable of reaching $\tau_{\rm sh}$ within the simulation time, so predictions of the GW signal from turbulence based on simulations are not available. However, analytical evaluations of the SGWB from MHD turbulence exist. 

The first analyses of the GW production by turbulence had some problems that led to an overestimate of the signal \cite{Kosowsky:2001xp,Dolgov:2002ra,Gogoberidze:2007an} (c.f. the discussion in \cite{Binetruy:2012ze} and references therein). The most recent analytical evaluation of GW emission from MHD turbulence generated during a first-order PT and freely decaying afterwards is the one of \cite{Caprini:2009yp}. This analytical evaluation maintains a certain level of intrinsic uncertainty, for example in that it has to rely on a theoretical turbulence model (usually, Kolmogorov turbulence is assumed) and on a model for the time decorrelation of the GW source. This uncertainty could only be addressed by numerical simulations of relativistic MHD turbulence. Furthermore, \cite{Caprini:2009yp} neglects the possibility of helical turbulence (see e.g.~\cite{Kahniashvili:2008pe}). Under these assumptions, the resulting contribution of MHD turbulence to the GW spectrum is \cite{Caprini:2009yp,Binetruy:2012ze} 
\begin{eqnarray}\label{eq:OmGWturb}
h^2\Omega_{\rm turb}(f) &=& 
3.35 \times 10^{-4} \, \left( \frac{H_*}{\beta} \right)
\left(\frac{\kappa_{\rm turb}\,\alpha}{1+\alpha}\right)^{\frac{3}{2}}\,
 \left( \frac{100}{g_*(T_*)}\right)^{1/3}\, v_w \\
&\times & \frac{(f/f_{\rm turb})^3}
{\left[ 1 + (f/f_{\rm turb}) \right]^{\frac{11}{3}} 
\left(1 + 8 \pi f/h_* \right)}\,, \nonumber
\end{eqnarray}
where 
\begin{equation}\label{eq:kappaturb}
\kappa_{\rm turb}=\epsilon \,\kappa_v\,,	
\end{equation}
 represents the (yet unknown) fraction of bulk kinetic energy associated to the vortical motions, as opposed to the compressional modes. Similarly to the case of sound waves, there is an amplification by a factor $\beta/H_*$\,, which is typical of sources that last longer than the average duration $1/\beta$ of the PT. In Eq.~\eqref{eq:OmGWturb}, $h_*=1.6 \cdot 10^{-4}(T_*/100 \,{\rm GeV})(g_*/100)^{1/6}$ mHz is the Hubble parameter red-shifted to today: it enters also as a consequence of the fact that turbulence acts as a source of GW for several Hubble times. Similarly to the sound waves case, the peak frequency is connected to the inverse characteristic length-scale of the source, the bubble size  $R_*$ towards the end of the PT: $f_{\rm turb}\simeq (3.5/2) (\beta/v_w)$, which becomes, after red-shifting,
\begin{equation}
f_{\rm turb} = 2.7 \times 10^{-2} \, {\rm mHz} \, \frac{1}{v_w}\, 
\left(\frac{\beta}{H_*} \right) \left(\frac{T_*}{100\,{\rm GeV}}\right) \left( 
\frac{g_*(T_*)}{100} \right)^{\frac{1}{6}} \, . 	
\end{equation}

\subsection{Examples of SGWB from a first order PT}

In order to predict the amplitude and peak frequency of the GW signal from a specific first order PT, one has to determine the value of the few parameters entering the GW spectrum, as shown in the previous subsections. These are the PT temperature $T_*$, the inverse duration of the PT $\beta/H_*$, the bubble wall velocity $v_w$, and the fraction of energy that contributes to the GW generation $(\Pi / \rho_{\rm tot})_*$. This latter becomes the factor $(\kappa\,\alpha/(1+\alpha))$ appearing in Eqs.~\eqref{eq:Omenv}, \eqref{eq:OmGWSound}, \eqref{eq:OmGWturb}, once translated into the two parameters representing the strength of the PT ($\alpha$, Eq.~\eqref{eq:alpha_thermal}) and the  fraction of vacuum energy that gets converted into gradient or kinetic energy ($\kappa$, Eq.~\eqref{eq:kappas}). 

These parameters can only be determined within a given model of the PT, and are not all independent among each other. In the case of a thermal phase transition, one first needs to find the bounce solution of the three-dimensional Euclidean action $S_3(T)$, which quantifies the probability of thermal jumping \cite{Huber:2007vva}. From this, one can then calculate the fraction of space that is covered by bubbles (neglecting overlap): $T_*$ can be defined as the temperature at which this fraction is equal to one. Moreover, knowing the action $S_3(T)$ as a function of temperature, one can calculate $\beta/\HH_*=T\,d(S_3/T)/dT$, which has to be evaluated towards the end of the PT to represent, as a matter of fact, the `duration' of the PT\footnote{Alternatively, it is possible to relate $\beta$ to the typical bubble size at the end of the PT through $v_b$, $\vev{R}\simeq 3v_b/\beta(T)$, where $\vev{R}$ can be estimated from the maximum of the bubble volume distribution \cite{Huber:2007vva}.}.

The bubble wall velocity $v_w$ and the fraction of energy that contributes to the GW generation $(\Pi / \rho_{\rm tot})_*$ cannot in general be evaluated solely from the action $S_3(T)$. Since these two parameters are connected to the dynamics of the bubble expansion in the primordial fluid, 
a knowledge of the total particle content and interactions of the theory is in principle necessary to determine them. The bubble wall velocity $v_w$ results from the balance among the driving force that makes the bubble expand (given by the pressure difference between the interior and the exterior of the bubble, which is connected to the latent heat) and the friction force due to the interaction of the bubble wall with the surrounding plasma, which slows down the bubble expansion. The friction can either be determined in a given particle theory model, for which all interactions are known, or it can be parametrised in terms of the independent parameter $\eta$, providing then a phenomenological description valid for several PT models \cite{Espinosa:2010hh}. Once $v_w$ is known, it gives the boundary condition for the hydrodynamical description of the bubble growth. 

The tensor anisotropic stress $\Pi$ sourcing GW is in general given by the the sum of the gradient energy in the Higgs-like field driving the phase transition, and by the bulk kinetic energy of the fluid set into motion by the bubble walls. This latter must be further divided into the component due to sound waves and the one due to MHD turbulence. If the PT is occurring in a thermal state and the friction is high, the bulk motion and the MHD turbulence are expected to dominate $\Pi$. The simulations of \cite{Hindmarsh:2013xza,Hindmarsh:2015qta,Hindmarsh:2017gnf} show that the contribution from the scalar field gradient energy is largely sub-dominant. In this case, the total GW signal is given by the sum of Eqs.~\eqref{eq:OmGWSound} and \eqref{eq:OmGWturb}. Moreover, it is shown in Ref.~\cite{Espinosa:2010hh} that the efficiency factor $\kappa_v$ in Eq.~\eqref{eq:kappas}, representing the fraction of vacuum energy that gets converted into bulk kinetic energy, can be related to $\alpha$ (Eq.~\eqref{eq:alpha_thermal}): in the limits of small and large $v_w$, Ref.~\cite{Espinosa:2010hh} finds
\begin{equation}
\label{eq:kappav}
\kappa_v \simeq 
\left\{\begin{array}{c c}
\alpha \left(0.73+0.083\sqrt{\alpha}+\alpha\right)^{-1} & v_w \sim 1\\
v_w^{6/5}6.9\, \alpha \left(1.36-0.037 \sqrt{\alpha}+\alpha \right)^{-1}& v_w 
\lesssim 0.1 \,.
\end{array}
\right.~
\end{equation}

If, on the contrary, the PT is very supercooled and friction is low, the role of the plasma is minor, and most of the energy remains in the form of gradient energy of the Higgs-like field. In this case, the total GW signal is given by Eq.~\eqref{eq:Omenv}, and furthermore, one can consistently approximate $\kappa_\phi\simeq 1$. Since the PT is very strongly first order, one has $\alpha \gg 1$, and the dependence on $\alpha$ of the SGWB in Eq.~\eqref{eq:Omenv} effectively drops. It is also important to point out that, in general, the strength of the PT $\alpha$ is connected with its duration $\beta/H_*$: strong PTs last longer, leading to small $\beta/H_*$. This increases the amplitude of the GW signal, but shifts the peak frequency to low values. 

\begin{figure}[t!]
\centering
\includegraphics[width=7.7cm]{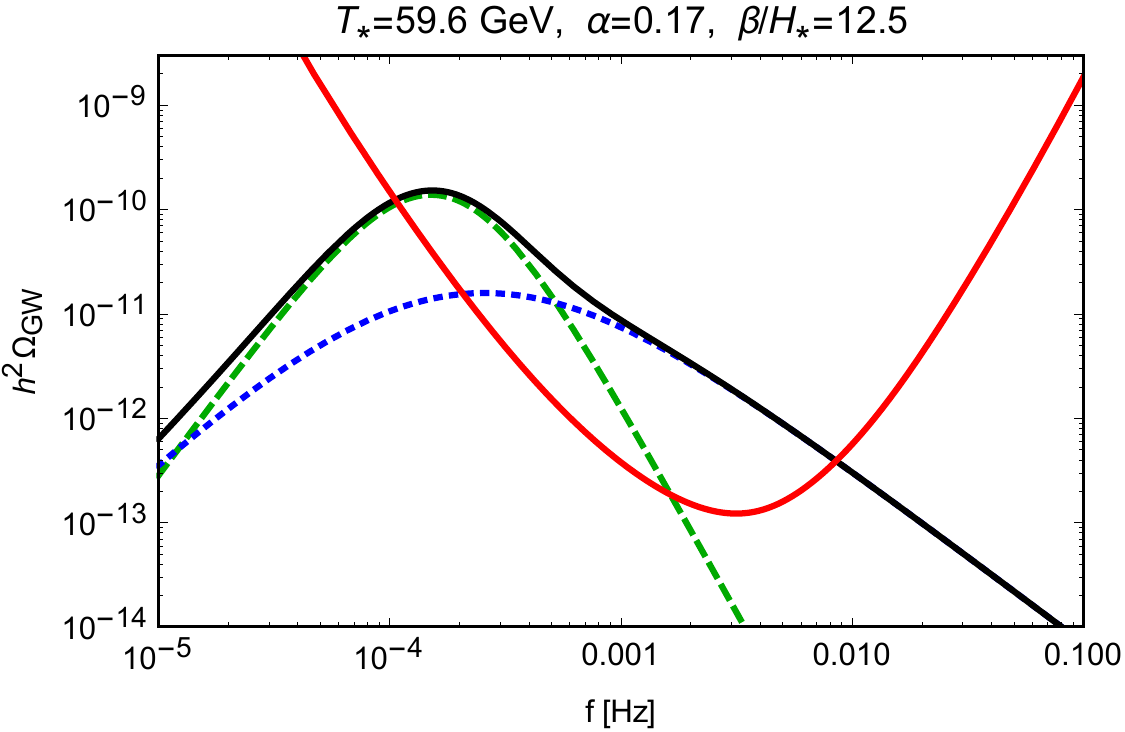}
\includegraphics[width=7.7cm]{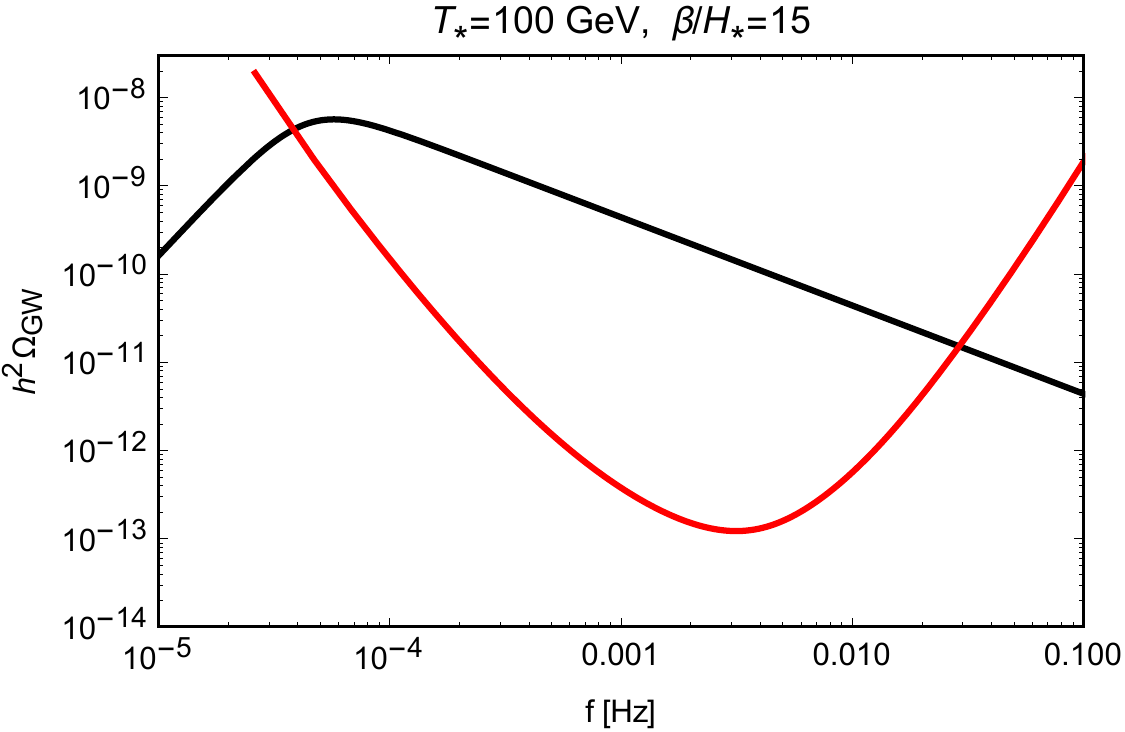}
\caption{SGWB spectra in two examples of first order PT, compared with the estimated sensitivity curve of the interferometer LISA calculated from~\cite{Audley:2017drz} (the red solid curve appearing in both pictures, c.f.~section \ref{sec:interferometers}). Left plot: the Higgs portal scenario, with parameters $\alpha=0.17\,,~\beta/H_*=12.5$ and $T_*=59.6$ GeV, see \cite{Caprini:2015zlo}. The green, dashed curve represents the GW signal from sound waves, while the blue, dotted curve represents the GW signal from MHD turbulence, where we have taken $\epsilon=1$ (c.f. Eq.~\eqref{eq:kappaturb}). Right plot: PT connected with the radion stabilization in the Randall Sundrum model, with $\beta/H_*=15$ and $T_*=100$ GeV, again see \cite{Caprini:2015zlo}. \label{Fig:PT}}
\end{figure}

Ref.~\cite{Caprini:2015zlo} presents a selection of PT scenarios, both related and not related to the EW symmetry breaking, that can produce SGWBs in the frequency range of the LISA interferometer. For each scenario there is a choice of benchmark values of $\alpha\,,~\beta/H_*$ and $T_*$ that can be realised within the model. This allows to give predictions for realistic GW signals, which can actually arise in well identified particle physics models. 

We close this section with two examples of SGWB from first order PTs, taken from Ref.~\cite{Caprini:2015zlo}: the Higgs portal scenario, with benchmark values $\alpha=0.17\,,~\beta/H_*=12.5$ and $T_*=59.6$ GeV, and the dilaton scenario, with benchmark values $\beta/H_*=15$ and $T_*=100$ GeV (as explained above, in this case the GW signal no longer depends on $\alpha \gg 1$). In the first case, we have set $v_w=0.95$ as done in \cite{Caprini:2015zlo}, while in the second case, since the PT is effectively happening in vacuum, we have set $v_w=1$. The resulting GW spectra are shown in Fig.~\ref{Fig:PT} together with the LISA sensitivity, taken from Ref.~\cite{Audley:2017drz}. In the Higgs portal scenario the SGWB is sourced by the plasma bulk motion. We therefore plot the two contributions: the one from sound waves, Eq.~\eqref{eq:OmGWSound}, with $\kappa_v$ given by the first line of Eq.~\eqref{eq:kappav}; and the one from turbulence, Eq.~\eqref{eq:OmGWturb}, with $\kappa_{\rm turb}=\epsilon \,\kappa_v$. Note that we have set $\epsilon=1$, since for the adopted benchmark point $\tau_{\rm sh} H_*\simeq 0.54$ and one therefore expects the formation of MHD turbulence. In the dilaton-like scenario, on the other hand, the PT is effectively happening in vacuum: we are therefore plotting only Eq.~\eqref{eq:Omenv}, with $\kappa_\phi=1$. It appears that both scenarios can provide a SGWB detectable by LISA, which could thereby help testing the occurrence of new physics beyond the standard model of particle physics, for models that are still viable notwithstanding LHC constraints.

\section{Cosmic defects}
\label{sec:CosmicDefects}

As we just described in Section~\ref{sec:FOPT}, a phase transition (PT) in the early Universe corresponds to a process of spontaneous symmetry-breaking from a symmetric phase (false vacuum) to a broken phase (true vacuum). This is typically driven by some scalar field(s) acquiring a non-zero vacuum expectation value within a certain vacuum manifold $\mathcal{M}$. If $\mathcal{M}$ meets certain conditions, cosmic defects may be produced as an aftermath product of a phase transition~\cite{Vilenkin:2000jqa,Hindmarsh:1994re}. In particular, if the vacuum manifold is topologically non-trivial, i.e.~has a non-trivial homotopy group $\pi_n(\mathcal{M}) \neq \mathcal{I}$, topological field configurations arise, producing strings ($n=1$), monopoles ($n=2$), or textures ($n=3$)~\cite{Kibble:1976sj}. For larger $n$, there is no topological obstruction for the symmetry-breaking field to reach the vacuum manifold at any point in space-time, and thus non-topological field configurations arise. Depending whether the symmetry broken is global or gauged, the defects generated are referred to as global or local defects, respectively. In all cases, topological or not, local or global, we refer to all of them as cosmic defects. 

Cosmic strings, no matter global or gauged, as well as any type of global defect (independently of their topology, or absence of it), present a $scaling$ behaviour, sufficiently long after the completion of the phase transition that created them~\cite{Turok:1991qq,Vilenkin:2000jqa,Hindmarsh:1994re,Durrer:2001cg}. The scaling regime is characterized by a self-similar evolution of the number density of defects, which is preserved within a causal volume, at every moment of cosmic history. In what follows we describe first the GW production by any network of cosmic defects in scaling (Section~\ref{sec:IrreduclibleGWCD}), and then we focus on the more significant and specific production of GWs from loops chopped off from a string network (Section~\ref{sec:GWcosmicStringLoops}).

\subsection{Irreducible emission from a cosmic defect network}
\label{sec:IrreduclibleGWCD}

As a network of cosmic defects evolves, its energy-momentum tensor adapts itself in order to maintain the scaling regime. Hence, the time evolution of the transverse-traceless part of the defect's network energy-momentum tensor, necessarily creates GWs. This was first appreciated in the context of a global phase transition in Ref.~\cite{Krauss:1991qu}, where it was argued, based on dimensional grounds and causality, that the field dynamics after a global phase transition should emit gravitational radiation. The amplitude of the GW backgrond was estimated using the quadrupole approximation, and it was concluded that the spectrum should be approximately scale-invariant. In the context of the large $N$ limit of a global phase transition~\cite{Turok:1991qq}, and using a full treatment of the tensor metric perturbation, it was demonstrated later on~\cite{JonesSmith:2007ne,Fenu:2009qf} that the self-ordering process of global textures after the phase transition, generates in fact an exact scale-invariant background of GWs. Numerical simulations in~\cite{Giblin:2011yh} further supported the idea of a scale invariant spectrum of GW, although only approximately, as the numerical spectrum of GWs was wiggly and slightly tilted. The origin of the scale-invariance of this background was finally clarified in~\cite{Figueroa:2012kw}, where it was demonstrated that any scaling source during radiation domination (RD), always radiates GWs with an exact scale-invariant energy density spectrum. In the case of cosmic defects, it is important to note that this is not related to their particular topology, nor to the order of the phase transition, or the global or local nature of the symmetry-breaking process that generates them. The scale invariance of the emitted GWs is just a consequence of the scaling regime during RD. These results can be further generalized~\cite{DaniMarkJonPRD}, extending the calculation of the GW emission by a scaling defect network to all cosmic history, including both radiation-dominated (RD) and matter-dominated (MD) eras. This introduces a new feature in the spectrum, which does not remain scale-invariant within the entire frequency range. 

This background represents an irreducible emission of gravitational waves from any type of cosmic string network, including global, Abelian, non-Abelian or semi-local strings, as well as from global defect networks, including domain walls, monopoles or non-topological textures. In what follows we review the analytical derivation of the resulting GW background emitted by the evolution of a defect network in scaling, and present some numerical results.

If we know the unequal-time-correlator (UTC) $\Pi^2(k,\eta_1,\eta_2)$ of a given system [recall Section~\ref{sec:spectrum_generic} and in particular Eq.~(\ref{eq:PIspec}), or alternatively Eq.~(\ref{eq:UTC}) in Section~\ref{sec:ParamRes}], we can compute the spectrum of GWs emitted by that system, by simply plugging $\Pi^2(k,\eta_1,\eta_2)$ into Eq.~(\ref{eq:rhogwPiRad}) [or, equivalently, into Eq.~(\ref{eq:GW_spectra(Pi)})]. Even before providing an explicit input for the UTC of a cosmic defect network, we can anticipate the form of the GW spectrum by simply recalling that any network of cosmic defects, after its formation in a phase transition, enters in a scaling regime. In scaling , the UTC can only depend on $k$ through the variables $x_1 = k\eta_1$ and $x_2=k\eta_2$, so from dimensional analysis the UTC must take the form
\begin{equation}\label{eq:UTCscaling}
\Pi^2(k,\eta_1,\eta_2) = {v^4\over\sqrt{\eta_1\eta_2}}\,\uetcT(k\eta_1,k\eta_2)\,,
\end{equation}
with $v$ the vacuum expectation value (VeV) of the symmetry breaking field. Using this form of the correlator, we can express the GW energy density spectrum, normalized to the critical energy density, as~\cite{Figueroa:2012kw,DaniMarkJonPRD}
\begin{eqnarray}\label{eq:GWspectrum3}
\hspace*{-1.0cm}\Omega_{\rm GW}(k,\eta) &\equiv & {1\over\rho_c}\frac{d\rhogw}{d\log k}(k,\eta) \\
\hspace*{-1.0cm} &=& {16\over3}\left({v\over \Mpl}\right)^4{k^2\over H^2a(\eta)^4}\int dx_1 dx_2 {a_1a_2\over\sqrt{x_1x_2}}\cos(x_1-x_2)\uetcT(x_1,x_2)\nn\,.
\end{eqnarray}
where $a_1 \equiv a(x_1/k)$, $a_2 \equiv a(x_2/k)$. Neglecting the late accelerated expansion due to dark energy, the energy budget of the Universe is dominated by non-relativistic matter since the moment of matter-radiation equality at redshift $(1+z_{\rm eq}) \equiv \Omega_{\rm m}^{(0)}/\Omega_{\rm rad}^{(0)} \simeq 3400$. Before, at redshift $z \gg z_{\rm eq}$, it was dominated by relativistic photons and neutrinos. Denoting by $\eta_{\rm eq}$ the moment of equality (in conformal time), the scale factor corresponding to a mixed radiation-matter fluid (ignoring dark energy), can be written as
\begin{equation}
\hspace*{-0.5cm} a(\eta) = a_{\rm eq}\left([(\sqrt{2}-1)(\eta/\eta_{\rm eq})+1]^2-1\right) = a_0^3\Omega_{\rm mat}^{(0)}{H_0^2\eta^2\over 4} + a_0^2\sqrt{\Omega_{\rm rad}^{(0)}}H_0\eta\,,
\end{equation}
with $a_{\rm eq}$ the scale factor at $\eta_{\rm eq}$. Using Eq.~(\ref{eq:GWspectrum3}) and the scale factor deep in radiation-dominaton, $a(\eta) \simeq \sqrt{\Omega_{\rm rad}^{(0)}}a_0^2H_0\eta$, we obtain the spectrum of GW at sub-horizon scales $x \equiv k\eta \gg 1$, for modes that entered inside the horizon during RD, as 
\begin{eqnarray}\label{eq:GWspectrum2}
\Omega_{\rm GW}(x,\eta) = \Omega_{\rm rad}(\eta)\left({v\over\Mpl}\right)^{\hspace*{-1mm}4}{\hspace*{-0.5mm}}\,{F}_{\rm RD}^{[\mathcal{U}]}(x)\,,\hspace*{1.15cm}\\
\label{eq:F_U}
{F}_{\rm RD}^{[\mathcal{U}]}(x) \equiv {16\over3}\int^x\hspace*{-0.3cm} dx_1 \int^x \hspace*{-0.3cm} dx_2~ \sqrt{x_1x_2} \cos(x_1-x_2)\,\uetcT(x_1,x_2)\,.
\end{eqnarray} 
At subhorizon scales, $\uetcT(x_1,x_2)$ is peaked at $x_1 = x_2 \equiv x$, and decays typically as $\propto x^{-p}$, with $p$ a positive real number $p > 2$~\cite{Figueroa:2012kw}. The convergence of the integration is thus guaranteed. We see that ${F}_{\rm RD}^{[\mathcal{U}]}(x)$ becomes progressively insensitive to the upper bound of integration, approaching asymptotically a constant value for $x \gg 1$. The function $F_{\uetcT}(x \gg 1)$ approaches rapid and asymptotically the constant ${F}_{\rm RD}^{[\mathcal{U}]}(\infty) \equiv {F}_{\rm RD}^{[\mathcal{U}]}(x\rightarrow \infty)$. As a consequence, the resulting GW spectrum at subhorizon scales, becomes exactly scale-invariant. The spectrum of the tensor modes emitted during RD, once such modes are well inside the horizon, becomes proportional to ${F}_{\rm RD}^{[\mathcal{U}]}(\infty)$,  and redhifts as $\propto {1\over a^4(\eta)}$, as it should for relativistic species such as GWs. For any type of cosmic defect network, there is always a function $\uetcT(x_1,x_2)$ that characterizes its scaling evolution. Hence, there is always a well determined value ${F}_{\rm RD}^{[\mathcal{U}]}(\infty)$, characterizing the amplitude of the GW background emitted. Redshifting the amplitude to today, one finally obtains
\begin{eqnarray}\label{eq:GWspectrumToday}
h^2\Omgw^{(0)}(k) \equiv {h^2\over\rho_{\rm c}}\left(\frac{d\rhogw}{d\log k}\right) = h^2\Omega_{\rm rad}^{(0)}\left({v\over\Mpl}\right)^{\hspace*{-1mm}4}{\hspace*{-0.5mm}}{F}_{\rm RD}^{[\mathcal{U}]}(\infty)\,.
\end{eqnarray}
The background of \gws\ produced during the radiation era by the evolution of any network of defects in scaling regime, is therefore exactly scale-invariant. The amplitude of this GW background today, is suppressed by the fraction $h^2\Omega_{\rm rad}$, and by the VeV as $(v/\Mpl)^4$. It also depends on the shape of the UTC, which ultimately determines the amplitude through $F_{\rm RD}^\infty[\mathcal{U}]$, which is different for each type of defect.

One can proceed to compute the GW spectrum emitted by a scaling network during MD\footnote{The results concerning the GWs emitted by a network of detects during MD, are based on work in progress~\cite{DaniMarkJonPRD}, and thus should be taken as preliminary only.}, analogously as how we did for RD. At times $\eta \gg \eta_{\rm eq}$, the scale factor can be approximated as $a(\eta) \simeq {1\over4}a^3_0\Omega_{\rm mat}^{(0)}H_0^2\eta^2$. Assuming that the UTC during MD is in scaling so it can be written as in Eq.~(\ref{eq:UTCscaling}), and using Eq.~(\ref{eq:GWspectrum3}), the spectrum of GW at sub-horizon scales $x \equiv k\eta \gg 1$ during MD, becomes
\begin{eqnarray}\label{eq:GWspectrum4}
\hspace*{-0.5cm} \Omega_{\rm GW}(x,\eta) = \Omega_{\rm rad}(\eta)\left({v\over\Mpl}\right)^{\hspace*{-1mm}4}{\hspace*{-0.5mm}}\,{k_{\rm eq}^2\over k^2}{F}_{\rm MD}^{[\mathcal{U}]}(x)\,,\\
\hspace*{-0.5cm}{F}_{\rm MD}^{[\mathcal{U}]}(x) \equiv {16\over3}(\sqrt{2}-1)^2\int_{x_{\rm eq}}^x\hspace*{-0.3cm} dx_1 \int_{x_{\rm eq}}^x \hspace*{-0.3cm} dx_2~ (x_1x_2)^{3/2}\cos(x_1-x_2)\,\uetcT(x_1,x_2)\,,
\label{eq:F_MD}
\end{eqnarray} 
where we have used $\Omega_{\rm mat}^{(0)}H_0a_0\eta_{\rm eq} = 2(\sqrt{2}-1)\sqrt{\Omega_{\rm rad}^{(0)}}$, $x_{\rm eq} \equiv k\eta_{\rm eq}$ and $k_{\rm eq} \equiv {1\over 2\eta_{\rm eq}}$. We define $2k_{\rm eq}\eta_{\rm eq} = 1$ as corresponding to the mode with half wavelength $\lambda_{\rm eq}/2 \equiv {\pi\over k_{\rm eq}}$ inside the horizon $1/\eta_{\rm eq}$ at the time of matter-radiation equality. Again, the integral becomes insensitive to the upper bound as $x \rightarrow \infty$. Redshifting this spectrum today, we obtain
\begin{eqnarray}\label{eq:GWspectrumTodayMD}
h^2\Omgw^{(0)}(k) \equiv {h^2\over\rho_{\rm c}}\left(\frac{d\rhogw}{d\log k}\right) = h^2\Omega_{\rm rad}^{(0)}\left({v\over\Mpl}\right)^{\hspace*{-1mm}4}{\hspace*{-0.5mm}}\left(k_{\rm eq}\over k\right)^2{F}_{\rm MD}^{[\mathcal{U}]}(\infty)\,,
\end{eqnarray}
indicating that the GW spectrum of modes emitted during MD and well inside the horizon, scales as $\propto 1/k^2$. The total spectrum today, spanning over MD and RD frequencies, is given by
\begin{eqnarray}
h^2\Omgw^{(0)}(k) = h^2\Omega_{\rm rad}^{(0)}\left({v\over\Mpl}\right)^{\hspace*{-1mm}4}\left({F}_{\rm RD}^{\infty}+{k_{\rm eq}^2\over k^2}{F}_{\rm MD}^{\infty}\right)\,.
\end{eqnarray}

\begin{figure*}[t]
\begin{center}
\includegraphics[width=7.5cm,height=5cm,angle=0]{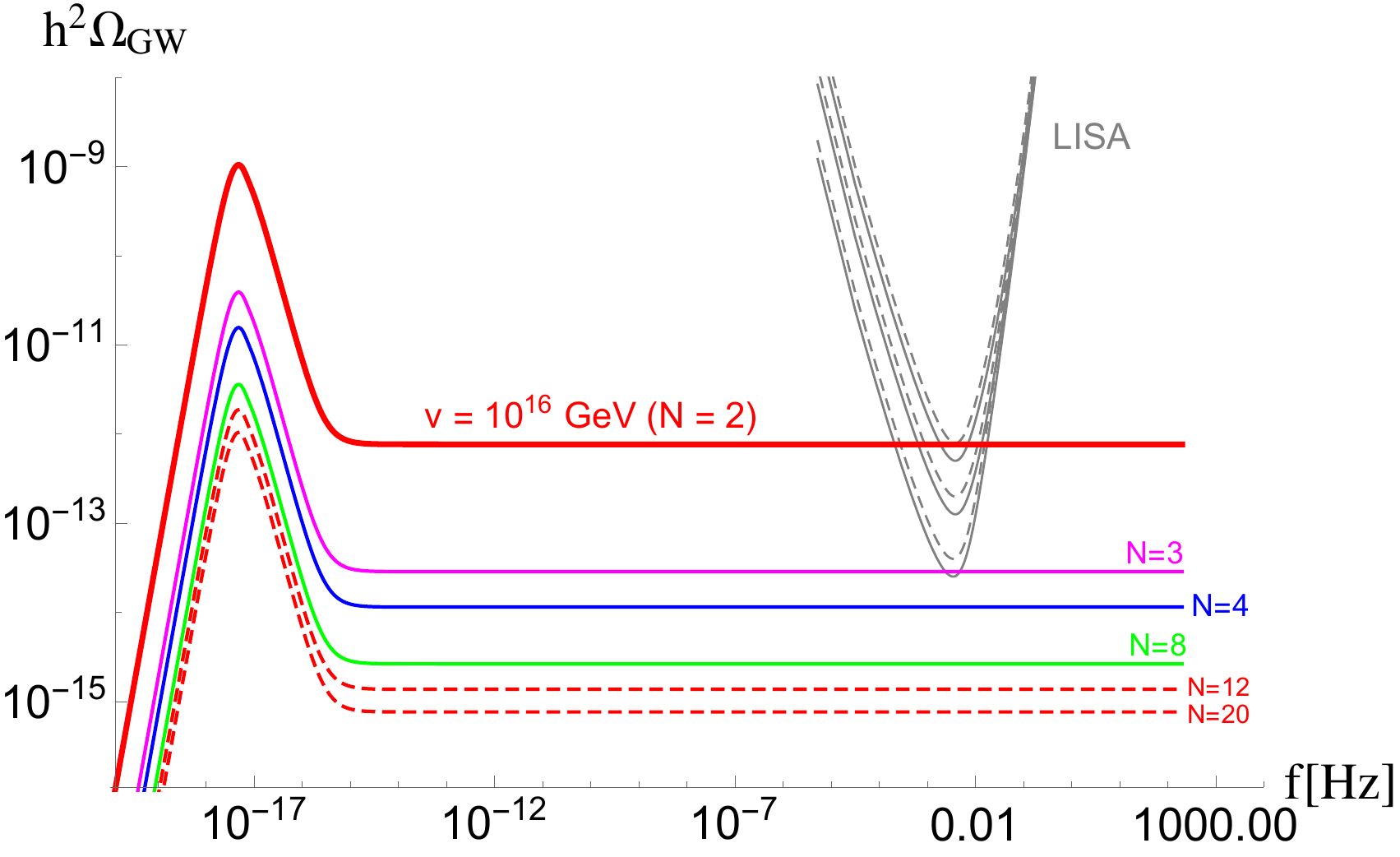}
\includegraphics[width=7.5cm,height=5cm,angle=0]{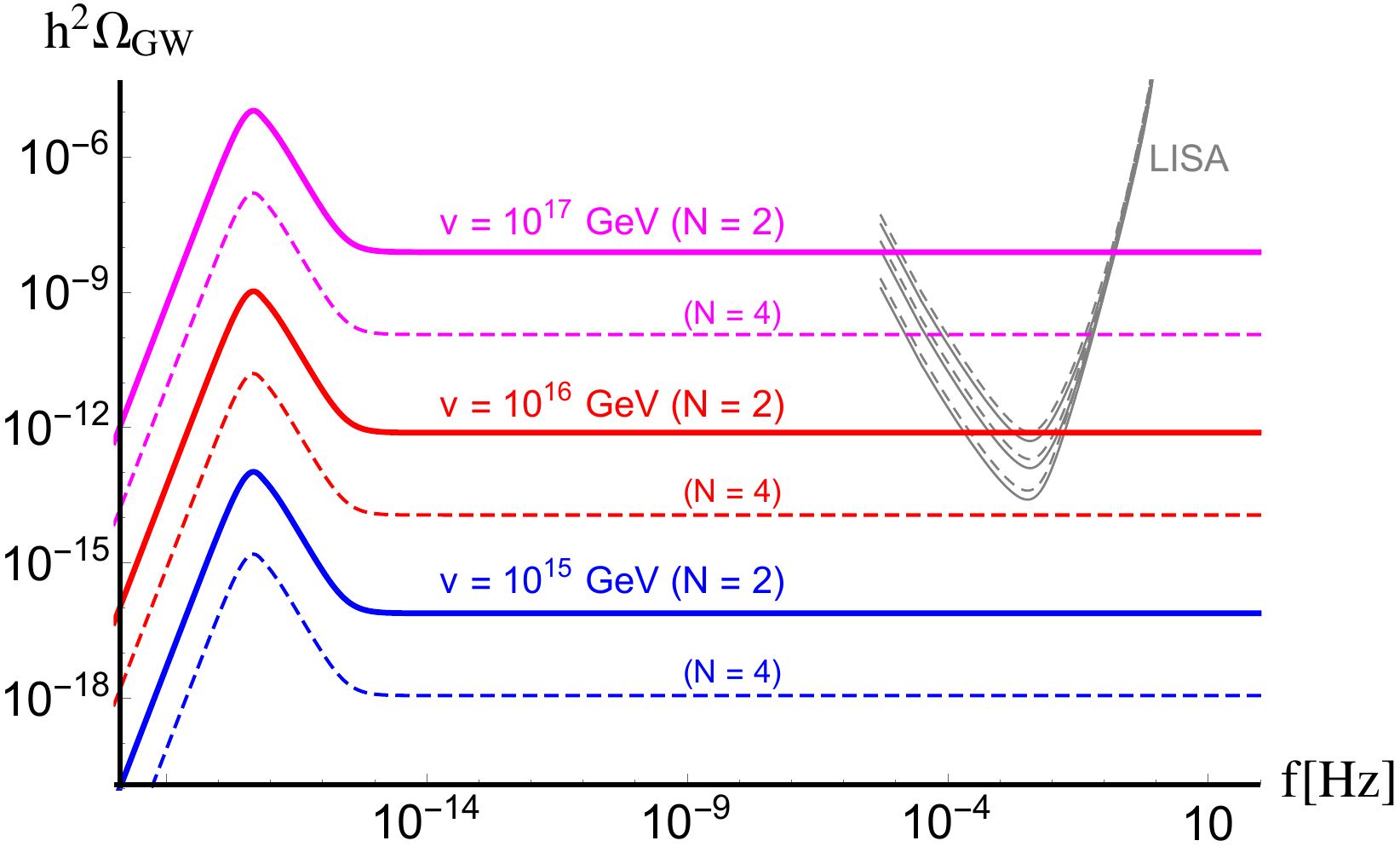}\\
\end{center}
\vspace*{-5mm}
\caption{Amplitude of the energy density spectrum today of the GW background emitted during the evolution of scaling networks resulting from a global symmetry breaking $O(N) \longrightarrow O(N-1)$. We show examples for $N = 2, 3, 4, 8, 12, 20$. The GW spectra are plotted against six proposed sensitivity curves for LISA (note that its final configuration has now been decided and it is given in \cite{Audley:2017drz}). The different spectra depict the scaling of the GW spectrum with the VeV and the number $N$. The case $v = 10^{-2}M_{\rm Pl}$, ruled out by CMB observations~\cite{Ade:2013xla,Lopez-Eiguren:2017dmc}, is shown nonetheless for illustrative purposes, to highlight the large amplitude of the background. For $v = 10^{-3}M_{\rm Pl}$, we see that the background from global strings ($N = 2$) could be marginally detectable by LISA, depending on the final configuration adopted.}
\vspace*{-3mm}
\label{fig:GWspectraScalingSources}
\end{figure*}

Let us finally note that if we evaluate the spectrum at super-horizon scales $x = k\eta < 1$, the integration in Eqs.~(\ref{eq:GWspectrum2}), (\ref{eq:GWspectrum4}) becomes sensitive to the upper bound $x$. For super-horizon scales, either during RD or MD, the signal actually scales as $\propto x^3$. The spectrum today does not remain, therefore, scale-invariant over all frequencies. The spectrum rather reaches a maximum at the frequency associated to the present horizon scale today $f_{0}$, decays as $\propto f^3$ for $f \ll f_{0}$, whereas it scales as $\propto 1/f^2$ at $f > f_0$. Eventually, it settles down to RD scale-invariant amplitude $\propto f^0$ for $f \gg f_{\rm eq}$, where $f_{\rm eq}$ is the frequency today corresponding to the horizon scale at the matter-radiation equality. We show the shape of the spectrum for various examples in Fig.~\ref{fig:GWspectraScalingSources}.

In the specific case of non-topological defects arising after the spontaneous symmetry breaking of a global O($N$) symmetry into O($N-1$), $\Pi^2(k,\eta_1,\eta_2)$ can be estimated analytically when $N \gg 1$. After some algebra and numerical evaluation, it is possible to arrive at an analytical expression for the GW background today, corresponding to the modes that entered during RD, as~\cite{JonesSmith:2007ne,Fenu:2009qf,DaniPhD}
\be\label{eq:GW_largeNfinal}
h^2\Omgw^{(0)}(f) \equiv {1\over\rho_{\rm c}^{(0)}}\left(\frac{d\rhogw}{d\log f}\right)_0 \simeq \frac{650}{N}\, \Omega^{(0)}_{\rm rad}\left(\frac{v}{{\Mpl}}\right)^{\!\!4}\,.
\ee
This result exhibits, in fact, the expected decreasing behaviour with the number of field components $N$: the larger $N$ is, the smaller the spatial gradients among the field components, and hence the smaller it is the source of GWs $\Pi \sim \partial\phi \partial\phi$ (see~\cite{Fenu:2009qf} for a detailed discussion on this). In the case of a general defect network, the correlator $\Pi^2(k,\eta_1,\eta_2)$ can only be obtained from field theory simulations, like those used for CMB studies~\cite{Lopez-Eiguren:2017dmc}. Using lattice simulations as an input, Ref.~\cite{Figueroa:2012kw} calculated numerically the GW amplitude from a system of global O($N$) defects, demonstrating that the numerical results converge well, as we increase $N$, to the large-$N$ analytical result Eq.~(\ref{eq:GW_largeNfinal}). Figure~\ref{fig:GWspectraScalingSources} depicts the numerical results corresponding to the simulations used in Ref.~\cite{Figueroa:2012kw} and~\cite{DaniMarkJonPRD}. In the case of global strings with $N = 2$, as this cannot be possibly interpreted as a large-N case, the numerical result deviates quite significantly from the prediction Eq.~(\ref{eq:GW_largeNfinal}) evaluated at $N = 2$. The numerical amplitude of the GW energy density spectrum for global strings is actually a factor $\sim 100$ larger than the analytical prediction~\cite{Figueroa:2012kw}. 

Unfortunately, there is no analysis available yet in the literature, about the detectability of this background by direct detection GW experiments. Given the current constraints from Planck $v \lesssim 10^{16}$ GeV on the VeV of a global phase transition~\cite{Ade:2013xla}, it seems that the GW background from global strings might be (for the largest possible VeV) marginally detectable by LISA. However, this may require re-assessment, based on the recent new analysis on the impact of global defects in the CMB~\cite{Lopez-Eiguren:2017dmc}. A re-evaluation of UTC's used as input in the numerical evaluation of the background, might also be desirable, given the limited dynamical range used in the simulations of~\cite{Figueroa:2012kw}. A proper assessment of the ability of GW direct detection experiments to detect this background requires therefore further work. It is clear however, in light of Fig.~\ref{fig:GWspectraScalingSources}, that the GW background from global strings with sufficiently large VeV (but below the CMB upper bound), will be accessible to futuristic experiments like DECIGO or BBO.

\subsection{Decay of cosmic string loops}
\label{sec:GWcosmicStringLoops}

So far we have discussed the irreducible emission of GWs by any network of cosmic defects, as long as the network is in scaling. We shall concentrate now on the particular case of cosmic strings, since $i)$ they represent (possibly) the best motivated case for defects from a the point of view of particle physics, and $ii)$ as we will explain in detail, they emit a significant extra amount of GWs (by a mechanism only proper to them), superseding the GW emission due to scaling discussed in Section~\ref{sec:IrreduclibleGWCD}. 

Cosmic strings correspond to one-dimensional topological defects that form during a phase transition in the early Universe, whenever the fundamental homotopy group of the corresponding vacuum manifold $\mathcal{M}$ is non-trivial, i.e.~$\pi_1(\mathcal{M}) \neq \mathcal{I}$. Cosmic strings arise naturally within well motivated inflationary models. For instance, local strings are always produced at the end of inflation in supersymmetric GUT models of Hybrid inflation~\cite{Jeannerot:2003qv}, as long as some reasonable assumptions are met. Cosmic strings can also be fundamental superstrings (as opposed to field theory configurations), which arise naturally in scenarios like brane inflation~\cite{Sarangi:2002yt}. Either type of strings are characterized by a linear energy density $\mu$, which in the Nambu-Goto picture (which describes an infinitely thin string) corresponds to its tension. This is typically expressed through the dimensionless quantity $G\mu$, where $G$ is Newton constant. In the case of standard field theory strings this quantity indicates the amplitude of the vacuum expectation value $v$ of the ordering field in the phase transition, typically as $G\mu = \pi(v/M_{Pl})^2$ (though the prefactor $\pi$ depends on conventions). Any network of cosmic (super-)strings is formed at any time by a population of `large' loops and `small' loops. The latter are loops with a diameter smaller than the causal horizon, whereas the former are loops so large, that only a fraction of their length lies within the causal horizon volume. As we will see shortly, this is precisely what makes special a string network in what concerns the emission of GWs, as the distribution of other cosmic defects cannot be split into `small' and `large' sub-groups. From now on, by loops we will be referring only to those of sub-horizon size, and we will label these with the subscript $<$. The larger ones will be referred to as the `infinite' (or long) strings, and will be labelled with the subscript $>$.

A crucial ingredient characterizing the evolution of any cosmic (super-)string network is the so called `intercommutation' property, whereby string loops intersect (or self-intersect), exchange `partners', and form new loops~\cite{Vilenkin:2000jqa}. A main difference between field strings and superstrings is that when the former collide and exchange partners, they do it with an intercommutation probability of $p = 1$. However, when cosmic superstrings collide in the three-dimensional space, they can intercommute with a reduced\footnote{This can be attributed to the extra dimensions in which the cosmic superstrings are moving, as a full intercommutation requires the collision in all the dimensions, and not just in the three macroscopic ones. As we will see, this aspect has a very relevant impact in the amplitude of the GWs emitted by the decay of cosmic superstring loops.} intercommutation probability $p < 1$. This property affects notably the evolution of the string network, and the production of the GWs expected from the decay of the loops. In general, as the loops have a significant tension, they will start oscillating relativistically after their formation. This causes them to decay releasing their energy via some `preferred' channel(s). In the case of local strings this channel is expected to be GWs, see e.g.~\cite{Vilenkin:1981bx,Vachaspati:1984gt,Caldwell:1991jj,Damour:2004kw}. Interestingly, real time simulations of the Abelian-Higgs (AH) scenario~\cite{Bevis:2006mj,Bevis:2010gj,Daverio:2015nva,Lizarraga:2016onn,Hindmarsh:2017qff} (which represents a simple field theory realization of local strings), show that the loops created tend to decay very fast, via classical emission of scalar and gauge fields. In the case of numerically simulated AH strings, the emission of GWs from the string network is therefore reduced only to the irreducible background described in section~\ref{sec:IrreduclibleGWCD} (which represents a much weaker background than that from the loop decay). It can be argued that the unexpected scalar and gauge emission in the numerical simulations might be due to the limited dynamical range used (given the limitation of our current computational capabilities): whereas the separation of scales between the string inverse curvature $L$ and the width of the cores $\delta$, cannot be larger than a factor $L/\delta \sim 10^{2}-10^{3}$ in present numerical simulations, in actual cosmological scenarios we expect a separation many orders of magnitude larger. Despite this limitation, the numerical simulations~\cite{Daverio:2015nva,Lizarraga:2016onn,Hindmarsh:2017qff} exhibit however, the development of a scaling regime, supporting the idea that their results should be, in principle, `scalable' to arbitrarily larger scale separations. Whether the decay of the loops into scalar and gauge bosons is an artefact or not due to the computational limitations, seems unlikely to be resolved any soon, unless some new ground-breaking computational strategy is envisaged. Acknowledging this caveat, in the rest of the section we will consider nonetheless, as usual, that cosmic loops from local strings are stable against scalar and gauge radiation, and decay only (or mostly) into GWs. 

The precise details of GW emission by cosmic string loops are very sensitive to various properties of the string network. The final spectrum will depend notably on $i)$ the string tension $G\mu$, $ii)$ the size of cosmic string loops relative to the horizon at birth $\alpha$, $iii)$ the spectral index $q$ of the emission spectrum, $iv)$ the cut-off $n_*$ in the emission spectrum (mimicking radiation backreaction~\cite{Caldwell:1996en}), and finally $v)$ the intercommutation probability $p$. The starting point to derive the spectrum from cosmic loops is the calculation of their number density $n(l,t)$, which counts how many loops with length between $l$ and $l+dl$ are present in the causal volume $V_H(t) = 1/H^3$, at a time $t$. Using on the one hand, the equation of state of infinite strings~\cite{Vilenkin:2000jqa}, $w_{>} \equiv p_{>}/\rho_{>} = {1\over3}(2\langle v^2\rangle-1)$, with $\langle v^2\rangle$ the mean squared velocity of the infinite strings, and on the other hand the conservation of the energy-momentum tensor of the whole system of infinite strings, cosmic loops and GWs, $\dot\rho_{\rm tot} = - 3H(\rho_{\rm tot}+p_{\rm tot})$, we can obtain the amount of energy (per unit of time) lost by the infinite string network, in order to create new cosmic loops: $\dot{E}_{<} = -V_H(t)\left(\dot{\rho}_{>} + 3H(t)(1+w_{>})\right)$. Assuming that loops are always `born' with a circumference size $l_b$ equal to a given fraction of the horizon scale at their time of birth $t_b$, $\alpha \equiv l_b(t_b)H(t_b)$, then it must be true that $\dot{E}_{<} = \alpha \mu H^{-1}(t) \dot{N}_{<}$, with $N_{<}$ the total number of loops (within the volume $V_H(t)$) formed since the creation of the network. Equating the two expressions for $\dot{E}_<$, one finds the formation rate of loops within the volume $V_H(t)$ as
\begin{eqnarray}
\dot{N}_{<} = {\rho_>\over \alpha\mu H(t)}\left(1-3w_{>}\right) \,.
\end{eqnarray}
The mean square velocity has been determined in numerical simulations, e.g.~in~\cite{Bennett:1989ak} as $\langle v^2\rangle = 0.43$ for RD and $\langle v^2\rangle = 0.37$ for MD (most recent evolution simulations~\cite{Ringeval:2005kr,BlancoPillado:2011dq} also support similar values). This makes the equation of state of the infinite strings $w_{>} = -0.047$ for RD and $w_{>} = -0.087$ for MD, and hence $\dot{N}_{<}\Big|_{\rm RD} \simeq 1.14\left({\rho_>\over \alpha\mu H(t)}\right)$ and $\dot{N}_{<}\Big|_{\rm MD} \simeq 1.26\left({\rho_>\over \alpha\mu H(t)}\right)$. The energy density of the infinite strings can be written as $\rho_> = A\mu H^2(t)$ with the proportionality constant determined again from numerical simulations, with Ref.~\cite{Bennett:1989ak} finding originally $A \simeq 52$ for RD and $A \simeq 31$ for MD, whereas more modern simulations~\cite{Ringeval:2005kr,BlancoPillado:2011dq} deviate from these values only by $\sim 10\%$. Using the quoted numbers, we obtain $\dot{N}_{<}\Big|_{\rm RD} \simeq 59.3 H(t)$ and $\dot{N}_{<}\Big|_{\rm MD} \simeq 39.1 H(t)$. In other words, at any time $t$ there is always a loop formation rate of the order of $few \times 10$ larger than the Hubble rate at that moment, $\dot{N}_{<} \sim \mathcal{O}(10) H(t)$.

In order to derive an expression for $n(l,t)$, we need to know the energy emission rate (assumed in the form of GWs) of a loop. This is simply given by $\dot{E}_{\rm GW} =  \Gamma G \mu^2$, with $\Gamma$ a constant parametrizing the efficiency of the energy loss mechanism. Based on numerical simulations~\cite{Casper:1995ub}, the value $\Gamma \simeq 50$ is normally used, with the most recent simulations~\cite{Blanco-Pillado:2017oxo} supporting this value. The length of a cosmic string loop at a time $t > t_b$ is simply $l(t,t_b) = \alpha H^{-1}(t) - \Gamma G\mu (t - t_b)$. From this we deduce that the differential change of length is $\dot{l} = -[{\alpha \dot H/H^2} + \Gamma G ]$, and from here we arrive at the wanted number of loops per volume $V_H$ and unit length, 
\begin{eqnarray}
n(l,t) &\equiv& {1\over V_{H}}{dN_{<}\over dl} = {1\over V_{H}}{\dot {N}_{<}\over \dot l} = {1\over V_{H}}{-\dot{N}_{<}\over(\alpha \dot H/H^2+\Gamma G\mu)}\,.
\end{eqnarray}

As a cosmic string loop of length $l$ oscillates under its tension $\mu$, it emits GWs in a series of harmonic modes. As the period of oscillation is $l/2$, the frequencies of the GWs emitted are harmonics of the inverse of this period, that is, $f_n \equiv 2n/l$. The power radiated in GWs into each harmonic mode is then $\dot{E}_{\rm GW}^{(n)} = P_n\Gamma G \mu^2$, with $\sum_{n=1}^{\infty} P_n = 1$, so that $\dot{E}_{\rm GW} \equiv \sum_n \dot{E}_{\rm GW}^{(n)} = \Gamma G \mu^2$. The fraction of energy emitted per harmonic can be parametrized in terms of a spectral index $q$ (characteristic of the type of loop), as $P_n \equiv D_q/n^{q+1}$. It can be argued that $q\approx 1/3$ for any string loop which has a cusp; something which is expected for string trajectories without a kink.
 
From the condition $\sum_n P_n = 1$ we obtain $D_q = 1/\zeta(q+1)$, with $\zeta(p)$ the Riemann Zeta function. We can approximate the discrete mode emission of GWs into a continuous emission, by considering $\sum_n P_n = D_q {l\over2}\sum_n (\Delta f_n) \times 1/(lf_n/2)^{q+1} \simeq l \int_{2/l}^{\infty} df \mathcal{P}(f)$, where $\mathcal{P}(f) \equiv C_q/(fl)^{q+1}$, with the normalization constant fixed as $C_q = 2^{q}q$ in order to guarantee the continuum normalization condition $l\int_{2/l}^{\infty}\mathcal{P}(f) = 1$. We can then write the `spectral power' emitted by a loop of size $l$, i.e.~the total energy emitted in GWs (per unit of time) within the frequency range $[f,f+df)$, as 
\begin{eqnarray}
dP_{\rm GW}(f) = \Gamma G\mu^2 l\mathcal{P}(f)df\,.
\end{eqnarray}
The effect of radiation backreaction is however ignored in the previous calculation, so in order to model this, Ref.~\cite{Caldwell:1996en} proposes to include a cut-off $n_*$, so that $P_n = 0$ for $n > n_*$. The radiation per mode is then modified into the fraction $P_n \equiv A_q^{\rm (br)} n^{-(q+1)}$, with the normalization constant fixed to $A_q^{\rm (br)} \equiv \sum_{n=1}^{n_*} n^{-(1+q)}$, in order to preserve the new normalization condition $\sum_{n=1}^{n_*} P_n = 1$. The value of $n_*$ is not known {\it a priori}, though in principle it can expected to be of the order of the ratio between the string curvature $R$ and the diameter of the string core $d_c$, $n_* \sim R/d_c$. Ref.~\cite{Sanidas:2012ee} suggests to consider it as a phenomenological parameter within the range $n_* \sim 10^{3}-10^5$. Translating this into the modelling of the GW emission in the continuum, we find  the spectral power emitted by a loop of size $l$ as $dP_{\rm GW}(f) = \Gamma G\mu^2 l\mathcal{P}(f)df$, where this time $\mathcal{P}(f) \equiv C_q^{\rm (br)}/(lf)^{q+1}$, with the constant $C_q^{\rm (br)} \equiv 2^qq/(1-1/n_*^{q})$ preserving the required normalization condition $\int_{2}^{2n_*}dx\mathcal{P}(x) = 1$.

Using the above results we can arrive at the spectrum of the stochastic background of GWs given by the superposition of the GW harmonics emitted by the distribution of loops present at different moments of time. Let us denote by $d\rho_{\rm GW}(t)$ the energy density in GWs emitted by loops of size within the length interval $[l, l + dl)$, during the time interval $[t, t + dt)$, and within the frequency range $[f_{\rm e}, f_{\rm e}+df_{\rm e})$. It follows that $d\rho_{\rm GW}(t) \equiv dP_{\rm GW}(f)dt \,n(l,t_e)\,dl$. Using the fact that the energy density of GWs redshifts as $1/a^4$ corresponding to relativistic species, and that the corresponding frequencies today are related to those at the time of emission by $f = f_{\rm e}a_{\rm e}/a_0$, we arrive at
\begin{eqnarray}
\hspace*{-0.75cm} d\rho^{\rm (0)} \equiv \left({a(t)\over a_0}\right)^4\,d\rho(t) = \left({a(t)\over a_0}\right)^3\,G\mu^2 \Gamma \,l \, \mathcal{P}((a_0/a(t))fl) \,df \,dt \,n(l,t)\,dl\,.
\end{eqnarray}
The final spectrum today of the GWs produced by the loops of a string network, emitted all through cosmic history from an initial time $t_*$ until today $t_0$, is then
\begin{eqnarray}\label{eq:GWfinalLoops}
{d\rho^{(0)}\over df} \equiv \Gamma{G\mu^2}\int_{t_*}^{t_0} dt \left({a(t)\over a_0}\right)^3 \int_0^{\alpha/H(t)} dl\, l \,n(l,t)\,\mathcal{P}((a_0/a(t))fl)\,.
\end{eqnarray}

Something that we have not mentioned yet is that, as pointed out in Refs.~\cite{Damour:2000wa,Damour:2001bk}, the GW signal from cosmic string loops includes not only the stationary and nearly Gaussian background that we have just described up to now, but also strong infrequent bursts that could be detected individually. These GW bursts are produced by string configurations known as {\it cusps}, corresponding to a highly boosted piece of a loop where the string folds up, and {\it kinks}, corresponding to shape-discontinuities that propagate along the strings at the speed of light. In this review we are rather interested on the stationary stochastic background that we have described so far, due to the continuous emission of GWs from the loops decay through all cosmic history\footnote{We will not consider the detection of individual bursts, but we point the interested reader on this to~\cite{gr-qc/0603115}.}. In principle, the strong infrequent GW bursts should not be included in the computation of the stationary background, in order not to over-estimate its amplitude. However, in practice, it has been found that for large initial loop sizes, removing the rare burst has practically no effect on the present-day GW spectrum~\cite{Siemens:2006yp,Olmez:2010bi,Binetruy:2012ze}, at least when the number of cusps and kinks per loop oscillation period is $\mathcal{O}(1)$. The latest simulations~\cite{BlancoPillado:2011dq,Blanco-Pillado:2013qja,Blanco-Pillado:2017oxo} clearly indicate that large initial loop sizes, of the order of $\mathcal{O}(0.1)$ times the size of the horizon at their time of birth, are preferred. We will thus not concern ourselves any further with the rare emissions from cusps and kinks. In order to evaluate the amplitude of the stationary background Eq.~(\ref{eq:GWfinalLoops}), it is however very relevant to asses whether there are cusps or kinks in the loops. This is simply because the presence of these determine essentially the spectral index $q$. For instance, $q = 1/3$ is obtained analytically for loops with cusps~\cite{Caldwell:1991jj,Vachaspati:1984gt} (and supported by numerical simulations in~\cite{Allen:1991bk}), while $q = 2/3$ is obtained for the contribution from kinks ($q = 1$ is also considered for a square loop with kinks in Ref.~\cite{Vilenkin:2000jqa}). As $n$ increases, the contribution from kinks decays faster than the one from cusps, we can expect generically that the total power at large $n$ will be dominated by the cusp contribution (unless the number of cusps is strongly suppressed compared to the number of kinks). We will then use $q = 1/3$ for loops with cusps, independently of the presence of kinks, and we will refer to these as `cuspy loops'. We will use $q = 2/3$ for loops that have kinks but no cusps, and will refer to these as  `kinky loops'. 

Let us note that the above spectral mode emission $P_n \propto 1/n^{q+1}$, or equivalently $\mathcal{P}(fl) \propto 1/(fl)^{q+1}$, is based on the asymptotic behavior expected for large $n$ (it assumes that the asymptotic behavior of the power is valid down to $n = 1$, which does not need to be the case). Hence, this modeling might be inaccurate for the lower harmonics, specially for $n = 1$. Since Eq.~(\ref{eq:GWfinalLoops}) was derived independently of the functional form of $\mathcal{P}(fl)$, it is interesting to compute the GW emission assuming that only the fundamental mode $n = 1$ of the GW harmonics is emitted. This can be modeled by simply considering a Dirac Delta distribution as $\mathcal{P}(fl) \delta^{(1)}(fl-2)$. 

\subsubsection{Observational constraints.}

\begin{figure*}[t]
\begin{center}
\includegraphics[width=5.3cm,height=7.5cm,angle=90]{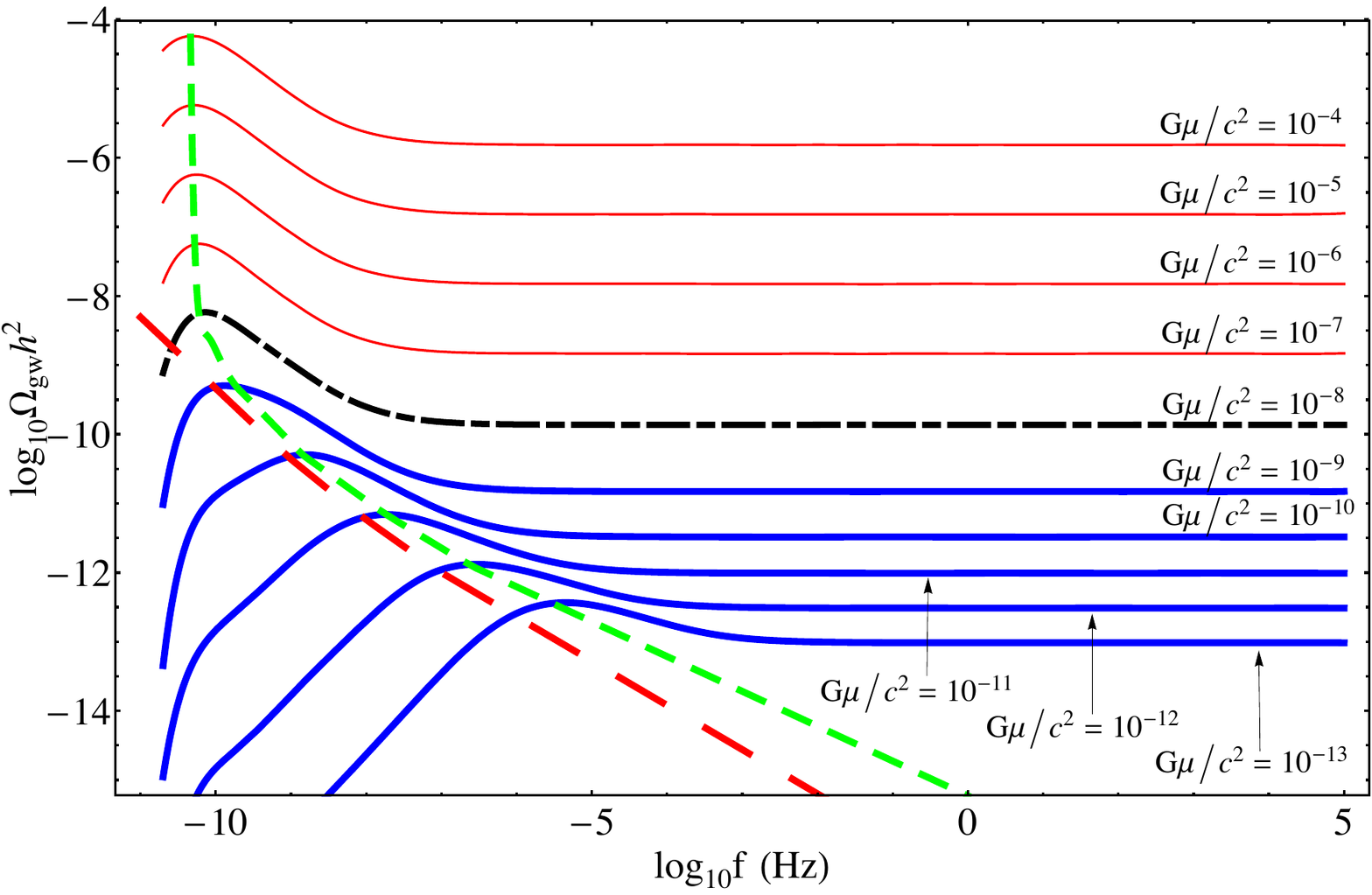}
~~~~\includegraphics[width=5.3cm,height=7.5cm,angle=90]{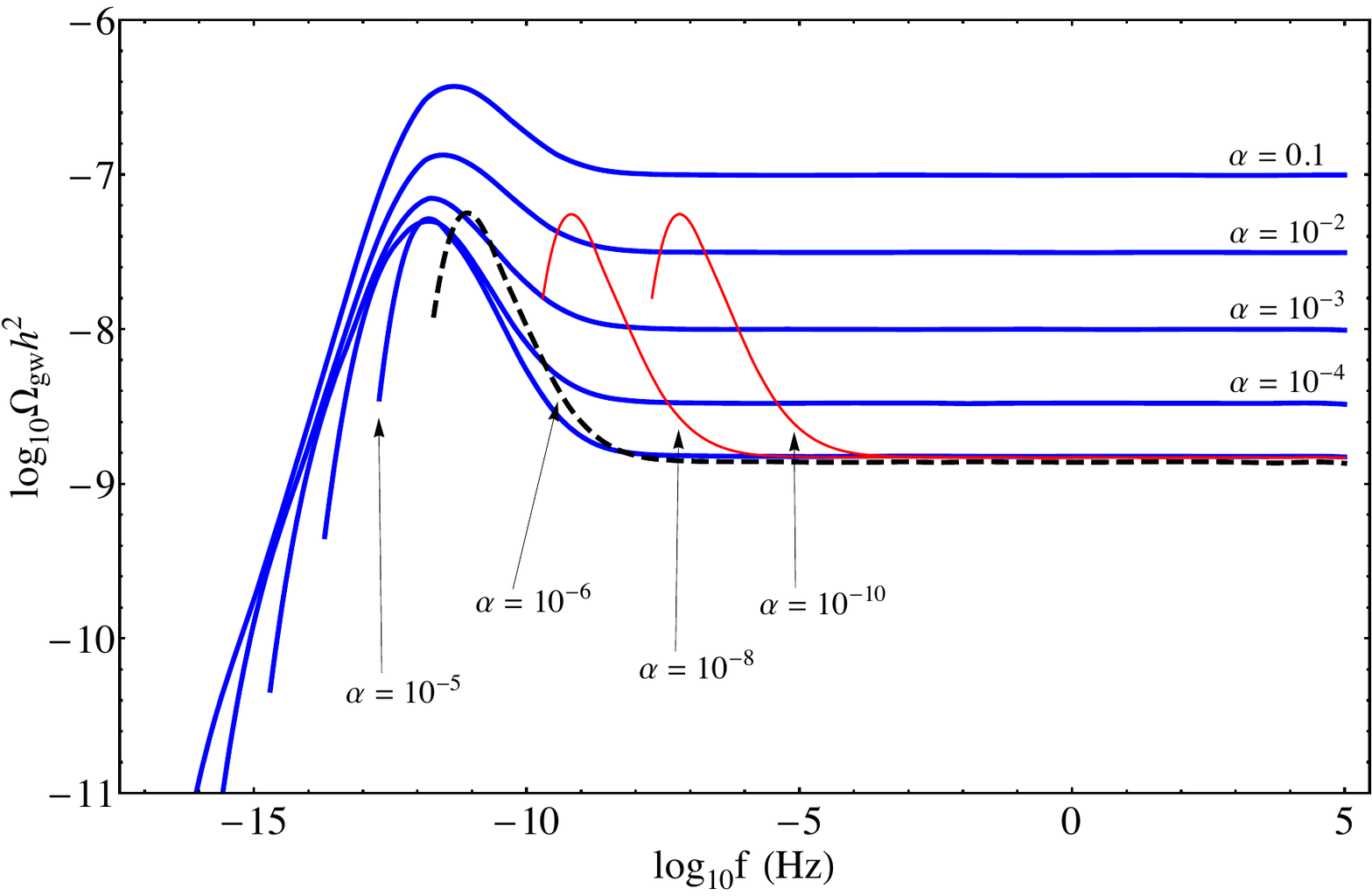}\\ 
\includegraphics[width=5.3cm,height=7.5cm,angle=90]{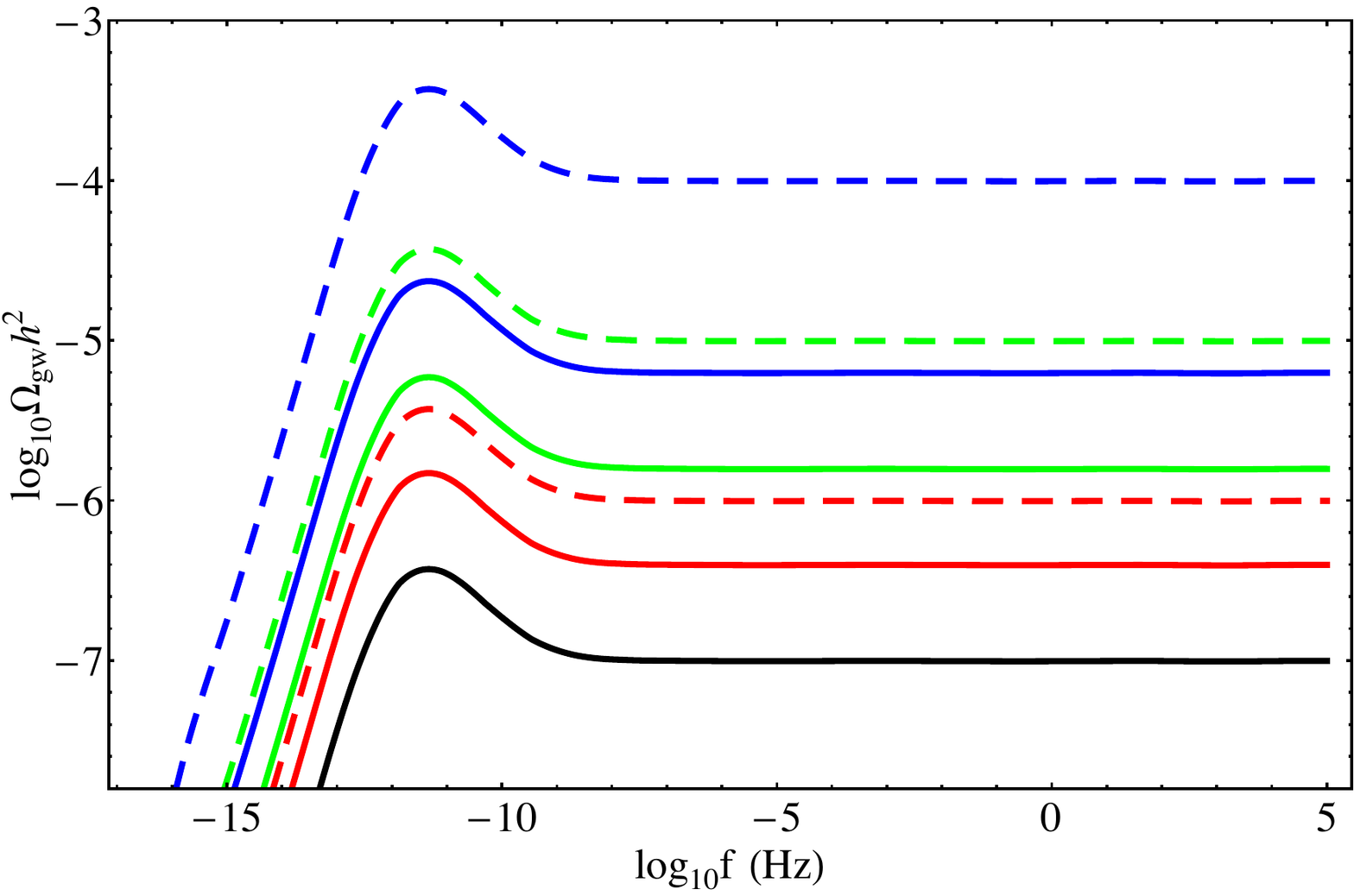}
~~~~\includegraphics[width=7.4cm,height=5.1cm,angle=0]{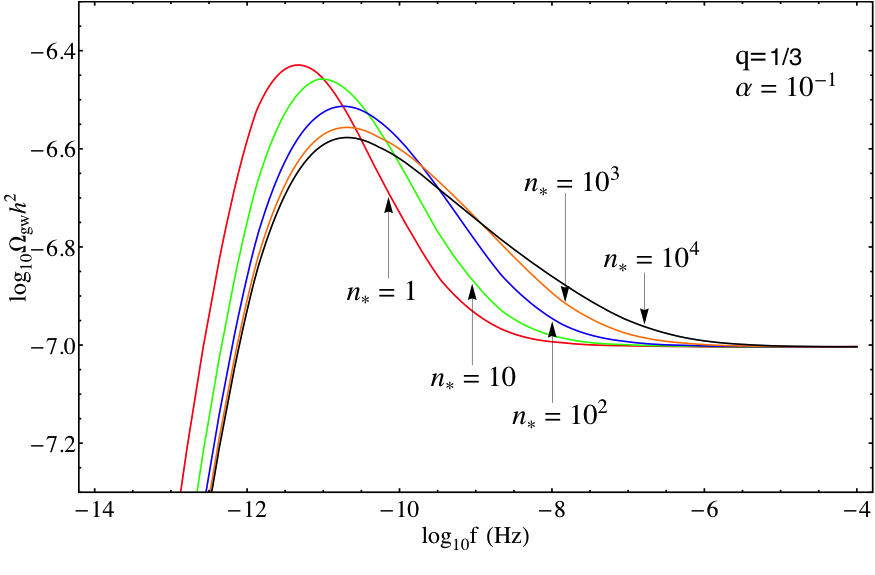}\\ 
\end{center}
\vspace*{-5mm}
\caption{All panels in this figure are taken from Ref.~\cite{Sanidas:2012ee}. They show the amplitude of the energy density spectrum today of the GW background emitted by the decay of loops chopped off from a string network all through cosmic history. Ref.~\cite{Sanidas:2012ee} considers the values $G\mu/c^2=10^{-7}$, $\alpha=10^{-7}$, $q=4/3$, $n_*=1$ and $p=1$, as a set of fiducial parameters. Thus, in each panel one parameter is varied while the others are fixed to their fiducial values (unless otherwise specified). The different panels exhibit the effect on the GW spectrum due to: variation of the tension $G\mu$ (top left panel, thick blue lines for large loops and thin red lines for small loop regime), variation of the loop birth size $\alpha$ (top right panel, thick blue lines for large loops and thin red lines for small loop regime), variation of the intercommutation probability $A(p,k) = Ap^{-k}$ [$A \simeq 52$ (RD), $31$ (MD)] (bottom left panel, $\alpha = 0.1$; red, green and blue solid lines are for $p=0.1,10^{-2},10^{-3}$ respectively, for $k=0.6$; red, green and blue dashed lines are for $k=1$. The black solid line is for $p=1$), and variation of the cut-off $n_*$ (bottom right panel, different colours for each value of $n_*$, as indicated in each curve).}
\vspace*{-3mm}
\label{fig:GWspectraLoops}
\end{figure*}

The sensitivity of the GW signal to different cosmic string parameters has been studied in multiple papers, see e.g.~\cite{Siemens:2006yp,DePies:2007bm,Olmez:2010bi,Binetruy:2012ze,Sanidas:2012ee}. The constraints on the different parameters characterizing the string network $\lbrace G\mu,\alpha,q,n_*, p\rbrace$, and in particular the constraint in the tension $G\mu$ (under various assumptions on the other parameters), have been re-assessed by successive analysis, see e.g.~\cite{Sanidas:2012ee,Blanco-Pillado:2013qja,Blanco-Pillado:2017rnf,Ringeval:2017eww,Blanco-Pillado:2017oxo}. In the following we review the results from Ref.~\cite{Sanidas:2012ee}, and compare them briefly against other recent results.

In Fig.~\ref{fig:GWspectraLoops} we show the amplitude of the energy density spectrum today of the GW background emitted by the loops from a string network under different assumptions (all panels are taken from Ref.~\cite{Sanidas:2012ee}). Defining a set of parameter values as fiducial, $G\mu/c^2=10^{-7}$, $\alpha=10^{-7}$, $q=1/3$, $n_*=1$ and $p=1$, each panel in Fig.~\ref{fig:GWspectraLoops} shows the effect on the GW spectrum due to varying one parameter only, while maintaining the others fixed to the fiducial values. The different panels exhibit the modification of the GW spectrum due to the variation of the tension $G\mu$ (top left panel), of the loop birth size $\alpha$ (top right panel), of the intercommutation probability $A(p,k) = Ap^{-k}$ [$A \simeq 52$ (RD), $31$ (MD)] (bottom left panel), and of the cut-off $n_*$ (bottom right panel). 

In order to calculate a constraint on the cosmic string tension in the most conservative manner, Ref.~\cite{Sanidas:2012ee} considered all the string network parameter combinations leading to a GW spectrum saturating the upper bound from PTA observations. They used, in particular, the GW stochastic background upper limit at a frequency of $f_1 = (1\,\rm yr)^{-1}$ from the EPTA data, see section~\ref{sec:PTA}. First, for each set of fixed $n_*,\, q$ and $p$ values, the amplitude $h^2\Omega_{\rm GW}(G\mu,\alpha)$ is obtained. The spectral index $n_{\Omega}(G\mu,\alpha)$ of the spectrum, assumed as $h^2\Omega_{\rm GW}(f) \propto (f/f_1)^{n_\Omega}$ around the reference frequency, is also obtained. This characterizes completely the SGWB as a function of $G\mu$ and $\alpha$ around the reference frequency of $f_1$. The $G\mu-\alpha$ parameter combinations which provide the constraint curve for each set of $n_*,\, q$ and $p$ in the $G\mu-\alpha$ parameter space, can then be calculated requiring $h^2\Omega_{\rm GW}(G\mu,\alpha;n_{\Omega}) = h^2\Omega_{\rm{GW,EPTA}}(f_1,n_{\Omega})$, where $h^2\Omega_{\rm{GW,EPTA}}(n_{\Omega};f_1)$ represents the EPTA limit in the form of a $\sim 95~\%$ exclusion curve of amplitude $h^2\Omega_{\rm GW}$ vs spectral index $n_\Omega$. Setting $p=1$, they obtain constraints for various values of $q$ and $n_*$ which satisfy the EPTA limit. For instance, the constraints for models with large cut-off $n_{*}$ are stronger than those with low $n_{*}$ for most of the $\alpha-G\mu$ combinations, except in the case of very small $G\mu$ or $\alpha$ (where the opposite takes place). The most conservative and generic constraint on the cosmic string tension can be set, therefore, by the parameter values where the constraint curve $G\mu$ vs $\alpha$ presents an absolute maximum, see Fig.~13 from~\cite{Sanidas:2012ee}. This corresponds to cosmic string networks with $\alpha\approx 10^{-5}$ and $n_*=1$. It gives~\cite{Sanidas:2012ee}
\begin{equation}
G\mu < 5.3\times10^{-7}\,,
\end{equation}
which represents a $95\%$ upper bound (for the specific set of parameters). The most recent simulations of cosmic string networks~\cite{Ringeval:2005kr,BlancoPillado:2011dq,Blanco-Pillado:2013qja,Ringeval:2017eww,Ringeval:2017eww} favour however, large loops at birth. For a loop size of $\alpha \sim 0.1$, the constraint obtained from the EPTA bound by Ref.~\cite{Sanidas:2012ee} becomes much stronger, of the order of $G\mu \lesssim 10^{-10}$. A recent analysis presented in Ref.~\cite{Blanco-Pillado:2017oxo} yields in fact a $95\%$ confidence level constraint as $G\mu < 1.5\times10^{-11}$. Ref.~\cite{Ringeval:2017eww} has also presented recently a new analysis, yielding $95\%$ confidence level constraints as $G\mu < 7.2\times10^{-11}$ and $G\mu < 1.0\times10^{-11}$, depending on the network modelling assumptions. In general, each of these works make different assumptions about cusps, kinks, gravitational back-reaction, and other aspects, so it is difficult to compare the results among them. What seems clear is that all of the most recent simulations prefer large loops, and that all analysis in such large-loop regime yield a strong constraint on the strings tension, of the order of $G\mu \lesssim 10^{-10}-10^{-11}$, depending on the details. Given this stringent upper bound, it is worth noting that unless future PTA observations observe soon a stochastic GW background, the limits on $G\mu$ will only improve marginally: a string network with tension much smaller than $G\mu \sim 10^{-11}$ will only sustain an amplitude for the GW background at frequencies above the observational range accessible to PTA's. If that was the case, a detection of the GW background from cosmic string networks will still be possible, but it will depend exclusively on observatories like LISA, or futuristic missions like DECIGO and BBO.

As a final comment, let us note that the GW signal from cosmic string loops can also probe the expansion rate of the universe at times before BBN. This is analogous to the case described in Section~\ref{sec:EvolTensInf}, about the distortion of the spectrum of the irreducible GW background from inflation, due to a non-standard equation of state before BBN. In the case of a cosmic string network, the effect is similar, so that the high frequency tail of the spectrum of the GWs emitted from the loops formed before BBN, can deviate from its usual flat shape, as long as the expansion rate of the Universe is other than radiation-dominated. See~\cite{Cui:2017ufi} for details.

\section{Conclusions}
\label{sec:Conclusion}

The recent direct detection of gravitational waves (GWs) from astrophysical binary systems by advanced LIGO and Virgo~\cite{Abbott:2016blz,Abbott:2016nmj,Abbott:2017vtc,Abbott:2017oio,TheLIGOScientific:2017qsa}, represents a milestone in astronomy and in physics. It has opened up a new window to explore the universe. Furthermore, the universe is actually expected to be permeated by various GW backgrounds of both astrophysical and cosmological origin. Several cosmological backgrounds could arise from a plethora of high energy phenomena that may have occurred in the primordial epochs of the universe evolution. GWs are, in fact, the most promising cosmic relic to probe directly the currently unknown physics of the early universe. In the present document, we have reviewed the actual status of our understanding of potential GWs sources in the early universe, that may have led to the production of cosmological GW backgrounds. 

In Sect.~\ref{sec:GWdef}, we review first the basic aspects, equations and conceptual difficulties, that emerge when defining the concept  of GWs itself. In Sect.~\ref{sec:general} we present a specialized discussion on the general properties of GW backgrounds of cosmological origin. We review the reasons to expect these backgrounds to be stochastic, describe some of their properties, and discuss how to characterize their spectrum. In Sect.~\ref{sec:bounds}, we review first, in Sects.~\ref{sec:BBN}-\ref{sec:PTA}, the present observational constraints on SGWBs. In Sect.~\ref{sec:interferometers}, we then quickly revise the basic principles for the detection of a SGWB using interferometry, and survey the features of current and planned GW direct detection experiments.

In Sects.~\ref{sec:inf} - \ref{sec:CosmicDefects} we describe in extensive detail the properties and origin of the SGWBs expected from the early universe. We classify them in five categories: irreducible background from inflation [Sect.~\ref{sec:inf}], beyond the irreducible background from inflation [Sect.~\ref{sec:infII}], preheating and similar non-perturbative phenomena [Sect.~\ref{sec:PreheatingAndOthers}], first order phase transitions [Sect.~\ref{sec:FOPT}] and cosmic defects [Sect.~\ref{sec:CosmicDefects}]. In Sect.~\ref{sec:inf} we first review, in Sect.~\ref{sec:PrimTensInf}, the irreducible GW background expected from any inflationary model, due to the amplification of initial quantum fluctuations of the tensor metric perturbations. In Sect.~\ref{sec:EvolTensInf} we discuss the evolution of this primordial background until the present, including post-inflationary effects that may affect its present-day amplitude. In Sect.~\ref{sec:infII} we describe how, under special circumstances, if new species or symmetries are at play during inflation, GWs with a large amplitude and a significant deviation from scale-invariance can also be produced during the inflationary period. We consider $i)$ the presence of fields during inflation leading to strong particle production [Sect.~\ref{sec:PartProdInf}], $ii)$ the enhancement of tensor perturbations at small scales due to spectator fields and effective field theory of inflation [Sect.~\ref{sec:BlueTilted}], $iii)$ alternative theories of gravity (other than General Relativity) driving the inflationary period [Sect.~\ref{sec:ModGrav}], and $iv)$ secondary GWs produced by the tensor anisotropic-stress or the formation primordial black-holes, from enhanced scalar perturbations at small scales [Sect.~\ref{sec:ScalarEnhancement}]. For completeness, we also consider the GW background produced from $v)$ alternative theories to inflation, namely Pre-Big-Bang, string gas and bounce cosmologies [Sect.~\ref{sec:InfAltern}]. 

In Sect.~\ref{sec:PreheatingAndOthers} we describe the GW production during preheating and related post-inflationary phenomena. We analyse the cases of standard parametric resonance of bosonic species [Sect.~\ref{sec:ParamRes}], parametric excitation of fermions [Sect.~\ref{sec:Fermionic}], symmetry breaking mediated by tachyonic instabilities [Sect.~\ref{sec:Tachyonic}], oscillations of flat-directions [Sect.~\ref{sec:susyQ}], and the dynamics of oscillons [Sect.~\ref{sec:Osci}]. In Sect.~\ref{sec:FOPT} we consider the GW production from first order phase transitions. We present all contributions to the signal, from bubble wall collisions [Sect.~\ref{sec:Bubbles}], to sound waves [Sect.~\ref{sec:SoundWaves}] and turbulent motions [Sect.~\ref{sec:Turbulence}]. Finally, in Sect.~\ref{sec:CosmicDefects}, we discuss GW production from cosmic defects, considering the irreducible GW emission expected from any network in scaling [Sect.~\ref{sec:IrreduclibleGWCD}], and the more specific but stronger GW emission due to the decay of the loops from a cosmic string network [Sect.~\ref{sec:GWcosmicStringLoops}].

Sects.~\ref{sec:inf} - \ref{sec:CosmicDefects} represent an exhaustive updated survey of the potential sources from the early universe, that produce a stochastic background of gravitational waves. We review in detail the physical origin and motivation of each source, the properties of the GW background they originate, and discuss the ability (if any) of GW detection experiments to place constraints on each scenario. Current GW detectors are not really optimized for the detection of a stochastic background of cosmological origin. The analysis presented in this review, indicates however that a fraction (in some cases significant) of the parameter space in realistic scenarios (e.g.~particle production during inflation, phase transitions, or cosmic string networks), is compatible with a detection by present or future experiments. There is therefore a real chance for a cosmological GW background to be detected in the near future. The benefits of a positive detection would be extraordinary, opening up a new observational window to fundamental high-energy physics that will never be reached by particle physics accelerators.

\section*{Acknowledgments}

CC greatest thank goes to Jean-Fran\c{c}ois Dufaux, with whom this project was initially started. We are very grateful to our collaborators from the LISA Cosmology Working Group for the work developed within the group, some of which is reviewed in this manuscript. We thank Pierre Auclair for contributing to section \ref{sec:EvolTensInf}, Oliver Gould for a general revision of the manuscript, Germano Nardini for corrections to section \ref{sec:FOPT},  
and Nicola Tamanini for his contribution to section \ref{subsec:GWcurved}. We acknowledge discussions with Michele Maggiore on section \ref{sec:CMB}, and with Chris Byrnes on section \ref{sec:SecGWScal}.


\newpage

\bibliography{tempBIBRefs}
\bibliographystyle{h-physrev4}


\end{document}